%% file: Thesis_Final.tex
\documentclass[a4paper,12pt,notitlepage]{report}

\usepackage{amsmath,amssymb,amsfonts,graphicx,slashed,fancyhdr,psfrag,setspace}
\usepackage{tcdThesis}

\numberwithin{equation}{section}% numbers equations per section

\begin{document}

\title{A theoretical study of spin filtering and its application to polarizing antiprotons}

\author{Donie O'Brien}
\prevdegrees{B.Sc.~(Hons.)}
\degree{Doctor of Philosophy}
\thesisdate{6}
\degreemonth{June}
\degreeyear{2008}
\department{Mathematics}
\supervisor{Nigel~H.~Buttimore}{Dr.}
\maketitle

\addcontentsline{toc}{section}{Declaration}
\makedeclaration
\input{dedication.tex}
%\summary{\include{tcdsummary}}
%\makesummary
%\acknowledgements{\include{acknowledgments}}
%\makeacknowledgements
\pagebreak
\addcontentsline{toc}{section}{Abstract}
\theabstract{\include{theabstract}}
\makeabstract
%\publications{\include{pubs}}
%\makepublications

%%%%Import the file summary.tex
\addcontentsline{toc}{section}{Summary}
\chapter*{\vspace*{-8ex}\begin{center}\Large{Summary}\vspace*{-1.7ex}\end{center}}
\input{summary.tex}

%%%%Import the file acknowledgments
\chapter*{\vspace*{-8ex}\begin{center}\Large{Acknowledgments}\vspace*{-1.7ex}\end{center}}
\input{acknowledgments.tex}
\addcontentsline{toc}{section}{Acknowledgments}

%Summary, where I outline the original parts of the thesis

\tableofcontents
%\addcontentsline{toc}{chapter}{Table of contents}

%\listoftables
%\addcontentsline{toc}{chapter}{List of tables}

%\listoffigures
%\addcontentsline{toc}{chapter}{List of figures}

\renewcommand{\tablename}{{\small Table}}
\renewcommand{\figurename}{{\small Figure}}  % Changes format for the label of the Figures.  I could also make it italics if needs be.

\pagebreak

\pagenumbering{arabic}
%Sets the page numbering to normal, and to start at 1 from page 1 of Chapter 1.

\chapter{Introduction}
\label{ch:Introduction}
% Maybe call it General Introduction????

\vspace*{7ex}
\begin{minipage}{6cm}
\end{minipage}
\hfill
\begin{minipage}{10cm}
\begin{quote}
The great Danish physicist Niels Bohr, it is said, had a good-luck horseshoe hanging in his office. \emph{\lq\lq You don't believe in that nonsense, do you?\rq\rq}\ a visitor once asked, to which Bohr replied, \emph{\lq\lq Of course not, but they say it brings you good luck whether you believe in it or not.\rq\rq}
\end{quote}
\end{minipage}
\vspace{10ex}

\begin{figure}[!h]
\centering
\includegraphics[height=5cm,width=5.5cm]{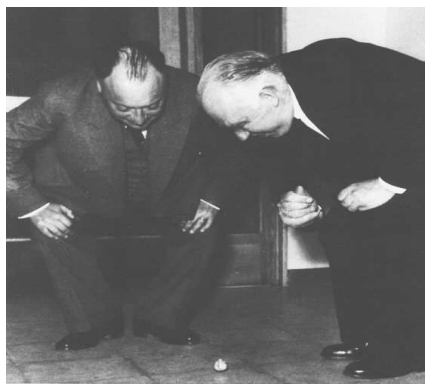}
\caption{\small{The great physicists Wolfgang Pauli (left) and Niels Bohr musing over the spin of a spinning top toy, trying to gain insight into the nature of spin in particle and nuclear physics.
}
}
\label{fig:Bohr_Pauli_Picture}
\end{figure}

\section{Spin and polarization}
\label{sec:Spin_and_polarization}

Spin is a fundamental property of elementary particles.  It was introduced theoretically by Wolfgang Pauli in the 1920's to explain how two electrons can exist in the ground state of an atom, while not violating his famous exclusion principle, for which he was awarded the 1945 Nobel Prize.  This principle states that two particles satisfying Fermi-Dirac statistics (later called fermions and defined by their half-integer spin) cannot exist in the same state at the same time.  Thus the two electrons in the inner shell of an atom must somehow be different, Pauli hypothesized that they have some differentiating characteristic called spin.  He theorized one electron to be in a \lq spin up' state and the other electron to be in \lq spin down' state, thus they do not violate the exclusion principle.  Uhlenbeck and Goudschmidt \cite{Goudschmidt:1926ea} also introduced the concept of spin around the same time as Pauli's work.  Pauli's theory of spin was non-relativistic, Paul Dirac developed the relativistic theory of spin in 1928 with the famous Dirac equation of a relativistic electron \cite{Dirac:1928hu}.

Spin was discovered experimentally by the famous Stern-Gerlach experiment in 1922 \cite{Gerlach:1922dv}.  Since then it has been an integral part of the Quantum Field Theories that describe particle interactions.

A beam of spin 1/2 particles will have the spins of each of the particles in either the \lq spin up' or \lq spin down' state.
For a beam of spin 1/2 particles the polarization is defined as 
\begin{equation}
\label{eq:Polarization_definition}
\mathcal{P} \ =\  \frac{N_+ \ -\  N_-}{N_+ \ +\  N_-}	\,,
\end{equation}
where $N_+$ and $N_-$ are the number of particles in the \lq spin up' and \lq spin down' states respectively.

\begin{figure}
\centering
\includegraphics[width=7cm]{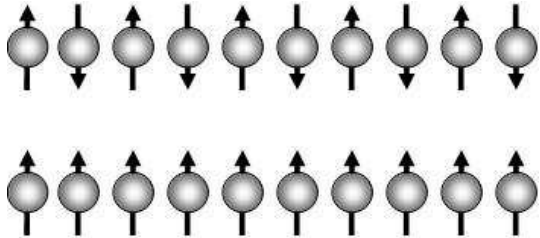}
\caption{\small{\it{The upper beam is unpolarized, with equal number of particles in the \lq spin up' and \lq spin down' states, which in reality are randomly distributed.  The lower beam is $100\%$ polarized, with all particles in the \lq spin up' state.  In practice the maximum beam polarizations achievable are about $90\%$.}}}
\label{fig:Beam_Polarization_Diagram}
\end{figure}

An unpolarized beam ($\mathcal{P} = 0$) is one where the spin states are randomly distributed, thus for a beam with a large number of spin 1/2 particles, half of the particles will be in the \lq spin up' state and half in the \lq spin down' state, as seen in Figure~\ref{fig:Beam_Polarization_Diagram}.  A polarized beam is one where more of the particles are in one spin state than the other\footnote{For further discussion on this see Figure~\ref{fig:N+N-Ratio_V_Polarization_Plot}.}.  For example, a $100 \, \%$ polarized beam ($\mathcal{P} = 1$) has all of the particles in one of the spin states.
%, a $0 \, \%$ polarized beam ($\mathcal{P} = 0$, completely unpolarized) will have half of the particles in the \lq spin up' state and half in the \lq spin down' state.  

Originally particle physics experiments used unpolarized beams and targets, thus completely overlooking the spins of the particles.  This way only total cross-sections instead of spin-dependent cross-sections can be measured.  Only a small portion of the reaction can thus be investigated.  In the words of the originator of the spin filtering method of polarization buildup, which much of the investigations in this thesis are based on, P.~L.~Csonka \cite{Csonka:1968}\,; \emph{\lq\lq One could, perhaps, say that the physicist who is able to measure only total cross-sections, is like the man in an art gallery who is only told the total weight of each statue, but is kept in ignorance of all other parameters specifying their shapes.  Most of us would agree that he is missing something\rq\rq}.  Nowadays highly polarized beams of certain particles are possible.  Polarized electrons and positrons have been used for many decades.  Baryons have proved more difficult to polarize.  The Relativistic Heavy Ion Collider (RHIC) at Brookhaven National Laboratory, New York is the first accelerator to use a high energy polarized proton beam.

%****Before the spin crisis there was a momentum crises.  Only about half the momentum of the proton was carried by its three constituent quarks, the rest must have been carried by chargeless particles; which led to the introduction of gluons and to the theory of QCD.

\section{A \lq\lq spin crisis'' in the parton model}
\label{sec:A_spin_crisis_in_the_parton_model}

\begin{figure}
\centering
\includegraphics{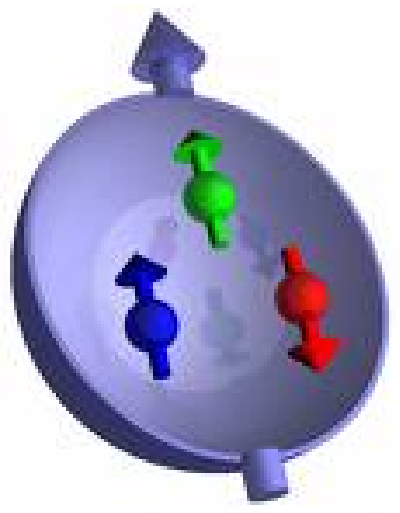}
%\hspace*{3em}
\hspace*{5em}
\includegraphics{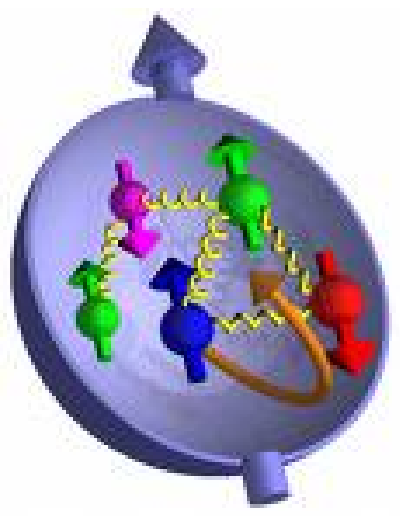}
\caption{\small{\it{These diagrams describe the internal structure of protons and neutrons, according to the theory of Quantum Chromo-Dynamics (QCD).  The diagram on the left shows the naive expectation that the spin of the nucleon is entirely constituted by the three valence quarks.  The EMC and SMC results proved that this was not correct.  The diagram on the right shows the current more complex view of nucleon spin structure, with contributions to the nucleon spin coming from the valence quarks, sea quarks, gluons and orbital angular momentum.
% (Diagrams courtesy of Hermes).
}}}
\label{fig:Proton_spin_structure}
\end{figure}

Prior to the European Muon Collaboration (EMC) \cite{EMC:1988,EMC:1989} and the Spin Muon Collaboration (SMC) \cite{Adeva:1993km} experiments at CERN it was assumed that all the spin of the nucleon was carried by its three constituent valence quarks.  The startling results of EMC in 1988 \cite{EMC:1988} and 1989 \cite{EMC:1989} showed that the three constituent valence quarks contribute very little to the spin of the nucleon.  This caused, what was dubbed ``The spin crisis in the parton model''\cite{Leader:1988vd,Anselmino:1988hn}, prompting a new theoretical investigation into the spin structure of the nucleon, which continues to this day.  The phrase \emph{``spin crisis''} which endures to this day was coined in a beautifully titled paper \emph{\lq\lq A crisis in the parton model: where, oh where is the proton's spin?''} by Mauro Anselmino and Elliot Leader \cite{Leader:1988vd}, and presented at the SPIN 1988 Symposium in Minneapolis, USA.  The fact that the two original EMC papers were the most cited experimental papers in the field for three years and have a combined total of over 2500 citations shows the immense effort that has been afforded to solving the ``spin crisis''.  It is now proposed that the spin of the nucleon is made up of the helicity of the constituent quarks $\Delta q$, the helicity of the gluons $\Delta G$, the orbital angular momentum of the quarks $L_q$, the orbital angular momentum of the gluons $L_g$ and the transversity of the quarks [referred to in the literature as either $\delta q$, $\Delta_T\,q$ or $h_{1\,q}$, we shall use the latter notation], as seen in Figure~\ref{fig:Proton_spin_structure}.  
Gluons, being massless spin-1 bosons, cannot be transversely polarized, hence there is no gluon transversity.  This gives us the longitudinal spin sum rule as follows:
\begin{equation}
\label{eq:Longitudinal_spin_sum_rule}
S_\mathrm{Nucleon}^{\,L} \ =\ \frac{1}{2} \ =\ \frac{1}{2}\,\Delta q \ +\  \Delta G \ +\  L^L_q \ +\  L^L_g\,,
\end{equation}
where the superscript $L$ refers to Longitudinal, and the transverse spin sum rule \cite{Bakker:2004ib}:
\begin{equation}
\label{eq:Transverse_spin_sum_rule}
S_\mathrm{Nucleon}^{\,T} \ =\  \frac{1}{2} \ = \ \frac{1}{2}\,h_{1\,q} \ +\  L^T_q \ +\  L^T_g\,,
\end{equation}
where the superscript $T$ refers to Transverse.
%****Not sure about the above equation, check Leader's (BLT) recent transverse spin sum rule.  Ensure our notation for transversity is consistent throughout the thesis, even in the diagrams.***
% I think this is correct, its a different notation than BLT use but I think its equivalent.

\begin{figure}[!h]
\centering
\includegraphics[height=5cm,width=5.5cm]{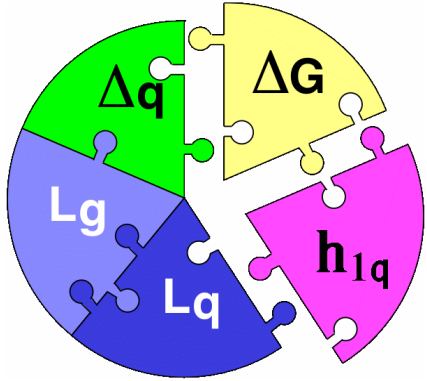}
\caption{\small{\it{The contributions to the spin of the nucleon: the helicity of the constituent quarks $\Delta q$,  the orbital angular momentum of the quarks $L_q$, the orbital angular momentum of the gluons $L_g$ all of which are known; the partly known helicity of the gluons $\Delta G$ and the unknown transversity of the quarks $h_{1\,q}$\,.
}}}
\label{fig:Proton_Structure_PieChart}
\end{figure}

The current knowledge of these constituents is summarized in Figure~\ref{fig:Proton_Structure_PieChart}.  The contributions $L_q$, $L_g$ and $\Delta q$ are known from experiment \cite{Ji:1996ek,Filippone:2001ux,Ji:2007zz,Thomas:2008ga} and from Lattice QCD studies \cite{Mathur:1999uf}.  There are currently many theoretical models \cite{Gluck:2000dy} and experimental programs obtaining information on $\Delta G$, these include HERMES, COMPASS, JLAB and RHIC.  But the last piece of the puzzle, the transversity distribution function $h_{1\,q}$ is to date almost completely unknown.  In order to best measure $h_{1\,q}$\,, a beam of polarized antiprotons would be required as we explain in the next section.  Much of this thesis is devoted to a theoretical investigation of possible methods to polarize an antiproton beam in a storage ring.

%***Explain that there was also a ``mass crises'' and a ``momentum crises'' in nucleon structure and how they were resolved.  Then explain the ``spin crisis'' in the nucleon spin structure.  Show the various possible solutions to it and say its still an open question, which will be investigated by the new generation of polarized experiments at GSI, J-PARC, BNL, COMPASS etc.  

\section{Antiprotons}
\label{sec:Antiprotons}

Antiprotons are the anti-particles of protons, which in turn are the core of the hydrogen atom, the most abundant element in the Universe.  The proton was shown to have an internal structure, {\it i.e.}\ not be a point particle, during seminal elastic electron-proton experiments in the Stanford Mark 3 accelerator, from 1954 to 1957 \cite{Hofstadter:1956qs}.  Robert Hofstadter was awarded the 1961 Nobel prize for this ground-breaking discovery which ushered in a new era of investigation into the structure of the nucleon.  A decade later, the much higher energy SLC accelerator was built at Stanford to investigate Deep Inelastic Scattering (DIS) experiments, showing the proton to be made up of point like quarks \cite{Breidenbach:1969kd}.  Again this achievement warranted the Nobel prize, in 1990 to Jerome Friedman, Henry Kendall and Richard Taylor.  The proton consists of $uud$ valence quarks, hence the antiproton consists of $\bar{u}\bar{u}\bar{d}$ valence quarks.  Above, as in the rest of the thesis, we denote antiparticles by an over-bar.  In shorthand notation protons are denoted $p$ and hence antiprotons are denoted $\bar{p}$.  Antiparticles have the opposite electromagnetic charge of their corresponding particle, thus for chargeless particles ({\it e.g.}\ the photon) the antiparticle is the same as the particle.  A proton has electromagnetic charge $+1$ in units where the electron charge is $-1$, thus an antiproton has electromagnetic charge $-1$.  While the concept of anti-matter often seems mysterious at first glance, it should be remembered that the positron (the anti-particle of the electron) was the third elementary particle discovered \cite{Anderson:1933mb}, after the electron and the photon.  Hence anti-matter has been an integral part of physical theories since 1932, the same year the neutron was discovered.
  
Antiprotons were discovered in 1955 by Owen Chamberlain and Emelio Segr\`e \cite{Chamberlain:1955ns}, who were awarded the 1959 Nobel prize for this ground-breaking discovery.  Professor Chamberlain, who recently passed away, spent the rest of his life contributing great efforts to the investigation of polarization phenomena and spin physics in general \cite{CERNcourier:05_2006}.  He was the first to investigate a possibility of polarizing antiprotons, and he co-organized the first workshop on polarizing antiprotons at Bodega Bay, California in 1985.  

The conclusions of this workshop \cite{Krisch:1986nt}, were that a high intensity polarized antiproton beam was not achievable at that time.  Another 22 years passed before the International community felt a sequel to this workshop was necessary, during which time interest in polarizing antiprotons grew steadily.  There has been much recent interest in producing a high intensity beam of polarized antiprotons, starting in 2004 with a proposal by the Polarized Antiproton eXperiments ($\mathcal{P}\mathcal{A}\mathcal{X}$) Collaboration at GSI Darmstadt \cite{Barone:2005pu}.  Since then many theories have been put forward on how to produce such a beam.  So in 2007 a sequel to the Bodega Bay workshop was organized in the newly founded Cockcroft Institute for accelerator research at the Daresbury Laboratory, UK \cite{Barber:2008a}.  A thorough investigation of the theoretical aspects of producing a polarized antiproton beam is presented in this thesis.

Antiproton-proton colliders have played an important role in the advancement of High Energy Physics.  In particular they led to the discovery of the $W$ and $Z$ bosons, and thus to the verification of the Weinberg, Glashow, Salam (1979 Nobel Prize) unified theory of electroweak interactions.  This was done by the UA1 and UA2 experiments at the SPS (Super Proton Synchrotron) collider in CERN in 1982 and led to the 1984 Nobel Prize to be awarded to Carlo Rubbia and Simon van der Meer.  One hopes that future polarized antiproton-proton colliders will lead to further epoch-making discoveries.

%\pagebreak

\section{Outline of the thesis}
\label{sec:Outline_of_the_thesis}

The major theme of the thesis is a theoretical investigation of the spin filtering method of polarization buildup, and an application of this to producing a high intensity polarized antiproton beam.  There is much debate in the International community as to the correct theoretical description of spin filtering.  We hope that the thorough analysis of spin filtering presented here will clarify some of this confusion.  No high intensity polarized antiproton beam has ever been achieved, and since a high intensity polarized antiproton beam could be used to measure many important quantities in particle physics, it is a main goal of the International community.  We first calculate all polarization dependent cross-sections in QED for the processes of interest, then we develop a set of differential equations using these polarization dependent cross-section to describe spin filtering; finally numerical results are obtained from this formalism.

In Chapter~\ref{ch:Motivation} the motivation for the thesis is outlined.  The benefits of a polarized antiproton beam are described, as are all possible methods to produce such a beam.  The methods to polarize bunches of other particles and atoms are also presented, such as electrons, positrons, protons, hydrogen and deuterium; and it is explained that none of these can be applied to the elusive case of polarizing antiprotons.  It is concluded that spin filtering is the most promising method to produce a high intensity polarized antiproton beam and the chapter concludes with an overview of spin filtering and a description of how it has been verified experimentally.

In Chapter~\ref{ch:Generic_helicity_amplitudes_and_spin_observables} all electromagnetic helicity amplitudes and spin observables, accounting for polarization effects in spin 1/2 - spin 1/2 elastic scattering are calculated.  Many of these results will be utilized in later chapters when providing a mathematical description of spin filtering, although their use is certainly not limited to this.  The spin 1/2 electromagnetic currents are introduced, both for point particles and particles with internal structure determined by electromagnetic form factors.  A generic equation is derived that can be used to calculate all polarization phenomena in elastic spin 1/2 - spin 1/2 electromagnetic scattering to first order in QED.  We then present results for all electromagnetic helicity amplitudes and spin observables for elastic spin 1/2 - spin 1/2 scattering.  The spin-averaged differential cross-section for spin 1/2 - spin 1/2 scattering is also presented in a new compact invariant form.  These results are then presented in Chapter~\ref{ch:Specific_helicity_amplitudes_and_spin_observables} for the specific cases of: antiproton-proton, antiproton-electron and positron-electron scattering.  Then the cross-sections and spin observables needed for spin filtering are explicitly presented, which will be utilized in the polarization evolution equations developed in Chapter~\ref{ch:Polarization_buildup_by_spin_filtering}.  The chapter concludes with a calculation of all spin 0 - spin 1 helicity amplitudes, which describe the scattering of deuterons off a carbon nucleus for example.

The theory of spin filtering is developed in Chapter~\ref{ch:Polarization_buildup_by_spin_filtering}.  A mathematical description of the related but simpler process of polarization buildup by the Sokolov-Ternov effect is first presented.  The ideas presented are utilized in the mathematical descriptions of spin filtering which follow.  The rates of change of the number of particles in each spin state are combined into a set of polarization evolution equations which describe the process of polarization buildup by spin filtering.  This set of polarization evolution equations is then analyzed and solved, emphasizing the physical implications of the dynamics.  The chapter concludes with an investigation of spin filtering of a stored beam.

Chapter \ref{ch:Various_scenarios_of_spin_filtering} presents a thorough investigation of spin filtering under various alternate scenarios, which would be of interest to any practical project to produce a high intensity polarized antiproton beam.  These scenarios are: 1) spin filtering while the beam is being accumulated, {\it i.e.}\ unpolarized particles are continuously being fed into the beam at a constant rate, 2)
unpolarized particles are continuously being fed into the beam at a linearly increasing rate, {\it i.e.}\ the particle input rate is ramped up, 3) the particle input rate is equal to the rate at which particles are being lost due to scattering beyond the ring acceptance angle, the beam intensity remaining constant, 4) increasing the initial polarization of a stored beam by spin filtering, and 5) the input of particles into the beam is stopped after a certain amount of time, but spin filtering continues.

As an application of the theoretical work presented throughout the thesis a possible method to produce a high intensity polarized antiproton beam by spin filtering off an opposing polarized electron beam is presented in Chapter~\ref{ch:Numerical_results}.  It is also outlined how this work can be applied to polarizing antiprotons by spin filtering off a polarized hydrogen target.  Firstly a description of the electron cooling technique to refocus the beam after scattering off the target each revolution in order to maintain high beam density is presented.  Then the various experimental input parameters, such as revolution frequency, target areal density, target polarization and the effective acceptance angle; needed to obtain realistic numerical estimates from our mathematical formalism are each described.  The benefits of using a lepton target are then described, before analyzing the case of spin filtering off an opposing polarized electron beam.  Finally spin filtering off a polarized hydrogen target is discussed, in the three cases of hydrogen with only electrons polarized, hydrogen with only protons polarized and finally hydrogen with both electrons and protons polarized.  It is shown that electromagnetic effects dominate hadronic effects in $\bar{p}\,p$ scattering in the region of low momentum transfer of interest in spin filtering. 

In Chapter~\ref{ch:Conclusions} some concluding remarks are presented.

\pagebreak

\section{Notation and conventions}
\label{sec:Notation_and_conventions}

The conventions will mainly follow the book of Peskin and Schroeder \cite{Peskin:1995}.  Rationalized units, where $\hbar =c=1$, will be used throughout the thesis unless otherwise stated.  Units in this system are as follows: $[\mathrm{length}] \,=\,[\mathrm{time}] \,=\,[\mathrm{energy}]^{\,-\,1} \,=\,[\mathrm{mass}]^{\,-\,1}$.  The usual convention of Greek characters representing four dimensional space-time indices $\left\{0,1,2,3\right\}$ and Latin characters representing three dimensional space indices $\left\{1,2,3\right\}$ is used throughout the thesis.  The end of a proof is represented by the symbol $\Box$, and $\equiv$ means that new objects are being defined. 

Lorentz 4-vectors are written as $x^\mu \,=\, \left(\,x^0,\, x^1,\, x^2,\, x^3\,\right)$, where $x^0$ is the time component
and $x^1$, $x^2$, and $x^3$ are the ${\bf \hat{x}}$, ${\bf \hat{y}}$, and ${\bf \hat{z}}$ space components respectively. Three dimensional vectors in Euclidean space are displayed in boldface, so that we can also write $x \,=\, \left(\,x^0, \,{\bf x}\,\right)$.  The Einstein summation convention, where repeated indices are summed over, is used throughout the thesis, unless otherwise specified.  Our conventions for Dirac spinors are presented in Appendix~\ref{Appendix:Dirac_algebra}.

The Feynman slash notation $\slashed{p} \,=\,\gamma^{\,\mu}\,p_\mu$, and the Minkowski metric tensor $\eta^{\,\mu\nu} \,=\,\mbox{diag}(+1,-1,-1,-1)$ are used.  The spin four vectors are normalized such that $S^{\,\mu} S_\mu \, = \, -\,1$.  We use the shorthand notation $A \cdot B$ for the scalar product in 4-dimensional Minkowski space, where $A \cdot B \,=\, A_{\,\mu}\,B\,^\mu \,=\, \eta_{\,\nu\mu}\, A^\nu B\,^\mu \,=\, A^0\,B^{\,0} \,-\, {\bf A} \cdot {\bf B}$.  The totally antisymmetric tensor $\epsilon^{\,\mu \nu \rho \sigma}$, also known as the Levi-Civita symbol, is defined such that $\epsilon^{\,0123} \,=\,+\,1$ and $\epsilon_{\,0123} \,=\,-\,1$, as seen in Appendix~\ref{Appendix:Dirac_algebra}.
% Ensure the above normalization of the Levi-Civita symbol is correct.

Antiparticles are denoted by an over-bar.  In shorthand notation protons are denoted $p$ and hence antiprotons are denoted $\bar{p}$.  Electrons and positrons are denoted by $e^-$ and $e^+$ respectively.  Time increases from left to right in all Feynman diagrams throughout the thesis.  Arrows on particle lines in Feynman diagrams denote the flow of particle number, which is forwards for particles and backwards for antiparticles.

We denote the time variable in each of the dynamical systems by $\tau$ to avoid confusion with the squared momentum transfer (Mandelstam $t$ variable) used throughout the thesis.

The scattering processes investigated in the thesis are always $2 \rightarrow 2$ elastic processes, with the momentum and spin 4-vectors of each particle labeled as:
\begin{equation*}
A\,\left(\,p_1,\,S_1\,\right) \ + \ B\,\left(\,p_2,\,S_2\,\right) \ \longrightarrow \ A\,\left(\,p_3,\,S_3\,\right) \ + \ B\,\left(\,p_4,\,S_4\,\right)\,,
\end{equation*}
with the particles above being the beam (1), target (2), scattered (3) and recoil (4) particles respectively.  The four momentum transfer is defined as 
\begin{equation*}
q \ =\ p_3 \,-\,p_1 \ =\  p_2\,-\,p_4\,.
\end{equation*}
The helicity amplitudes are represented by 
\begin{equation*}
\mathcal{M}\left(\,\mbox{scattered},\,\mbox{recoil}\,;\,\mbox{beam},\,\mbox{target}\,\right) \ = \ \mathcal{M}\left(\,\lambda_3,\,\lambda_4\,;\,\lambda_1,\,\lambda_2\,\right)\,.
\end{equation*}
The arguments are to be read from right to left, as $\lambda_1$ and $\lambda_2$ correspond to the incoming particles in the reaction and $\lambda_3$ and $\lambda_4$ correspond to the outgoing particles in the reaction.  For spin 1/2 - spin 1/2 scattering the helicities $\lambda_i \,=\,\pm$ for $i \in \left\{\,1,2,3,4\,\right\}$ are $+$ if the particles spin vector points in the direction of its momentum vector and $-$ if the particles spin vector points in the opposite direction to its momentum vector.  The $\pm$ in the helicity amplitudes are shorthand for $\pm 1/2$\,, the helicity of a spin 1/2 particle.  

The spin observables are represented by $K_{ab}$ for the polarization transfer observable and $\left(\,1\,-\,D_{ab}\,\right)$ for the depolarization observable, where $a,b \in \left\{\,X,Y,Z\,\right\}$ for the direction of the particles spin vector where its momentum is along the $Z$ direction.  The subscripts are read from right to left, in {\it e.g.}\ $K_{ab}$ where $b$ is the direction of the spin vector of the incoming particle and $a$ is the direction of the spin vector of the outgoing particle.

\pagebreak
  
\chapter{Motivation}
\label{ch:Motivation}

\vspace*{5ex}
\begin{minipage}{6cm}
\end{minipage}
\hfill
\begin{minipage}{10cm}
\begin{quote}
\emph{\lq\lq Polarization data has often been the graveyard of
fashionable theories. If theorists had their way, they
might just ban such measurements altogether out of
self-protection.\rq\rq}
\flushright{J.~D.~Bjorken}
\end{quote}
\end{minipage}
\vspace{8ex}

In this chapter the motivation for the present work is discussed.  The benefits to the high energy physics community of a high intensity polarized antiproton beam are first presented in section~\ref{sec:Motivation_for_a_polarized_antiproton_beam}, by describing the important parameters in particle physics that could be measured and investigated with such a beam.  Section~\ref{sec:Polarizing_bunches_of_particles_or_atoms} describes how bunches of other particles and atoms are polarized, such as electrons, positrons, protons, hydrogen and deuterium.  Unfortunately none of these tried and tested techniques can be applied to the elusive case of polarizing antiprotons.  Section~\ref{sec:Methods_to_polarize_an_antiproton_beam} describes and compares some possible methods to polarize antiprotons.  It is concluded that spin filtering is the most promising technique to produce a polarized antiproton beam as it is the only technique that has been experimentally verified.  The chapter concludes with an overview of the theory of spin filtering, and a section showing how spin filtering was verified for polarizing a proton beam by repeated scattering off a polarized hydrogen target in a storage ring by the FILTEX experiment in 1993.

\section{Motivation for a polarized antiproton beam}
\label{sec:Motivation_for_a_polarized_antiproton_beam}

A high intensity polarized antiproton beam would provide the unique possibility to measure many very important quantities in particle physics.  The most important quantity that could be measured is the transversity distribution of quarks inside protons, which has eluded direct measurement thus far.  Two other important investigations, into single spin asymmetries and nucleon electromagnetic form factors, can also be greatly advanced if a high intensity polarized antiproton beam was available.  These motivations for producing a high intensity polarized antiproton beam are described in detail below.  
%Only the three most important motivations are described here, for a more detailed list of motivations see Ref.~\cite{Barone:2005pu}.

\subsection{The transversity distribution function}
\label{subsec:The_transversity_distribution_function}

The transversity distribution function is the last leading twist\footnote{Leading twist means that in the factorization of a physical process the parton distribution function appears in the leading order of $1/Q^{\,2}$.} piece of the QCD description of the partonic structure of the nucleon, in the collinear limit\footnote{The collinear limit is where the intrinsic transverse motion of the quarks is averaged over.
%(${\bf k}_\perp \rightarrow 0$).
}, that has not been directly measured.  It describes the quark transverse polarization inside a transversely polarized nucleon.  In fact, to date, almost nothing is known about the transversity distribution, except for the recent work of Anselmino {\it et al.}~\cite{Anselmino:2007fs,Anselmino:2007zr}.  Unlike the other leading twist distributions [the unpolarized quark distribution $q\left(x,Q^2\right)$ and the helicity distribution  $\Delta q\left(x,Q^2\right)$] which have been measured, the transversity  $h_{1\,q}\left(x,Q^2\right)$ [sometimes referred to in the literature as $\Delta_T\,q\left(x,Q^2\right)$ or $\delta q\left(x,Q^2\right)$] can neither be accessed in deep inelastic scattering of leptons off nucleons, nor can it be reconstructed from the knowledge of $q\left(x,Q^2\right)$ and $\Delta q\left(x,Q^2\right)$ \cite{Barone:2005pu}.  In a transversely polarized hadron, $h_{1\,q}\left(x,Q^2\right)$ is the number density of quarks with momentum fraction $x$ and polarization parallel to that of the hadron, minus the number density of quarks with the same momentum fraction and antiparallel polarization, {\it i.e.}\ $h_{1\,q}\left(x,Q^2\right) = q^{\uparrow}\left(x,Q^2\right) - q^{\downarrow}\left(x,Q^2\right)$ \cite{Barone:2001sp}.   One cannot claim to understand the spin structure of the nucleon until all three leading twist structure functions have been measured.

\begin{figure}
\centering
\includegraphics[height=4cm]{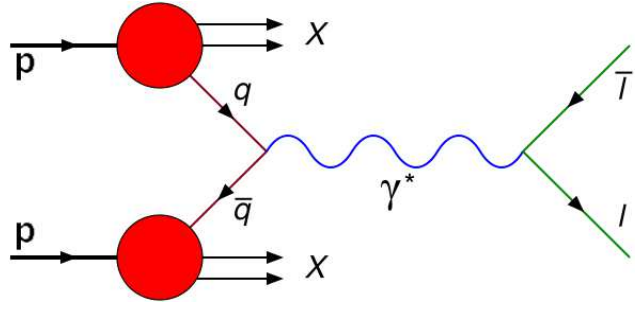}
\caption{\small{\it{The Drell\,-Yan lepton pair production process $p\,p \rightarrow \bar{l}\,l\,X$, via a virtual photon $\gamma^*$.}}}
\label{fig:Drell_Yan_Diagram}
\end{figure}

In order to best access the transversity distribution function, the double spin asymmetry $A_\mathrm{TT}$ in the Drell\,-Yan production of lepton pairs must be measured; thus both initial particles in a reaction must be transversely polarized.  It could in future be done for $p^{\uparrow}\, p^{\uparrow}$ scattering at RHIC, but this asymmetry is expected to be small from theory \cite{Anselmino:2004ki}, as explained below.  Also the cross-section for Drell\,-Yan lepton pair production is much higher for $\bar{p}\,p$ scattering than for $p\,p$ scattering ($\sigma_\mathrm{DY}^{\ \bar{p}\,p} \ \gg \sigma_\mathrm{DY}^{\ p\,p}$); because in the former case valance quarks in the proton annihilate with valance antiquarks in the antiproton, as opposed to with sea antiquarks in the second proton in the latter case.   The Drell\,-Yan lepton pair production process \cite{Drell:1970wh} is shown in Figure~\ref{fig:Drell_Yan_Diagram}.
The double spin asymmetry, an experimentally measurable quantity, is defined as
\begin{equation}
\label{eq:Double_spin_asymmetry}
A_\mathrm{ij} \ \equiv \ \frac{\mathrm{d}\sigma^{\uparrow\uparrow} \,-\, \mathrm{d}\sigma^{\uparrow\downarrow}}{\mathrm{d}\sigma^{\uparrow\uparrow} \,+\, \mathrm{d}\sigma^{\uparrow\downarrow}}\,,
\end{equation}
where $i$ and $j$ can be either $\mathrm{L}$ for longitudinal, or $\mathrm{T}$ for transverse.
For $p^{\uparrow}\, p^{\uparrow}$ Drell\,-Yan processes the transverse double spin asymmetry is \cite{Anselmino:2004ki}
\begin{eqnarray}
\label{eq:Transverse_double_spin_asymmetry_pp}
A_\mathrm{TT}^{p\,p}  & = & \displaystyle{\frac{\mathrm{d}\Delta\hat{\sigma}}{\mathrm{d}\hat{\sigma}}\ \frac{\sum_q\, e_q^2\left[\,h_{1\,q}^{\,p} \left(x_1,M^{\,2}\right) \ h_{1\,\bar{q}}^{\,p} \left(x_2,M^{\,2}\right) \ +\  h_{1\,\bar{q}}^{\,p} \left(x_1,M^{\,2}\right)\ h_{1\,q}^{\,p}\left(x_2,M^{\,2}\right)\,\right]}{\sum_q\, e_q^2\left[\,q^{\,p}\left(x_1,M^{\,2}\right)\ \bar{q}^{\,p}\left(x_2,M^{\,2}\right) \ +\  \bar{q}^{\,p}\left(x_1,M^{\,2}\right)\ q^{\,p}\left(x_2,M^{\,2}\right)\,\right]}} \,, \nonumber\\[2ex]
& \approx & \displaystyle{ \frac{\mathrm{d}\Delta\hat{\sigma}}{\mathrm{d}\hat{\sigma}}\ \frac{h_{1\,u}^{\,p}\left(x_1,M^{\,2}\right)\ h_{1\,\bar{u}}^{\,p}\left(x_2,M^{\,2}\right)\ + \ h_{1\,\bar{u}}^{\,p}\left(x_1,M^{\,2}\right)\ h_{1\,u}^{\,p}\left(x_2,M^{\,2}\right)}{u^{\,p}\left(x_1,M^{\,2}\right)\ \bar{u}^{\,p}\left(x_2,M^{\,2}\right) \ + \ \bar{u}^{\,p}\left(x_1,M^{\,2}\right)\ u^{\,p}\left(x_2,M^{\,2}\right)}\,} ,
\end{eqnarray}
where $\mathrm{d}\Delta\hat{\sigma}$ and $\mathrm{d}\hat{\sigma}$ are the polarized and unpolarized cross-sections of the elementary QED process $q\,\bar{q}\rightarrow l^-\,l^+$ respectively, $M$ is the invariant mass of the lepton pair, $e_q$ is the electromagnetic charge of the quarks and $x_1$ and $x_2$ are the fraction of their respective nucleon momentum carried by each of the interacting partons.  The leading term in the approximation comes from the fact that the $u$ quark dominates at large $x$ \cite{Barone:2001sp}.  Whereas for $p^{\uparrow}\,\bar{p}^{\,\uparrow}$ Drell\,-Yan processes
\begin{eqnarray}
\label{eq:Transverse_double_spin_asymmetry_ppbar}
A_\mathrm{TT}^{p\,\bar{p}} & = &  \displaystyle{\frac{\mathrm{d}\Delta\hat{\sigma}}{\mathrm{d}\hat{\sigma}}\ \frac{\sum_q\, e_q^2\left[\,h_{1\,q}^{\,p} \left(x_1,M^{\,2}\right)\  h_{1\,\bar{q}}^{\,\bar{p}} \left(x_2,M^{\,2}\right)\ + \ h_{1\,\bar{q}}^{\,p} \left(x_1,M^{\,2}\right)\ h_{1\,q}^{\,\bar{p}}\left(x_2,M^{\,2}\right)\,\right]}{\sum_q\, e_q^2\left[\,q^{\,p}\left(x_1,M^{\,2}\right)\ \bar{q}^{\,\bar{p}}\left(x_2,M^{\,2}\right) \ +\  \bar{q}^{\,p}\left(x_1,M^{\,2}\right)\ q^{\,\bar{p}}\left(x_2,M^{\,2}\right)\,\right]}} \,,\nonumber \\[2ex]
& \approx & \frac{\mathrm{d}\Delta\hat{\sigma}}{\mathrm{d}\hat{\sigma}}\ \frac{h_{1\,u}^{\,p} \left(x_1,M^{\,2}\right)\ h_{1\,\bar{u}}^{\,\bar{p}} \left(x_2,M^{\,2}\right)}{u^{\,p}\left(x_1,M^{\,2}\right)\ \bar{u}^{\,\bar{p}}\left(x_2,M^{\,2}\right)}\, ,
\end{eqnarray}
the latter of which is much larger since there are more antiquarks in antiprotons making $h_{1\,u}^{\,p} \left(x,Q^2\right) \,=\,h_{1\,\bar{u}}^{\,\bar{p}} \left(x,Q^2\right) \gg h_{1\,\bar{u}}^{\,p}\left(x,Q^2\right) \,=\,h_{1\,u}^{\,\bar{p}}\left(x,Q^2\right)$ \cite{Anselmino:2004ki}.  
%(***Note also $\bar{u}^{\bar{p}} \gg \bar{u}^p$ but the transversity distribution functions are still easier to measure in the second case.***)  
Thus $A_\mathrm{TT}^{p\,\bar{p}}$, which can only be measured using a polarized antiproton beam, is expected to be much bigger than $A_\mathrm{TT}^{p\,p}$.  Note in the literature all quantities in these equations are often written with respect to the proton using the fact that the distribution of antiquarks in a proton is equal to the distribution of quarks in an antiproton etc.\ but here we want to keep the antiproton distribution functions explicit.  Using the above fact, and at $x_1\,=\,x_2$, eq.~(\ref{eq:Transverse_double_spin_asymmetry_ppbar}) reduces to 
\begin{equation}
A_\mathrm{TT}^{p\,\bar{p}} \ = \ \frac{\mathrm{d}\Delta\hat{\sigma}}{\mathrm{d}\hat{\sigma}}\ \left[\,\frac{h_{1\,u}^{\,p} \left(x_1,M^{\,2}\right)}{u\left(x_1,M^{\,2}\right)} \,\right]^{\,2}\, ,
\end{equation}
providing a unique direct way to measure a single transversity distribution function~\cite{Anselmino:2004ki}. 

Also $q\left(x,Q^2\right)$ and $\bar{q}\left(x,Q^2\right)$ decrease with increasing $x$, so to measure $A_\mathrm{TT}$ large $x_1$ and $x_2$ is favoured \cite{Anselmino:2004ki,Efremov:2004qs}.  Interestingly this happens for lower energy scattering again making a low/medium energy facility, such as that proposed by the $\mathcal{P}\mathcal{A}\mathcal{X}$ Collaboration, more suited than RHIC\footnote{Note there is a proposal to run RHIC at $\sqrt{s} = 50 \ \mbox{GeV}$ instead of their usual $\sqrt{s} = 200 \ \mbox{GeV}$ which would make it suitable in this regard, but the problem of no antiproton beam would still remain.  In Run 6 (2006) RHIC used $\sqrt{s} = 62.4\ \mbox{GeV}$.}.  At RHIC energies, even though $A_\mathrm{TT}$ could be detected it only measures the transversity of the sea quarks, at the lower $\mathcal{P}\mathcal{A}\mathcal{X}$ energies we could investigate the transversity of the valence quarks \cite{Anselmino:2007fs,Efremov:2004qs}.

\subsection{Single spin asymmetries}
\label{subsec:Single_spin_asymmetries}

Single Spin Asymmetries (SSA), where one of the initial particles in the reaction is polarized in the direction of the arrows below, are defined as
\begin{equation}
\label{eq:Single_spin_asymmetry}
A_\mathrm{j} \ \equiv \ \frac{\mathrm{d}\sigma^{\uparrow} \,-\, \mathrm{d}\sigma^{\,\downarrow}}{\mathrm{d}\sigma^{\uparrow} \,+\, \mathrm{d}\sigma^{\,\downarrow}}\,,
\end{equation}
where $j$ can be either $\mathrm{L}$ for longitudinal, or $\mathrm{T}$ for transverse.

Some data on SSA in $\bar{p}^{\,\uparrow}\, p$ Drell-Yan lepton pair production was obtained by the E704 experiment at Fermilab \cite{Adams:2003fx,Bravar:1996ki}, but because the collisions were in the energy region of $J/\psi$ production\footnote{The $J/\psi$ particle is the first excited state of charmonium, a meson consisting of one charm quark and one charm antiquark.  Two papers by separate experiments announcing its discovery were published on the same day, one group naming it the $J$ particle and the other group naming it the $\psi$ particle.  It has come to be known as the $J/\psi$ particle.} it was difficult to distinguish the Drell-Yan signal from the large $J/\psi$ production background.  The low intensity polarized antiproton beam used in E704 is described in section~\ref{subsec:Antihyperon_decay}.

Importantly analyzing charm production in $\bar{p}^{\,\uparrow}\, p$ scattering will make it possible to disentangle the Sivers \cite{Sivers:1989cc,Sivers:1990fh} and the Collins mechanisms \cite{Collins:1992kk}, of which there is great theoretical interest.  In general, both effects contribute to the measured SSA, but in the case of charm production the Collins mechanism drops out.

A polarized antiproton beam would allow further analysis of single spin asymmetries in $\bar{p}^{\,\uparrow}\, p$ scattering, augmenting the brief Fermilab data on this \cite{Adams:2003fx}, and adding to the current data on single spin asymmetries which have been observed in $\bar{p}\ p^{\uparrow}$ and $p^{\uparrow}\,p$ reactions \cite{Krueger:1998hz,Bravar:1996ki,Boglione:2007dm} and the double spin asymmetries observed in $p^{\uparrow}\,p^{\uparrow}$ reactions at RHIC \cite{Abelev:2006uq}.  These observed asymmetries are very large, up to $40\%$ \cite{Krueger:1998hz}, prompting Stan Brodsky to call them \emph{\lq\lq the greatest asymmetries ever seen by a human being"} constituting  \emph{\lq\lq one of the unsolved mysteries of hadron physics"}.  There is much current interest in the theoretical community to try to achieve a satisfactory understanding of these large single spin asymmetries \cite{Boglione:2007dm,Brodsky:2002cx,Anselmino:2005ea,Anselmino:2005sh,Bacchetta:2007sz,Brodsky:2005yw}.

\subsection{Electromagnetic form factors of the proton}
\label{subsec:Electromagnetic_form_factors_of_the_proton}

The fact that nucleons (protons and neutrons) are not point particles and have an internal structure, is parameterized into electromagnetic form factors, as treated later in section~\ref{sec:Spin_1/2_electromagnetic_currents}.  The Sachs electric and magnetic form factors $G_E$ and $G_M$ contain information on the finite charge radius of the proton, thus are very important components of a complete understanding of particle physics.  They can be measured experimentally but there is not, to date, complete agreement between the experimental results and theoretical models of the form factors \cite{Dubnickova:1992ii,Punjabi:2005wq}. 

There is much current theoretical interest in nucleon time-like form factors \cite{Brodsky:2003gs}.  A polarized antiproton beam would enable the first measurement of the moduli and the relative phase of the time-like electric and magnetic form factors $G_E$ and $G_M$ of the proton.  An unexpected $Q^{\,2} = -\,q^{\,2}$ dependence of the $G_E(q^{\,2})\,/\,G_M(q^{\,2})$ ratio of the electric and magnetic form factors of the proton, has been observed at the Jefferson laboratory (JLAB), the ratio decreasing monotonically with increasing $Q^{\,2}$ \cite{Jones:1999rz,Gayou:2001qd}.  It would be possible to clarify this unexpected $Q^{\,2}$ dependence by a measurement of the relative phases of $G_E(q^{\,2})$ and $G_M(q^{\,2})$ in the time-like region, which would constrain and discriminate strongly between the models for the form factors.  This phase can be measured for the first time in the reactions $\bar{p}^{\,\uparrow}\,p \rightarrow e^+ e^-$ and $\bar{p}\,p^\uparrow \rightarrow e^+ e^-$ \cite{Brodsky:2003gs}, the former of which is uniquely possible with a polarized antiproton beam.  The JLAB data was obtained by analyzing the polarization transfer reaction $p\,e^{-\,\uparrow} \rightarrow p^{\,\uparrow} \,e^-$, this data could be augmented and checked by analyzing the polarization transfer reaction $\bar{p}\,e^{-\,\uparrow} \rightarrow \bar{p}^{\,\uparrow}\, e^-$, which is at the heart of the spin filtering technique discussed in detail throughout this thesis.

The relative phase ambiguity can also be addressed by measuring the double spin asymmetry in the reaction $\bar{p}^{\,\uparrow}\,p^\uparrow \rightarrow l^+\,l^-$, where $l$ is any lepton.  This reaction can also be used to analyze the $G_E \,-\,G_M$ separation, thus serving as a check of the Rosenbluth separation in the time-like region \cite{Buttimore:2007cv}.

\section{Polarizing bunches of particles or atoms}
\label{sec:Polarizing_bunches_of_particles_or_atoms}

High intensity beams of polarized electrons, positrons and protons, as well as polarized atomic gas targets have been used in high energy physics laboratories throughout the world.  We now briefly describe how they are polarized.

\begin{itemize}

\item Sokolov-Ternov effect (\lq radiative' or \lq self'-polarization): A beam of charged particles circulating in a storage ring is automatically polarized because of a difference in the spin-flip transition rates due to emission of photons by synchrotron radiation induced by bending in the magnetic field of the ring.  This method works well for polarizing electrons and positrons, but not for heavier particles such as protons or antiprotons, as explained in section~(\ref{subsec:Spontaneous_synchrotron_radiation_emission}).

\item Atomic hydrogen and deuterium are polarized by removing atoms in certain hyperfine states, and inducing angular momentum conserving transitions between hyperfine states.

\item Once hydrogen is polarized the electrons can be stripped off in a magnetic field leaving polarized protons.

\end{itemize}

\noindent
Unfortunately it is not possible to produce a high intensity polarized antiproton beam using any of these tried and tested methods, as to do so one would need a large supply of antihydrogen atoms. 

The Sokolov-Ternov effect will be described in section~(\ref{subsec:Spontaneous_synchrotron_radiation_emission}) and we describe how polarized hydrogen is obtained in section~(\ref{subsec:Polarizing_hydrogen_gas}).

\subsection{Polarizing hydrogen gas}
\label{subsec:Polarizing_hydrogen_gas}

Unpolarized hydrogen atoms in a strong magnetic field equally populate each of four hyperfine states:
\begin{eqnarray*}
|\uparrow_{p}\,\downarrow_{e}\,\rangle \hspace*{3em} |\downarrow_{p}\,\downarrow_{e}\,\rangle \hspace*{3em} |\downarrow_{p}\,\uparrow_{e}\,\rangle \hspace*{3em} |\uparrow_{p}\,\uparrow_{e}\,\rangle 
\end{eqnarray*}
An inhomogeneous magnetic field acts as a Stern-Gerlach apparatus separating the atoms in the states $|\uparrow_{p}\,\uparrow_{e}\,\rangle$ and $|\downarrow_{p}\,\uparrow_{e}\,\rangle$ from those in the states $|\downarrow_{p}\,\downarrow_{e}\,\rangle$ and $|\uparrow_{p}\,\downarrow_{e}\,\rangle$.  A sextupole magnet focuses the atoms in one pair of states while defocusing the others.  Thus one can extract atoms in the states $|\uparrow_{p}\,\uparrow_{e}\,\rangle$ and $|\downarrow_{p}\,\uparrow_{e}\,\rangle$, {\it i.e.}\ hydrogen atoms in which the electrons are totally polarized, but protons unpolarized.  If one then requires hydrogen atoms in which the protons are totally polarized but the electrons unpolarized, angular momentum conserving transitions from $|\downarrow_{p}\,\uparrow_{e}\,\rangle$ to $|\uparrow_{p}\,\downarrow_{e}\,\rangle$ can be induced by a radio frequency field.

Hydrogen with both the electrons and protons polarized can be obtained by isolating the $|\uparrow_{p}\,\uparrow_{e}\,\rangle$ state, but with only half the intensity of hydrogen with {\it either} electrons {\it or} protons polarized.

In summary there are three types of polarized hydrogen, with all atoms in the hyperfine states as follows
\begin{eqnarray}
\label{eq:Getting_Hydrogen_polarization_state1}
|\uparrow_{p}\,\uparrow_{e}\,\rangle \ + \ |\uparrow_{p}\,\downarrow_{e}\,\rangle \hspace*{1em} \implies \hspace*{1em} \mathcal{P}_p \ = \ 1 \hspace*{1em} \mbox{and} \hspace*{1em} \mathcal{P}_e \ = \ 0 \\[2ex]
\label{eq:Getting_Hydrogen_polarization_state2}
|\uparrow_{p}\,\uparrow_{e}\,\rangle \ + \ |\downarrow_{p}\,\uparrow_{e}\,\rangle \hspace*{1em} \implies \hspace*{1em} \mathcal{P}_p \ = \ 0 \hspace*{1em} \mbox{and} \hspace*{1em} \mathcal{P}_e \ = \ 1 \\[2ex]
\label{eq:Getting_Hydrogen_polarization_state3}
|\uparrow_{p}\,\uparrow_{e}\,\rangle \hspace*{1em} \implies \hspace*{1em} \mathcal{P}_p \ = \ 1 \hspace*{1em} \mbox{and} \hspace*{1em} \mathcal{P}_e \ = \ 1 
\end{eqnarray}
where we denote the polarization of the protons in the hydrogen by $\mathcal{P}_p$ and the polarization of the electrons in the hydrogen by $\mathcal{P}_e$\,.  In practice the atoms are not perfectly isolated in certain hyperfine states, thus the electron and proton polarizations in polarized hydrogen are less than one.  The HERMES Collaboration at DESY have utilized polarized hydrogen and polarized deuterium targets with $\mathcal{P}_p \,=\,0.9$ and/or $\mathcal{P}_e\,=\,0.9$ \cite{Airapetian:2004yf}, and these targets are now being used by the $\mathcal{P}\mathcal{A}\mathcal{X}$ Collaboration in COSY J\"ulich for preliminary tests on spin filtering \cite{Steffens:2007zz,Nass:2007zz}.   

Other atoms, such as deuterium, can be polarized analogously.  Stripping these atoms of electrons in a magnetic field leaves a polarized ion beam.  

It is not possible to generate a beam of polarized antiprotons by this means as it is, thus far, not possible to accumulate large numbers of antihydrogen atoms.

The RHIC collider in Brookhaven National Laboratory, New York is the world's first high intensity polarized proton accelerator.  A high intensity polarized antiproton collider would greatly supplement and add to results obtained at RHIC.

%\section{Accelerating polarized beams}
%\label{subsec:Accelerating_polarized_beams}
%
%***Write this section***
%
%***Add short section on accelerating polarized beams, depolarizing resonances and Siberian snakes somewhere***
%
%***This section could possibly be omitted, if I felt there was too much experimental/technical detail in the Motivation Chapter of the thesis.***  Just briefly mention this in the text, perhaps with a footnote.

\section{Methods to polarize an antiproton beam}
\label{sec:Methods_to_polarize_an_antiproton_beam}

Now that we have demonstrated the incredible potential of a high intensity polarized antiproton beam, let us investigate the various methods of generating such a beam.  Physicists have been trying to produce beams of polarized antiprotons for over 25 years, a great summary of proposed methods is given in Ref.~\cite{Krisch:1986nt}.  Atomic beam sources, used in the production of polarized protons and heavy ions will not work because of the annihilation of antiprotons with matter.  The E704 experiment at Fermilab has produced polarized antiprotons from the decay of polarized $\Lambda$ hyperons, but the intensities achieved were too low for current needs.  Storing antiprotons in a storage ring would help build up to a high luminosity beam.

Spin filtering has been proven to work for protons scattering off a polarized internal hydrogen target in the FILTEX experiment at the TSR ring in Heidelberg in 1992-1993 \cite{Rathmann:1993xf}.  Thus spin filtering is the only plausible experimentally tested technique for generating a high intensity polarized antiproton beam.  In light of this we devote much of this thesis to the theoretical understanding of spin filtering in general.  As an application of our theoretical work we propose a method to polarize antiprotons by spin filtering off an opposing polarized electron beam, and calculate the polarization buildup time and maximum polarization possible in this case.

\subsection{Antihyperon decay}
\label{subsec:Antihyperon_decay}

Antihyperons are produced when a Multihundred-GeV proton beam strikes a target.  The antihyperons decay into antiprotons, which should have the same polarization as the protons from hyperon decay.  A polarized antiproton beam of this type was produced at Fermilab's E704 experiment \cite{Grosnick:1989qv}. 
%The antiprotons had a polarization of about $40 \, \%$.  
The low intensity (because it is a tertiary beam) and large phase space made it difficult to store and accelerate these polarized antiprotons; however it was possible to scatter them off a polarized or unpolarized proton target.  The polarization of the antiprotons comes from parity violating decays of antilambdas, and the measured polarization was as high as $64 \, \%$.  The target they used to produce the antilambdas was Beryllium, and their polarized antiproton beam intensities were up to $1.5 \times 10^{\,5} \ \mbox{s}^{-1}$ \cite{Grosnick:1989qv}.  This method of producing polarized antiprotons is not suitable for high luminosity experiments, such as the high intensity beam a storage ring could provide, which are needed to access transversity and other measurements as outlined in section~\ref{sec:Motivation_for_a_polarized_antiproton_beam}.

\subsection{Stern-Gerlach separation}
\label{subsec:Stern_Gerlach_separation}

A possible method to produce a polarized antiproton beam from an unpolarized antiproton beam is based on the Stern-Gerlach effect.  In an inhomogeneous magnetic field the spins of particles, aligned parallel or antiparallel to the field, are deflected in opposite directions and become spatially separated.  For this reason this method, proposed by Niinikoski and Rossmanith in 1985 \cite{Niinikoski:1985bm}, is also called the {\it spin-splitter} technique.  A major advantage of this method is that the beam can first be accelerated to any desired energy and then polarized, thus avoiding the loss of polarization associated with accelerating polarized beams\footnote{Beams tend to lose some of their polarization at certain depolarization resonance energies during acceleration.  This problem can be circumvented by utilizing Siberian Snakes \cite{Derbenev:1975hd}, devices which flip the polarization vector of each beam particle by $180$ degrees each revolution.  Thus any deflections from the polarization axis are canceled out every two revolutions.}. 

In a typical storage ring inhomogeneous magnetic fields are provided by the quadrupoles.  It was hoped that the spatial separation of the particles in the two spin states would add up on passing through many quadrupoles, and further over many revolutions in the storage ring; eventually leading to a macroscopic separation of the particles in opposite spin states \cite{Niinikoski:1985bm}.  One spin state can then be dumped, or flipped, and one is left with a polarized antiproton beam.

Unfortunately, after much interest in this technique \cite{Cameron:2003cg,Cameron:2003ch}, the International Community has doubts as to whether effects in successive quadrupoles will add up coherently \cite{Barber:2008}.  The effects may continuously cancel each other out and one will be left with no net separation of particles in the two spin states.  At the very least this method would have to be experimentally verified before being considered a practical method of producing a polarized antiproton beam.

\subsection{Spontaneous synchrotron radiation emission}
\label{subsec:Spontaneous_synchrotron_radiation_emission}

\psfrag{G}{{\bf {\huge $\gamma$}}}
\begin{figure}
\centering
\includegraphics[width=11cm]{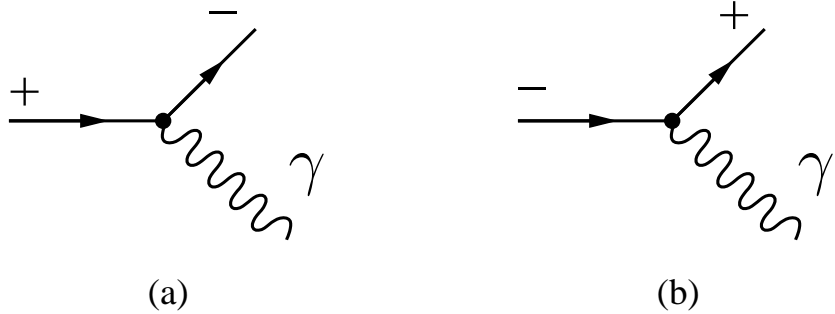}
\caption{\small{\it{The two spin-flip Feynman diagrams that contribute to the Sokolov-Ternov effect.  In (a) an electron in the \lq spin up' state gets flipped to the \lq spin down' state, while in (b) an electron in the \lq spin down' state gets flipped to the \lq spin up' state; after emitting a photon due to synchrotron radiation induced by bending in a magnetic field.  The cross-sections for these two processes are not equal and as such there will be a gradual buildup of polarization in the beam, known as the Sokolov-Ternov effect.}}}
\label{fig:Sokolov-Ternov_Effect__Diagrams}
\end{figure}

Charged particles emit synchrotron radiation, in the form of photons, when bent in a magnetic field.  There is a slight difference in the spin-flip transition cross-sections due to this photon emission: $\sigma\left(\,e^-_\uparrow \rightarrow e^-_\downarrow \, \gamma \,\right)\, \neq \, \sigma\left(\,e^-_\downarrow \rightarrow e^-_\uparrow \, \gamma\,\right)$, thus over time the beam of charged particles acquires some polarization.  The cross-section for a particle in the \lq spin up' state to flip to the \lq spin down' state on emitting a photon by synchrotron radiation is different to the cross-section for a particle in the \lq spin down' state to flip to the \lq spin up' state.  The \lq spin up' and \lq spin down' states are defined as the particle's spin being aligned parallel and antiparallel to the magnetic field of the storage ring respectively.  This \lq self polarization' is called the Sokolov-Ternov effect after the Russian theorists who discovered it around 1963 \cite{Sokolov:1963}.  

In a perfect storage ring an equilibrium polarization of $\mathcal{P}_\mathrm{ST} = 8\,/ \left(\,5\,\sqrt{3}\,\right) \approx 0.924$ is reached \cite{Lee:1997}.  In practice the maximum polarization achievable is slightly less than this ideal value due to imperfections of the magnetic fields in the storage ring.  However in less than one hour electron beams at TRISTAN in Japan and HERA in Germany acquired polarizations of about $80 \, \%$ or more \cite{Lee:1997}.  The Sokolov-Ternov radiative polarization is along the vertical direction perpendicular to the storage ring plane.  

% TRISTAN is an electron-positron collider in KEK.

\begin{table}
\begin{center}
\begin{tabular}{|c||c|c|c||c|c|}
\hline
%\multicolumn{6}{|c|}{} \\*[-2.2ex]
  & \multicolumn{3}{|c||}{} & \multicolumn{2}{|c|}{}\\[-2ex]
 & \multicolumn{3}{|c||}{Electron storage rings} & \multicolumn{2}{|c|}{Proton storage rings}\\[0.5ex]
\hline
 & & & & &  \\*[-2ex]
                        &  LEP     & TRISTAN &  HERA    & LHC      & SSC \\[0.5ex]
\hline
 & & & & &  \\*[-2ex]
$E$\ [GeV]              & $46.5$   & $30$    & $27.521$ & $7000$   & $20000$\\[0.5ex]
$\gamma 
= \sqrt{1-\beta^{\,2}}$ & $90999$  & $58710$ & $53858$  & $7462$   & $21317$\\[0.5ex]
$R$\ [m]                & $4243$ & $480$     & $1008$   & $4243$   & $12096$\\[0.5ex]
$\rho$\ [m]             & $3096.2$ & $246.5$ & $575$    & $3096.2$ & $10108$\\[0.5ex]
$N_\gamma$              & $6039$   & $3896$  & $3574$   & $567$    & $1429$\\[0.5ex]
\hline
 & & & & &  \\*[-2ex]
$\tau_\mathrm{\,ST}$\ [min] & $308$    & $2$     & $35$  & $2.8 \times 10^{14}$     & $4.5 \times 10^{13}$\\[0.5ex]
\hline
\end{tabular}
\end{center}
\caption{\small{\it{Properties of some high energy electron storage rings, and proposed proton storage rings.  $E$ is the kinetic energy of the beam, $\gamma = \sqrt{1 - \beta^{\,2}}$ is the relativistic Lorentz factor, $R$ the mean radius of the storage ring consisting of identical bending magnets of bending radius $\rho$ separated by straight sections combining to give a total circumference $2\,\pi R$, $N_\gamma$ is the average number of photons emitted per particle per revolution and $\tau_\mathrm{\,ST}$ is the time taken to reach the equilibrium Sokolov-Ternov polarization.  Parts of this table are reproduced from Refs.~\cite{Lee:1997} and \cite{Lee:2004}.  As one can see this method of polarization buildup takes too long for (anti)proton rings.}}}
\label{table:Properties_of_Synchrotrons}
\end{table}

The effect is much stronger for electrons than for protons as the rate of synchrotron radiation (number of photons emitted per second) is related to the velocity of the particle not its energy.  Because (anti)protons are approximately $1800$ times more massive than electrons, at a given energy electrons are traveling at a much higher velocity, {\it i.e.}\ much closer to the speed of light ($\gamma_e = (\,m_p\,/\,m_e\,)\,\gamma_p \approx 1800\,\gamma_p$).  If (anti)protons were moving this close to the speed of light they too would become self polarized by the Sokolov-Ternov effect.  Thus at a given energy the time taken to polarize electrons by the Sokolov-Ternov effects is much shorter than the time taken to polarize (anti)protons.  Even at the $20 \ \mbox{TeV}$ of the proposed Superconducting Super Collider (SSC) it would take antiprotons or protons about $10^7 \ \mbox{years}$ to acquire a useful polarization\footnote{See Table~\ref{table:Properties_of_Synchrotrons}.}, and much higher kinetic energies would be required to provide a practical method of polarizing antiprotons by the Sokolov-Ternov effect.
%
%It has been calculated that (anti)protons must have a kinetic energy of $70 \ \mbox{TeV}$ in order for the polarization to buildup in a matter of hours or less \cite{Krisch:1986nt}, 

The time taken to reach the equilibrium polarization is given by \cite{Sokolov:1963,Lee:1997,Jackson:1975qi}:
%
%\begin{equation}
%\label{eq:Sokolov_Ternov_Polarization_Time2}
%\tau_\mathrm{ST} \ = \ \frac{8}{5\,\sqrt{3}}\frac{m^{\,2}\,c^{\,2}\,\rho^{\,2}\,R}{e^{\,2}\,\hbar\,\gamma^{\,5}}
%\end{equation}
%%
%****This equation is only dimensionally correct in units where $e^2 \rightarrow e^2/4\,\pi\,\epsilon_0$, where $\epsilon_0$ is the permittivity of free space.  **No there is no $\epsilon_0$ in HEP, they use $r_e = e^2/mc^2$, think about this but it may actually be best to change back to the above equation as in Leader and Jackson** We could change $e^2$ to the classical electron radius $r_e$ which would solve the confusion as the $4\,\pi\,\epsilon_0$ confusion is subsumed into the definition of $r_e$ in different systems of units, cgs and MKS.  Problem is that one then needs to change to the classical proton radius for protons or antiprotons.So we would have****
%
\begin{equation}
\label{eq:Sokolov_Ternov_Polarization_Time}
\tau_\mathrm{ST} \ = \ \frac{8}{5\,\sqrt{3}}\,\frac{m\,\rho^{\,2}\,R}{r\,\hbar\,\gamma^{\,5}}\,,
\end{equation}
where $m$ is the particle's mass, $r$ the classical radius of the particle (electron or proton), $\hbar = h\,/\,2\,\pi$ the reduced Planck's constant, $R$ the mean radius of the storage ring consisting of identical bending magnets of bending radius $\rho$ separated by straight sections combining to give a total circumference $2\,\pi R$ and $\gamma = 1\,/\,\sqrt{1 - \beta^{\,2}}$ is the relativistic Lorentz factor, where $\beta = v\,/\,c$ is the ratio of the particles velocity to that of light.  Some properties of current and proposed future synchrotrons are presented in Table~\ref{table:Properties_of_Synchrotrons}.
%Note that the classical radius of the proton is 1800 times smaller than the classical radius of the electron.  This is somewhat counterintuitive and I should be very aware of it.

The Sokolov-Ternov effect is similar to systems investigated later in this thesis, and it can be described by systems of differential equations similar to ones we develop to describe spin filtering.  Hence to provide a comparison we present and solve a set of differential equations describing the Sokolov-Ternov effect in section~\ref{sec:The_Sokolov-Ternov_Effect}.
%
%
%The results from the table are calculated in the mathematica notebook Sokolov_Ternov_Effect.nb
%
Synchrotron radiation is the physical principle behind the antenna, emitting photons in the form of radio waves, and some lasers generated by wiggler magnets.  The light produced by synchrotron radiation is used by biologists and many storage rings have a second life as intense light sources after high energy physics experiments have ceased.  Electrons lose much of their energy in a storage ring due to synchrotron radiation, typically emitting hundreds to thousands of photons per revolution\footnotemark[5],  and as a consequence very high energy electron/positron accelerators, such as the International Linear Collider (ILC), must be linear to avoid this problem.  

We conclude that this method of polarization buildup would take too long for an antiproton beam to be considered practical at present energies.

%*** Ensure the equation is also valid for protons not just electrons.  This section is then edited and complete for the Thesis.***

\subsection{Polarization of directly produced antiprotons}
\label{subsec:Polarization_of_directly_produced_antiprotons}

It is well known that the particles produced when a high energy proton beam strikes a target have some polarization at certain production angles.  Some of the particles produced will be antiprotons; in fact this is how antiproton beams are obtained \cite{Eades:1999ff}.  Unfortunately the polarization generally seems to be larger at larger production angles where the cross-sections are smaller \cite{Krisch:1986nt}.  Thus it appears difficult to simultaneously obtain antiprotons with a high polarization and a high beam intensity using this method.

\subsection{The theory of spin filtering} 
\label{subsec:The_theory_of_spin_filtering}

\begin{figure}
\centering
\includegraphics[width=15cm]{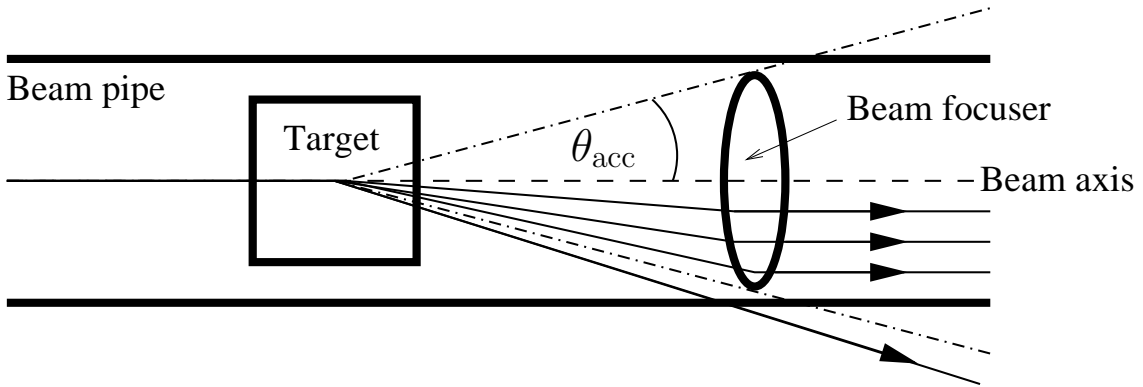}
\caption{\small{\it{This diagram describes the spin filtering technique.  Beam particles travel along the beam axis and scatter off the target.  Particles scattered at angles greater than the acceptance angle $\theta_\mathrm{acc}$ are lost from the beam, while particles scattered at angles less than $\theta_\mathrm{acc}$ pass through a beam focuser and remain in the beam.  In this simplistic diagram the beam focuser is represented by a lens, but in reality the beam is focused by electron cooling as explained in section~\ref{sec:Beam_Cooling}.}}}
\label{fig:Spin_Filtering}
\end{figure}

\begin{figure}
\caption{\it{The following two diagrams provide a schematic representation of the two physical processes, selective scattering out of the ring (left) and selective spin-flip (right), that contribute to polarization buildup by spin filtering in a storage ring.   Particles in the \lq spin up' state are represented by blue squares and particles in the \lq spin down' state are represented by yellow squares, while the grey box represents a polarized target.  In both cases the beam is initially unpolarized with equal numbers of particles in the \lq spin up' and \lq spin down' states.}}
\begin{minipage}{7cm}
\vspace*{1ex}
\includegraphics[width=7cm]{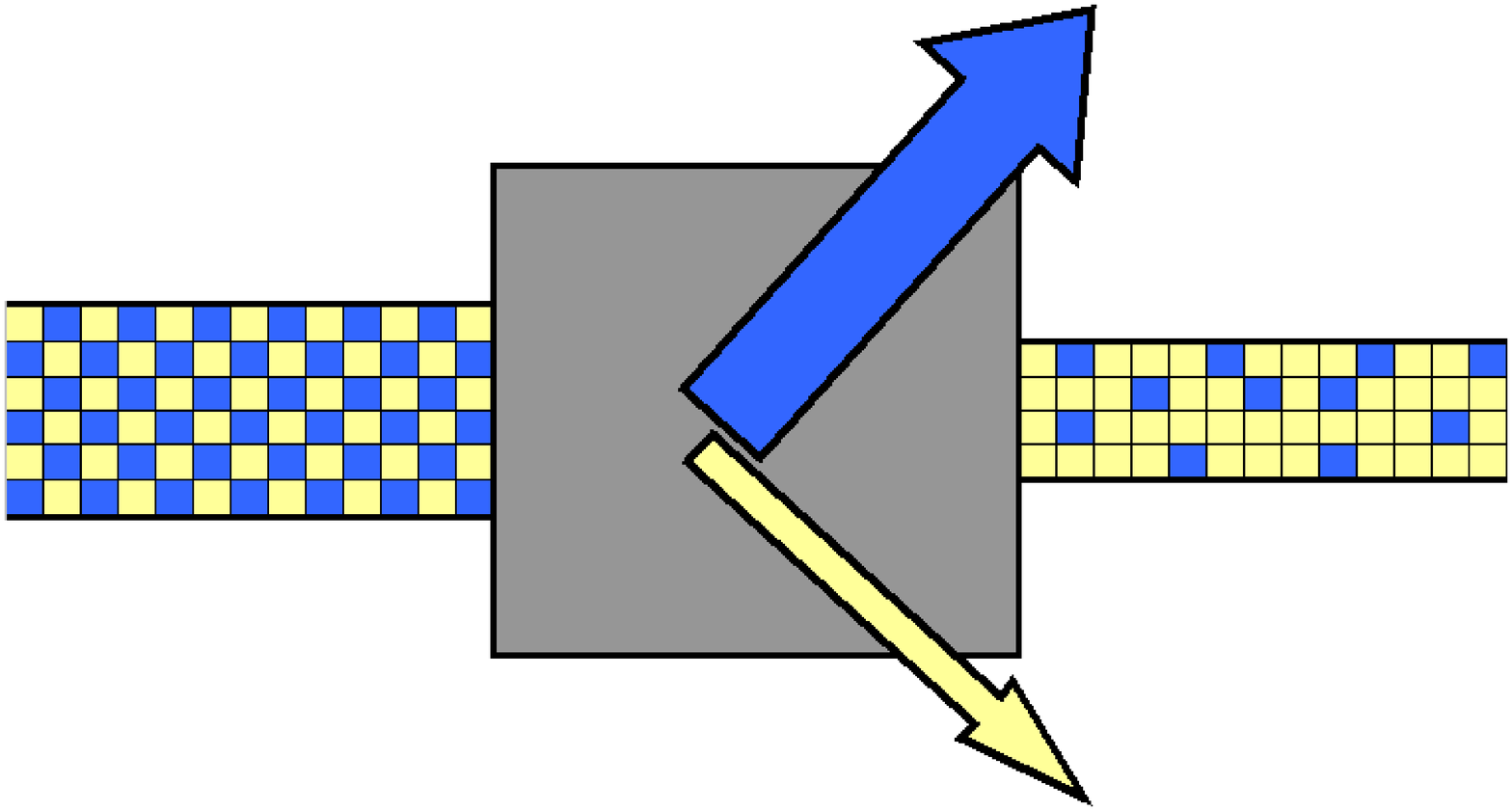}
\vspace*{1ex}
\end{minipage}
%\vspace{2ex}
\hfill
\begin{minipage}{7cm}
\vspace*{3.5ex}
\includegraphics[width=7cm]{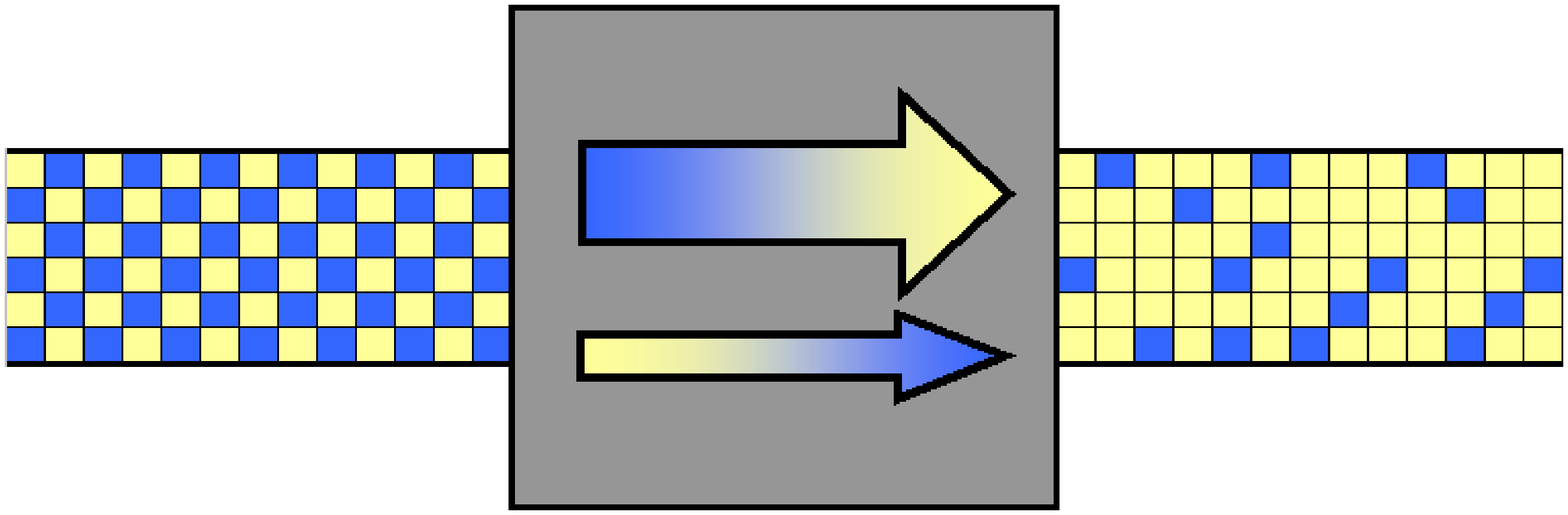}
\vspace*{1ex}
\end{minipage}
%\small{
{\it {\bf Selective scattering out of the ring (left):} When interacting with the polarized target at certain energies particles in the \lq spin up' state are scattered out of the beam at a higher rate than particles in the \lq spin down' state, hence the larger blue arrow than yellow arrow.  Thus one is left with a beam that has more particles in the \lq spin down' state, {\it i.e.}\ the beam is now polarized, represented by the excess of yellow squares in the final beam.  Note that since particles have been scattered out of the ring there are less particles in the beam after interaction than were in the beam initially, this is represented by the smaller final beam.  If the target was unpolarized particles in both spin states would be scattered out of the beam at equal rates, thus no polarization buildup would occur via this process.}\\[1ex]  {\it {\bf Selective spin-flip (right):} On interaction with the polarized target at certain energies, the \lq spin up' to \lq spin down' spin-flip cross-section is larger than the \lq spin down' to \lq spin up' spin-flip cross-section.  We represent this by different size arrows with colours fading from blue to yellow and from yellow to blue respectively.  Thus after interaction with the target the beam will have more particles in the \lq spin down' state than in the \lq spin up' state, {\it i.e.}\ the beam is now polarized, represented by the excess of yellow squares in the final beam.  Note that the beam intensity is the same after interaction with the polarized target in this process since particles are not lost from the beam, they are just flipped from one spin state to the other. If the target was unpolarized particles in both spin states would have their spins flipped at equal rates, thus no polarization buildup would occur via this process.}
%}
%\end{center}
\label{fig:Spin_Filtering_Diagrams}
\end{figure}

The spin filtering method of polarization buildup \cite{Csonka:1968,Rathmann:1993xf,Rathmann:2004pm}, described schematically in Figure~\ref{fig:Spin_Filtering}, consists of a circulating beam repeatedly interacting with a polarized internal target in a storage ring.  Originally proposed by P.~L.~Csonka in 1968 \cite{Csonka:1968}, it is based on the selective removal of particles from the beam, and selective spin-flip while remaining in the beam, due to spin-dependent scattering off a polarized target.
%It has been shown experimentally that, at certain energies, there is a difference between the total cross-sections for the scattering of particles with their spins aligned parallel or antiparallel to their direction of motion, {\it i.e.}\ $\Delta\,\sigma_L \,\equiv\, \sigma_\mathrm{tot}( \rightleftarrows ) \,-\, \sigma_\mathrm{tot}( \rightrightarrows ) \,\neq\, 0$, for both proton-proton and antiproton-proton scattering \cite{Grosnick:1996sy}.  
Many particles are scattered at small angles but remain in the beam after refocusing each revolution.  This introduces a characteristic laboratory frame acceptance angle $\theta_\mathrm{acc}$, scattering above which causes particles to be lost from the beam.  There is also a minimum laboratory frame scattering angle $\theta_\mathrm{min}$, corresponding to the Bohr radius of the atoms in the target, below which scattering is prevented by Coulomb screening \cite{Jackson:1999}, as described in section~\ref{subsec:Minimum_scattering_angle}.  The two physical processes that contribute to polarization buildup by spin filtering, as described in Figure~\ref{fig:Spin_Filtering_Diagrams}, are: 
\\[2ex]
\hspace*{7em}{\bf (a)} spin selective scattering out of the beam, and 
\\[2ex]
\hspace*{7em}{\bf (b)} selective spin-flip while remaining in the beam  
\\[2ex]

Thus particles in one spin state may be scattered out of the beam, or have their spin-flipped while remaining in the beam, at a higher rate than particles in the other spin state.  Hence over time one spin state is depleted more than the other leading to a beam polarization.  The beam will diverge slightly after many interactions with the target, but can be refocused by beam cooling, as explained in section~\ref{sec:Beam_Cooling}.  We prove later in the thesis that beam cooling does not depolarize a stored antiproton beam.

As the beam polarization increases the beam intensity decreases, when there is scattering out of the beam.  So one can obtain beam polarization at the expense of losing beam intensity.  Low beam intensity means low event rate, hence low statistics in a measurement, which is never desired.  This trade-off between beam polarization and beam intensity is characteristic of spin filtering and must be optimized to produce a sufficient beam polarization while maintaining reasonable beam intensity.

%\psfrag{theta}{{\Large $\theta$}}
\psfrag{Tacc}{{\Large $\theta_\mathrm{acc}$}}

%In the presence of polarized protons of magnetic number $m=1/2$ in the target, beam protons with $m=1/2$ are scattered less often than those with $m=-1/2$, which eventually causes the stored beam to acquire a polarization parallel to the proton spin of the hydrogen atoms during spin filtering \cite{Rathmann:2004pm}.  
%
%

An advantage for the spin filtering method is that polarized hydrogen and deuterium jet targets have already been developed for other projects.  Highly polarized high density gas jet targets have been used in the HERMES and COMPASS experiments.  The HERMES experiment has been decommissioned since the shutdown of the HERA accelerator complex in DESY, and the polarized gas target has been transfered to COSY in J\"ulich, Germany to be used in spin filtering studies.  It is likely that the HERMES polarized gas target will be used by the $\mathcal{P}\mathcal{A}\mathcal{X}$ Collaboration in a future spin filtering Antiproton Polarizer Ring at FAIR, GSI Darmstadt.  

There has been much debate amongst theorists as to what mechanisms are responsible for the polarization buildup in spin filtering.  Contributions come from the electromagnetic scattering of beam antiprotons off the electrons in the hydrogen target and from the electromagnetic and hadronic scattering of beam antiprotons off the protons in the hydrogen target.  Horowitz and Meyer, in 1994, were the first to highlight the importance to spin filtering of the electrons in the hydrogen target \cite{Horowitz:1994,Meyer:1994}.  They claim that electrons in a hydrogen target are not massive enough to deflect antiprotons beyond the acceptance angle of any storage ring, a fact which we demonstrate later in the thesis.  Thus scattering of the antiprotons off the electrons in a hydrogen target causes no beam losses and any polarization buildup must be due to spin-flip transitions \cite{MacKay:2006}.  In 2005 two groups from the Budker Institute for Nuclear Physics, Russia and the Institute for Nuclear Physics, J\"ulich, Germany claimed that such spin-flip effects are small thus spin filtering off polarized electrons in a hydrogen target will lead to a negligible rate of polarization buildup \cite{Milstein:2005bx,Nikolaev:2006gw,Dmitriev:2007ms}.  An experiment has been proposed to test this claim \cite{PAX:2006a}, by investigating the converse case of whether unpolarized electrons in a helium--4 target depolarize a stored polarized proton beam.  The helium--4 target is chosen because its nuclei is spin - 0, hence any polarization transfer must come from scattering off its electrons.  There are currently two schools of thought regarding spin filtering of antiprotons off a polarized hydrogen target, (1) proposal \cite{Rathmann:2004pm} building on the work of Horowitz and Meyer, which advocates using a hydrogen target with high electron polarization and low proton polarization; and (2) the Budker/J\"ulich proposal \cite{Milstein:2005bx,Nikolaev:2006gw} to use a hydrogen target with low electron polarization and high proton polarization.  As is often the case the matter must be resolved by an experiment \cite{PAX:2006a} to see which method is preferable.

There are many advantages of using a lepton target instead of an atomic target, the foremost of which is that antiprotons cannot be absorbed by a lepton target as they are in a atomic target due to annihilation with the protons in the atomic target.  This fact has led to two proposals for spin filtering off polarized lepton beams: one off a co-moving polarized positron beam \cite{Walcher:2007sj} by a group in Mainz, Germany and the other, presented in this thesis, off an opposing polarized electron beam \cite{O'Brien:2007hu}.  The momentum of an opposing electron beam causes antiprotons to be scattered beyond acceptance, hence allowing contributions from both of the physical process, selective scattering out of the beam and selective spin-flip, of spin filtering.  A thorough treatment of the dynamics of spin filtering has been presented recently by the present author \cite{O'Brien:2007sw,O'Brien:2007jz} and forms much of the later chapters of this thesis.

\pagebreak

Spin filtering is the only method to produce a polarized antiproton beam in a storage ring that has been successfully tested, by the FILTEX experiment in 1993 \cite{Rathmann:1993xf}, as described in section~\ref{sec:Verification_of_spin_filtering}.  As a result, much of this thesis is devoted to a theoretical understanding of the spin filtering process, under various scenarios.

\section{Verification of spin filtering} 
\label{sec:Verification_of_spin_filtering}

Polarization buildup by spin filtering has been proven to work in the FILTEX experiment at the Test Storage Ring (TSR) at the Max Planck Institute for Nuclear Physics in Heidelberg, Germany \cite{Rathmann:1993xf}.  We summarize their results below.  

\begin{figure}[!h]
\centering
\includegraphics[height=9cm,width=13cm]{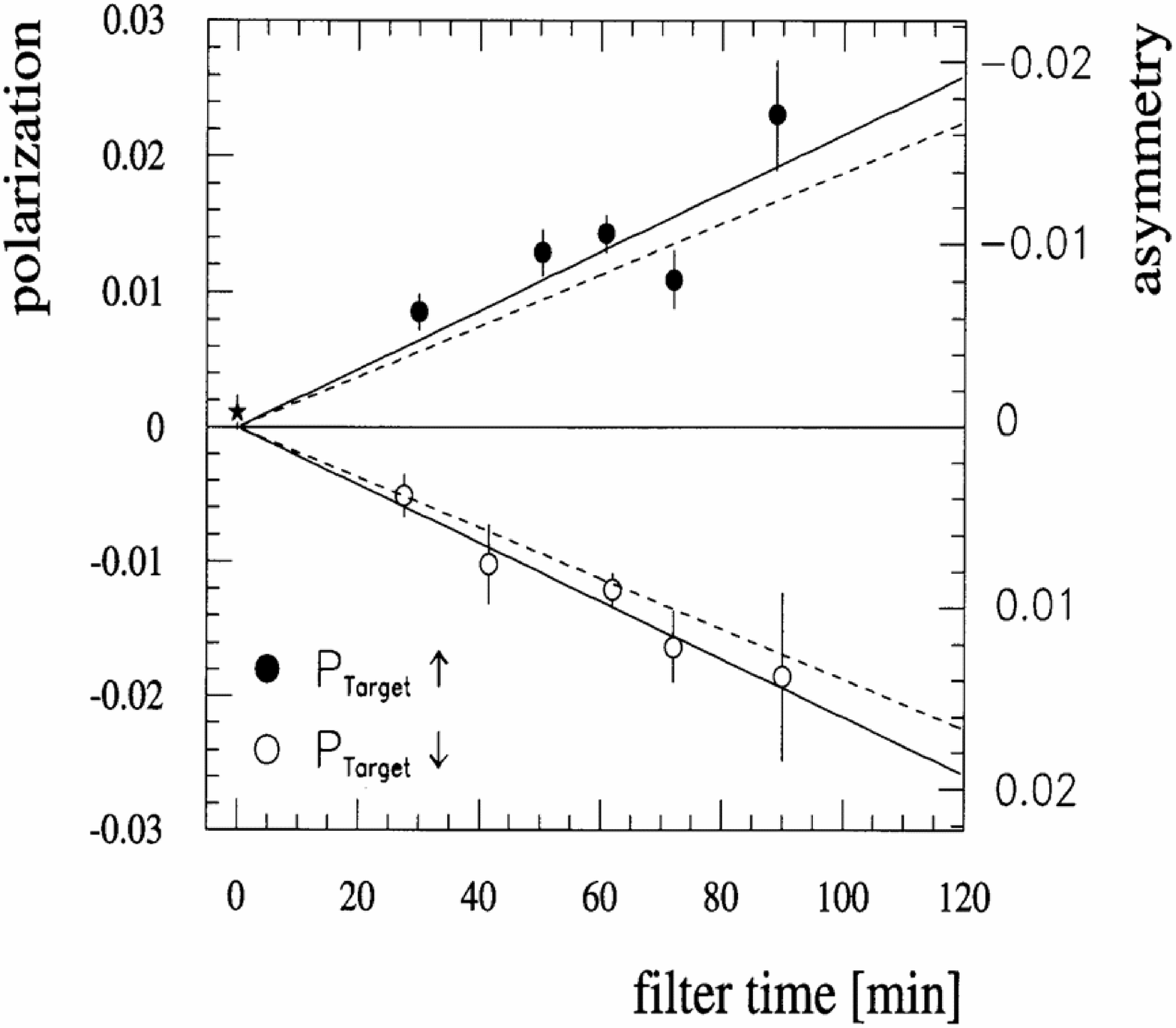}
\caption{\small{\it{The results of the FILTEX experiment, showing polarization buildup over time.  The solid lines show the best fit to the data with a rate of polarization buildup of $1.24 \times 10^{\,-2}\ \mbox{h}^{-1}$.  The dashed lines are based on the expected buildup rate from the model presented in Ref.~\cite{Rathmann:1993xf}, from where this plot has been reproduced with permission from the authors. 
%***$\tau_1$ is the time required to reach a 100\% polarized beam $100/1.24 \approx 80$.  Read through Rathmann's paper to find out what exactly the polarization lifetime means.***
}}}
\label{fig:FILTEX_Results}
\end{figure}
In the TSR a $23 \ \mbox{MeV}$ proton beam was stored and repeatedly made to interact with a polarized internal hydrogen gas target.  A polarized hydrogen target with atoms in the hyperfine state $|\uparrow_{p}\,\uparrow_{e}\,\rangle$ was used, {\it i.e.}\ where both the protons and electrons are polarized.  The target density was $6 \times 10^{\,13}$ polarized hydrogen atoms per $\mbox{cm}^2$, and the frequency of revolution was $1.177 \ \mbox{MHz}$ \cite{Rathmann:1993xf}.  The beam was left to orbit in the ring passing through the target each revolution for times between $30$ and $90$ minutes; then the polarization was measured.  The proton beam was initially unpolarized and over time it gained a small amount of polarization as shown in Figure~\ref{fig:FILTEX_Results}.

The polarization buildup rate of the proton beam at FILTEX was \cite{Rathmann:1993xf}\,:
\begin{equation}
\label{eq:FILTEX_result}
\frac{\mathrm{d}\,\mathcal{P}_\mathrm{\,beam}}{\mathrm{d}\,\tau} \ = \ 0.0124 \ \pm \ 0.0006 \ \ \ \mbox{per hour.}
\end{equation}
\noindent
After $90 \ \mbox{minutes}$ the polarization had increased to $1.86 \, \%$ and the beam intensity had decreased to $5 \, \%$ of its original value \cite{Rathmann:1993xf}.  But with a better configuration of the experiment, and a dedicated spin filtering polarizer ring, the rate of polarization buildup could be greatly increased.  The TSR ring had an acceptance angle measured to be $\theta_\mathrm{acc} = 4.4\,\pm \,0.5 \ \mbox{mrad}$, which we could optimize for our needs.

The beam lifetime $\tau_*$, which we discuss later in the thesis, is the time taken for the number of particles in the beam to decrease by a factor of $e = 2.78$.  The beam lifetime in the TSR during the FILTEX experiment, with the polarized internal target in the ring, was $30 \ \mbox{minutes}$.  We show in section~\ref{subsec:Beam_lifetime_and_figure_of_merit} that the polarization achieved after two beam lifetimes is an important measure of a spin filtering scheme.  At FILTEX this value was measured to be $\mathcal{P}_\mathrm{beam}\left(\,2\,\tau_*\,\right) \,=\, 0.0124$.

This was just a feasibility test for spin filtering, and while it verifies that the method works, the polarization buildup rate was small.  In order to maximize the effect of spin filtering a dedicated spin filtering polarizing ring would need to be built.  The $\mathcal{P}\mathcal{A}\mathcal{X}$ Collaboration has recently proposed the construction of such a ring called the Antiproton Polarizer Ring (APR) inside the HESR at FAIR in GSI Darmstadt, Germany.

\pagebreak

\chapter{Generic helicity amplitudes and spin observables}
\label{ch:Generic_helicity_amplitudes_and_spin_observables}

\vspace*{5ex}
\begin{minipage}{6cm}
\end{minipage}
\hfill
\begin{minipage}{10cm}
\begin{quote}
\emph{\lq\lq The most incomprehensible thing about the world is that it is comprehensible.\rq\rq}
\flushright{Albert Einstein}
\end{quote}
\end{minipage}
\vspace{8ex}

In this chapter all electromagnetic helicity amplitudes and spin observables, accounting for polarization effects in spin 1/2 - spin 1/2 elastic scattering are calculated.  Many of these results will be utilized in later chapters when providing a mathematical description of spin filtering, although their use is certainly not limited to this.  We begin in section~\ref{sec:Spin_1/2_electromagnetic_currents} by introducing the spin 1/2 electromagnetic currents, both for point particles and particles with internal structure determined by electromagnetic form factors.  A generic equation is derived in section~\ref{sec:Generic_elastic_spin_1/2_spin_1/2_calculation} that can be used to calculate all polarization phenomena in elastic spin 1/2 - spin 1/2 electromagnetic scattering to first order in QED.  The spin-averaged differential cross-section for spin 1/2 - spin 1/2 scattering is presented in a new compact invariant form in section \ref{sec:Spin_averaged_cross-section}.  We then present results for all electromagnetic helicity amplitudes for elastic spin 1/2 - spin 1/2 scattering in section~\ref{sec:Helicity_amplitudes}, and for electromagnetic spin observables for elastic spin 1/2 - spin 1/2 scattering in section~\ref{sec:Spin_observables}.

\section{Spin 1/2 electromagnetic currents}
\label{sec:Spin_1/2_electromagnetic_currents}

In this section the spin 1/2 electromagnetic currents of both point-like particles and non-point-like particles, with internal structure defined by electromagnetic form factors, are introduced.  We investigate $2 \ \mbox{particle}  \rightarrow 2 \ \mbox{particle}$ elastic scattering processes in the space-like region, with the mass and the momentum and spin 4-vectors of each particle labeled as:
\begin{equation*}
A\,\left(\,M,\,p_1,\,S_1\,\right) \ + \ B\,\left(\,m,\,p_2,\,S_2\,\right) \ \longrightarrow \ A\,\left(\,M,\,p_3,\,S_3\,\right) \ + \ B\,\left(\,m,\,p_4,\,S_4\,\right)\,,
\end{equation*}
with the particles above being the beam (1), target (2), scattered (3) and recoil (4) particles respectively.  The four momentum transfer is defined as 
\begin{equation}
q \ =\ p_3 \,-\,p_1 \ =\  p_2\,-\,p_4\,.
\end{equation}
For $2 \ \mbox{particle}  \rightarrow 2 \ \mbox{particle}$ elastic scattering the spin-averaged differential cross section is related to the helicity amplitudes $\mathcal{M}(\lambda_3\,\lambda_4;\lambda_1\,\lambda_2)$
 by
\begin{equation}
\label{eq:Cross-section_to_helicity_amplitudes}
s\,\frac{\mathrm{d}\,\sigma}{\mathrm{d}\,\Omega} \ = \ \frac{1}{\left(\,8\,\pi\,\right)^{2}}\, \sum_{\lambda_1\,\lambda_2\,\lambda_3\,\lambda_4}  \frac{1}{\left(\,2\,s_A \,+\,1\,\right)\ \left(\,2\,s_B\,+\,1\,\right)}\ |\,\mathcal{M}(\,\lambda_3,\,\lambda_4\,;\,\lambda_1,\,\lambda_2\,)\,|^{\,2} \,,
\end{equation}
where $\lambda_1,\lambda_2$ and $\lambda_3,\lambda_4$ are the helicities of the initial and final particles respectively, $s_A$ and $s_B$ are the spins of the two particles in the elastic process, and the $s$ and $t$ are Mandelstam variables \cite{Mandelstam:1958xc} defined in Appendix~\ref{Appendix:Relations_between_Mandelstam_variables}.  We label the mass of particle $A$, taken to be an antiproton, as $M$ and the mass of particle $B$, taken to be an electron or a proton, as $m$.  Define electromagnetic form factors $F_1\!\left(q^2\right)$ and $F_2\!\left(q^2\right)$, with normalization $F_1(0)=1$ and $F_2(0)=\kappa_p = \mu_\mathrm{p} - 1$, the anomalous magnetic moment of the proton, where $q^2 = t$ in the $t$-channel case that we are solely interested in.  Form factors are empirical quantities, obtained from experiment, which describe the fact that protons are not point-like particles and have an internal structure.  They include all effects of the strong nuclear interaction inside the proton, hence are very difficult to calculate theoretically.  The Sachs electric $G_E(t) = F_1(t) + F_2(t) \ t/\left(4\,M^{\,2}\right)$ and magnetic $G_M(t) = F_1(t) + F_2(t)$ form factors are used.  In the $t$-channel, also known as the space-like region, the form factors are real functions of $t$.  Although not treated in this thesis it is worth mentioning that this is not true in the $s$-channel, also known as the time-like region, where the form factors are complex functions of $s$.  For a treatment of polarization observables in the time-like region see Ref.~\cite{Buttimore:2006mq}.
\psfrag{jmu}{$j^{\mu}$}
\psfrag{p1}{$p_1$}
\psfrag{p2}{$p_2$}
\psfrag{p3}{$p_3$}
\psfrag{p4}{$p_4$}
\psfrag{Jmu}{$J_p^{\mu}$}
\psfrag{Jamu}{$J_{\bar p}^{\mu}$}
\psfrag{Gmu}{$\gamma^{\,\mu}$}
\begin{figure}
\centering
\includegraphics{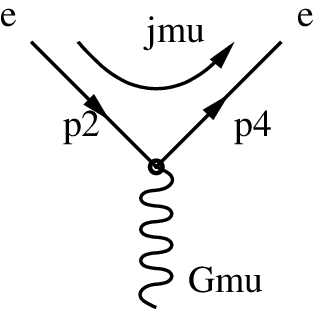}
\hspace*{3em}
\includegraphics{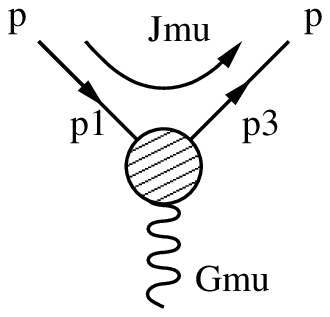}
\hspace*{3em}
\includegraphics{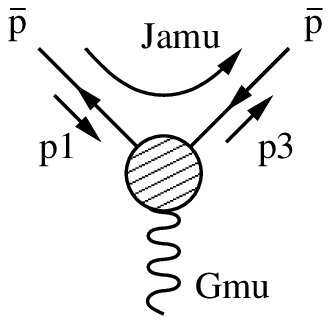}
\caption{\small{\it{The electron, proton and antiproton electromagnetic currents; $j^{\mu}$, $J_p^{\mu}$ and $J_{\bar p}^{\mu}$ respectively.  Time increases from left to right.  These Feynman diagrams are converted into the mathematical expressions for the currents presented in this section using the Feynman rules of Appendix~\ref{Appendix:Feynman_rules_for_QED}.  The shaded circle in the proton and antiproton currents describe that these are not point particles, and have an internal structure described by form factors.}}
}
\label{fig:Electromagnetic_currents}
\end{figure}
From Figure~\ref{fig:Electromagnetic_currents}, and using the Feynman rules for QED presented in Appendix~\ref{Appendix:Feynman_rules_for_QED}, the electron current for particle $B$ is
\begin{equation}
\label{eq:Electron_current}
j_{B}^{\,\mu}\ = \ -\,i\,e\,\bar{u}\left(\,p_4,\,\lambda_4\,\right)\,\gamma^{\,\mu}\,u\left(\,p_2,\,\lambda_2\,\right)\,,
\end{equation}
where $e$ is the electron charge and $u\left(\,p_2,\,\lambda_2\,\right)$ and $\bar{u}\left(\,p_4,\,\lambda_4\,\right)$ are the spinors of the incoming and outgoing electron respectively.

Generalizing this to a non-point-like particle, such as the proton, with a finite extent defined by the Pauli and Dirac electromagnetic form factors $F_1\!\left(q^2\right)$ and $F_2\!\left(q^2\right)$, one obtains the most general (anti)proton\footnote{We are only interested in $t$-channel elastic scattering, {\it i.e.}\ not $s$-channel antiproton-proton annihilation which is suppressed in comparison to the $t$-channel amplitude in the low $|\,t\,|$ region of interest in a storage ring, as explained in Figure~\ref{fig:Bhabha_Feynman_diagram}.  In this case an antiproton can be treated as a negatively charged proton, with internal electromagnetic structure described by the same form factors as the proton.  Hence, as is customary in the phenomenology literature \cite{Horowitz:1994,Block:1996jd}, $u$ and $\bar{u}$ spinors can be used for the antiproton current in the $t$-channel instead of the $v$ and $\bar{v}$ anti-spinors required in the treatment of annihilation.  The treatment presented here is in exact agreement with a treatment where the proton current involves spinors and the antiproton current involves anti-spinors.  The minus signs introduced into eq.~(\ref{eq:Generic_calculation_with_epsilons}) because of the anti-spinor completeness relations of eqs.~(\ref{eq:Completeness_relations1} and \ref{eq:Completeness_relations3}) only contribute to terms that vanish when the traces are evaluated using the trace theorems of eq.~(\ref{eq:Gamma_matrix_trace_theorems}).  Hence all cross-sections and spin observables for $t$-channel elastic antiproton-antiproton, antiproton-proton and proton-proton scattering are equal to first order in QED.  The anti-spinor formalism is less general then the one presented here as one is forced, due to the Gordon decomposition identity for anti-spinors derived in Appendix~\ref{Appendix:Derivation_of_the_Gordon_decomposition_identities}, to specify that the particle and antiparticle have opposite anomalous magnetic moments.  Hence the results could not be applied to antiproton-neutron scattering for example, as they can be in the present formalism.} electromagnetic current 
%, that is consistent with Time-Reversal and Parity Invariance and current conservation $\partial_{\mu}\,J^{\,\mu} = 0$,
 %\footnote{We introduce the form factor $\hat{F}_2$ briefly in this section, to keep the dependence on the particles anomalous magnetic moment $\kappa$ explicit.  The anomalous magnetic moment will then be factored into the definition of the form factor $F_2$, which will be used in the remainder of the thesis.}
for particle $A$:
\begin{equation}
\label{eq:Proton_current_before_Gordon_decomposition}
\hspace*{-0.4em}
 J_{\!A}^{\,\mu} \, = \, \pm\,i\,e_{p} \ \bar u(p_3,\lambda_3) \left[\,F_1\!\left(q^2\right)  \, \gamma^{\,\mu}  \,+\,   \frac{F_2\!\left(q^2\right)}{2\,M}\,i\,\sigma^{\,\mu\,\nu} \left(\,p_3 \,-\,p_1\,\right)_\nu \,\right]u(p_{\,1},\lambda_1) \,,
\end{equation}
where $e_p = -\, e_{\bar p} = -\,e$ is the charge on the proton and consequently the upper sign is for the antiproton current and the lower sign is for the proton current, and where
\begin{equation}
\label{eq:Gamma_Matrix_anticommutation_relation}
\sigma^{\,\mu\,\nu} \ \equiv \ \frac{i}{2}\left[\,\gamma^\mu,\gamma^\nu \,\right]  \ = \ \frac{i}{2}\left(\,\gamma^\mu \,\gamma^\nu \,-\, \gamma^\nu \gamma^\mu \,\right)\,.
\end{equation}
Using the Gordon decomposition identity \cite{Gordon:1928,Bjorken:1964}\,:
\begin{eqnarray}
\label{eq:Gordon_decomposition_identity}
\bar{u}(p')\,\gamma^{\,\mu}\,u(p) & = & \bar{u}(p')\,\left[\frac{\left(\,p \,+\, p' \,\right)^{\,\mu}}{2\,M} \ +\  \frac{i\,\sigma^{\,\mu\,\nu}\left(\,p'\,-\, p\,\right)_\nu}{2\,M}\,\right]u(p)\,,
%\\[2ex]
%\bar{v}(p')\,\gamma^{\,\mu}\,v(p) & = & \bar{v}(p')\,\left[\,-\,\frac{\left(\,p \,+\, p' \,\right)^{\,\mu}}{2\,M} \ -\  \frac{i\,\sigma^{\,\mu\,\nu}\left(\,p'\,-\, p\,\right)_\nu}{2\,M}\,\right]v(p)  \,,
\end{eqnarray}
which is derived in Appendix~\ref{Appendix:Derivation_of_the_Gordon_decomposition_identities}, the (anti)proton electromagnetic current can be written as
\begin{eqnarray}
\label{eq:Proton_current_after_Gordon_decomposition}
 J_{\!A}^{\,\mu} & = & \pm \,i\,e_{p} \ \bar u(p_3,\lambda_3) \left( G_M \, \gamma^{\,\mu} \ -\  F_2\,\, \frac{p_1^{\,\mu} \,+\, p_3^{\,\mu}}{2\,M} \,\right)u(p_{\,1},\lambda_1) \,,
%\\[2ex]
%\label{eq:Antiproton_current_after_Gordon_decomposition}
%J_{\!A\,\bar p}^{\,\mu} & = & -\,i\,e_{\bar{p}} \ \bar v(p_3,\lambda_3) \left( G_M \, \gamma^{\,\mu} \ +\  F_2\,\, \frac{p_{\,1}^{\,\mu} \,+\, p_3^{\,\mu}}{2\,M} \,\right)v(p_{\,1},\lambda_1) \,,
\end{eqnarray}
which has a simpler gamma matrix structure than eq.~(\ref{eq:Proton_current_before_Gordon_decomposition}), hence will allow for easier computations.

It is important to note that the structureless limit (point-like particles) is obtained when $F_1\!\left(q^2\right) \,=\,1$ and\footnote{More correctly for a point-like electron $F_2\!\left(0\right) \,=\, \alpha / (2\,\pi) \,+\, \mathcal{O}(\alpha^2)$ \cite{Weinberg:1995}.  This is a result of the anomalous magnetic moment of the electron and was first calculated by Julian Schwinger shortly after the famous Shelter Island Conference in 1947 \cite{Schwinger:1948iu}.  The anomalous magnetic moment of the electron has now been calculated to order $\alpha^4$ in QED and agrees with the experimentally measured value to 10 decimal places \cite{Maggiore:2005}, making QED one of the most accurately verified theories in the history of physics.} $F_2\!\left(q^2\right) \,=\,0$, hence $G_M\!\left(q^2\right) \,=\,G_E\!\left(q^2\right)\,=\,1$.  Applying this condition to the proton current in eq.~(\ref{eq:Proton_current_after_Gordon_decomposition}) one obtains the electron current presented in eq.~(\ref{eq:Electron_current}).  Because we are interested in both antiproton-proton and antiproton-electron scattering, we shall calculate the generic case of the elastic scattering of two structured spin 1/2 particles.  One can then take the one particle point-like limit to account for antiproton-electron scattering, or the two particle point-like limit to account for positron-electron scattering.  Hence we generalize the electron current to account for non-point-like particles by using
\begin{equation}
\label{eq:Structured_electron_current_after_Gordon_decomposition}
 J_{B}^{\,\mu} \ = \ -\,i\,e \ \bar u(p_4,\lambda_4) \left( g_M \, \gamma^{\,\mu} \ -\  f_2\,\, \frac{p_{\,2}^{\,\mu} \,+\, p_4^{\,\mu}}{2\,m} \,\right)u(p_{\,2},\lambda_2) \,,
\end{equation}
where we label the Dirac, Pauli and Sachs electromagnetic form factors of particle $B$ by lowercase $f_1(t)$, $f_2(t)$, $g_M(t) = f_1(t)+ f_2(t)$ and $g_E(t) = f_1(t) + f_2(t)\,t\,/\,(4\,m^2)$ respectively, with the usual normalizations $f_1(0)=1$ and $f_2(0)=\kappa = \mu -1$ therefore $g_M(0)= \mu$ and $g_E(0) =1$.  

The above electromagnetic currents, and hence all results derived using them, can be applied to electrically neutral particles which nevertheless have an anomalous magnetic moment, such as the neutron, by encompassing the charge $e$ into the definitions of the form factors and using the new normalizations $F_1^{\,n}(0)\,=\,0$ and $F_2^{\,n}(0)\,=\, \kappa_n \,=\, -1.913$ in units of nuclear magnetons.

\section{Generic elastic spin 1/2 - spin 1/2 calculation}
\label{sec:Generic_elastic_spin_1/2_spin_1/2_calculation}

The helicity amplitudes can be related to the currents, to first order in QED, by the relation
\begin{equation}
\label{eq:Amplitude_in_terms_of_currents}
i \,\mathcal{M}(\lambda_3\,\lambda_4\,;\lambda_1\,\lambda_2) \ = \ J_{\!A}^{\,\mu} \left(\,M,\,\lambda_3,\,\lambda_1\,\right)\, \left(\,\frac{-\,i\,\eta_{\mu \nu}}{q^{\,2}}\,\right) J_{B}^{\,\nu} \left(\,m,\,\lambda_4,\,\lambda_2\,\right) \,.
\end{equation}
%
%where the $i\,\epsilon$ term does not contribute to first order in QED.
Therefore the generic electromagnetic amplitude for a structured spin 1/2 particle of mass $M$ scattering elastically off a structured spin 1/2 particle of mass $m$, via single $t$-channel photon exchange, is
\begin{eqnarray}
\label{eq:Antiproton_proton_Amplitude}
i \,\mathcal{M}(\lambda_3\,\lambda_4\,;\lambda_1\,\lambda_2) & = & \pm \,(-\,i)^3\,e^{\,2} \ \bar u(p_3,\lambda_3) \left( G_M \, \gamma^{\,\mu} \ -\ \frac{F_2}{2\,M}\,R^{\,\mu} \,\right)u(p_{\,1},\lambda_1) \\[2ex]
 & & \times  \left(\,\frac{\eta_{\mu \nu}}{q^{\,2}}\,\right) \bar u(p_4,\lambda_4) \left( g_M \, \gamma^\nu \ -\ \frac{f_2}{2\,m}\,r^\nu \,\right)u(p_{\,2},\lambda_2) \,, \nonumber \\[3ex]
 & & \hspace*{-10.5em} \ =  \frac{\pm\,i\,e^{\,2}}{q^{\,2}}\,\underbrace{\bar u(p_3,\lambda_3) \!\left(\!G_M \, \gamma^{\,\mu} - \frac{F_2}{2\,M}\,R^{\,\mu} \!\right)\!u(p_{\,1},\lambda_1)}_{\displaystyle{A}} \,\underbrace{\bar u(p_4,\lambda_4)\! \left(\!g_M \, \gamma_\mu  - \frac{f_2}{2\,m}\,r_\mu \!\right)\!u(p_{\,2},\lambda_2)}_{\displaystyle{B}} \nonumber
\end{eqnarray}
where the upper sign is for $p\,p$ or $\bar{p}\,\bar{p}$ scattering and the lower sign is for $\bar{p}\,p$ scattering,  and where we have defined 
\begin{equation}
\label{eq:Definition_of_R_and_r}
R^{\,\mu} \ \equiv \ p_1^{\,\mu} \ +\  p_3^{\,\mu}
\hspace*{2em} \mbox{and} \hspace*{2em}
r^\nu \ \equiv \ p_2^{\,\nu}  \ +\  p_4^{\,\nu} \,,
\end{equation} 
%
%and the $i\,\epsilon$ term only contributes to higher order loop amplitudes which we are not calculating here.
%
and used the fact that $e_{\bar{p}} = - e_p = e$.  Note that in the $t$-channel the hermiticity of the electromagnetic currents implies that the form factors are real functions of $q^2 = t$ ({\it i.e.}\ $F_1 = F_1^{\,*}$, $f_1 = f_1^{\,*}$, $F_2 = F_2^{\,*}$, $f_2 = f_2^{\,*}$ and hence $G_M = G_M^{\,*}$ and $g_M = g_M^{\,*}$) \cite{Weinberg:1995}.  
\psfrag{jmu}{$j^{\mu}$}
\psfrag{pl1}{$p_1,\lambda_1$}
\psfrag{pl2}{$p_2,\lambda_2$}
\psfrag{pl3}{$p_3,\lambda_3$}
\psfrag{pl4}{$p_4,\lambda_4$}
%\psfrag{Jmu}{$J_p^{\mu}$}
%\psfrag{Jamu}{$J_{\bar p}^{\mu}$}
\psfrag{Gmu}{$\gamma^{\,\mu}$}
\begin{figure}
\centering
\includegraphics{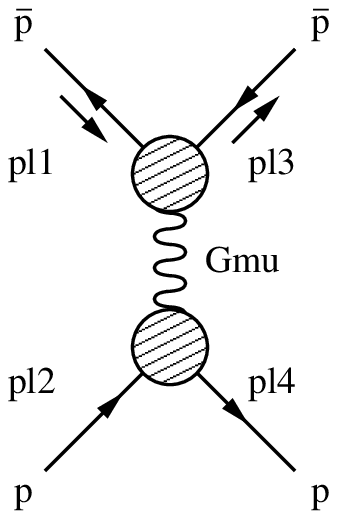}
\hspace*{4em}
\includegraphics{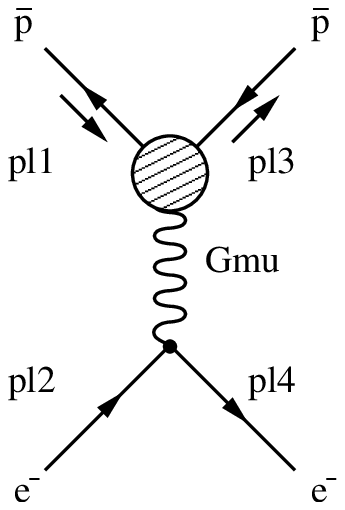}
\hspace*{4em}
\includegraphics{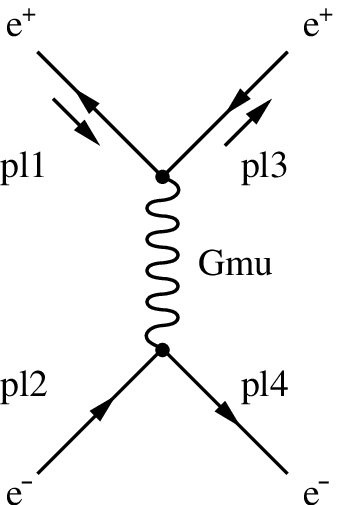}
\caption{\small{\it{The Feynman diagrams for single photon exchange in antiproton-proton, antiproton-electron and positron-electron scattering respectively.  As elsewhere in the thesis the time axis increases from left to right.  The antiproton-proton case is generic and encompasses the other two cases.  In this section we calculate all spin-dependent phenomena for electromagnetic spin 1/2 - spin 1/2 elastic scattering, of particles with internal structure described by form factors, to first order in QED.}}
}
\label{fig:Generic_Feynman_Diagrams}
\end{figure}
Now one must obtain
\begin{equation}
\label{eq:Amplitude_squared}
|\mathcal{M}|^{\,2} \ = \ \mathcal{M}\,\mathcal{M}^{\,*} \ = \ \frac{e^{\,4}}{q^{\,4}} \,\left(A \,B\right)\,\left(A\,B\right)^{\,*} \ = \ \frac{e^{\,4}}{q^{\,4}} \ A\,B\,B^{\,*} A^{\,*} \,,
\end{equation}
where, using relations in Appendix~\ref{Appendix:Dirac_algebra}, the complex conjugates of $A$ and $B$, defined in eq.~(\ref{eq:Antiproton_proton_Amplitude}), are
\begin{eqnarray}
\label{eq:Conjugate_currents}
A^{\,*} & = & \bar u(p_{\,1},\lambda_1) \left( G_M \, \gamma^{\,\nu} \ -\  \frac{F_2}{2\,M}\,R^{\,\nu} \right) u(p_3,\lambda_3)\,,\\[2ex]
B^{\,*} & = & \bar u(p_{\,2},\lambda_2)\left( g_M \, \gamma_\nu  \ -\  \frac{f_2}{2\,m}\,r_\nu  \right) u(p_4,\lambda_4) \,.
\end{eqnarray}
Substituting the above into eq.~(\ref{eq:Amplitude_squared}), requires using the completeness relations:
\begin{eqnarray}
\label{eq:Completeness_relations4}
u(p_i,\lambda_i)\, \bar{u}(p_i,\lambda_i) & = & \frac{1}{2}\left(\,\slashed{p}_i \,+\, m_i \,\right)\left(\,1\,+\,\gamma_5\,\slashed{S}_i\,\right) \,,
%\\[2ex]
%u(p_4, \lambda_4)\, \bar{u}(p_4,\lambda_4) & = & \frac{1}{2}\left(\,\slashed{p}_4 \,+\, m \,\right)\left(\,1\,+\,\gamma_5\,\slashed{S}_4\,\right) \nonumber\,,\\[2ex]
%u(p_1, \lambda_1)\, \bar{u}(p_1, \lambda_1) & = & \frac{1}{2}\left(\,\slashed{p}_1 \,+\, M \,\right)\left(\,1\,+\,\gamma_5\,\slashed{S}_1\,\right) \,,\\[2ex]
%u(p_3, \lambda_3)\, \bar{u}(p_3, \lambda_3) & = & \frac{1}{2}\left(\,\slashed{p}_3 \,+\, M \,\right)\left(\,1\,+\,\gamma_5\,\slashed{S}_3\,\right) \nonumber\,,
%
\end{eqnarray}
where $\lambda_i$ and $S_i$ are the helicity and spin four vector of the particle with momentum $p_i$ where $i \in \left\{1,2,3,4\right\}$, normalized such that $S_i^{\,\mu} \,{S_i}_\mu \,=\,-\,1$ and constrained by the orthogonality condition $p_i^{\,\mu} \,{S_i}_\mu \,=\,0$.  The mass of the particle with momentum $p_i$ is denoted by $m_i$ where we have that $m_1 = m_3 = M$ and $m_2 = m_4 = m$.

The result is a generic equation for all polarization phenomena in elastic spin 1/2 - spin 1/2 electromagnetic scattering to first order in QED:
\begin{eqnarray}
\label{eq:Generic_calculation}
 & &  16\ \displaystyle{\left(\ \frac{q}{e}\ \right)^4}\, |\,\mathcal{M}\,|^{\,2} = \\[1ex]
& & \hspace*{-1.5em}\mbox{Tr}\left[\left(\slashed{p}_4 + m \right) \!\left(1
  +
  \gamma_5 \, \slashed{S}_4 \right)\! \left(g_M \gamma^\nu - \displaystyle{\frac{f_2\, r^\nu}{2\,m}} \right)\! \left( \slashed{p}_2
  +
  m\right)\! \left(1+\gamma_5 \,  \slashed{S}_2 \right)\! \left(g_M \gamma^\mu - \displaystyle{\frac{f_2\, r^\mu}{2\,m}} \right)\right] \times \nonumber \\[1ex] 
 & &\hspace*{-1.6em}
  \mbox{Tr}\left[\left(\slashed{p}_1 \!+ M \right)\! \left(1
  +
  \gamma_5 \,\slashed{S}_1 \right)\! \left(\!G_M \gamma_\mu
  - \displaystyle{\frac{F_2\,R_\mu}{2\,M}} \right) \!\left( \slashed{p}_3 \!+ M \right)\left(1 + \gamma_5 \,  \slashed{S}_3\right) \!\left(\!G_M \gamma_\nu - \displaystyle{\frac{F_2\,R_\nu}{2\,M}} \right)
  \right] \nonumber
\end{eqnarray}
%
%The first trace above corresponds to a structured particle (such as a proton), and the second trace corresponds to a structured antiparticle (such as an antiproton).  
%Note that one can transform between a particle trace and an antiparticle trace, or vice versa, by setting $M \rightarrow -\,M$. 

This generic equation can thus be used to calculate all helicity amplitudes and spin observables by substituting specific values for the spin ($S_i$) and momenta ($p_i$) four vectors, and can describe equal particle scattering in the case $f_1 \rightarrow F_1$, $f_2 \rightarrow F_2$ and $m \rightarrow M$.  This also applies to antiproton-electron scattering by setting  one particle to be point-like using $f_1 \rightarrow 1$ and $f_2 \rightarrow 0$ and hence $g_M \rightarrow 1$, in which case the first trace simplifies to the familiar electron trace:
\begin{equation}
\label{eq:Electron_trace_complete}
\mbox{Tr}\left[\left(\slashed{p}_4 \, +\,  m \right)\left(1
  \, +\, 
  \gamma_5 \, \slashed{S}_4 \right) \gamma^\nu \left(\, \slashed{p}_2
  \, +\, 
  m\right) \left(1 \, +\, \gamma_5 \,  \slashed{S}_2 \right) \gamma^\mu \right]\,,
\end{equation}
which when polarization effects are averaged over gives the familiar spin-averaged electron trace
\begin{equation}
\label{eq:Electron_trace}
\mbox{Tr}\left[\left(\,\slashed{p}_2 \,+\, m\,\right)\,\gamma_\mu\,\left(\,\slashed{p}_4 \,+\, m\,\right)\,\gamma_\nu\,\right] \ = \ 4\,\left(\,{p_2}_\mu\, {p_4}_\nu \,+\, {p_2}_\nu\, {p_4}_\mu \,+\, \frac{t}{2}\,\eta^{\mu\nu}\,\right)\,,
\end{equation}
which has been evaluated using the trace theorems presented in Appendix~\ref{Appendix:Dirac_algebra}. 

Equation~(\ref{eq:Generic_calculation}) can be generalized to directly evaluate the squares of all electromagnetic helicity amplitudes.  We introduce the constants $\epsilon_i$ multiplying each $S_i$, where $i \in \{1,2,3,4\}$.
\begin{eqnarray}
\label{eq:Generic_calculation_with_epsilons}
& &  16\ \displaystyle{\left(\ \frac{q}{e}\ \right)^4\, |\,\mathcal{M}\,|^{\,2}} = \\[1ex]
& & \hspace*{-1.65em}\mbox{Tr}\!\left[\!\left( \slashed{p}_4 \!+ m \right)\!\! \left(1
  +
  \epsilon_4\,\gamma_5 \, \slashed{S}_4 \right)\!\! \left(\!g_M \gamma^\nu - \displaystyle{\frac{f_2\, r^\nu}{2\,m}} \right)\!\! \left( \slashed{p}_2\!
  +
  m\right)\!\! \left(1+\epsilon_2\,\gamma_5 \, \slashed{S}_2 \right)\!\! \left(\!g_M \gamma^\mu -  \displaystyle{\frac{f_2\, r^\mu}{2\,m}} \right)\!\right] \!\times \nonumber \\[1ex] 
& &  \hspace*{-1.65em}
  \mbox{Tr}\!\left[\!\left( \slashed{p}_1\! +\! M \right)\!\!\left(1
  +
  \epsilon_1\,\gamma_5 \, \slashed{S}_1 \right)\!\!\left(\!G_M \gamma_\mu
  -  \displaystyle{\frac{F_2\,R_\mu}{2\,M}} \right) \!\!\left( \slashed{p}_3 \!+\! M \right) \!\!\left(1 +
  \epsilon_3\,\gamma_5 \, \slashed{S}_3\right)\!\!\left(\!G_M \gamma_\nu -  \displaystyle{\frac{F_2\,R_\nu}{2\,M}} \right)\!\right] \nonumber
\end{eqnarray}
Later we will set $\epsilon_i = \pm 1$ to account for different helicity states.  Equation~(\ref{eq:Generic_calculation_with_epsilons}) is used to derive the helicity amplitudes and spin observables throughout the remainder of this chapter.
 
The spin four vectors are now normalized so that all $\epsilon_i= +1$ corresponds to the helicity amplitude $\phi_1 = \mathcal{M}(\,+,+\,;\,+,+\,)$, and the $\pm 1$ in the helicity amplitudes now relate to the signs of the $\epsilon_i$.

The momenta and longitudinal, transverse\footnote{Transverse to the direction of motion but still in the scattering plane, sometimes called {\it Sideways} and denoted $S$.} and normal spin four vectors in the Centre-of-Mass frame are presented in Table~\ref{table:CM_frame_momenta_and_spin_4vectors}.  Substituting combinations of these four vectors into eq.~(\ref{eq:Generic_calculation_with_epsilons}), and computing the traces, will provide expressions for all spin-dependent cross-sections for elastic spin 1/2 - spin 1/2 scattering to first order in QED.  The traces were computed using the computer algebraic program {\tt Mathematica} and its add on package {\tt Tracer} \cite{Jamin:1991dp}.  The {\tt Mathematica} code for a typical calculation is presented in Appendix~\ref{Appendix:Sample_Mathematica_code}.  

\begin{table}[!h]
\begin{center}
\begin{tabular}{|ccc|ccc|} \hline
\multicolumn{6}{|c|}{} \\*[-2ex]
\multicolumn{6}{|c|}{Centre-of-Mass Momenta vectors} \\[0.5ex] \hline
& & & & &\\[-1.8ex]
$ p_1 $&$  = $&$ \left(\,E_A,\,0,\,0,\,k\,\right) $& $ p_3 $&$  = $&$ \left(\,E_A,\,k\,\sin\theta,\,0,\,k\,\cos\theta\,\right)$\\[1ex]
$ p_2 $&$  =$&$  \left(\,E_B,\,0,\,0,\,-\,k\,\right) $&$ p_4 $&$  = $&$ \left(\,E_B,\,-\,k\,\sin\theta,\,0,\,-\,k\,\cos\theta\,\right)$\\[1ex]
 \hline
\multicolumn{6}{|c|}{} \\*[-2ex]
\multicolumn{6}{|c|}{Centre-of-Mass Normal spin vectors}\\[0.5ex]\hline
& & & & & \\[-1.8ex]
$ S_1^{\,N} $&$  =$&$ \left(\,0,\,0,\,1,\,0\,\right)$& $ S_3^{\,N} $&$ =$ &$ \left(\,0,\,0,\,1,\,0\,\right)$\\[1ex]
$ S_2^{\,N} $&$ = $&$\left(\,0,\,0,\,1,\,0\,\right)$&$ S_4^{\,N} $&$ = $&$ \left(\,0,\,0,\,1,\,0\,\right)$\\[1ex]
 \hline
\multicolumn{6}{|c|}{} \\*[-2ex]
\multicolumn{6}{|c|}{Centre-of-Mass Transverse spin vectors}\\[0.5ex]\hline
& & & & & \\[-1.8ex]
$  S_1^{\,T} $&$  =$&$ \,\left(\,0,\,1,\,0,\,0\,\right)$& $ S_3^{\,T} $&$ =$ &$ \left(\,0,\,\cos\theta,\,0,\,-\sin\theta\,\right)$\\[1ex]
$ S_2^{\,T} $&$ = $&$ \left(\,0,\,1,\,0,\,0\,\right)$&$ S_4^{\,T} $&$ = $&$ \left(\,0,\,-\cos\theta,\,0,\,\sin\theta\,\right)$\\[1ex]
\hline
\multicolumn{6}{|c|}{} \\*[-2ex]
\multicolumn{6}{|c|}{Centre-of-Mass Longitudinal spin vectors} \\[0.5ex]\hline
& & & & & \\[-1.8ex]
$  S_1^{\,L} $&$  =$&$ \displaystyle{\frac{1}{M}}\,\left(\,k,\,0,\,0,\,E_A\,\right)$& $ S_3^{\,L} $&$ =$ &$ \displaystyle{\frac{1}{M}}\,\left(\,k,\,E_A\,\sin\theta,\,0,\,E_A\,\cos\theta\,\right)$\\[2ex]
$ S_2^{\,L} $&$ = $&$ \displaystyle{\frac{1}{m}}\,\left(-\,k,\,0,\,0,\,E_B\,\right)$&$ S_4^{\,L} $&$ = $&$ \displaystyle{\frac{1}{m}}\,\left(-\,k,\,E_B\,\sin\theta,\,0,\,E_B\,\cos\theta\,\right)$\\[2ex]
 \hline
\end{tabular}
\end{center}
\caption{\small{\it{Momenta and spin 4--vectors in the Centre-of-Mass frame.  The Centre-of-Mass energies of particles $A$ and $B$ are $E_A=\sqrt{k^{\,2}\,+\,M^{\,2}}$ and $E_B=\sqrt{k^{\,2}\,+\,m^{\,2}}$ respectively, where $k$ is the modulus of the Centre-of-Mass 3--momentum.  The Centre-of-Mass scattering angle is denoted by $\theta$.}}}
\label{table:CM_frame_momenta_and_spin_4vectors}
\end{table}
%
%**********To go from proton trace to antiproton trace we should change the sign of $M$ everywhere, including in the definition of $F$.  But doing this we find that both traces are the same, but only for the spin-averaged case.  We need to flip the sign of the mass only of the second particle, which is the antiproton.  So all of our generic equations with $f_1$, $f_2$, $m$ and $F_1$, $F_2$ and $M$ just changing the sign of $M$ only (not $m$) transforms from proton-proton results to antiproton-proton results.   {\bf NNNBBB}**************
%
\noindent
The spin 4-vectors of Table~\ref{table:CM_frame_momenta_and_spin_4vectors} satisfy the general expression:
\begin{equation}
S^{\,\mu} \ = \  \frac{1}{M}\left(\,{\bf p} \cdot {\bf \hat{s}}\,,\,M\,{\bf \hat{s}} \ + \ \frac{{\bf p} \cdot {\bf \hat{s}}}{E\,+\,M}\,{\bf p}\,\right) \,,
\end{equation}
where ${\bf p}$, $E$ and $M$ are the momentum 3--vector, the energy and the mass of the particle in question and ${\bf \hat{s}}$ is a unit 3--vector identifying a generic spatial direction \cite{Arenhovel:2007gi,Barone:2003}.  When ${\bf \hat{s}}$ is parallel (or antiparallel) to ${\bf p}$ then the particle is longitudinally polarized while if ${\bf \hat{s}}$ is perpendicular to ${\bf p}$ then the particle is transversely polarized.  From this equation one can easily verify that $S^{\,\mu} \, S_{\,\mu} = -1$ and $p^{\,\mu} S_{\,\mu} =0$, where $p^{\,\mu} = (E\,,\,{\bf p})$, as is seen to be satisfied by all of the vectors in Table~\ref{table:CM_frame_momenta_and_spin_4vectors}.  

%\pagebreak

\section{Spin-averaged cross-section}
\label{sec:Spin_averaged_cross-section}

The spin-averaged differential cross-section for $t$-channel elastic spin 1/2 - spin 1/2 scattering can be obtained by setting each $S_i = 0$ in eq.~(\ref{eq:Generic_calculation_with_epsilons}), and multiplying by $2^{\,4}=16$ to counter the four factors of $1/2$ from the spin-dependent completeness relations that are absent in the spin-averaged completeness relations.  One obtains
\begin{eqnarray}
\label{eq:Spin_averaged_generic_calculation}
|\,\mathcal{M}\,|^{\,2} & =& \displaystyle{\left(\, \frac{e}{q}\, \right)^4}\,\mbox{Tr}\left[\left( \slashed{p}_4 + m \right) \left(g_M \gamma^\nu - \displaystyle{\frac{f_2\, r^\nu}{2\,m}} \right) \left( \slashed{p}_2
  +
  m\right) \left(g_M \gamma^\mu -  \displaystyle{\frac{f_2\, r^\mu}{2\,m}} \right)\right] \times \nonumber \\[1ex] 
& & 
  \mbox{Tr}\left[\left( \slashed{p}_1 + M \right)\left(G_M \gamma_\mu
  -  \displaystyle{\frac{F_2\,R_\mu}{2\,M}} \right) \left( \slashed{p}_3 + M \right)\left(G_M \gamma_\nu - \displaystyle{\frac{F_2\,R_\nu}{2\,M}} \right)\right]
\end{eqnarray}
%
%Note that, because the trace of a product with an odd number of $\gamma$ matrices is zero, the antiproton and proton 
These traces can be evaluated using the trace theorems of Appendix~\ref{Appendix:Dirac_algebra} to obtain
\begin{eqnarray}
% & & \mbox{Tr}\left[\left(\slashed{p}_1 - M \right) \left(G_M \gamma_\mu
%  + \displaystyle{\frac{F_2\,R_\mu}{2\,M}} \right) \left( \slashed{p}_3 - M \right)\left(G_M \gamma_\nu + \displaystyle{\frac{F_2\,R_\nu}{2\,M}} \right)
%  \right] \nonumber\\[2ex]
& & \mbox{Tr}\left[\left(\slashed{p}_1 + M \right) \left(\!G_M \gamma_\mu
  - \displaystyle{\frac{F_2\,R_\mu}{2\,M}} \right)  \left( \slashed{p}_3 + M \right) \left(G_M \gamma_\nu - \displaystyle{\frac{F_2\,R_\nu}{2\,M}} \right) \right] \\[2ex]
& & \hspace*{-1.5em}= \displaystyle{4\,G_M^{\,2}\left({p_1}_\mu \, {p_3}_\nu \,+\, {p_1}_\nu \, {p_3}_\mu \,+\, \frac{t}{2}\,\eta^{\mu\nu}\,\right)}  
- 4\,G_M\,F_2\,R_\mu \,R_\nu + 2\,F_2^{\,2}\,R_\mu \,R_\nu\left(\,1 - \frac{t}{4\,M^{\,2}}\,\right) \nonumber
\end{eqnarray}
%
%and as a consistency check notice that transforming $M \rightarrow -\,M$ converts each of the above traces to the other one, but does not alter the computed result.  
%
%Therefore the spin-averaged cross-section for $t$-channel elastic antiparticle-antiparticle, antiparticle-particle and particle-particle scattering to first order in QED are all equal to each other.  
The result follows from eq.~(\ref{eq:Spin_averaged_generic_calculation}), after including a factor of $1/4$ from averaging over initial spin states and summing over final spin states, as presented in our recent paper \cite{O'Brien:2006zt}\,:
\begin{eqnarray}
\label{eq:Generic_spin_averaged_cross-section}
\frac{s}{\alpha^{\,2}}\,\frac{\mathrm{d}\,\sigma}{\mathrm{d}\,\Omega}
& =
&
\left(\frac{4\,m^2\,g_E^{\,2} \,-\,t\,g_M^{\,2}}{4\,m^2 \,-\, t}\right)\left(\frac{4\,M^{\,2}\,G_E^{\,2} \,-\,t\,G_M^{\,2}}{4\,M^{\,2} \,-\, t}\right)\frac{\left(\,M^{\,2}\,-\,m^2\,\right)^2 \ - \ s\,u}{t^2} \nonumber\\[2ex]
& & +\ \left(\frac{2\,m\,Mg_E\,G_E}{t}\right)^2 \ + \ \frac{1}{2}\,\,g_M^{\,2}\,G_M^{\,2}\,,
\end{eqnarray}
where we have used the \emph{electromagnetic coupling constant} (\emph{fine structure constant}) $\alpha = e^{\,2}/4\,\pi$ and the Sachs electric form factor of each particle:
\begin{equation}
G_E\!\left(q^2\right) \ = \ F_1\!\left(q^2\right) \,+\, \frac{t}{4\,M^{\,2}}\,F_2\!\left(q^2\right)
\hspace*{0.7em} \mbox{and} \hspace*{0.7em}
g_E\!\left(q^2\right) \ = \ f_1\!\left(q^2\right) \,+\, \frac{t}{4\,m^2}\,f_2\!\left(q^2\right) .
\end{equation}
Equation~(\ref{eq:Generic_spin_averaged_cross-section}), describing the spin-averaged elastic scattering of any two spin 1/2 particles or antiparticles, is a generalization of the famous Rosenbluth formula for elastic electron-proton scattering \cite{Rosenbluth:1950yq}.  It corresponds to a twofold generalization of the Rosenbluth formula in that the mass of neither particle has been neglected and the internal structure of both particles is included, and is expressed here in a new invariant form.

%\pagebreak

\section{Helicity amplitudes}
\label{sec:Helicity_amplitudes}

Helicity is defined as the projection of the particles spin 3-vector in the direction of its momentum 3-vector.  Helicity is a discretized quantity, having values of either $\pm \hbar\,/\,2$ for a spin 1/2 particle, because the spin of a particle with respect to an axis is quantized.  Helicity states are always longitudinally polarized, {\it i.e.}\ either along direction of motion, which we denote by $+$ for $+\,1/2$, or opposite direction of motion, which we denote by $-$ for $-\,1/2$ in spin 1/2 - spin 1/2 scattering\footnote{Remembering we set $\hbar = 1$ throughout the thesis.}.

The helicity amplitudes for $2 \ \mbox{particle} \rightarrow 2 \ \mbox{particle}$ scattering processes were introduced by Jacob and Wick \cite{Jacob:1959at}, and are represented by 
\begin{equation}
\label{eq:Helicity_amplitudes_definition}
\mathcal{M}\left(\,\mbox{scattered},\,\mbox{recoil}\,;\,\mbox{beam},\,\mbox{target}\,\right) \ = \ \mathcal{M}\left(\,\lambda_3,\,\lambda_4\,;\,\lambda_1,\,\lambda_2\,\right)\,.
\end{equation}
The arguments are to be read from right to left, as $\lambda_1$ and $\lambda_2$ correspond to the incoming particles in the reaction and $\lambda_3$ and $\lambda_4$ correspond to the outgoing particles in the reaction.  For spin 1/2 - spin 1/2 scattering the helicities $\lambda_i \,=\,\pm$ for $i \in \left\{\,1,2,3,4\,\right\}$ are $+$ if the particles spin vector points in the direction of its momentum vector and $-$ if the particles spin vector points in the opposite direction to its momentum vector.  The $\pm$ in the helicity amplitudes are shorthand for $\pm 1/2$\,, the helicity of a spin 1/2 particle.  For $2 \ \mbox{particle} \rightarrow 2 \ \mbox{particle}$ spin 1/2 - spin 1/2 scattering there are 16 helicity amplitudes:
\begin{eqnarray}
\label{eq:16_spin1/2-spin1/2_helicity_amplitudes}
\begin{tabular}{ccccccc}
1. & $\mathcal{M}(+,+\,;+,+)$ & $\equiv$  & $\phi_1$ & \hspace*{3em} & 9.  & $\mathcal{M}(-,-\,;+,+)$ \\
2. & $\mathcal{M}(+,+\,;-,-)$ & $\equiv$  & $\phi_2$ & \hspace*{3em} & 10. & $\mathcal{M}(-,+\,;-,+)$ \\
3. & $\mathcal{M}(+,-\,;+,-)$ & $\equiv$  & $\phi_3$ & \hspace*{3em} & 11. & $\mathcal{M}(-,+\,;+,-)$ \\
4. & $\mathcal{M}(+,-\,;-,+)$ & $\equiv$  & $\phi_4$ & \hspace*{3em} & 12. & $\mathcal{M}(+,-\,;-,-)$ \\
5. & $\mathcal{M}(+,+\,;+,-)$ & $\equiv$  & $\phi_5$ & \hspace*{3em} & 13. & $\mathcal{M}(-,-\,;+,-)$ \\
6. & $\mathcal{M}(+,+\,;-,+)$ & $\equiv$  & $\phi_6$ & \hspace*{3em} & 14. & $\mathcal{M}(-,-\,;-,+)$ \\
7. & $\mathcal{M}(+,-\,;+,+)$ & $\equiv$  & $\phi_7$ & \hspace*{3em} & 15. & $\mathcal{M}(-,+\,;-,-)$ \\
8. & $\mathcal{M}(-,+\,;+,+)$ & $\equiv$  & $\phi_8$ & \hspace*{3em} & 16. & $\mathcal{M}(-,-\,;-,-)$   
\end{tabular}
\end{eqnarray}
Parity Invariance and Time-reversal Invariance, which are strictly satisfied in both electromagnetic and hadronic reactions, provide relations between these 16 helicity amplitudes.  Parity changes the direction of motion but does not change the spin, hence it flips the helicity of a particle.  Therefore Parity Invariance acts on the helicity amplitudes as follows\footnote{In the general case there would be a quantity $\eta \,=\,\frac{\eta_C\,\eta_D}{\eta_A\,\eta_B}\,(-\,1)^{\,s_A + s_B - s_C - s_D}$, where $s$ and $\eta$ correspond to the spin and intrinsic parity of each particle in the reaction, as a factor on the right hand side of eq.~(\ref{eq:Parity_invariance}).  But for elastic scattering, which we are solely interested in, particles $A$ and $C$ are the same and particles $B$ and $D$ are the same; therefore $\eta \,=\,(-1)^0\,=\,1$.} \cite{Jacob:1959at,Goldberger:1960md}:
\begin{equation}
\label{eq:Parity_invariance}
\mathcal{M}\left(\,-\lambda_3,\,-\lambda_4\,;\,-\lambda_1,\,-\lambda_2\,\right) \ =\ (-1)^{\,\lambda - \mu}\,\mathcal{M}\left(\,\lambda_3,\,\lambda_4\,;\,\lambda_1,\,\lambda_2\,\right)\,,
\end{equation}
where $\lambda \,\equiv \, \lambda_1 \,-\, \lambda_2$ and $\mu \,\equiv\, \lambda_3 \,-\, \lambda_4$\,, keeping in mind that each $\lambda_i$ is $\pm 1/2$\,.  Thus this reduces the 16 helicity amplitudes to eight independent ones:
\begin{eqnarray}
\hspace*{-1em}
\begin{tabular}{cclcc}
$\mathcal{M}(+,+\,;+,+)$ & $=$ & $(-1)^{0-0} \    \mathcal{M}(-,-\,;-,-)$  & $=$ & $\ \ \mathcal{M}(-,-\,;-,-)$\\[1ex]
$\mathcal{M}(+,+\,;+,-)$ & $=$ & $(-1)^{-1-0}\  \mathcal{M}(-,-\,;-,+)$  & $=$ & $-   \mathcal{M}(-,-\,;-,+)$\\[1ex]
$\mathcal{M}(+,+\,;-,+)$ & $=$ & $(-1)^{1-0} \    \mathcal{M}(-,-\,;+,-)$  & $=$ & $-   \mathcal{M}(-,-\,;+,-)$\\[1ex]
$\mathcal{M}(+,-\,;+,+)$ & $=$ & $(-1)^{0-(-1)}\    \mathcal{M}(-,+\,;-,-)$  & $=$ & $-   \mathcal{M}(-,+\,;-,-)$\\[1ex]
$\mathcal{M}(-,+\,;+,+)$ & $=$ & $(-1)^{0-1}\     \mathcal{M}(+,-\,;-,-)$  & $=$ & $-   \mathcal{M}(+,-\,;-,-)$\\[1ex]
$\mathcal{M}(+,+\,;-,-)$ & $=$ & $(-1)^{0-0}\     \mathcal{M}(-,-\,;+,+)$  & $=$ & $\ \ \mathcal{M}(-,-\,;+,+)$\\[1ex]
$\mathcal{M}(+,-\,;-,+)$ & $=$ & $(-1)^{-1-1}\    \mathcal{M}(-,+\,;+,-)$  & $=$ & $\ \ \mathcal{M}(-,+\,;+,-)$\\[1ex]
$\mathcal{M}(+,-\,;+,-)$ & $=$ & $(-1)^{-1-(-1)}\ \mathcal{M}(-,+\,;-,+)$  & $=$ & $\ \ \mathcal{M}(-,+\,;-,+)$
\end{tabular}
\end{eqnarray}
%
% These have all been checked in my notes 9-1-08
%
Time-reversal Invariance means that the amplitude for a reaction is unchanged if the direction of time is reversed.  It acts on the helicity amplitudes by interchanging incoming and outgoing particles as follows \cite{Jacob:1959at,Goldberger:1960md}:
\begin{equation}
\label{eq:Time_reversal_invariance}
\mathcal{M}\left(\,\lambda_1,\,\lambda_2\,;\,\lambda_3,\,\lambda_4\,\right) \ = \ (-1)^{\,\lambda - \mu}\,\mathcal{M}\left(\,\lambda_3,\,\lambda_4\,;\,\lambda_1,\,\lambda_2\,\right)\,.
\end{equation}
This reduces the eight remaining helicity amplitudes to six independent ones, by the relations,
\begin{eqnarray}
\begin{tabular}{cclcc}
$\mathcal{M}(+,-\,;+,+)$ & $=$ & $(-1)^{1-0}\ \mathcal{M}(+,+\,;+,-)$    & $=$  & $\ -\mathcal{M}(+,+\,;+,-)$ \\[1ex]
$\mathcal{M}(-,+\,;+,+)$ & $=$ & $(-1)^{-1-0}\ \mathcal{M}(+,+\,;-,+)$ & $=$  & $\ -\mathcal{M}(+,+\,;-,+)$
\end{tabular}
\end{eqnarray}
Hence there are six independent helicity amplitudes, $\phi_1,\,\phi_2,\,\phi_3,\,\phi_4,\,\phi_5$ and $\phi_6\,$, for $2 \ \mbox{particle}  \rightarrow 2\ \mbox{particle}$ spin 1/2 - spin 1/2 scattering, as follows:
\begin{eqnarray}
\label{eq:Spin1/2_spin1/2_Helicity_amplitudes_final}
\phi_1 & \equiv & \ \ \ \ \mathcal{M}(+,+\,;+,+)  \ \ \ =\ \ \ \ \ \mathcal{M}(-,-\,;-,-) \nonumber \\[1ex]
\phi_2 & \equiv & \ \ \ \ \mathcal{M}(+,+\,;-,-)  \ \ \ = \ \ \ \ \ \mathcal{M}(-,-\,;+,+) \nonumber\\[1ex]
\phi_3 & \equiv & \ \ \ \ \mathcal{M}(+,-\,;+,-)  \ \ \ = \ \ \ \ \ \mathcal{M}(-,+\,;-,+) \nonumber\\[1ex]
\phi_4 & \equiv & \ \ \ \  \mathcal{M}(+,-\,;-,+)  \ \ \ = \ \ \ \ \ \mathcal{M}(-,+\,;+,-) \\[1ex]
\phi_5 & \equiv & \ \ \ \  \mathcal{M}(+,+\,;+,-)  \ \ \ = \ \ -\,\mathcal{M}(-,-\,;-,+) \nonumber \\[1ex] & = &  \ -\,\mathcal{M}(+,-\,;+,+)  \ \ \ =\ \ \ \ \ \mathcal{M}(-,+\,;-,-) \nonumber\\[1ex]
\phi_6 & \equiv & \ \ \ \  \mathcal{M}(+,+\,;-,+)  \ \ \ = \ \ -\,\mathcal{M}(-,-\,;+,-) \nonumber \\[1ex] & = &  \ -\,\mathcal{M}(-,+\,;+,+) \ \ \ = \ \ \ \ \ \mathcal{M}(+,-\,;-,-) \nonumber
\end{eqnarray}
As can be seen from eq.~(\ref{eq:Spin1/2_spin1/2_Helicity_amplitudes_final}), $\phi_1$ and $\phi_3$ are non-spin-flip amplitudes, $\phi_2$ and $\phi_4$ are double-spin-flip amplitudes and $\phi_5$ and $\phi_6$ are single-spin-flip amplitudes.  By double-spin-flip we mean that both particles in the reaction have their spins flipped, {\it i.e.}\ their helicities reversed, and by single-spin-flip only one particle in the reaction has its spin flipped, while non-spin-flip means none of the particles in the reaction have their spins flipped.  The non-spin-flip amplitudes, $\phi_1$ and $\phi_3$, are also known as {\it spin-elastic} amplitudes.

Note that for identical particle scattering one also has that $\phi_6 = - \,\phi_5$ and hence there are only five independent helicity amplitudes in this case, as will be shown.

The multiplicity of an independent helicity amplitude is defined as the number of times it appears in the set of total helicity amplitudes, as seen in eq.~(\ref{eq:Spin1/2_spin1/2_Helicity_amplitudes_final}).  Hence $\phi_1$, $\phi_2$, $\phi_3$ and $\phi_4$ each have multiplicity $2$ while $\phi_5$ and $\phi_6$ have multiplicity $4$.  Thus one can write the spin-averaged differential cross-section in terms of the independent helicity amplitudes as
\begin{eqnarray}
\label{eq:Cross-section_to_independent_helicity_amplitudes_invariant}
\hspace*{-2em} s\,\frac{\mathrm{d}\,\sigma}{\mathrm{d}\,\Omega} & = & \frac{1}{\left(8\,\pi\right)^2} \ \sum_{\lambda_1\,\lambda_2\,\lambda_3\,\lambda_4} \frac{1}{4}\ |\,\mathcal{M}(\,\lambda_3,\,\lambda_4\,;\,\lambda_1,\,\lambda_2\,)\,|^{\,2} \,,\nonumber \\[2ex] 
& = & \frac{1}{2 \left(8\,\pi\right)^2} \, \left(\,|\phi_1|^{\,2} \,+\,|\phi_2|^{\,2} \,+\,|\phi_3|^{\,2} \,+\,|\phi_4|^{\,2} \,+\,2\,|\phi_5|^{\,2} \,+\,2\,|\phi_6|^{\,2} \,\right) .
\end{eqnarray}
We now calculate the modulus of all six independent helicity amplitudes for elastic scattering of spin 1/2 - spin 1/2 particles or antiparticles to first order in QED,  by substituting the four longitudinal spin vectors (four pure helicity states) from Table~\ref{table:CM_frame_momenta_and_spin_4vectors} into eq.~(\ref{eq:Generic_calculation_with_epsilons}) and setting specific values ($\pm 1$) for the $\epsilon$'s to obtain
\begin{eqnarray}
\label{eq:Generic_helicity_amplitudes}
\frac{\phi_1}{\pm\alpha} \!& = &\!\! \frac{s - m^2 - M^{\,2}}{t}\left(\!1\!+\!\frac{t}{4\,k^2}\right) f_1\,F_1 - f_1\,F_1 - f_2\,F_1 - f_1\,F_2 - \frac{1}{2}\,f_2\,F_2\left(\!1 - \frac{t}{4\,k^2}\right) \nonumber \\[3ex]
\frac{\phi_2}{\pm\alpha} \!& = & \! \frac{1}{2}\left(\frac{m}{k}\,f_1 - \frac{k}{m}\,f_2\right)\!\left( \frac{M}{k}\,F_1- \frac{k}{M}\,F_2\right)\!+ \frac{s - m^2\! - M^{\,2}\! - 2\,k^2}{4\,m\,M}\left(\!1+\!\frac{t}{4\,k^2}\right) f_2\,F_2\nonumber \\[3ex]
\frac{\phi_3}{\pm\alpha} \!& = &\! \left(\frac{s - m^2 - M^{\,2}}{t}\, f_1\,F_1 + \frac{f_2\,F_2}{2}\,\right)\left(1 + \frac{t}{4\,k^2}\right)  \\[3ex]
\phi_4 \!& = &\! -\phi_2 \nonumber \\[3ex]
\frac{\phi_5}{\pm\alpha} \!& = &\! \sqrt{\frac{s\left(4\,k^2 + t\right)}{-t}}\left[\frac{f_1\,F_1\,m}{4\,k^2}\left(\!1 \!- \frac{m^2\! - M^{\,2}}{s}\right) \!- \frac{F_1\,f_2}{2\,m} \!+\frac{t\,f_2\,F_2}{16\,m\,k^2}\left(\!1 \!+ \frac{m^2\! - M^{\,2}}{s}\right)\!\right]\nonumber \\[3ex]
\frac{\phi_6}{\pm\alpha} \!& = &\!\! \sqrt{\frac{s\left(4\,k^2 + t\right)}{-t}}\!\left[\frac{f_1\,F_1\,M}{4\,k^2}\!\left(\frac{M^{\,2}\! - m^2}{s}-1\!\right) \!+\frac{F_2\,f_1}{2\,M} -\frac{t\,f_2\,F_2}{16\,M\,k^2}\left(\!1 \!+ \frac{M^{\,2}\! - m^2}{s}\right)\!\right]\nonumber 
\end{eqnarray}
in agreement with those found by other methods \cite{Bystricky:1976jr,Buttimore:1978ry,LaFrance:1980}.  The linear combinations $\phi_+ \equiv \phi_1 + \phi_3$ and $\phi_- \equiv \phi_1 - \phi_3$ appear often in the observables
\begin{equation}
\label{eq:Phi+}
\frac{\phi_1 + \phi_3}{\pm\alpha} \ = \ -\,g_M\,G_M \ +\ \left(1\,+\,\frac{t}{4\,k^2}\right)\left(2\,f_1\,F_1\,\frac{s - m^2 - M^{\,2}}{t} \,+ \,f_2\,F_2\,\right)\,,
\end{equation}
\begin{equation}
\label{eq:Phi-}
\hspace*{-25.4em}\frac{\phi_1 - \phi_3}{\pm\alpha} \ = \ -\,g_M\,G_M\,,
\end{equation}
along with the simpler combinations $\phi_2 \,+\, \phi_4 \,=\, 0$ and $\phi_2 \,-\, \phi_4 \,=\, 2\,\phi_2$\,.

%As presented above the helicity amplitudes correspond to antiparticle-particle scattering, but can be transformed to particle-particle scattering by setting $M\rightarrow -\,M$ or to antiparticle-antiparticle scattering by setting $m\rightarrow -\,m$, for these results see our paper~\cite{O'Brien:2006zt}.  This will be the case for all of the expressions derived in the next few sections, but to avoid repetition we shall only state this explicitly here.  Transforming to equal particle-particle scattering by setting $M\rightarrow -\,M$ and then $f_1=F_1$, $f_2=F_2$ and $m=M$ one sees that $\phi_5 = -\,\phi_6$. 

On comparison to previously published proton-proton helicity amplitudes \cite{Bystricky:1976jr,Buttimore:1978ry,LaFrance:1980}, and using the fact that $e_{\bar p} = - e_p$, one finds that the $\pm\,\alpha$ factors above are plus for like-charged particles ({\it e.g.}\ proton-proton or antiproton-antiproton scattering) and minus for unlike-charged particles ({\it e.g.}\ antiproton-proton scattering).

The above $t$-channel helicity amplitudes can be transformed into the corresponding $s$- and $u$-channel helicity amplitudes using crossing symmetry \cite{Buttimore:2007cv,Trueman:1964}.

%\pagebreak

\section{Spin observables}
\label{sec:Spin_observables}

The spin observables for a $2 \ \mbox{particle} \rightarrow 2 \ \mbox{particle}$ elastic reaction, as introduced in \cite{Bystricky:1976jr,LaFrance:1980}, are described in Table~\ref{tab:Spin_Observables}.  In this section we present expressions for them to first order in QED for $t$-channel electromagnetic scattering.  In particular the spin transfer and depolarization observables will play an important role in the remainder of the thesis.

\begin{table}%[!h]
\begin{tabular}{rcl}
\small{{\bf Unpolarized:}} & \begin{minipage}{10em} \centering \includegraphics[width=1.6cm]{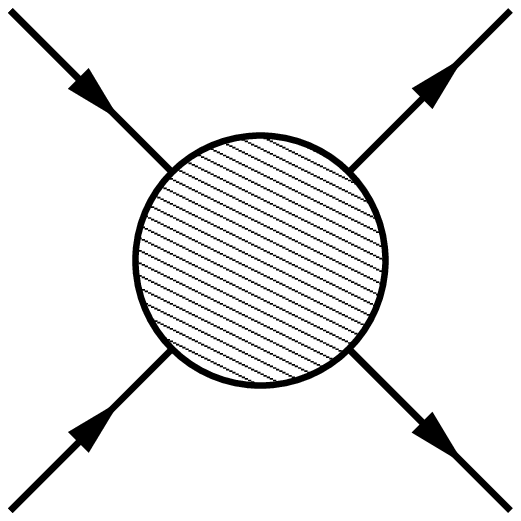} \end{minipage}  &  $\displaystyle{\frac{\mathrm{d}\,\sigma}{\mathrm{d}\,\Omega}}$\\[6ex]
\small{{\bf  1 particle polarized:}}  & & \\[2ex]
\small{Single spin asymmetries} &  \begin{minipage}{21em} \centering \includegraphics[width=1.7cm]{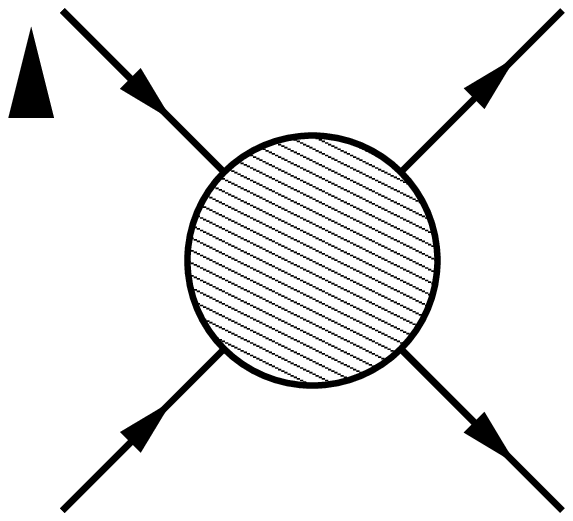} \hspace*{0.3em} \includegraphics[width=1.7cm]{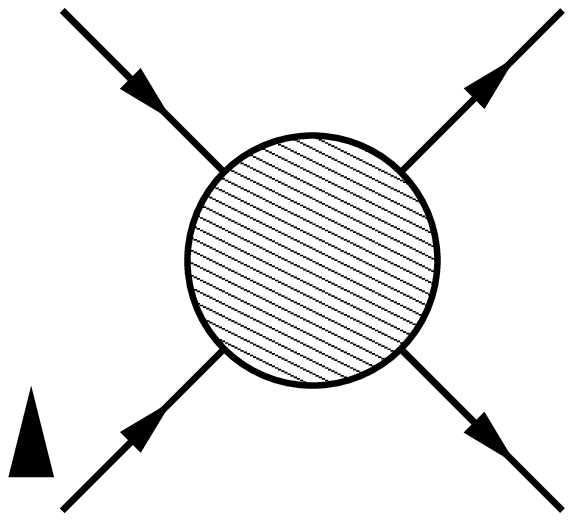} \hspace*{0.3em} \includegraphics[width=1.7cm]{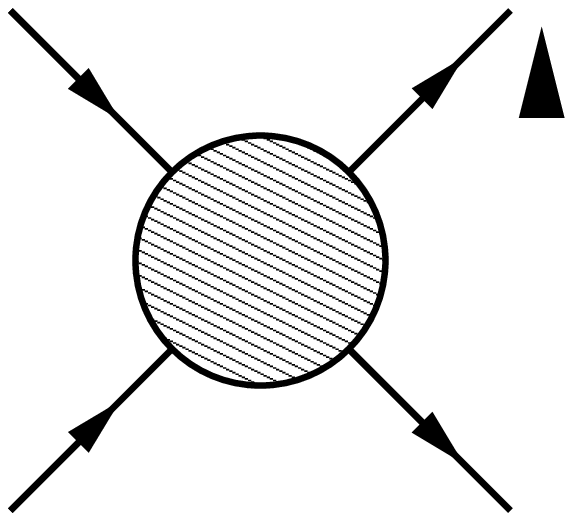} \hspace*{0.3em} \includegraphics[width=1.7cm]{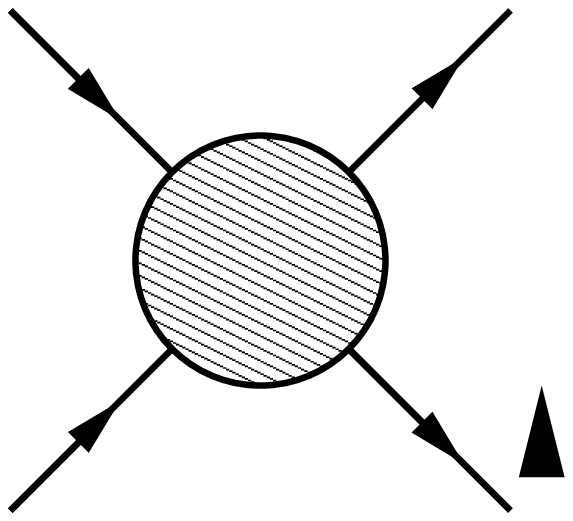} \end{minipage} &  $A_{i}$\\[6ex]
\small{{\bf 2 particles polarized:}}  & & \\[2ex]
\small{Polarization transfer:}
&  \begin{minipage}{15em} \centering \includegraphics[width=1.8cm]{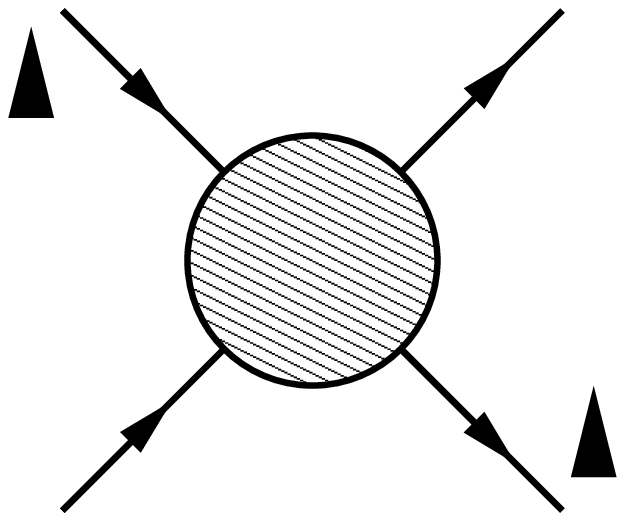} \hspace*{1em} \includegraphics[width=1.8cm]{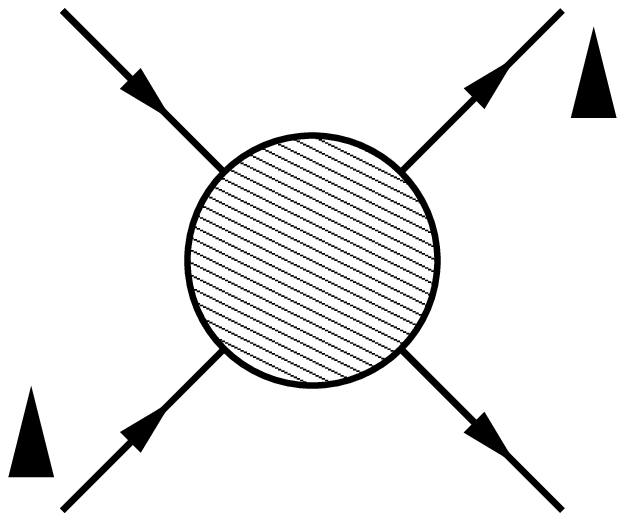}\end{minipage}
 &  $K_{ij}$ \\[7ex]
\small{Depolarization:} & \begin{minipage}{15em} \centering \includegraphics[width=1.8cm]{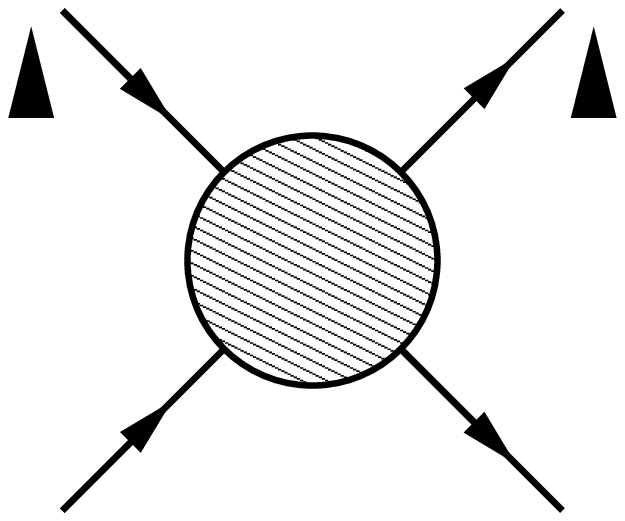} \hspace*{1em} \includegraphics[width=1.8cm]{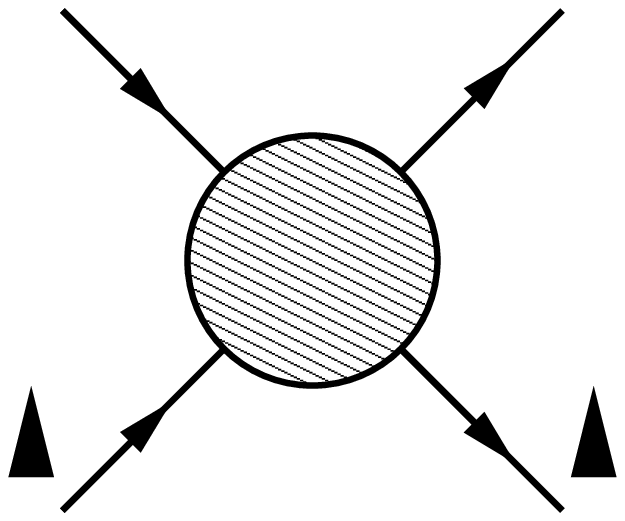}\end{minipage}  &  $D_{ij}$ \\[7ex]
\small{Double spin asymmetries:} & \begin{minipage}{15em} \centering \includegraphics[width=1.8cm]{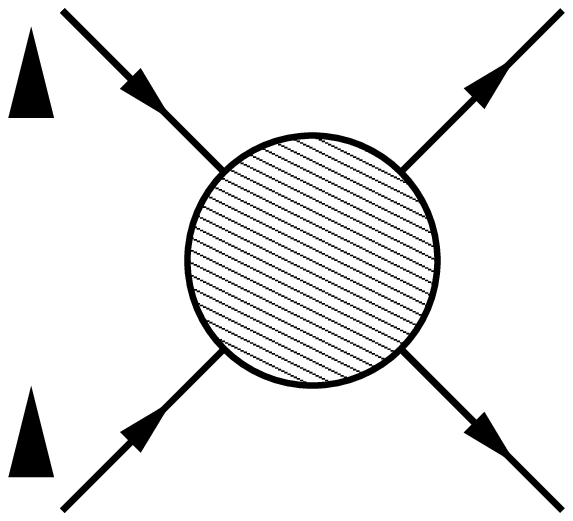} \hspace*{1em} \includegraphics[width=1.8cm]{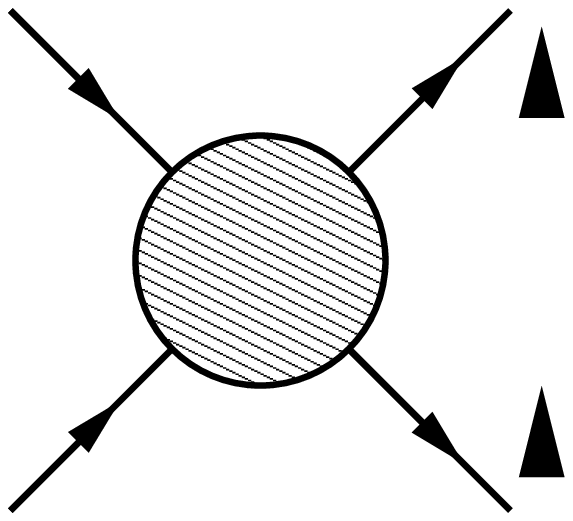}\end{minipage} & $A_{ij}$\\[9ex]
\small{{\bf 3 particles polarized:}} &  \hspace*{-2em}\begin{minipage}{22em} \centering \includegraphics[width=1.9cm]{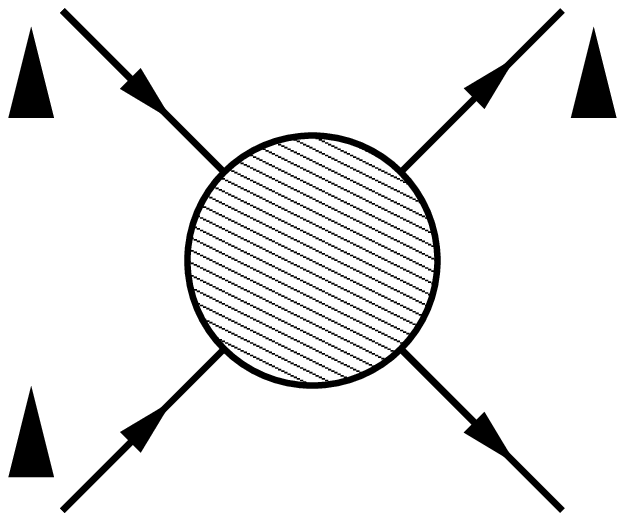} \hspace*{0.3em} \includegraphics[width=1.9cm]{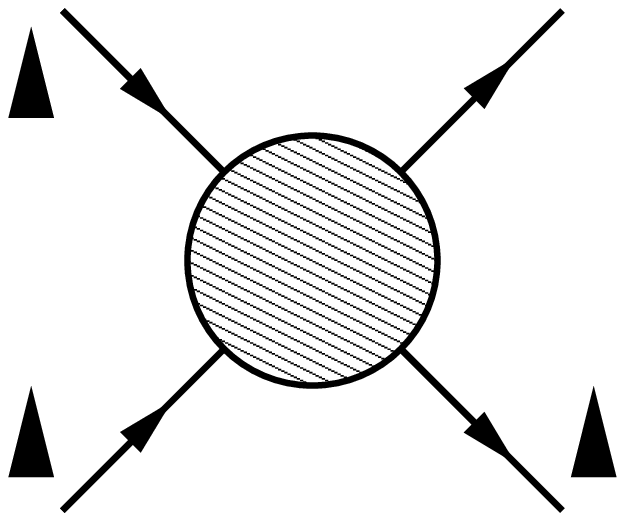} \hspace*{0.3em} \includegraphics[width=1.9cm]{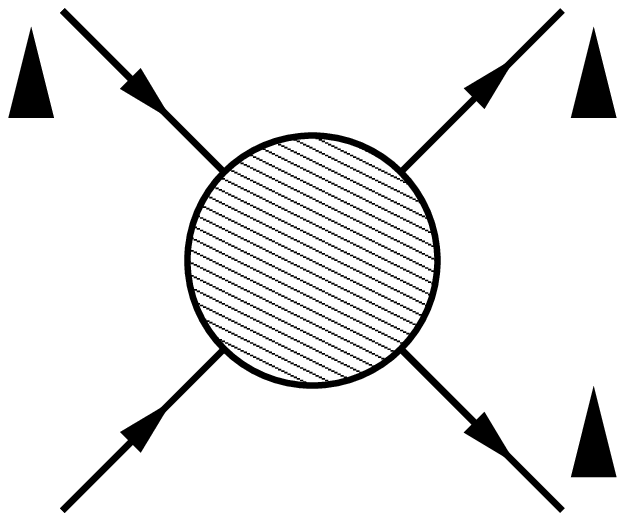} \hspace*{0.3em} \includegraphics[width=1.9cm]{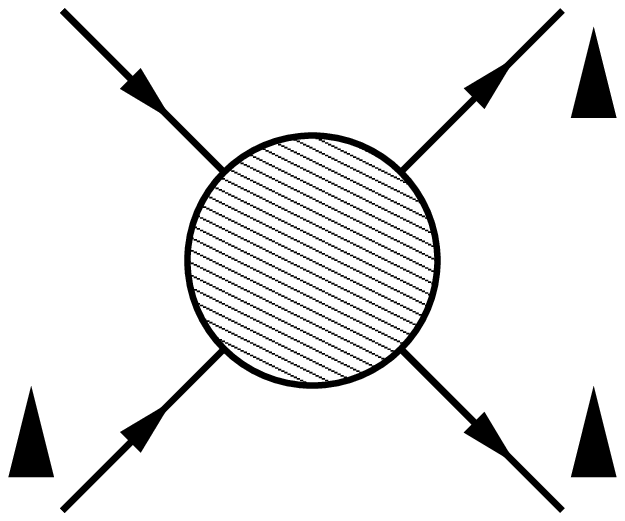}  \end{minipage}  \hspace*{-2em} & $C_{ijk}$  \\[9ex]
\small{{\bf 4 particles polarized:}}      & \begin{minipage}{6em} \centering \includegraphics[width=1.9cm]{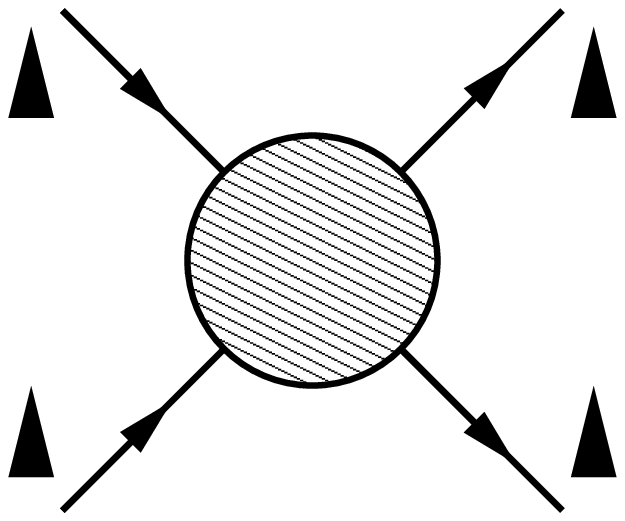} \end{minipage} & $C_{ijkl}$ \\[5ex]
\end{tabular}
\caption{\small{\it{The $16$ spin observables of a $2 \ \mbox{particle} \rightarrow 2 \ \mbox{particle}$ elastic reaction.  An upward pointing triangle represents that the beam is polarized, while the absence of a triangle represents an unpolarized beam.  The right hand column shows the symbols for the spin observables that will be used throughout the thesis, as defined in Ref.~\cite{Bystricky:1976jr}.  The subscripts $i,j,k,l \, \in \, \left\{\,X,Y,Z\,\right\}$ correspond to the direction of the polarization of each particle.  Time increases from left to right as elsewhere in the thesis.}}} 
\label{tab:Spin_Observables}
\end{table}

\subsection{Polarization transfer observables}
\label{subsec:Polarization_transfer_observables}
\noindent
Setting $S_1 = S_4 =0$ in eq.~(\ref{eq:Generic_calculation_with_epsilons}) and subtracting the spin-averaged differential cross-section gives a generic equation for spin transfer from initial particle $B$ to final antiparticle $A$\,:
\begin{eqnarray}
\label{eq:Generic_spin_transfer_equation}
\frac{\mathrm{d}\,\sigma}{\mathrm{d}\,\Omega}\,K_\mathrm{ij} \!&\! = \!&\! \frac{\alpha^2\,g_M\,G_M}{8\,m\,M\,s\,t^2}\left\{F_2\, t\  q \cdot S_3\, \left[\,4\, m^2\, f_1\ p_3 \cdot S_2 \, +\,  f_2 \left(\,2\left(\,m^2 \,+\, M^{\,2} -\, s\,\right) q \cdot S_2  \right.\right.\right.  \nonumber \\[2ex]
& & \left.\left.\left. +\ p_3 \cdot S_2\ t \,-\, q \cdot S_2\ t\,\right)\,\right] \ -\ 16\,m^2\, M^{\,2}\, F_1\, g_E\ p_1 \cdot S_2\ q \cdot S_3\right. \nonumber \\[2ex]
& & \left. \,+\ 4\, M^{\,2}\, G_E\,\left[\,4\, m^2\, f_1\, \left(\,p_3 \cdot S_2\ q \cdot S_3 \,-\, S_2 \cdot S_3\, t\,\right) \right.\right. \nonumber \\[2ex] 
& & \left.\left.\hspace*{0.1em} +\ t\,f_2\, \left(\,p_3 \cdot S_2\ q \cdot S_3 \ +\  q \cdot S_2\ \left(\,p_2\,+\,p_4\,\right) \cdot S_3 \ -\  S_2 \cdot S_3\ t\,\right)\,\right]\,\right\}\,,
\end{eqnarray}
into which specific vectors $S_2$ and $S_3$ will be inserted to give the various polarization transfer observables.  Scattering is in the $XZ$ plane, so the coordinates are $X$ (Transverse), $Y$ (Normal) and $Z$ (Longitudinal), where in the above equation $i,j \in \{X,Y,Z\}$.
%, which are related to  the Argonne LAB coordinates $S$, $N$ and $L$ \cite{Leader:2005}.  
%
\subsubsection{The spin observables $K_\mathrm{XX}$, $K_\mathrm{YY}$ and $K_\mathrm{ZZ}$}
\noindent
The spin observable $K_\mathrm{XX}$ is obtained by inserting the transverse polarized spin vectors $S_2 = S_2^{\,T}$ and $S_3 = S_3^{\,T}$ from Table~\ref{table:CM_frame_momenta_and_spin_4vectors} into the generic spin transfer equation~(\ref{eq:Generic_spin_transfer_equation}):
\begin{eqnarray}
\label{eq:Generic_K_XX}
\frac{\mathrm{d}\,\sigma}{\mathrm{d}\,\Omega}\,K_\mathrm{XX} & = & \alpha^2\frac{g_M\,G_M}{8\,k^2\,m\,M\,s}\,\left\{\,4\,m^2\,f_1\,\left(\,M^{\,2}\,F_1 \,-\, k^2\,F_2\,\right) \ + \ f_2\,\left[\,-\,4\,k^2\,M^{\,2}\,F_1 \right. \right. \nonumber \\[2ex]
 & &  \left. \left. \quad +\ \left(\,4\, k^4 \ +\ \left(\,4\,k^2\,+\, t\,\right)\sqrt{k^2 \,+\, m^2}\,\sqrt{k^2 \,+\, M^{\,2}}\,\right)\, F_2\,\right]\right\}\,.
\end{eqnarray}
Inserting the normal polarized spin vectors $S_2 = S_2^{\,N}$ and $S_3 = S_3^{\,N}$ from Table~\ref{table:CM_frame_momenta_and_spin_4vectors} into the generic spin transfer equation~(\ref{eq:Generic_spin_transfer_equation}) gives
\begin{equation}
\label{eq:Generic_K_YY}
\frac{\mathrm{d}\,\sigma}{\mathrm{d}\,\Omega}\,K_\mathrm{YY} \ \,=\, \ \left(\frac{2\,\alpha^2}{s\ t}\right)\,m\,M \,g_E\,g_M\,G_E\,G_M \,.
\end{equation}
Inserting the longitudinally polarized spin vectors $S_2 = S_2^{\,L}$ and $S_3 = S_3^{\,L}$ from Table~\ref{table:CM_frame_momenta_and_spin_4vectors} into the generic spin transfer equation~(\ref{eq:Generic_spin_transfer_equation}) gives
\begin{eqnarray}
\label{eq:Generic_K_ZZ}
\frac{\mathrm{d}\,\sigma}{\mathrm{d}\,\Omega}\,K_\mathrm{ZZ} & = & \alpha^2\,\frac{g_M\,G_M\,}{8\ k^2\,s^2\, t}\left\{\left[\,s^2 \,-\, \left(M^{\,2} \,-\, m^2\right)^2\right]\,\left(4\,k^2\,+\, t\right)\, f_1\, F_1 \right. \nonumber\\[2ex]
& &  \left. 
\phantom{\sqrt{k^2 \,+\, M^{\,2}}} \quad
+\,s\,\left(4\,k^2\,f_1 \,-\, t\, f_2\right)\,\left(\,4\,k^2\, F_1 \,-\, t\,F_2\,\right)\right\}\,.
\end{eqnarray}

\subsubsection{The spin observables $K_\mathrm{XZ}$ and $K_\mathrm{ZX}$}
\noindent
When the spin four vectors $S_2 = S_2^{\,L}$ and $S_3 = S_3^{\,T}$ from Table~\ref{table:CM_frame_momenta_and_spin_4vectors} are substituted into the generic spin transfer equation~(\ref{eq:Generic_spin_transfer_equation}) we obtain
\begin{eqnarray}
\label{eq:Generic_K_ZX}
\frac{\mathrm{d}\,\sigma}{\mathrm{d}\,\Omega}\,K_\mathrm{XZ} & = &  \frac{\alpha^2\,g_M\,G_M}{16\,M\,s^{\,3/2}\,t}\sqrt{\frac{-\,t\,\left(\,4\,k^2 \,+\, t\,\right)}{k^{\,4}}}\,\left\{\left(\,s \,+\, M^{\,2} \,-\, m^2\,\right)\,t\,f_2\,F_2 \right.\nonumber \\[2ex]
& & 
\left. \qquad +\  4\,f_1\,\left[\,M^2\,\left(\,s \,+\, m^2 \,-\, M^{\,2}\,\right)F_1 \ -\ 2\,k^2\,s\,F_2\,\right]\right\}\,.
\end{eqnarray}
Inserting the spin four vectors $S_2 = S_2^{\,T}$ and $S_3 = S_3^{\,L}$ from Table~\ref{table:CM_frame_momenta_and_spin_4vectors} we obtain
\begin{eqnarray}
\label{eq:Generic_K_XZ}
\frac{\mathrm{d}\,\sigma}{\mathrm{d}\,\Omega}\,K_\mathrm{ZX} & = & \frac{\alpha^2\,g_M\,G_M}{16\,m\,\,s^{\,3/2}\,t}\sqrt{\frac{-\,t\,\left(\,4\,k^2 \,+\, t\,\right)}{k^{\,4}}}\,\left\{\,\left(\,s \,+\, m^2 \,-\, M^{\,2}\,\right)\,t\,f_2\,F_2 \right. \nonumber \\[2ex]
& & \left. \qquad  +\ 4\,F_1\,\left[\,m^2\,\left(\,s \,+\, M^{\,2} \,-\, m^2\,\right)\,f_1 \ -\ 2\,k^2\,s\,f_2\,\right] \right\}\,.
\end{eqnarray} 
As expected by parity and time-reversal invariance we confirm that
\begin{equation}
\label{eq:Generic_Ks_equal_zero}
K_\mathrm{XY} \ = \ K_\mathrm{YX} \ = \ K_\mathrm{YZ} \ = \ K_\mathrm{ZY} \ = \ 0\,.
\end{equation}

%\pagebreak

\subsection{Depolarization spin observables}
\label{subsec:Depolarization_spin_observables}
\noindent
The observable $D_{ij}$ is the polarization remaining after interaction with the target, while of interest here is the loss of polarization after interaction with the target, {\it i.e.}\ $\left(1 - D_{ij}\right)$.  We present results to leading order in small $|\,t\,|$.  Here setting $S_2 = S_4 = 0$ in the generic eq.~(\ref{eq:Generic_calculation_with_epsilons}) and subtracting the spin-averaged differential cross-section, then subtracting the result from the spin-averaged equation gives a generic depolarization equation for antiparticle $A$, into which the various vectors $S_1$ and $S_3$ can be substituted.  The {\tt Mathematica} code to derive this generic depolarization equation is included in Appendix~\ref{Appendix:Sample_Mathematica_code}, but the equation itself is too long to be practical to include here.  The $\approx$ sign means to first order in small $|\,t\,|$.
\subsubsection{The spin observables $\left(\,1 - D_\mathrm{XX}\,\right)$, $\left(\,1 - D_\mathrm{YY}\,\right)$ and $\left(\,1 - D_\mathrm{ZZ}\,\right)$}
\noindent
Substituting the transverse spin vectors $S_1 = S_1^{\,T}$ and $S_3 = S_3^{\,T}$ from Table~\ref{table:CM_frame_momenta_and_spin_4vectors} into the generic depolarization equation gives
\begin{eqnarray}
\label{eq:Generic_1_minus_D_XX}
\frac{\mathrm{d}\,\sigma}{\mathrm{d}\,\Omega}\,\left(\,1 - D_\mathrm{XX}\,\right) & \approx & \frac{-\,2\,\alpha^2\,F_1^2}{k^2\, m^2\,t\,s}\, \left\{ \left[\,m^4\, \left(\,k^2 \,+\, M^{\,2}\,\right)\, f_1^2  \ +\ s\,k^{\,4}\,f_2^{\,2} \right.\right.\\[2ex] 
& & \qquad \qquad \qquad \left. \left.-\ k^2\ m^2\, \left(\,s \,+\, M^{\,2} \,-\, m^2\,\right)\, f_1\ f_2 \nonumber
\right]\right\}\,.
\end{eqnarray}
Inserting the normal polarized spin vectors $S_1 = S_1^{\,N}$ and $S_3 = S_3^{\,N}$ from Table~\ref{table:CM_frame_momenta_and_spin_4vectors} into the generic depolarization equation gives
\begin{equation}
\label{eq:Generic_1_minus_D_YY}
\frac{\mathrm{d}\,\sigma}{\mathrm{d}\,\Omega}\,\left(\,1 - D_\mathrm{YY}\,\right)  \ \,=\, \ \frac{\alpha^2}{2\,s}\ g_M^{\,2}\ G_M^{\,2}\,, \ \ \ \ \ \ \ \ \mbox{complete to all orders in $t$.}
\end{equation}
Inserting the longitudinal spin vectors $S_1 = S_1^{\,L}$ and $S_3 = S_3^{\,L}$ from Table~\ref{table:CM_frame_momenta_and_spin_4vectors} into the generic depolarization equation gives
\begin{eqnarray}
\label{eq:Generic_1_minus_D_ZZ}
\frac{\mathrm{d}\,\sigma}{\mathrm{d}\,\Omega}\,\left(\,1 - D_\mathrm{ZZ}\,\right) & \approx & \frac{-\,2\,\alpha^2\,F_1^2}{k^{\,2}\, M^{\,2}\,t\,s}\, \left\{ \left[\,M^4\, \left(k^2 \,+\, m^2\right)\, f_1^2 \ +\  s\,k^{\,4}\,f_2^{\,2} \right.\right.\\[2ex] 
& & \qquad \qquad \qquad \left. \left.-\ k^2\, M^2\, \left(\,s\,+\,m^2 \,-\, M^{\,2}\,\right)\, f_1\ f_2 \nonumber
\right]\right\}\,.
\end{eqnarray}
\subsubsection{The spin observables $\left(\,1 - D_\mathrm{XZ}\,\right)$ and $\left(\,1 - D_\mathrm{ZX}\,\right)$}
\noindent
Substituting the spin vectors $S_1 = S_1^{\,L}$ and $S_3 = S_3^{\,T}$ from Table~\ref{table:CM_frame_momenta_and_spin_4vectors} into the generic depolarization equation gives
\begin{equation}
\label{eq:Generic_1_minus_D_XZ}
\frac{\mathrm{d}\,\sigma}{\mathrm{d}\,\Omega}\,\left(\,1 - D_\mathrm{XZ}\,\right)  \  \,\approx \,  \ \frac{4\, \alpha^2\, f_1^{\,2}\, F_1^{\,2}}{s\ t^{\,2}}\, \left[\ 2\ k^{\,4} \,+\, m^2\ M^{\,2}  \,+\, \  k^{\,2}\, \left(\,s \,-\, 2\,k^{\,2}\,\right)\,\right]\,,
\end{equation}
and $\displaystyle{\frac{\mathrm{d}\,\sigma}{\mathrm{d}\,\Omega}\,\left(\,1 - D_\mathrm{ZX}\,\right)}$, found by inserting into the generic depolarization equation the spin vectors $S_1 = S_1^{\,T}$ and $S_3 = S_3^{\,L}$ from Table~\ref{table:CM_frame_momenta_and_spin_4vectors}, is the same as above to leading order in $t$.  Note this is just the leading $t$ part of the spin-averaged case.

As expected by parity and time-reversal invariance we confirm that
\begin{equation}
\label{eq:Generic_1_minus_D_equals_zero}
D_\mathrm{XY} \ = \ D_\mathrm{YX} \ = \ D_\mathrm{YZ} \ = \ D_\mathrm{ZY} \ = \ 0\,.
\end{equation}

\subsection{Spin asymmetries}
\label{subsec:Spin_asymmetries}
\noindent
For electromagnetic scattering, to first order in QED, all single and triple spin asymmetries are zero,
\begin{equation}
A_{i} \ = \ C_{ijk} \ =\  0\ \ \ \ \ \ \ \ \mbox{where $i$, $j$, $k$}\, \in \{\mathrm{\,X,Y,Z\,}\} \,,
\end{equation}
and the double spin asymmetries equal the polarization transfer spin observables:
\begin{equation}
A_{ij} \ =\  K_{ij}\ \ \ \ \ \ \ \ \mbox{where $i$, $j$}\, \in \{\mathrm{\,X,Y,Z\,}\} \,.
\end{equation}
as a consequence of the tree-level electromagnetic helicity amplitudes all being real, and the fact that $\phi_2 = -\,\phi_4$ to first order in QED\footnote{See expressions for the spin observables in terms of helicity amplitudes presented Ref.~\cite{Bourrely:1980mr} and in Table A.10.5 of Ref.~\cite{Leader:2005}.}.
There are also \lq four spin measurement' spin observables $C_{ijkl}$ as described in Table~\ref{tab:Spin_Observables} where the polarization of the beam, target, scattered and recoil particles are measured.  These \lq four spin measurement' spin observables are not needed for spin filtering as the polarization of the recoil particle will not be measured, hence they will not be treated in this thesis.  For a treatment of this see Refs.~\cite{LaFrance:1980,Leader:2005,Artru:2008cp}.

Radiative corrections are not treated in this thesis.  In elastic $e^-\,p$ scattering radiative corrections are estimated to give corrections of 1\% - 3\%, and are only considered in very high precision experiments.  This is treated in Refs.~\cite{Mo:1968cg,Maximon:2000hk,Maximon:2000hm,Afanasev:2001nn}.

% Perhaps omit this comment about radiative corrections, ask Nigel whether to include it or not.  Including it may prompt a question on it which is outside the scope of the thesis.  Know it for defence anyway.
\chapter{Specific helicity amplitudes and spin observables}
\label{ch:Specific_helicity_amplitudes_and_spin_observables}

\vspace*{5ex}
\begin{minipage}{6cm}
\end{minipage}
\hfill
\begin{minipage}{10cm}
\begin{quote}
\emph{\lq\lq Mathematics, rightly viewed, posses not only truth, but a supreme beauty - a beauty cold and austere, like that of a sculpture.\rq\rq}
\flushright{Bertrand Russell}
%\emph{\lq\lq All of my life through, the new sights of Nature made me rejoice like a child.\rq\rq}
%\flushright{Marie Curie}
\end{quote}
\end{minipage}
\vspace{8ex}

The expressions for the generic spin 1/2 - spin 1/2 electromagnetic helicity amplitudes and spin observables derived in Chapter~\ref{ch:Generic_helicity_amplitudes_and_spin_observables} are presented in this chapter for the specific cases of: antiproton-proton scattering in section~\ref{sec:Antiproton_proton_helicity_amplitudes_and_spin_observables}, antiproton-electron scattering in section~\ref{sec:Antiproton_electron_helicity_amplitudes_and_spin_observables}  and positron-electron scattering in section~\ref{sec:Positron_electron_helicity_amplitudes_and_spin_observables}.  Then the specific cross-sections and spin observables needed for spin filtering are explicitly presented in section~\ref{sec:Observables_needed_for_spin_filtering}.  These results are of importance to many areas of particle physics, and will be utilized throughout the remainder of this thesis.  The chapter concludes with a calculation of all spin 0 - spin 1 helicity amplitudes in section~\ref{sec:Spin_0_spin_1_helicity_amplitudes}, which describe the scattering of deuterons off a carbon nucleus for example.

\pagebreak

\section{Introduction}
\label{sec:QEDCalculations_Introduction}

In this chapter we calculate all spin-averaged and spin-dependent electromagnetic cross-sections for elastic antiproton-proton, antiproton-electron and positron-electron scattering, to first order in QED.  These are required in the descriptions of spin filtering that follow later in the thesis.  In the region of low momentum transfer of importance in storage rings electromagnetic effects dominate over hadronic effects.  So we calculate the electromagnetic contribution to these cross-sections, and focus on the low $|\,t\,|$ behaviour.  As explained in section~\ref{subsec:Minimum_scattering_angle} there is a minimum momentum  transfer $q_\mathrm{\,min}$ (and correspondingly a minimum scattering angle $\theta_\mathrm{min}$), below which scattering is prevented by Coulomb screening, corresponding to an impact parameter of the Bohr radius $a_B \,=\,52900 \ \mbox{fm}$ of the atom \cite{Jackson:1999}.  Antiprotons flying past an atom at impact parameters greater than $a_B$ see the atom as neutral and do not interact with the atom \cite{Nikolaev:2006gw}.  For momentum transfers of $q > q_\mathrm{\,min} = 1/a_B$ the energy transfered is greater than the binding energy of the hydrogen atom, hence to a good approximation the antiproton scatters from free protons and electrons in the hydrogen atom \cite{Horowitz:1994,Meyer:1994}, {\it i.e.}\
\begin{equation}
\label{eq:Atom_as_free_constituents_approximation}
\left(\frac{\mathrm{d}\,\sigma}{\mathrm{d}\,\Omega}\right)_{\bar{p}\,H} \ \approx \ \ \,\left(\frac{\mathrm{d}\,\sigma}{\mathrm{d}\,\Omega}\right)_{\bar{p}\,p} \ + \ \ \,\left(\frac{\mathrm{d}\,\sigma}{\mathrm{d}\,\Omega}\right)_{\bar{p}\,e^-}
\end{equation}
where $H$ represents a hydrogen atom.

Since the objective is to polarize the antiproton beam the transfer of polarization from initial target particles to the final antiproton beam is of principal importance.  Hence a thorough investigation of the elastic polarization transfer reactions
\begin{eqnarray}
\label{eq:Pbar_p_polarization_transfer_reaction}
\bar{p} \,+\, p^{\,\uparrow} & \longrightarrow & \bar{p}^{\,\uparrow} \,+\, p \\[2ex]
\label{eq:Pbar_e_polarization_transfer_reaction}
\bar{p} \,+\, e^{-\,\uparrow} & \longrightarrow & \bar{p}^{\,\uparrow} \,+\, e^-
\end{eqnarray}
is presented.  The effects of depolarization on the antiproton beam are also calculated.  

We are also interested in the interactions of an antiproton beam with an opposing polarized electron beam, which eq.~(\ref{eq:Pbar_e_polarization_transfer_reaction}) equally describes.  Coulomb screening also introduces a minimum scattering angle, $\theta_\mathrm{min}$, in this case; due to the impact parameter being limited by half of the average separation of electrons in the beam, as explained in section~\ref{subsec:Minimum_scattering_angle}.

\section{$\bar{p}\,p$ helicity amplitudes and spin observables}
\label{sec:Antiproton_proton_helicity_amplitudes_and_spin_observables}

In this section we present the expressions for the electromagnetic helicity amplitudes and spin observables in the specific case of antiproton-proton scattering.  We have only derived expressions for $t$-channel scattering, while for antiproton-proton scattering there is also a contribution from $s$-channel (annihilation) scattering, as shown in Figure~\ref{fig:Bhabha_Feynman_diagram}.  However the $t$-channel results are dominant in the low $|\,t\,|$ region of interest in a storage ring, as explained in Figure~\ref{fig:Bhabha_Feynman_diagram}, and electromagnetic effects also dominate over the hadronic effects in this low $|\,t\,|$ region.  The results in this section are important in the region where $|\,t\,| \,<\, |\,t_c\,|$ for antiproton-proton collisions with total cross section $\sigma^{\bar{p}\,p}_{\mathrm{tot}}$, defined by \cite{Buttimore:1978ry,Kopeliovich:1974ee}
\begin{equation}
\label{eq:t_c_defined}
t_c \ =\  - \,\frac{8\, \pi \, \alpha}{\beta_\mathrm{lab}\,\sigma^{\bar{p}\,p}_\mathrm{tot}\,\sqrt{1\,+\,\rho^2}} 
 \ \approx \  
- \,0.001 \ \ \mbox{(GeV/}c)^{\,2} \,,
\end{equation}
where the electromagnetic interaction dominates the hadronic interaction, as derived in section~\ref{subsec:Electromagnetic_and_hadronic_scattering}.  Here the laboratory velocity is $\beta_\mathrm{lab} =\sqrt{s\,(s-4\,M^{\,2})}\ /\,(s-2\,M^2)$ and $\rho = \mathcal{R}e \{\phi^h_+\}\,/\,\mathcal{I}m \{\phi^h_+\}$ the ratio of real to imaginary parts of the hadronic non-flip amplitude\footnote{The total $\bar{p}\,p$ cross-section behaves like $\sigma^{\bar{p}\,p}_{\mathrm{tot}} \,\approx\, 75.5/\beta_\mathrm{lab} \ \mbox{mb}$ for energies up to $1 \ \mbox{GeV}$ \cite{Rathmann:2004pm,Amsler:2008zz}, and $\rho^2 \approx 0$ at all energies measured thus far (see Figure~4-12 of Ref.~\cite{Klempt:2002ap}).  Therefore $t_c$ is energy independent, at least up to energies of $1 \ \mbox{GeV}$.}.

The electromagnetic helicity amplitudes and spin observables for $\bar{p}\,p$ scattering can be obtained by setting equal masses and form factors ($f_1 = F_1$, $f_2 = F_2$, $g_E = G_E$, $g_M = G_M$ and $m = M$) in the expressions provided in sections~(\ref{sec:Helicity_amplitudes} and \ref{sec:Spin_observables}).  These are required by the  $\mathcal{PAX}$ Collaboration to analyze the buildup of polarization of an antiproton beam by interactions with the protons in a hydrogen target.  The helicity amplitudes in section \ref{sec:Helicity_amplitudes} now become
\begin{eqnarray}
\label{eq:Pbar_p_helicity_amplitudes}
\frac{\phi_1}{-\,\alpha} & = & \left(\frac{s + 4\,k^2}{2\,t} + \frac{M^{\,2}}{2\,k^2}\right) F_1^{\,2} - 2\,F_1\,F_2 + \left(\frac{t - 4\,k^2}{8\,k^2}\right) F_2^{\,2} \nonumber \,,\\[2ex]
\frac{\phi_2}{-\,\alpha} & = &  \frac{-\,\phi_4}{-\,\alpha}  \ = \  \left(\frac{M^{\,2}}{2\,k^2}\right) F_1^{\,2} \,-\, F_1\,F_2 \,+\, \left[\frac{s\left(\ t+8\,k^2\ \right)}{32\,M^{\,2}\,k^2}-\frac{1}{2}\right] F_2^{\,2} \nonumber \,,\\[2ex]
\frac{\phi_3}{-\,\alpha} & = & \left(\frac{s \,-\, 2\,M^{\,2}}{t}\,F_1^{\,2}\,+\,\frac{F_2^{\,2}}{2} \right)\left(\,1 \,+\, \frac{t}{4\,k^2} \right) \,,\\[2ex]
\frac{\phi_5}{-\,\alpha} & = & \frac{1}{2\,M}\sqrt{\frac{\,s\,\left(4\,k^2 \,+\,t\right)}{-t}}\,\left[\left(\frac{M^{\,2}}{2\,k^2}\right) F_1^{\,2} \,-\, F_1\,F_2 \,+\, \left(\frac{t}{8\,k^2}\right) F_2^{\,2}\, \right]\nonumber \,,
\end{eqnarray}
where $\phi_6 = -\,\phi_5$ in this case, as expected.  The generic spin transfer equation for this case is
\begin{eqnarray}
\label{eq:Generic_spin_transfer_Pbar_P} 
\frac{\mathrm{d}\,\sigma}{\mathrm{d}\,\Omega}\,K_\mathrm{ij} & = & \frac{\alpha^2 \,G_M^{\,2}}{8\, M^{\,2}\, s\ t^2} \left\{\,q \cdot S_2\ q \cdot S_3 \left(\,4\, M^{\,2} \,-\, 2\, s \,-\, t\,\right) t\, F_2^{\,2}  
      \right.  \\[2ex]
& & \left. \qquad \qquad \quad + \ 4\,\left[\,2\, p_3 \cdot S_2\ q \cdot S_3 \ +\ 
              q \cdot S_2\, \left(\,p_2 \,+\,p_4\,\right) \cdot S_3\,\right] M^{\,2}\, t\, F_2\, G_E 
\right. \nonumber \\[2ex]
& & \left. \qquad \qquad \qquad \qquad \qquad \quad + \ 16\, M^{\,4}\, G_E \left(\,q \cdot S_2\  q \cdot S_3 \ -\  S_2 \cdot S_3\ t\, G_E\,\right)
\,\right\} \nonumber \,, 
\end{eqnarray}
and the spin transfer observables now become
\begin{eqnarray}
\label{eq:Pbar_p_spin_transfer_observables}
\frac{\mathrm{d}\,\sigma}{\mathrm{d}\,\Omega}\,K_\mathrm{XX} & = & \frac{\alpha^2\,G_M^{\,2}}{8\,s\,k^{\,2}\,M^2}\left\{\!4\,M^{\,4}F_1^{\,2}\,-\, 8\,k^2 M^{\,2}F_1\,F_2 
\,+\,\left[4\, k^4 \,+\,\left(k^2\,+\,\frac{t}{4}\right) s\right] F_2^{\,2}\right\} \,,\nonumber\\[2ex]
\frac{\mathrm{d}\,\sigma}{\mathrm{d}\,\Omega}\,K_\mathrm{YY} & = & \left(\frac{2\,\alpha^2}{s\,t}\right) M^{\,2} \,G_E^{\,2}\,G_M^{\,2} \,, \\[2ex]
\frac{\mathrm{d}\,\sigma}{\mathrm{d}\,\Omega}\,K_\mathrm{ZZ} & = & \frac{\alpha^2\,G_M^{\,2}}{8\ k^2\,s\,\,t}\left[\,s\left(4\,k^2
+ t\,\right)F_1^{\,2}
+\left(\,4\,k^2\, F_1 - t\,F_2\,\right)^2\,\right] \nonumber\,.
\end{eqnarray}
For $\bar{p}\,p$ elastic scattering $K_\mathrm{XZ} = K_\mathrm{ZX}$, and we obtain
\begin{eqnarray}
\label{eq:Pbar_p_K_XZ}
\!\frac{\mathrm{d}\,\sigma}{\mathrm{d}\,\Omega}\,K_\mathrm{XZ} & = &  \frac{\mathrm{d}\,\sigma}{\mathrm{d}\,\Omega}\,K_\mathrm{ZX} \,,\\[2ex]  
 & = & \frac{\alpha^2\,G_M^{\,2}}{2\,M\,t\,\sqrt{s}}\,\sqrt{\frac{-\,t\,\left(\,4\,k^2 + t\,\right)}{k^{\,4}}}\,\left(\frac{M^2\,F_1^{\,2}}{2}  \,-\, k^2\,F_1\,F_2 \,+\, \frac{t\,F_2^{\,2}}{8} \right) \,. \nonumber
\end{eqnarray}
The depolarization spin observables to leading order in small $|\,t\,|$ become
\begin{eqnarray}
\label{eq:Pbar_p_depolarization_observables}
\frac{\mathrm{d}\,\sigma}{\mathrm{d}\,\Omega}\,\left(\,1 - D_\mathrm{XX}\,\right) & \approx &  \frac{-\,2\,\alpha^2\,F_1^{\,2}}{k^{\,2}\,M^{\,2}\,s\,t}\,\left(\,k^{\,2} \,+\, M^{\,2}\,\right)\left(\,M^{\,2}\,F_1 \,-\, 2\,k^2\,F_2\,\right)^2 \nonumber \,,\\[2ex]
\frac{\mathrm{d}\,\sigma}{\mathrm{d}\,\Omega}\,\left(\,1 - D_\mathrm{YY}\,\right) & = & \frac{\alpha^2}{2\,s} \, G_M^{\,4}\,  \ \ \ \ \ \ \ \ \mbox{complete to all orders in $t$} \nonumber \,,\\[2ex]
\frac{\mathrm{d}\,\sigma}{\mathrm{d}\,\Omega}\,\left(\,1 - D_\mathrm{ZZ}\,\right) & \approx & \frac{-\,2\,\alpha^2\,F_1^{\,2}}{k^{\,2}\,M^{\,2}\,s\,t}\,\left(\,k^{\,2} \,+\, M^{\,2}\,\right)\left(\,M^{\,2}\,F_1 \,-\, 2\,k^2\,F_2\,\right)^2 \,,\\[2ex]
\frac{\mathrm{d}\,\sigma}{\mathrm{d}\,\Omega}\,\left(\,1 - D_\mathrm{XZ}\,\right) & \approx & \frac{\mathrm{d}\,\sigma}{\mathrm{d}\,\Omega}\,\left(\,1 - D_\mathrm{ZX}\,\right)  \ \approx \ \frac{\mathrm{d}\,\sigma}{\mathrm{d}\,\Omega} \ \approx \ \frac{4\,\alpha^2\, F_1^{\,4}}{s\, t^{\,2}}\left(\,2\, k^{\,2} \,+\, M^{\,2}\,\right)^{\,2} \nonumber\,.
\end{eqnarray}
Note that to first order in small $|\,t\,|$ the antiproton and proton form factors can be approximated as $F_1 \approx 1$, $F_2 \approx \mu_p - 1$ ({\it i.e.}\ $G_M \approx \mu_p$ and $G_E \approx 1$), where $\mu_p$ is the magnetic moment of the proton.

The $t$-channel spin-averaged differential cross-section for this case simplifies to
\begin{equation}
\label{eq:Pbar_p_spin_averaged_cross_section}
\frac{s}{\alpha^2}\,\frac{\mathrm{d}\,\sigma}{\mathrm{d}\,\Omega} \ = \ 
\left(\frac{4\,M^{\,2}\,G_E^{\,2} -t\,G_M^{\,2}}{4\,M^{\,2} - t}\right)^2 \left(\frac{- s\,u}{t^2}\right) \ +\  \left(\frac{2\,M^{\,2}\,G_E^{\,2}}{t}\right)^2 \ +\  \frac{1}{2}\,\,G_M^{\,4}\,.
\end{equation}
The above helicity amplitudes and spin observables are seen to be correct by a comparison to the relations between the helicity amplitudes and spin observables presented in Ref.~\cite{Bourrely:1980mr} and in Table A.10.5 of Ref.~\cite{Leader:2005}.

In addition to antiproton-proton scattering the results of this section equally apply to any spin 1/2 (anti)baryon-(anti)baryon elastic electromagnetic scattering in the $t$-channel.  Also, given that quarks are spin 1/2 particles, the above expressions could be applied to $t$-channel (anti)quark-(anti)quark electromagnetic scattering if in future quarks are found to have an internal structure.

%\pagebreak

\section{$\bar{p}\,e^-$ helicity amplitudes and spin observables}
\label{sec:Antiproton_electron_helicity_amplitudes_and_spin_observables}
%See notes 19-2-2008 where these are derived from the generic relations

The spin observables for $\bar{p}\,e$ scattering can be obtained by setting $f_1 = 1$ and  $f_2 = 0$ ({\it i.e.}\ $g_E = g_M = 1$) in the expressions from sections~(\ref{sec:Helicity_amplitudes} and \ref{sec:Spin_observables}), to account for the elastic scattering of a structured particle off a point-like particle.  These are required by the $\mathcal{PAX}$ Collaboration to analyze the buildup of polarization of an antiproton beam by interactions with the electrons in a hydrogen target, or interactions with a polarized electron beam as treated in Chapter~\ref{ch:Numerical_results}.
The helicity amplitudes in section~\ref{sec:Helicity_amplitudes} for antiproton-electron elastic collisions become
\begin{eqnarray}
\label{eq:Pbar_e_helicity_amplitudes}
\frac{\phi_1}{\alpha} & = & \left(\,s - m^2 - M^{\,2}\,\right) \left(1 \,+\, \frac{t}{4\ k^{\,2}}\right) \frac{F_1}{t} \,-\, F_1 \,-\, F_2 \,,\nonumber\\[2ex]
\frac{\phi_2}{\alpha} & = &   \frac{-\,\phi_4}{\alpha}  \ = \ \frac{m\ M\ F_1}{2\ k^{\,2}} \,-\, \frac{m\ F_2}{2\ M}\,,\nonumber \\[2ex]
\frac{\phi_3}{\alpha} & = &\left(\,s - m^2 - M^{\,2}\,\right) \left(1 \,+\, \frac{t}{4\ k^{\,2}}\right) \frac{F_1}{t} \,,\\[2ex]
\frac{\phi_5}{\alpha} & = & \sqrt{\frac{s\, \left(\,4\ k^2 \,+\, t\,\right)}{-\,t}}\ \left[\frac{m\, \left(\,s - m^2 + M^{\,2}\,\,\right)
 F_1}{4\ k^{\,2}\ s} \right]\,,\nonumber\\[2ex]
\frac{\phi_6}{\alpha} & = & -\,\sqrt{\frac{s\, \left(\,4\ k^2 + t\,\right)}{-\,t}}\ \left[\frac{M\, \left(\,s +m^2 - M^{\,2}\,\right) 
F_1}{4\ k^{\,2}\ s} - \frac{F_2}{2\, M}\right]\nonumber\,.
\end{eqnarray}
The generic spin transfer equation for this case is
\begin{equation}
\label{eq:Generic_spin_transfer_Pbar_E}
\hspace*{-0.64em}\frac{\mathrm{d}\,\sigma}{\mathrm{d}\,\Omega}\, K_{ij} \, = \,  \frac{-\,2\,\alpha^2\,m\,M\,G_M}{s\ t}\left[\,G_E\,S_2 \cdot S_3 + \!\left( \frac{F_2}{2\,M^2}\,p_3 \cdot S_2 - \frac{F_1}{t}\, q \cdot S_2\right)  q \cdot S_3\,\right] \!, 
\end{equation}
which is a generalization of equation (3) of Ref.~\cite{Horowitz:1994}, which was at the root of the initial interpretation of the FILTEX results on spin filtering in 1994.  We do not neglect the $1/M^{\,2}$ term, or make the non-relativistic approximations of eqs.~(\ref{eq:Non_relativistic_approximations}), as done in Ref.~\cite{Horowitz:1994}.

\noindent
The spin transfer observables for antiproton-electron elastic scattering are
\begin{eqnarray}
\label{eq:Pbar_e_spin_transfer_observables}
\frac{\mathrm{d}\,\sigma}{\mathrm{d}\,\Omega}\,K_\mathrm{XX} & = &  \alpha^2\frac{m\ G_M}{2\,k^{\,2}\,M\,s}\,\left(\,M^{\,2}\,F_1 \,-\, k^{\,2}\,F_2\,\right)\,,\nonumber \\[2ex]
\frac{\mathrm{d}\,\sigma}{\mathrm{d}\,\Omega}\,K_\mathrm{YY} & = &  \frac{2\,\alpha^2\,m\,M \,G_M\,G_E}{s\ t}  \,,\nonumber\\[2ex]
\frac{\mathrm{d}\,\sigma}{\mathrm{d}\,\Omega}\,K_\mathrm{ZZ} & = & \frac{\alpha^2\,G_M\,}{8\, k^{\,2}\,s^2\, t} \left\{\left[s^2 - \left(M^{\,2} \,-\, m^2\right)^2\,\right]\,\left(4\,k^2\,+\, t\,\right) F_1 \right.\nonumber\\[2ex]
 & & \left. \quad
\phantom{\left[s^2 - \left(M^{\,2} \,-\, m^2\right)^2\,\right]} +\ 4\,k^2\,s\,\left(4\,k^2 F_1 \,-\, t\,F_2\right)\right\}\,,\\[2ex]
\frac{\mathrm{d}\,\sigma}{\mathrm{d}\,\Omega}\,K_\mathrm{XZ} & = & \frac{\alpha^2\,G_M}{4\,M\,s^{\,3/2}\ t}\sqrt{\frac{-\,t\left(4\,k^2 + t\,\right)}{k^{\,4}}}\,\left[M^2\left(\,s + m^2\!\! - M^2\right)F_1\, -\, 2\,k^2s\,F_2\,\right] \,, \nonumber\\[2ex]
\frac{\mathrm{d}\,\sigma}{\mathrm{d}\,\Omega}\,K_\mathrm{ZX} & = &  \frac{\alpha^2\,m\,F_1\,G_M}{4\,s^{\,3/2}\ t}\,\sqrt{\frac{-\,t\,\left(\,4\,k^2 \,+\, t\,\right)}{k^{\,4}}}\,\,\left(\,s  - m^2 + M^{\,2}\,\right) \nonumber \,.
\end{eqnarray}
The depolarization spin observables to leading order in small $|\,t\,|$ for this case are
\begin{eqnarray}
\label{eq:Pbar_e_depolarization_observables}
\frac{\mathrm{d}\,\sigma}{\mathrm{d}\,\Omega}\,\left(\,1 - D_\mathrm{XX}\,\right) & \approx & \frac{-\,m^2\,\alpha^2\,F_1^{\,2}}{2\ k^2\ s^2\ t}\ \left(\,s - m^2 + M^{\,2}\,\right)^2 \,,\nonumber\\[2ex]
\frac{\mathrm{d}\,\sigma}{\mathrm{d}\,\Omega}\,\left(\,1 - D_\mathrm{YY}\,\right) & = & \frac{\alpha^2}{2\,s} \, G_M^{\,2}\,  \ \ \ \ \ \ \ \ \mbox{complete to all orders in $t$}\,, \nonumber\\[2ex]
\frac{\mathrm{d}\,\sigma}{\mathrm{d}\,\Omega}\,\left(\,1 - D_\mathrm{ZZ}\,\right) & \approx & \frac{-M^2\,\alpha^2\,F_1^{\,2}}{2\ k^2\ s^2\ t}\ \left(\,s + m^2 - M^{\,2}\,\right)^2 \,,\\[2ex]
\frac{\mathrm{d}\,\sigma}{\mathrm{d}\,\Omega}\,\left(\,1 - D_\mathrm{XZ}\,\right) & \approx & \frac{\mathrm{d}\,\sigma}{\mathrm{d}\,\Omega}\,\left(\,1 - D_\mathrm{ZX}\,\right)  \ \approx \ \frac{\mathrm{d}\,\sigma}{\mathrm{d}\,\Omega} \ \approx \ \frac{4\, \alpha^2\,  F_1^{\,2}}{s\ t^{\,2}}\ \left(\, s\, k^{\,2} \,+\, m^2\, M^{\,2}\,\right)  \nonumber\,.
\end{eqnarray}

\pagebreak
\noindent
The spin-averaged differential cross-section from section~\ref{sec:Spin_averaged_cross-section} for this case simplifies to
\begin{equation}
\label{eq:Pbar_e_spin_averaged_cross_section}
\frac{s}{\alpha^2}\,\frac{\mathrm{d}\,\sigma}{\mathrm{d}\,\Omega}
 \ = \ 
\left(\frac{4\,M^{\,2}\,G_E^{\,2} -t\,G_M^{\,2}}{4\,M^{\,2} - t}\right)\,\frac{\left(\,M^{\,2}-m^2\,\right)^2 - s\,u}{t^2} \,+\, \left(\frac{2\,m\,M\,G_E}{t}\right)^2 \,+\, \frac{1}{2}\,\,\,G_M^{\,2} \,,
\end{equation}
the familiar Rosenbluth formula \cite{Rosenbluth:1950yq}, where we have not neglected the mass of the electron.

The low energy (non-relativistic) limit of the above equations can be obtained using
\begin{eqnarray}
\label{eq:Non_relativistic_approximations}
s & = & \left(\,E^\mathrm{\,cm}_1 \,+\, E^\mathrm{\,cm}_2\,\right)^{\,2} \ \approx \ \left(\,m \,+\, M \,\right)^{\,2} \ \approx \ M^{\,2} \,, \nonumber \\[2ex]
G_E & \approx & 1 \,, \\[2ex]
G_M & \approx & \mu_p \ = \ 1 \,+\, \kappa_p \,. \nonumber
\end{eqnarray}
Of particular importance is the non-relativistic limit of $\left(\mathrm{d}\,\sigma/\mathrm{d}\,\Omega\right)K_\mathrm{YY}$  in eq.~(\ref{eq:Pbar_e_spin_transfer_observables}) for polarization transfer in $\bar{p}\,e^{-\,\uparrow} \rightarrow \bar{p}^{\,\uparrow}\,e^-$ scattering, which using $t = -\,4\,k^2\,\sin^2\frac{\theta}{2}$ where $\theta$ is the Centre-of-Mass scattering angle gives
\begin{equation}
\frac{\mathrm{d}\,\sigma}{\mathrm{d}\,\Omega} \ K_\mathrm{YY} \ \approx \ -\,\frac{\alpha^2\,m\,\left(\,1 \,+\, \kappa_p\,\right)}{2\,k^2\,M\,\sin^2\!\left(\frac{\theta}{2}\right)} \,.
\end{equation}
This is precisely equation (4) of Ref.~\cite{Horowitz:1994}, used extensively throughout their paper and early $\mathcal{PAX}$ calculations.  The equations presented in this thesis generalize this work to relativistic energies.  Non-relativistic expressions for the spin transfer observables have also been presented recently in Ref.~\cite{Arenhovel:2007gi,Milstein:2008tc}, in the context of a proposal to polarize antiprotons by repeated interaction with a co-moving polarized positron beam at very low relative velocities \cite{Walcher:2007sj}. 

In addition to antiproton-electron elastic scattering the results of this section apply also to any (anti)baryon-(anti)lepton elastic scattering.  In particular they apply to antiproton-muon elastic scattering which may be of use in spin filtering, where the fact that muons have approximately 200 times the mass of electrons will greatly enhance the polarization transfer cross-sections $K_\mathrm{XX}$ and $K_\mathrm{YY}$ in eq.~(\ref{eq:Pbar_e_spin_transfer_observables}).

\section{$e^+e^-$ helicity amplitudes and spin observables}
\label{sec:Positron_electron_helicity_amplitudes_and_spin_observables}

The electromagnetic helicity amplitudes and spin observables for $t$-channel elastic positron-electron scattering can be obtained by making the transformations $f_1 = F_1 \rightarrow 1$, $f_2 = F_2 \rightarrow 0$, $g_E = G_E \rightarrow 1$, $g_M = G_M \rightarrow 1$ and $M \rightarrow m$ in the expressions provided in sections~(\ref{sec:Helicity_amplitudes} and \ref{sec:Spin_observables}).  The helicity amplitudes from section \ref{sec:Helicity_amplitudes} now become
\begin{eqnarray}
\label{eq:Ebar_e_helicity_amplitudes}
\frac{\phi_1}{-\,\alpha} & = & \left(\frac{s + 4\,k^2}{2\,t} \ +\  \frac{m^{\,2}}{2\,k^2}\right) \nonumber \,,\\[2ex]
\frac{\phi_2}{-\,\alpha} & = &  \frac{-\,\phi_4}{-\,\alpha}  \ = \  \frac{m^{\,2}}{2\,k^2} \nonumber \,,\\[2ex]
\frac{\phi_3}{-\,\alpha} & = & \left(\frac{s \,-\, 2\,m^2}{t}\,\right)\left(\,1 \,+\, \frac{t}{4\,k^2} \right) \,,\\[2ex]
\frac{\phi_5}{-\,\alpha} & = & \frac{m}{4\,k^2}\,\sqrt{\frac{\,s\,\left(4\,k^2 \,+\,t\right)}{-t}}\nonumber \,,
\end{eqnarray}
where again $\phi_6 = -\,\phi_5$ in this case, as was found for ${\bar p}\,p$ scattering.

\begin{figure}
\centering
\includegraphics{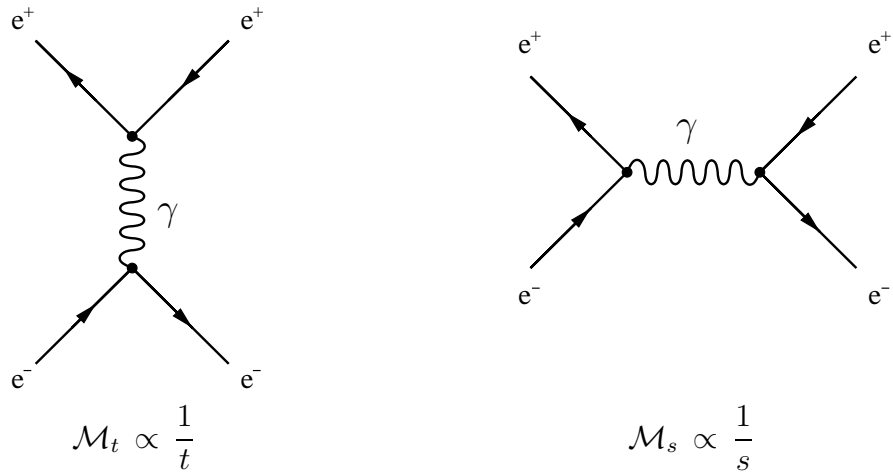}
\vspace*{3ex}
\caption{\small{\it{The two tree-level Feynman diagrams for elastic positron-electron scattering, also known as Bhabha scattering.  The antiproton-proton case is analogous.  Contributions to the full amplitude $\mathcal{M} = \mathcal{M}_t + \mathcal{M}_s$ come from the $t$-channel (left diagram), being proportional to $1/\,t$, and $s$-channel (right diagram), being proportional to $1/s$, as shown.  For low momentum transfer (small $|\,t\,|$), and also at high energies (large $s$), the $t$-channel contributions dominate.  As always time increases from left to right.
}}}
\label{fig:Bhabha_Feynman_diagram}
\end{figure}

The generic spin transfer equation for this case is
\begin{equation}
\label{eq:Generic_spin_transfer_Ebar_E}
\frac{\mathrm{d}\,\sigma}{\mathrm{d}\,\Omega}\, K_{ij}\  = \ \frac{-\,2\,\alpha^2\,m^2}{s\ t} \,\left[\, S_2 \cdot S_3 \ - \ \frac{q \cdot S_2 \ q \cdot S_3}{t}\,\right]\,,
\end{equation}
and the spin transfer observables for $e^+\,e^-$ elastic scattering are
\begin{eqnarray}
\label{eq:Ebar_e_spin_transfer_observables}
\frac{\mathrm{d}\,\sigma}{\mathrm{d}\,\Omega}\,K_\mathrm{XX} & = & \frac{\alpha^2\,m^2}{2\,s\,k^{\,2}} \,,\nonumber\\[2ex]
\frac{\mathrm{d}\,\sigma}{\mathrm{d}\,\Omega}\,K_\mathrm{YY} & = & \frac{2\,\alpha^2\,m^2}{s\,t} \,, \\[2ex]
\frac{\mathrm{d}\,\sigma}{\mathrm{d}\,\Omega}\,K_\mathrm{ZZ} & = & \frac{\alpha^2}{8\ k^2\,s\,\,t}\,\left[\,s\left(4\,k^2 \,+\, t\,\right) \,+\,16\,k^4\,\right] \nonumber\,.
\end{eqnarray}
For $e^+\,e^-$ elastic scattering $K_\mathrm{XZ} = K_\mathrm{ZX}$ hence one obtains
\begin{equation}
\label{eq:Ebar_e_K_XZ}
\!\frac{\mathrm{d}\,\sigma}{\mathrm{d}\,\Omega}\,K_\mathrm{XZ} \ = \ \frac{\mathrm{d}\,\sigma}{\mathrm{d}\,\Omega}\,K_\mathrm{ZX} \ = \ \frac{\alpha^2\,m}{4\ t\,\sqrt{s}}\,\sqrt{\frac{-\,t\,\left(\,4\,k^2 + t\,\right)}{k^{\,4}}} \,. \nonumber
\end{equation}
The depolarization spin observables to leading order in small $|\,t\,|$ for this case are
\begin{eqnarray}
\label{eq:Ebar_e_depolarization_observables}
\frac{\mathrm{d}\,\sigma}{\mathrm{d}\,\Omega}\,\left(\,1 - D_\mathrm{XX}\,\right) & \approx &  \frac{-\,2\,\alpha^2\,m^2}{k^{\,2}\,s\,t}\,\left(\,k^{\,2} \,+\, m^2\,\right) \nonumber \,,\\[2ex]
\frac{\mathrm{d}\,\sigma}{\mathrm{d}\,\Omega}\,\left(\,1 - D_\mathrm{YY}\,\right) & = & \frac{\alpha^2}{2\,s} \,  \ \ \ \ \ \ \ \ \mbox{complete to all orders in $t$} \nonumber \,,\\[2ex]
\frac{\mathrm{d}\,\sigma}{\mathrm{d}\,\Omega}\,\left(\,1 - D_\mathrm{ZZ}\,\right) & \approx & \frac{-\,2\,\alpha^2\,m^2}{k^{\,2}\,s\,t}\,\left(\,k^{\,2} \,+\, m^2\,\right) \,,\\[2ex]
\frac{\mathrm{d}\,\sigma}{\mathrm{d}\,\Omega}\,\left(\,1 - D_\mathrm{XZ}\,\right) & \approx & \frac{\mathrm{d}\,\sigma}{\mathrm{d}\,\Omega}\,\left(\,1 - D_\mathrm{ZX}\,\right)  \ \approx \ \frac{\mathrm{d}\,\sigma}{\mathrm{d}\,\Omega} \ \approx \ \frac{4\,\alpha^2\,}{s\ t^{\,2}}\left(\,2\ k^{\,2} \,+\, m^2\,\right)^{\,2} \nonumber\,.
\end{eqnarray}
The $t$-channel spin-averaged differential cross-section for this case simplifies to
\begin{equation}
\label{eq:Ebar_e_spin_averaged_cross_section}
\frac{s}{\alpha^2}\,\frac{\mathrm{d}\,\sigma}{\mathrm{d}\,\Omega} \ = \ 
\left(\frac{\,- s\,u\,}{t^2}\right) \ +\  \left(\frac{\,2\,m^2\,}{t}\right)^2 \ +\  \frac{1}{2}\,,
\end{equation}
which can be written in the more familiar form \cite{O'Brien:2006tp}
\begin{equation}
\label{eq:Ebar_e_spin_averaged_cross_section2}
\frac{\mathrm{d}\,\sigma}{\mathrm{d}\,\Omega} \ = \ \frac{\alpha^2}{s\,t^2}\left[\,\left(\,2\,m^2\ -\ s\,\right)^{\,2}  \ +\  s\,t \ +\ \frac{t^2}{2}\,\right] \,.
\end{equation}
In elastic electron-positron scattering, also known as Bhabha scattering \cite{Bhabha:1936zz}, for very low momentum transfer the $t$-channel part of the spin-averaged differential cross-section dominates over the $s$-channel part.  But the complete spin-averaged Bhabha differential cross-section must include both $t$-channel and $s$-channel Feynman diagrams in the amplitude, as shown in Figure~\ref{fig:Bhabha_Feynman_diagram}, to give
\begin{eqnarray}
\label{eq:Bhabha_cross-section}
s \,\frac{\mathrm{d}\,\sigma}{\mathrm{d}\,\Omega}& = &\,\overbrace{\frac{\alpha^2}{t^2}\,\left[\,\left(\,2\,m^2 \,-\,s\,\right)^{\,2} \ + \ s\,t \ +\ \frac{t^2}{2}\ \right]}^{\mbox{$t$-channel part}}  \\[2ex]
& &  + \ \underbrace{\frac{\alpha^2}{s^2}\,\left[\,\left(\,2\,m^2 \,-\,t\,\right)^{\,2} \ +\ s\,t \ + \ \frac{s^2}{2}\ \right]}_{\mbox{$s$-channel part}} \ + \ \underbrace{\frac{\alpha^2}{s\,t}\,\left[\,\left(\,s\, +\, t\,\right)^{\,2} \ - \ 4\,m^4\,\right]}_{\mbox{cross-term part}} \nonumber \,.
\end{eqnarray}
where one sees that the $t$-channel part transforms into the $s$-channel part, and vice-versa, by the well known crossing symmetry obtained by interchanging $s$ and $t$.
\psfrag{G2}{\large{$\gamma$}}
\psfrag{TT}{$\displaystyle{\mathcal{M}_t} \,\propto \,\displaystyle{\frac{1}{t}}$}
\psfrag{SS}{$\displaystyle{\mathcal{M}_s} \,\propto \,\displaystyle{\frac{1}{s}}$}

The results of this section also apply to any combination of $t$-channel elastic (anti)lepton-(anti)lepton, (anti)lepton-(anti)quark and (anti)quark-(anti)quark electromagnetic scattering.  This is particularly useful since the masses of each particle have been retained in the equations.
 
%\section{Other reference frames}
%\label{sec:Other_reference_frames}

\section{Observables needed for spin filtering}
\label{sec:Observables_needed_for_spin_filtering}

In this section we present the leading $t$ approximation of all spin observables needed for spin filtering.  These expressions will be integrated over a range of $t$ later in the thesis.  All expressions are written in terms of invariants using eqs.~(\ref{eq:Invariant_lambda_defined} and \ref{eq:CMFrame_to_invariants}) above.  The low $|\,t\,|$ approximation of the form factors\footnote{Writing $F_1(t)$ and $F_2(t)$ in terms of $G_E(t)$ and $G_M(t)$ and then using a Taylor expansion of eq.~(\ref{eq:Dipole_model_for_Sachs_form_factors}) one obtains the Taylor expansions $F_1(t) \, = \, 1 \,+\, 2.30777\ t \,+\, 4.37246 \ t^{\,2} \,+\, \mathcal{O}\left(t^{\,3}\right)$ and $F_2(t) \, = \, 1.79285 \,+\, 5.5594\ t \,+\, 12.2483\ t^{\,2} \,+\, \mathcal{O}\left(t^{\,3}\right)$.}, $F_1(t) \approx 1$ and $F_2(t) \approx \mu_p \,-\,1$ therefore $G_E(t) \approx 1$ and $G_M(t) \approx \mu_p$, are used, as is seen to be valid in the dipole model for the Sachs form factors 
\begin{equation}
\label{eq:Dipole_model_for_Sachs_form_factors}
G_E(t)\ = \ \frac{G_M(t)}{\mu_p} \ = \ \frac{1}{\left(\,1 \,-\, t/\Lambda^2\,\right)^{2}} \,,
\end{equation}
with $\Lambda^2 \,=\, 0.71 \ \mbox{(GeV/}c)^{\,2}$ obtained from a best fit to experimental data \cite{Sill:1992qw,Walker:1993vj}.

Spin filtering is an azimuthally symmetrical process, as will be explained in Chapters~\ref{ch:Polarization_buildup_by_spin_filtering} and \ref{ch:Various_scenarios_of_spin_filtering}, hence the transverse contributions will be averaged in what follows.

\pagebreak
\noindent
The Centre-of-Mass expressions presented throughout this chapter and the previous chapter can be transformed into invariants, or into the Laboratory reference frame using the following relations \cite{Byckling:1973}:
\begin{eqnarray}
\label{eq:Invariant_lambda_defined}
\lambda \ \equiv \ 4\,k^{\,2}\,s & = & \left[\,s\,-\,\left(\,m\,+\,M\,\right)^{\,2}\,\right]\left[\,s\,-\,\left(\,m\,-\,M\,\right)^{\,2}\,\right]  \,,\\[2ex]
\label{eq:Invariant_lambda_m_equals_M}
& = & s\,\left(\,s \,-\, 4\,M^{\,2}\,\right) \ \ \ \ \ \ \mbox{when} \ m\,=\,M \,,\\[2ex]
\label{eq:CMFrame_to_invariants}
\frac{\mathrm{d}\,\sigma}{\mathrm{d}\,t} & = & \frac{\pi}{k^{\,2}}\ \frac{\mathrm{d}\,\sigma}{\mathrm{d}\,\Omega} \ = \ \frac{4\,\pi\,s}{\lambda}\ \frac{\mathrm{d}\,\sigma}{\mathrm{d}\,\Omega} \,,\\[2ex]
\label{eq:CM_to_LAB_momenta}
k & = & \frac{M}{\sqrt{s}}\ p_\mathrm{lab}\,,\\[2ex]
\label{eq:t_to_CM_theta}
t & = & -\,2\,k^{\,2}\,\left(\,1 - \cos\theta\,\right) \ = \ -\,4\,k^{\,2}\,\sin^{\,2}\frac{\theta}{2} \,,
\end{eqnarray}
where $\theta$ is the Centre-of-Mass scattering angle and $\lambda$ is a Lorentz invariant.  Of interest in eq.~(\ref{eq:t_to_CM_theta}) are the particular cases that $t\,=\,0$ when $\theta \,=\, 0$ and $t\,=\,-\,4\,k^{\,2}\,$ when $\theta \,=\, \pi$, corresponding to total backward scattering.  One sees that for elastic scattering $t\,<\,0$ for all Centre-of-Mass scattering angles $\theta \,\neq\,0$.

\subsection{Antiproton - proton scattering}
\label{subsec:Antiproton_proton_scattering}
\noindent
In this section, where the masses of the two particles are equal, $\lambda$ is the invariant defined in eq.~(\ref{eq:Invariant_lambda_m_equals_M}).   The $\approx$ sign refers to the first term in the expansion in $t$.  To leading order in small $|\,t\,|$ the relevant spin observables for single photon exchange antiproton-proton scattering are:
\begin{eqnarray}
\label{eq:Pbar_p_leading_t_observables}
\frac{K_\mathrm{XX} \,+\, K_\mathrm{YY}}{2}\ \frac{\mathrm{d}\,\sigma}{\mathrm{d}\,t} &\approx &\displaystyle{\frac{4\,\pi\,\alpha^2\ M^{\,2}\ \mu_p^{\,2}}{\lambda\ t}} \,,\nonumber\\[1ex]
\frac{\left(1 - D_\mathrm{XX}\right) \,+\, \left(1 - D_\mathrm{YY}\right)}{2}\ \frac{\mathrm{d}\,\sigma}{\mathrm{d}\,t} &\approx &\!\!\!\displaystyle{ \frac{-\, \pi\,\alpha^2 \left(\lambda \,+\, 4\,M^{\,2}s\right)}{\lambda^2\ M^{\,2}\ s^2\ t}\,\left[\,2\,M^{\,2}s \,+\, \lambda \left(1 \,-  \,\mu_p\right)\, \right]^{\,2}} \,,\nonumber\\[1ex]
K_\mathrm{ZZ}\ \frac{\mathrm{d}\,\sigma}{\mathrm{d}\,t} & \approx & \displaystyle{\frac{4\ \pi \ \alpha^2 \ \mu_p^2}{s\ t\ \lambda}\ \left(\lambda \,+\, 2\,M^{\,2}\,s\right)} \,,\\[1ex]
\left(1 - D_\mathrm{ZZ}\right)\ \frac{\mathrm{d}\,\sigma}{\mathrm{d}\,t} & \approx &\!\!\!\displaystyle{ \frac{-\,2\,\pi\,\alpha^2 \left(\lambda \,+\, 4\,M^{\,2}s\right)}{\lambda^2\ M^{\,2}\ s^2\ t}\,\left[\,2\,M^{\,2}s \,+\, \lambda \left(1 \,-  \,\mu_p\right)\, \right]^{\,2}}.\nonumber
\end{eqnarray}
The leading $t$ approximation of the spin-averaged cross-section for this case is:
\begin{equation}
\label{eq:Pbar_p_spin_averaged_leading_t}
\frac{\mathrm{d}\,\sigma}{\mathrm{d}\,t} \ \approx \ \frac{4\, \pi \, \alpha^{\,2}}{\lambda}\ \frac{\left(\,s \,-\,2\,M^{\,2}\,\right)^{\,2}}{t^{\,2}} \,.
\end{equation}

\subsection{Antiproton - electron scattering}
\label{subsec:Antiproton_electron_scattering}
\noindent
In this section, where the masses of the two particles are not equal, $\lambda$ is the invariant defined in eq.~(\ref{eq:Invariant_lambda_defined}).  The leading $t$ terms of the relevant observables for antiproton-electron scattering are:
\begin{eqnarray}
\label{eq:Pbar_e_leading_t_observables}
\frac{K_\mathrm{XX} \,+\, K_\mathrm{YY}}{2}\ \frac{\mathrm{d}\,\sigma}{\mathrm{d}\,t} &\approx &\displaystyle{ \frac{\,4\, \pi \, \alpha^2\, m\, M\, \mu_p}{\lambda\ t}} \,,\nonumber\\[1ex]
\frac{\left(1 - D_\mathrm{XX}\right) \,+\, \left(1 - D_\mathrm{YY}\right)}{2}\ \frac{\mathrm{d}\,\sigma}{\mathrm{d}\,t} &\approx &\displaystyle{\frac{-\,4\, \pi \, m^2\,\alpha^2\,\left(\,s \,-\, m^2 \,+\, M^{\,2}\,\right)^2}{\lambda^2\ t}} \,,\nonumber\\[1ex]
K_\mathrm{ZZ}\ \frac{\mathrm{d}\,\sigma}{\mathrm{d}\,t} &\approx & \displaystyle{\frac{4\, \pi \, \alpha^2\, \mu_p}{\lambda\ t}\,\left(\,s \,-\, m^2 \,-\, M^{\,2}\,\right)} \,,\\[1ex]
\left(1 - D_\mathrm{ZZ}\right)\ \frac{\mathrm{d}\,\sigma}{\mathrm{d}\,t} &\approx &\displaystyle{\frac{-\,8\ \pi \ M^{\,2}\,\alpha^2\,\left(\,s \,+\, m^2 \,-\, M^{\,2}\,\right)^2}{\lambda^2\ t}} \,.\nonumber
\end{eqnarray}
The leading $t$ approximation of the spin-averaged cross-section for this case is:
\begin{equation}
\label{eq:Pbar_e_spin_averaged_leading_t}
\frac{\mathrm{d}\,\sigma}{\mathrm{d}\,t} \ \approx \ \frac{4\, \pi \, \alpha^{\,2}}{\lambda}\ \frac{\left(\,s\,-\,m^2\,-\,M^{\,2} \,\right)^{\,2}}{t^{\,2}}\,.
\end{equation}

\section{Spin 0 - spin 1 helicity amplitudes}
\label{sec:Spin_0_spin_1_helicity_amplitudes}

In this section we calculate the electromagnetic helicity amplitudes for spin 0 - spin 1 scattering\footnote{This is the only section of the thesis where the particles involved are not spin 1/2.}, generalizing the spin 0 - spin 1 helicity amplitudes presented to leading order in the low $|\,t\,|$ approximation in Ref.~\cite{Buttimore:2004vi}.  While not being directly used later in the thesis this calculation uses the formalism developed earlier, and would be applicable to the scattering of deuterons\footnote{The nucleus of a deuterium atom, a bound state consisting of a proton and a neutron, is called a deuteron.} (which are spin 1) off a carbon nucleus (spin 0).

A spin 1 particle has three possible spin states, represented by $-1$, $0$ and $+1$, while a spin 0 particle has only one spin state.  We represent the helicity amplitudes for spin 0 - spin 1 scattering as $\mathcal{M}(\,\lambda_b\,;\,\lambda_a\,)$ where $\lambda_a$ and $\lambda_b$ are the helicities of the incoming and outgoing spin 1 particle respectively, hence $\lambda_a\,,\lambda_b \in \left\{-,\,0,\,+\right\}$.
%Since the helicity of the spin 0 particle cannot change we omit it in the notation, which is shorthand for $\mathcal{M}(\,\lambda_0,\,\lambda_b\,;\,\lambda_0,\,\lambda_a\,)$, where $\lambda_0$ is the helicity of the spin 0 particle.   
Hence the complete set of spin 0 - spin 1 helicity amplitudes are:
\begin{eqnarray}
\label{eq:9_spin0-spin1_helicity_amplitudes}
\begin{tabular}{cccccccc}
1. & $\mathcal{M}(+\,;+)$ & \hspace*{1em} & 4. & $\mathcal{M}(-\,;+)$ & \hspace*{1em} & 7. & $\mathcal{M}(0\,;+)$ \\[1ex]
2. & $\mathcal{M}(+\,;-)$ & \hspace*{1em} & 5. & $\mathcal{M}(-\,;-)$ & \hspace*{1em} & 8. & $\mathcal{M}(0\,;-)$ \\[1ex]
3. & $\mathcal{M}(+\,;0)$ & \hspace*{1em} & 6. & $\mathcal{M}(-\,;0)$ & \hspace*{1em} & 9. & $\mathcal{M}(0\,;0)$ 
\end{tabular}
\end{eqnarray}
All definitions from section~\ref{sec:Helicity_amplitudes} are equally valid here and one obtains $\lambda - \mu \ =\  \left(\,\lambda_0 - \lambda_a\,\right) - \left(\,\lambda_0 - \lambda_b\,\right) \ = \ \lambda_b - \lambda_a$\,.  Hence Parity Invariance defined in eq.~(\ref{eq:Parity_invariance}) gives the following relations between the helicity amplitudes
\begin{equation}
\begin{tabular}{cclcc}
$\mathcal{M}(-\,;-)$ & $=$ & $(-1)^{\,1-1}\ \mathcal{M}(+\,;+)$    & $=$ & $\ \ \ \mathcal{M}(+\,;+)$ \\[1ex]
$\mathcal{M}(-\,;0)$ & $=$ & $(-1)^{\,1-0}\ \mathcal{M}(+\,;0)$    & $=$ & $-\,\mathcal{M}(+\,;0)$ \\[1ex]
$\mathcal{M}(0\,;-)$ & $=$ & $(-1)^{\,0-1}\ \mathcal{M}(0\,;+)$    & $=$ & $-\,\mathcal{M}(0\,;+)$ \\[1ex]
$\mathcal{M}(-\,;+)$ & $=$ & $(-1)^{\,1-(-1)}\ \mathcal{M}(+\,;-)$ & $=$ & $\ \  \mathcal{M}(+\,;-)$
\end{tabular}
\end{equation}
Thus reducing the nine helicity amplitudes to five independent ones.  Applying Time-reversal Invariance, defined in eq.~(\ref{eq:Time_reversal_invariance}), further reduces to four independent helicity amplitudes by the following relations:
\begin{equation}
\begin{tabular}{cclcc}
$\mathcal{M}(+\,;0)$ & $=$ & $(-1)^{\,1-0}\ \mathcal{M}(0\,;+)$    & $=$ & $-\,\mathcal{M}(0\,;+)$ \\[1ex]
$\mathcal{M}(-\,;0)$ & $=$ & $(-1)^{\,-1-0}\ \mathcal{M}(0\,;-)$    & $=$ & $-\,\mathcal{M}(0\,;-)$
\end{tabular}
\end{equation}
Now we can present the four independent helicity amplitudes for spin 0 - spin 1 scattering:
\begin{eqnarray}
\label{eq:4_independent_spin0-spin1_helicity_amplitudes}
H_1 & \equiv & \mathcal{M}(+\,;+) \ = \ \mathcal{M}(-\,;-) \nonumber \\[1ex]
H_2 & \equiv & \mathcal{M}(+\,;0) \ \ = \ \mathcal{M}(0\,;-) \ = \ -\,\mathcal{M}(0\,;+) \ = \ -\,\mathcal{M}(-\,;0) \nonumber\\[1ex]
H_3 & \equiv & \mathcal{M}(+\,;-) \ = \ \mathcal{M}(-\,;+) \\[1ex]
H_4 & \equiv & \mathcal{M}(0\,;0) \nonumber \,,
\end{eqnarray}
where it is seen that $H_1$ and $H_4$ are the non-spin-flip amplitudes, while $H_2$ and $H_3$ are the spin-flip amplitudes.  The non-spin-flip amplitudes, $H_1$ and $H_4$, are also known as {\it spin-elastic} amplitudes.  It can be seen from eq.~(\ref{eq:4_independent_spin0-spin1_helicity_amplitudes}) that $H_1$ and $H_3$ have multiplicity $2$, $H_2$ has multiplicity $4$ and $H_4$ has multiplicity $1$.  Thus one can write the spin-averaged differential cross-section in terms of the independent helicity amplitudes as
\begin{eqnarray}
\label{eq:Cross-section_to_independent_helicity_amplitudes_Spin0-Spin1}
s\,\frac{\mathrm{d}\,\sigma}{\mathrm{d}\,\Omega} & = & \frac{1}{\left(8\,\pi\right)^2} \ \sum_{\lambda_a\,\lambda_b}\  \frac{1}{\left(\,2\,s_A \,+\,1\,\right)\left(\,2\,s_B\,+\,1\,\right)}\ |\,\mathcal{M}(\,\lambda_b\,;\,\lambda_a\,)\,|^{\,2}\,, \nonumber \\[2ex] 
& = &  \frac{1}{\left(8\,\pi\right)^2}\ \sum_{\lambda_a\,\lambda_b} \ \frac{1}{3}\ |\,\mathcal{M}(\,\lambda_b\,;\,\lambda_a\,)\,|^{\,2}\,, \nonumber \\[2ex] 
& = & \frac{1}{3\left(8\,\pi\right)^2} \, \left(\,2\,|H_1|^{\,2} \,+\,4\,|H_2|^{\,2} \,+\,2\,|H_3|^{\,2} \,+\,|H_4|^{\,2}\,\right) \,,
\end{eqnarray}
where $s_A\,=\,0$ and $s_B\,=\,1$ are the spins of the two particles in the elastic scattering process.  The helicity amplitudes for elastic spin 0 - spin 1 electromagnetic scattering can be found by calculating
\begin{equation}
\label{eq:Calculating_Spin0-Spin1_helicity_amplitudes}
\mathcal{M}(\,\lambda_b\,;\,\lambda_a\,) \ = \ \frac{j_\mu \, J^\mu (\,\lambda_b\,;\,\lambda_a\,)}{q_\nu \,q^{\,\nu}} \,,
\end{equation}
where $j_\mu$ and $J^{\,\mu}$ are the spin 0 and spin 1 electromagnetic currents respectively, as defined below.

The spin 0 current is very simple as it has no helicity structure.  The electromagnetic current for a spin 0 particle of charge $Z\,e$ and form factor $F_0\!\left(q^2\right)$ is simply \cite{Weinberg:1995,Aitchison:2003}
\begin{equation}
\label{eq:Spin0_current}
j_\mu \ = \ Z\,e\,F_0\!\left(q^2\right)\, \left(\,p \,+\,p'\,\right)_\mu \,.
\end{equation}
%
%This is derived in Weinberg 1 page 543
%
Since the deuteron is a spin 1 object its electromagnetic structure is described by three form factors, charge monopole $G_C$, charge quadrupole $G_Q$ and magnetic dipole $G_M$, assuming parity and time-reversal invariance.

The most general form of the deuteron electromagnetic current, assuming Parity and Time-reversal invariance is \cite{Gourdin:1963,Gourdin:1974iq,Gilman:2001yh}:
\begin{eqnarray}
\label{eq:Deuteron_current}
J^{\,\mu}(\,\lambda_b\,;\,\lambda_a\,) & = & e \,\epsilon^*_\rho (\,p_b,\,\lambda_b\,) \left[\,F_1^{\,\mathrm{d}}\!\left(q^2\right)\,R^{\,\mu} \, \eta^{\,\rho\,\sigma} \ - \ \frac{F_2^{\,\mathrm{d}}\!\left(q^2\right)}{2\, M^{\,2}_d} \,R^{\,\mu}\,q^{\,\rho} \,q^{\,\sigma} \right. \nonumber \\[2ex]
& & \quad 
\left.- \ G_1^{\,\mathrm{d}}\!\left(q^2\right)\left(\,\eta^{\,\mu\,\rho}\,\eta^{\,\nu\,\sigma} \,-\, \eta^{\,\mu\,\sigma}\,\eta^{\,\nu\,\rho}\,\right)\,q_\nu 
\phantom{\frac{1}{2}} \hspace*{-0.7em}
\,\right] \epsilon_\sigma (\,p_a,\,\lambda_a\,) \,,
\end{eqnarray}
where $M_d$ is the mass of the deuteron, $R^{\,\mu} \,=\, p_a^{\,\mu} \,+\,p_b^{\,\mu}$ and $\epsilon_\sigma (\,p_a,\,\lambda_a\,)$ is a polarization 4-vector restricted to three independent components by the condition $p_a^{\,\mu} \,\epsilon_\mu (\,p_a,\,\lambda_a\,) \,=\,0$.  The quantities $F_1^{\,\mathrm{d}}$, $F_2^{\,\mathrm{d}}$ and $G_1^{\,\mathrm{d}}$ are the electromagnetic form factors of the deuteron with normalizations
\begin{equation}
\label{eq:Deuteron_form_factor_normalizations}
F_1^{\,\mathrm{d}}(0) \ = 1, \hspace*{3em} F_2^{\,\mathrm{d}}(0) \ = \ Q \,+\,\mu_d \,-\,1, \hspace*{3em} Q_1^{\,\mathrm{d}}(0) \ = \ \mu_d \,,
\end{equation}
where $Q$ is the quadrupole moment of the deuteron in units of $e/M_d^{\,2}$ and $\mu_d$ is the magnetic dipole moment of the deuteron in units of $e/2\,M_d$\,.

%Since we are interested in the region of low momentum transfer where electromagnetic effects dominate over hadronic effects, we will neglect the sub-leading $t$ behaviour of the deuteron current as:
%%
%\begin{eqnarray}
%\label{eq:Deuteron_current_leading_behaviour}
%J^{\,\mu}(\,\lambda_b\,;\,\lambda_a\,) & \approx & e \,\epsilon^*_\rho (\,p_b,\,\lambda_b\,) \left[\, - \ \frac{F_2^{\,\mathrm{d}}(q^2)}{2\, M^{\,2}_d} \,R^{\,\mu}\,q^{\,\rho} \,q^{\,\sigma} \,\right] \epsilon_\sigma (\,p_a,\,\lambda_a\,) \,,
%\end{eqnarray}
%
The complete set of initial and final deuteron polarization vectors, in the Centre-of-Mass (CM) frame, where the initial momentum of the deuteron is $p_a \,=\,\left(\,E_d,\,0,\,0,\,k\,\right)$ where $E_d \,=\,\sqrt{k^{\,2} \,+\,M_d^{\,2}}$ is the energy of the deuteron, are as follows \cite{Corbett:1984}:
\begin{eqnarray}
\label{eq:Initial_deuteron_polarization_vectors}
\hspace*{-4em}
 \mbox{Initial} \ \left\{ \ \ \ \ \begin{array}{ccl} \displaystyle{\epsilon^{\,\mu} (\,p_a\,,+)\ } & = & \displaystyle{\frac{1}{\sqrt{2}} \ \left(\,0,\,-1,\,-\,i,\,0\,\right)} \\[3ex]
\displaystyle{\epsilon^{\,\mu} (\,p_a\,,\, 0\, )\ } & = & \displaystyle{\frac{1}{M_d}\ \left(\,k,\,0,\,0,\,E_d\,\right)} \\[3ex]
\displaystyle{\epsilon^{\,\mu} (\,p_a\,,-)\ } & = & \displaystyle{\frac{1}{\sqrt{2}} \ \left(\,0,\,1,\,-\,i,\,0\,\right) }\end{array} \right.
\end{eqnarray}
\begin{eqnarray}
\label{eq:Final_deuteron_polarization_vectors}
\mbox{Final\ } \ \left\{ \ \ \ \  \begin{array}{ccl}
\displaystyle{{\epsilon^*}^\mu (\,p_b\,,+)} & = & \displaystyle{\frac{1}{\sqrt{2}} \  \left(\,0,\,-\cos\theta,\,i,\,\sin\theta\,\right)} \\[3ex]
\displaystyle{{\epsilon^*}^\mu (\,p_b\,,\, 0\, )} & = & \displaystyle{\frac{1}{M_d} \ \left(\,k,\,E_d \sin\theta,\,0,\,E_d \cos\theta\,\right)} \\[3ex]
\displaystyle{{\epsilon^*}^\mu (\,p_b\,,-)} & = & \displaystyle{\frac{1}{\sqrt{2}} \ \left(\,0,\,\cos\theta,\,i,\,-\sin\theta\,\right)} \end{array} \right.
\end{eqnarray}
where $\theta$ is the CM scattering angle and $i\,=\,\sqrt{-1}$.  The momentum transfer is
\begin{equation}
\label{eq:Spin0_Spin1_momentum_transfer}
q_{\,\mu} \ = \ \eta_{\,\mu\,\nu}\ q^{\,\nu} \ = \ \left(\,0,\,-\,k\sin\theta,\,0,\,k\,-\,k\cos\theta\,\right)\,.
\end{equation}
Combining all of the above into eq.~(\ref{eq:Calculating_Spin0-Spin1_helicity_amplitudes}) gives the helicity amplitudes for a deuteron of mass $M_d$
 colliding with a spin zero nucleus of charge $Z\,e$
 and electromagnetic form factor $F_0(t)$, as follows:
\begin{eqnarray}
\label{eq:H1_spin0-spin1_helicity_amplitude}
    \frac{H_1}{Z\, e^2 \,F_\mathrm{0}}
  & = &  
   \left[\, (s - u) \left( \frac{ -F^\mathrm{\,d}_1}{t} 
   \,+\,
   \frac{F^\mathrm{\,d}_2}{4\,M_d^{\,2}} \right) 
   \, + \, 
   G^\mathrm{\,d}_1 \,\right]
   \left( 1 \,+\, \frac{t}{4k^2} \right)\,, \\[7ex]
\label{eq:H2_spin0-spin1_helicity_amplitude}
   \frac{H_2}{Z\, e^2\, F_\mathrm{0}}
 &  = &
  \left\{ \sqrt{ \frac{1}{M_d^{\,2}} + \frac{1}{k^2} }\  \left[\,(s - u) \left( 		  -
   F^\mathrm{\,d}_1 \, 
   \,+\,
   F^\mathrm{\,d}_2 \,\frac{t}{4\,M_d^{\,2}} \right)
   \,+\,
   \,t \, G^\mathrm{\,d}_1 \,\right]  \right. \nonumber
\\[4ex] 
& & \left.   \qquad \qquad \qquad
\phantom{\sqrt{ \frac{1}{M_d^{\,2}} + \frac{1}{k^2} }}
+ \ 
   \frac{2\,k \,\sqrt{s}}{M_d}\  G^\mathrm{\,d}_1 \right\}
   \sqrt{ \,\frac{2}{-t} \, - \frac{1}{2\,k^2} }\,, \\[7ex]
\label{eq:H3_spin0-spin1_helicity_amplitude}
   \frac{H_3}{Z\, e^2\, F_\mathrm{0}}
   & = &
   F^\mathrm{\,d}_1 \, \frac{(s - u)}{4\,k^2}
   \ - \
   \left[\,F^\mathrm{\,d}_2 \frac{(s - u)}{4\,M_d^{\,2}} \ + \ G^\mathrm{\,d}_1 \,\right]
   \left(1 \,+\, \frac{t}{4\,k^2} \right) \,, \\[7ex]
\label{eq:H4_spin0-spin1_helicity_amplitude}
  \frac{H_4}{Z\, e^2\, F_\mathrm{0}}
 & = &
  -\,F^\mathrm{\,d}_1 \, \frac{(s - u)}{2}
 \left[\,
       \frac{2}{t} \,+\, \frac{1}{M_d^{\,2}} \,+\, \frac{1}{k^2}
 \,\right] \ + \ F^\mathrm{\,d}_2\,\frac{t \,(s - u)}{8\,M_d^{\,2}}
   \left( \frac{1}{M_d^{\,2}} \,+\, \frac{1}{k^2} \right)\nonumber
\\[4ex]
& &
  \qquad \qquad \qquad \qquad \qquad \!  + \ 
   2\,G^\mathrm{\,d}_1 \left[\,
%  
%   \phantom{\left(\frac{t^2}{k^2}\right)} \right.
%
   \frac{(s - u)}{4\,M_d^{\,2}} \,+\, \frac{t}{4\,k^2} \,+\, 1\, \right] \,.
%\;
%   \approx
%\;
%   F^\mathrm{\,d}_1 \, \frac{s}{t}
%\, .
\end{eqnarray}
In the above expressions the dependence on the mass of the spin-0 nucleus is in the $(\,s\,-\,u\,)$ terms.
%Note that many of these are off by factors of 2 or 4 or -1 from Nigel and Trueman's paper.  Perhaps try to find out which are correct.

%\pagebreak

\chapter{Polarization buildup by spin filtering}
\label{ch:Polarization_buildup_by_spin_filtering}

\vspace*{5ex}
\begin{minipage}{6cm}
\end{minipage}
\hfill
\begin{minipage}{10cm}
\begin{quote}
\emph{\lq\lq If, as I have reason to believe, I have disintegrated the nucleus of the atom, this is of greater significance than the war.\rq\rq}\\[2ex]
Ernest Rutherford, apologizing for absence from a meeting of the International Anti-submarine Warfare Committee.
\end{quote}
\end{minipage}
\vspace{8ex}

The theory of spin filtering is developed in this chapter.  A mathematical description of the related but simpler process of polarization buildup by the Sokolov-Ternov effect is first presented in section~\ref{sec:The_Sokolov-Ternov_Effect}.  The ideas presented are utilized in the mathematical descriptions of spin filtering which follow.  In section~\ref{sec:Polarization_evolution_equations} the rates of change of the number of particles in each spin state are combined into a set of polarization evolution equations which describe the process of polarization buildup by spin filtering.  This set of polarization evolution equations is then analyzed and solved in section~\ref{sec:Solving_the_polarization_evolution_equations}, emphasizing the physical implications of the dynamics.  

\pagebreak

\section{The Sokolov-Ternov effect}
\label{sec:The_Sokolov-Ternov_Effect}

The fact that an electron beam acquires a \lq self-polarization' due to the emission of synchrotron radiation in a storage ring is called the Sokolov-Ternov effect \cite{Sokolov:1963}, and has been described in section~\ref{subsec:Spontaneous_synchrotron_radiation_emission} of this thesis.  It turns out that one can describe the Sokolov-Ternov effect by a system of polarization evolutions equations very similar to a scenario of spin filtering when there is no scattering out of the ring.  The physical principles behind both systems are identical, the polarization buildup in both being induced by a discrepancy in the spin-flip transition rates.  The spin-flip in the Sokolov-Ternov effect is induced by synchrotron radiation as a result of the charged particle being bent in a magnetic field, whereas in spin filtering the spin-flip is induced by interactions with a polarized internal target.

In order to introduce the mathematics of  systems of polarization evolution equations we now present and solve a set of polarization evolution equations that describe the Sokolov-Ternov effect.  This will provide a stepping-stone to the mathematical description of spin filtering which follows later in the chapter.  The intermediate details of the calculations will be presented here, but omitted in later sections.  Please note as described in section~\ref{subsec:Spontaneous_synchrotron_radiation_emission} that this effect is much stronger for electrons than (anti)protons and the Sokolov-Ternov \lq self-polarization' is not a practical method to produce a polarized antiproton beam at present energies.

We denote the transition rates in this section by $W_{ab}$ instead of $\sigma_{ab}$, where $a,b \in \{+,-\}$, to distinguish from the spin filtering spin-flip case.  It is important to note that there is no target in the Sokolov-Ternov effect, which makes it physically different to spin filtering, although it can be described similarly.  The number of particles in the \lq spin up' and \lq spin down' states can change by two mechanisms while emitting synchrotron radiation: (1) \lq spin up' particles can be flipped to \lq spin down' particles, the cross-section for which we label as $W_{+-}$\,; and (2) \lq spin down' particles can be flipped to \lq spin up' particles, the cross-section for which we label as $W_{-+}$\,.  Mechanism (1) constitutes a decrease in the number of \lq spin up' particles ($N_+$) and an increase in the number of \lq spin down' particles ($N_-$), while mechanism (2) constitutes a decrease in the number of \lq spin down' particles and an increase in the number of \lq spin up' particles.  This explains the signs of the coefficients in eq.~(\ref{eq:Sokolov_Ternov_System}) which follows.

%Let $W_{+-}$ denote the spin-flip transition rate from the $+$ state to the $-$ state and $W_{-+}$ denote the spin-flip transition rate from $-$ state to the $+$ state.  
\pagebreak
\noindent
Therefore the Sokolov-Ternov effect can be described by the polarization evolution equations
\begin{eqnarray}
\label{eq:Sokolov_Ternov_System}
\frac{\mathrm{d}}{\mathrm{d}\,\tau}\left[\begin{array}{c} N_+ \\[2ex] N_-\end{array}\right] & = &\left[\begin{array}{ccc} - \,W_{+-} & W_{-+}\\[2ex] W_{+-} & - \,W_{-+}\end{array}\right] \ \left[\begin{array}{c} N_+ \\[2ex] N_-\end{array}\right] \,.
\end{eqnarray}
where $\tau$ is the time variable\footnote{We denote the time variable in each of the dynamical systems by $\tau$ to avoid confusion with the squared momentum transfer (Mandelstam $t$ variable) used throughout the thesis.}.

As long as the two spin-flip transition rates $W_{+-}$ and $W_{-+}$ are not equal, {\it i.e.}\ $W_{+-} \,\neq\,W_{-+}$, there will be a buildup of beam polarization over time.  It has been found that there is a slight difference between these rates, and this is the basis of the original Sokolov-Ternov idea \cite{Sokolov:1963}.

This system is identical to the system presented later in eq.~(\ref{eq:Meyers_No_Loss_System}), which describes spin filtering when there is no scattering out of the ring, except for the values of the matrix entries.  Solving this system gives eigenvalues $\lambda_1 = 0$ and $\lambda_2 = -\,\left(\,W_{-+} \,+\,W_{+-}\,\right)$, leading to eigenvectors
\begin{eqnarray*}
{\bf v}_1 = \left[\,\begin{array}{c} W_{-+} \\[2ex] W_{+-} \end{array}\,\right] 
\hspace*{3em} \mbox{and} \hspace*{3em}
{\bf v}_2 = \left[\,\begin{array}{c} 1 \\[2ex] -\,1 \end{array}\,\right] \,.
\end{eqnarray*}
And the solution to the system is
\begin{eqnarray*}
\left[\begin{array}{c} N_+(\tau) \\[2ex] N_-(\tau) \end{array}\right] & = & c_1\,{\bf v}_1\,e^{\lambda_1\,\tau} \ + \ c_2\,{\bf v}_2\,e^{\lambda_2\,\tau} \,,\\[2ex]
& = & c_1\,\left[\,\begin{array}{c} W_{-+} \\[2ex] W_{+-} \end{array}\,\right]\,e^{0} \ + \ c_2\,\left[\,\begin{array}{c} 1 \\[2ex] -\,1 \end{array}\,\right]\,e^{-\,\left(\,W_{-+} \,+\,W_{+-}\,\right)\,\tau} \,,
\end{eqnarray*}
where $c_1$ and $c_2$ are constants to be determined from the initial conditions, hence
\begin{eqnarray}
N_+(\tau) & = & c_1\,W_{-+} \ + \ c_2\,e^{-\,\left(\,W_{-+} \,+\,W_{+-}\,\right)\,\tau} \,, \nonumber \\[2ex]
N_-(\tau) & = & c_1\,W_{+-} \ - \ c_2\,e^{-\,\left(\,W_{-+} \,+\,W_{+-}\,\right)\,\tau} \,,
\end{eqnarray}
where one sees that $N_+(\tau) \, + \, N_-(\tau) \ = \ c_1\,\left(\,W_{-+} \,+\,W_{+-}\,\right) = \mbox{constant} \equiv N_0$, and also that $\mathrm{d}\,N_+\,/\,\mathrm{d}\,\tau = -\, \mathrm{d}\,N_-\,/\,\mathrm{d}\,\tau$ as it must be if there is no scattering out of the ring.  Imposing the initial conditions $N_+(0) \,=\, N_-(0) \,=\, N_0\,/\,2$, corresponding to a beam that is initially unpolarized, gives the constants
\begin{equation*}
c_1 \ = \ \frac{N_0}{W_{-+} \,+\,W_{+-} }
\hspace*{3em} \mbox{and} \hspace*{3em}
c_2 \ = \ \frac{N_0}{2}\,\frac{W_{+-} \,-\,W_{-+}}{W_{-+} \,+\,W_{+-}} \,,
\end{equation*}
and thus the complete solution
\begin{equation}
\label{eq:Sokolov_Ternov_solution}
\mathcal{P}(\tau) \ = \ \frac{N_+ \,-\,N_-}{N_+ \,+\,N_-} \ = \ \frac{W_{-+} \,-\,W_{+-}}{W_{-+} \,+\,W_{+-}}\ \left[\,1 \ - \ e^{-\,\left(\,W_{-+} \,+\,W_{+-}\,\right)\,\tau}\,\right] \,.
\end{equation}
\begin{figure}
\centering
\input{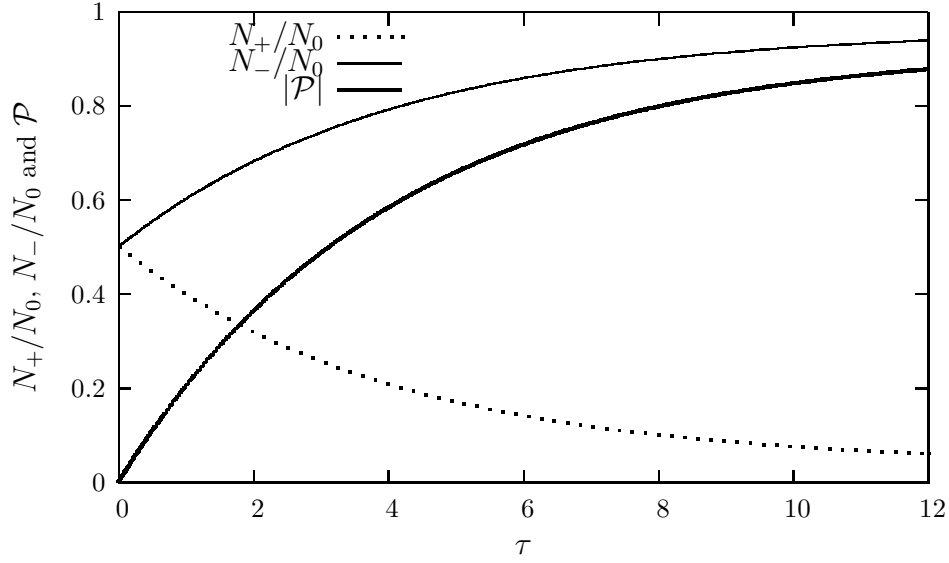} %x-range to 15
\caption{\small{\it{A schematic graph of the Sokolov-Ternov effect, showing that the number of particles in the \lq spin up' state ($N_+$) decreases with time while the number of particles in the \lq spin down' state ($N_-$) increases with time.  Hence $|\mathcal{P}| = |(N_+ - N_-)/(N_+ + N_-)|$, the absolute value of the beam polarization, increases with time.  One sees that the relations $\mathrm{d}N_+/\mathrm{d}\tau =-\,\mathrm{d}N_-/\mathrm{d}\tau$ and $N_+ + N_- = N_0$ are satisfied throughout.  The graph just shows general trends, therefore we do not specify the units of the time axis.
%The scaling factors for the N's cancel in the polarization hence dont effect the polarization curve.  See notes 1-2-08 on eps graph of this. 
}}}
\label{fig:Sokolov_Ternov_Plot}
\end{figure}
The spin-flip transition rates $W_{+-}$ and $W_{-+}$ are defined from the theory of synchrotron radiation as \cite{Lee:1997,Sokolov:1963}
\begin{equation}
\label{eq:Sokolov_Ternov_spin_flip_transition_rates}
W_{+-} \ = \ \frac{W_0}{2}\,\left(\,1 + \frac{8}{5\,\sqrt{3}}\,\right)
\hspace*{2em} \mbox{and} \hspace*{2em}
W_{-+} \ = \ \frac{W_0}{2}\,\left(\,1 - \frac{8}{5\,\sqrt{3}}\,\right) \,,
\end{equation}
where \cite{Sokolov:1963,Jackson:1975qi}
\begin{equation}
\label{eq:Sokolov_Ternov_time}
W_0 \ = \ \frac{5\,\sqrt{3}}{8}\,\frac{m\,\rho^{\,2}\,R}{r\,\hbar \, \gamma^5} \,,
\end{equation}
all quantities defined as in eq.~(\ref{eq:Sokolov_Ternov_Polarization_Time}) which in fact, as we shall show, is the reciprocal of the above equation.  Adding and subtracting the spin-flip transition rates gives
\begin{eqnarray}
W_{-+} \,+\,W_{+-} & = & W_0 \,, \nonumber \\[2ex]
W_{-+} \,-\,W_{+-} & = & \frac{-\,8}{5\,\sqrt{3}}\,W_0 \,.
\end{eqnarray}
One can now present the complete solutions for the number of particles in both the \lq spin up' and \lq spin down' states as a function of time $\tau$\,:
\begin{eqnarray}
N_+(\tau) & = & \frac{N_0}{2} \ + \ \frac{8\,N_0}{10\,\sqrt{3}}\,\left(\,e^{-\,W_0\,\tau} \ - \ 1\,\right) \,,\\[2ex]
N_-(\tau) & = & \frac{N_0}{2} \ - \ \frac{8\,N_0}{10\,\sqrt{3}}\,\left(\,e^{-\,W_0\,\tau} \ - \ 1\,\right) \,,
\end{eqnarray}
which can trivially be seen to satisfy $N(\tau) \,\equiv \,N_+(\tau) \,+\,N_-(\tau) \,=\,N_0 \,=\,\mbox{constant}$.

The steady state polarization (Sokolov-Ternov polarization) $\mathcal{P}_\mathrm{ST}$ is reached when $\tau > 1\,/\,\left(\,W_{-+} \,+\,W_{+-}\,\right) = W_0^{\,-1} \equiv \tau_\mathrm{\,ST}$ where $\tau_\mathrm{\,ST}$ is presented in eq.~(\ref{eq:Sokolov_Ternov_Polarization_Time})\footnote{For times well above $\tau = \tau_\mathrm{\,ST}$, $\left(\,1 \,-\, e^{-\,\tau\,/\,\tau_\mathrm{\,ST}}\,\right) \approx 1$.  The time $\tau_\mathrm{\,ST}$ depends strongly on the Lorentz $\gamma$ factor and the mean radius $R$ of the storage ring but is typically of the order of minutes or hours for electron storage rings, see Table~\ref{table:Properties_of_Synchrotrons}.}
\begin{equation}
\label{eq:Sokolov_Ternov_maximum_polarization}
\mathcal{P}_\mathrm{ST} \ = \ \frac{W_{-+} \,-\,W_{+-}}{W_{-+} \,+\,W_{+-}}  \ = \ \frac{-\,8}{5\,\sqrt{3}} \ \approx \ -\,0.924 \,,
\end{equation}
and one has the complete solution
\begin{equation}
\label{eq:Sokolov_Ternov_complete_solution}
\mathcal{P}(\tau) \ = \ \mathcal{P}_\mathrm{ST} \,\left[\,1 \ - \ e^{-\,W_0\,\tau}\,\right] \ = \ \frac{-\,8}{5\,\sqrt{3}}\,\left[\,1 \ - \ e^{-\,\tau\,/\,\tau_\mathrm{\,ST}}\,\right] \,.
\end{equation}
In a perfect ring one obtains an equilibrium polarization of $92.4\%$ after time $\tau_\mathrm{\,ST}$.  In practice the maximum polarization achieved is slightly less than this due to imperfections in the magnetic fields of real synchrotrons.  Some parameters, including $\tau_\mathrm{\,ST}$, of current and proposed future synchrotrons are presented in Table~\ref{table:Properties_of_Synchrotrons}.  The general behaviour of the number of particles in the \lq spin up' and \lq spin down' states, along with the polarization buildup, due to the Sokolov-Ternov effect are plotted in Figure~\ref{fig:Sokolov_Ternov_Plot}.

\pagebreak

\section{Polarization evolution equations}
\label{sec:Polarization_evolution_equations}

%Should I include ``for spin filtering'' in this section title????

\begin{figure}[!h]
\setlength{\unitlength}{1cm}
\begin{minipage}{7cm}
\includegraphics[width=7cm]{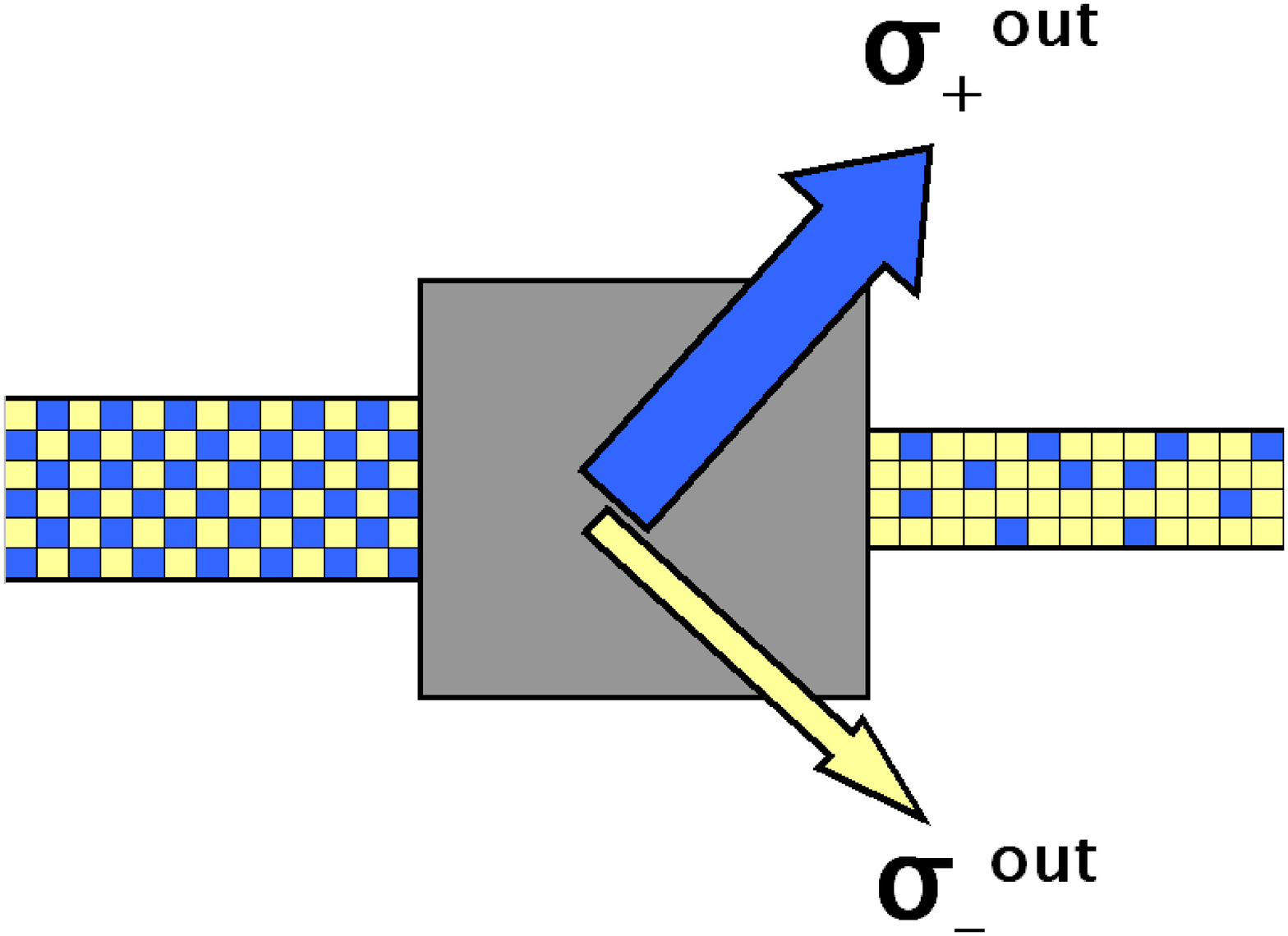}
\end{minipage}
\hfill
\begin{minipage}{7cm}
\includegraphics[width=7cm]{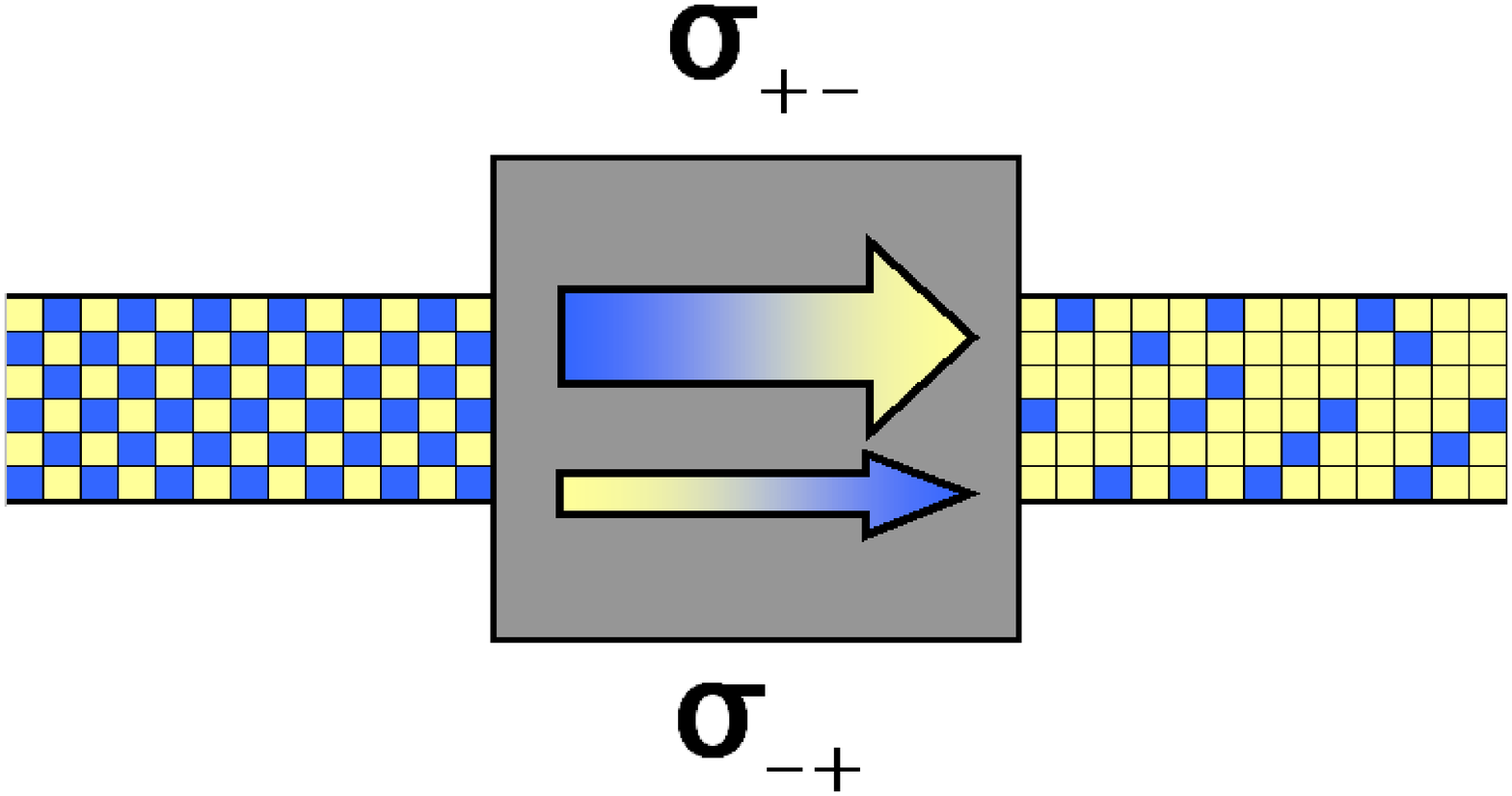}
\end{minipage}
\caption{\small{\it{These diagrams describing the two physical processes, selective scattering out of the beam (left) and selective spin-flip while remaining in the beam (right), that contribute to polarization buildup by spin filtering in a storage ring have been explained in Figure~\ref{fig:Spin_Filtering_Diagrams}.  Here we label the cross-section for a particle in the \lq spin up' state to be scattered out of the beam as $\sigma^\mathrm{\,out}_+$, the cross-section for a particle in the \lq spin down' state to be scattered out of the beam as $\sigma^\mathrm{\,out}_-$, the cross-section for a particle in the \lq spin up' state to be flipped to the \lq spin down' state while remaining in the beam as $\sigma_{+-}$\, and the cross-section for a particle in the \lq spin down' state to be flipped to the \lq spin up' state while remaining in the beam as $\sigma_{-+}$.  In order for each of these processes to contribute to polarization buildup of the beam we must have $\sigma^\mathrm{\,out}_+ \, \neq \, \sigma^\mathrm{\,out}_-$ and $ \sigma_{+-}\, \neq \, \sigma_{-+}$ respectively.  These cross-sections will be used below in the mathematical evolution equations to describe the rate of buildup of polarization by spin filtering.}}}
\label{fig:Spin_Filtering_Diagrams_With_Symbols}
\end{figure}
In this section we develop sets of differential equations that describe the buildup of polarization of an antiproton beam by spin filtering.  Consistency checks are then performed on the systems of equations, which provides a chance to highlight the underlying physical phenomena under investigation.  The method of polarization buildup by spin filtering has been outlined in section~\ref{subsec:The_theory_of_spin_filtering}, which the reader may wish to read again before continuing here.  Since the cross-sections for an interaction between a beam particle and a particle in the target are low we neglect the effects of multiple scattering, which has a cross-section orders of magnitude smaller than single scattering.  Hence we consider two possibilities each time the beam passes through the target: (1) a beam particle can pass through the target without interaction, or (2) a beam particle can scatter off at most one of the target particles.  
%The former case is dominant as the single interaction case is suppressed by the coupling constant, for example in QED the single scattering case is suppressed by a factor of the {\it electromagnetic coupling constant} (also called the {\it fine structure constant}) $\alpha \,\approx\,1\,/\,137$.  

Interactions with residual gas due to a non-perfect vacuum in the storage ring can also be neglected as the density of the internal target is much higher than the density of the residual gas.  The high density of the target, and the fact that it is a constantly replenished gas jet, ensures that there is no significant target depolarization.

Recall the two physical processes that contribute to spin filtering shown in Figure~\ref{fig:Spin_Filtering_Diagrams_With_Symbols}.  The number of particles in the \lq spin up' state can change by three means: (1) \lq spin up' particles being scattered out of the beam, the cross-section for which we label as $\sigma^\mathrm{\,out}_+$, (2) \lq spin up' particles being flipped to \lq spin down' particles while remaining in the beam, the cross-section for which we label as $\sigma_{+-}$, and (3) \lq spin down' particles being flipped to \lq spin up' particles while remaining in the beam, the cross-section for which we label as $\sigma_{-+}$.  Mechanisms (1) and (2) constitute a decrease in the number of \lq spin up' particles and (3) constitutes an increase in the number of \lq spin up' particles.  This explains the signs of the coefficients in eq.~(\ref{eq:Meyers_System}) which follows.  Correspondingly the number of particles in the \lq spin down' state can also change by three means: (1) \lq spin down' particles being scattered out of the beam, the cross-section for which we label as $\sigma^\mathrm{\,out}_-$, (2) \lq spin down' particles being flipped to \lq spin up' particles while remaining in the beam ($\sigma_{-+}$) and (3) \lq spin up' particles being flipped to \lq spin down' particles while remaining in the beam ($\sigma_{+-}$).  All of this can be expressed in the following set of polarization evolution equations:
\begin{eqnarray}
\label{eq:Meyers_System}
\frac{\mathrm{d}}{\mathrm{d}\,\tau}\left[\begin{array}{c} N_+ \\[2ex] N_-\end{array}\right] & = & - \,n\,\nu \ \left[\begin{array}{ccc} \sigma^\mathrm{\ out}_+ \,+ \,\sigma_{+-} & -\,\sigma_{-+}\\[2ex] - \,\sigma_{+-} & \sigma^\mathrm{\ out}_- \,+\, \sigma_{-+}\end{array}\right] \ \left[\begin{array}{c} N_+ \\[2ex] N_-\end{array}\right] \,,
\end{eqnarray}
where $\tau$ is the time variable, $n$ is the areal density of the target, $\nu$ is the revolution frequency of the beam and $N_+(\tau)$ and $N_-(\tau)$ are the number of beam particles in the \lq spin up' and \lq spin down' states at time $\tau$ respectively.

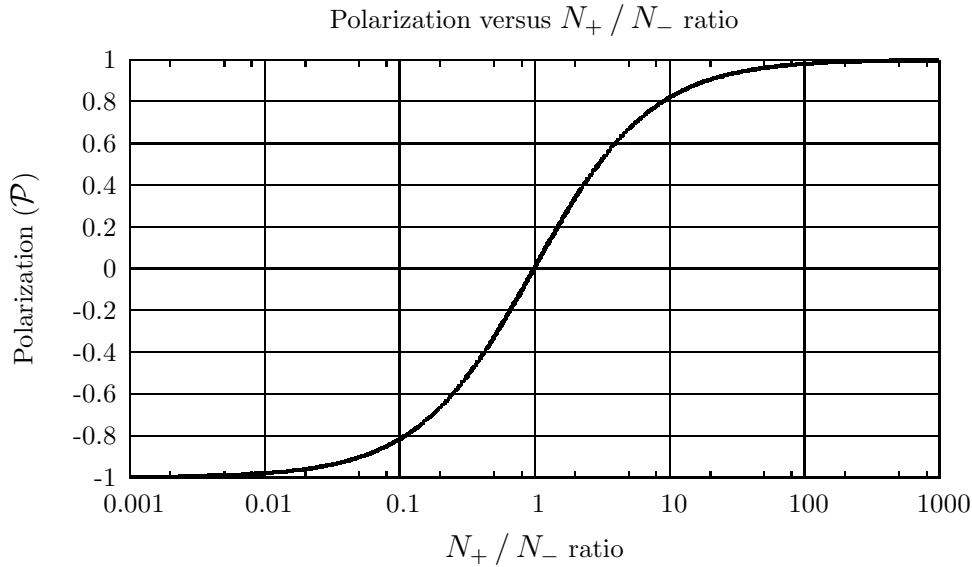
\begin{figure}[!h]
\begin{minipage}{14cm}
\begin{center}
\input{N+N-Ratio_V_Polarization_Plot.tex}
\end{center}
\vspace*{2ex}
\end{minipage}
\vfill
\begin{minipage}{14cm}
\begin{center}
\hspace*{1em}
\begin{tabular}{|c||c|c|c|c|c|c|}
\hline
 & & & & & &  \\*[-2ex]
 $N_+ \ =$                         &  $N_-$  &  $2\,N_-$  & $3\,N_-$  & $10\,N_-$ &  $19\,N_-$ & $100\,N_-$ \\[0.5ex]
\hline
 & & & & & &  \\*[-2ex]
 Polarization ($\mathcal{P}$) $=$  &  $0$    & $0.33$   & $0.5$  &  $0.818$   & $0.9$      &  $0.98$ \\[0.5ex]
\hline
\end{tabular}
\end{center}
\vspace*{1ex}
\end{minipage}
\caption{\small{\it{This graph, from equation~\ref{eq:Polarization_definition_J/N}, shows how the polarization ($\mathcal{P}$) changes as the ratio ($N_+\,/\,N_-$) of the number of particles in the \lq spin up' state to the number of particles in the \lq spin down' state changes.  The horizontal axis is a log scale.  The change in polarization as the number of particles in each of the spin states changes is very important throughout the thesis, so we highlight a few points in the table.  The second last entry in the table is particularly relevant as the target used in our numerical calculations is $90\%$ polarized.  Note that $0\,\leq\, |\mathcal{P}|\,\leq \, 1$ and when $N_- \,>\,N_+$ the polarization is defined to be negative, {\it i.e.}\ $-1\,\leq\,\mathcal{P}\,<\,0$.}}}
\label{fig:N+N-Ratio_V_Polarization_Plot}
\end{figure}

For a beam that is initially unpolarized one imposes the following initial conditions
\begin{equation}
\label{eq:Unpolarized_initial_conditions}
N_+(0) \ = \ N_-(0) \ = \ \frac{N_0}{2} \,,
\end{equation}
where $N_0$ is the total number of particles in the beam initially.  We define the {\it beam intensity} $N(\tau) \equiv N_+(\tau) \, + \, N_-(\tau)$ as the total number of particles in the beam at time $\tau$, and the {\it beam total spin} at time $\tau$ as $J(\tau) \equiv N_+(\tau) \, - \, N_-(\tau)$ so that the polarization of the beam at time $\tau$ is simply given by
\begin{equation}
\label{eq:Polarization_definition_J/N}
\mathcal{P}(\tau) \ = \ \frac{N_+(\tau) \, - \, N_-(\tau)}{N_+(\tau) \, + \, N_-(\tau)} \ = \ \frac{J(\tau)}{N(\tau)} \,.
\end{equation}
The change in beam polarization as the number of particles in each of the spin states changes is very important throughout the thesis, so we plot this dependence and highlight a few points in Figure~\ref{fig:N+N-Ratio_V_Polarization_Plot}.

\begin{figure}
\centering
\input{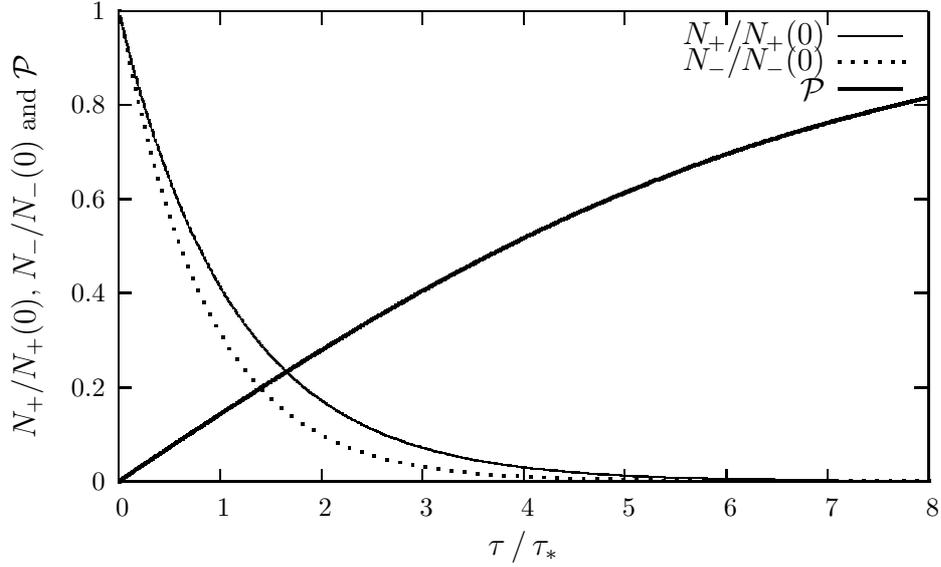} %x-range only to 8, which is better in this case
\caption{\small{\it{A schematic graph showing that the number of particles in the \lq spin up' ($N_+$) and \lq spin down' ($N_-$) states each decrease with time, but at different rates.  Hence the beam polarization $\mathcal{P} =(N_+ -N_-)/(N_+ +N_-)$ increases with time.  The time axis is scaled by the beam lifetime $\tau_*$, as described in section~\ref{subsec:Beam_lifetime_and_figure_of_merit}.
%The scaling factors for the N's cancel in the polarization hence dont effect the polarization curve.  See notes 1-2-08 on eps graph of this. 
}}
}
\label{fig:N+N-P_Plot}
\end{figure}

A problem with spin filtering where particles are scattered out of the beam is that while the beam polarization increases the beam intensity decreases.  We propose possible solutions to this problem in Chapter~\ref{ch:Various_scenarios_of_spin_filtering}.  The behaviour of the number of particles in the \lq spin up' and \lq spin down' states, along with the polarization, as time increases is shown in Figure~\ref{fig:N+N-P_Plot}.  

Note some treatments of spin filtering investigate a scenario where no particles are scattered out of the beam, {\it i.e.}\ the maximum scattering angle for the process is less than the ring acceptance angle, which is the case for antiprotons scattering off electrons in an atomic target \cite{Horowitz:1994,Meyer:1994,Milstein:2005bx,Nikolaev:2006gw,O'Brien:2007hu} and for antiprotons scattering off a co-moving beam of electrons or positrons \cite{Walcher:2007sj}.  In these scenarios only selective spin-flip can contribute to polarization buildup, and one avoids the problem of decreasing beam intensity.  The low density of the targets currently available causes the rate of polarization buildup using these methods to be slow, but the enhanced cross-sections at low energies suggested in Refs.~\cite{Walcher:2007sj,Arenhovel:2007gi,Milstein:2008tc} may counteract this difficulty.  We analyze such systems later in the thesis.

Before solving this system of polarization evolution equations we shall prove a number of short lemma's providing a consistency check that the equations accurately describe the physical phenomena we wish to model.  This also provides a chance to highlight the dynamical properties of the physical system, as this plays a major role in the rest of the thesis.
\\
\\
{\bf Lemma 1}
\\
\\
{\it If $\sigma^\mathrm{\,out}_+ \, = \, \sigma^\mathrm{\,out}_-$ and $\sigma_{+-}\, = \, \sigma_{-+}$ there will be no buildup of beam polarization, but there will still be loss of beam intensity $N(\tau)$.}

%\pagebreak
%
\noindent
{\bf \emph{Proof:}}
\\
\\
When $\sigma^\mathrm{\,out}_+ \, = \, \sigma^\mathrm{\,out}_-$ and $\sigma_{+-}\, = \, \sigma_{-+}$ the polarization evolution equations reduce to
\begin{eqnarray}
\label{eq:System_with_Unpolarized_target}
\frac{\mathrm{d}}{\mathrm{d}\,\tau}\left[\begin{array}{c} N_+ \\[2ex] N_-\end{array}\right] & = & - \,n\,\nu \ \left[\begin{array}{ccc} \sigma^\mathrm{\ out}_+ \,+ \,\sigma_{+-} & -\,\sigma_{+-}\\[2ex] - \,\sigma_{+-} & \sigma^\mathrm{\ out}_+ \,+\, \sigma_{+-}\end{array}\right] \ \left[\begin{array}{c} N_+ \\[2ex] N_-\end{array}\right] \,,
\end{eqnarray}
{\it i.e.}\
\begin{eqnarray}
%\label{eq:Meyers_No_Loss_System2}
\frac{\mathrm{d}\,N_+}{\mathrm{d}\,\tau} & = & - \,n\,\nu \ \left[\,\left(\,\sigma^\mathrm{\ out}_+ \,+\sigma_{+-}\,\right)\,N_+ \ - \ \sigma_{+-}\,N_-\,\right] \,,\nonumber \\[2ex]
\frac{\mathrm{d}\,N_-}{\mathrm{d}\,\tau} & = & - \,n\,\nu \ \left[\,-\,\sigma_{+-}\,N_+ \ + \ \left(\,\sigma^\mathrm{\ out}_+ \,+\, \sigma_{+-}\,\right)\,N_-\,\right] \,,
\end{eqnarray}
which can be added and subtracted to give
\begin{eqnarray}
\label{eq:System_with_Unpolarized_target1}
\frac{\mathrm{d}\,N(\tau)}{\mathrm{d}\,\tau} & = &  \, - \, n \, \nu \,\sigma^\mathrm{\ out}_+ \,N(\tau) \,,\nonumber \\[2ex]
\frac{\mathrm{d}\,J(\tau)}{\mathrm{d}\,\tau} & = &  \, - \, n \, \nu \,\left(\,\sigma^\mathrm{\ out}_+ \, + \,2\,\sigma_{+-} \,\right)\,J(\tau)\,,
\end{eqnarray}
two uncoupled first order separable ODE's which can be integrated to give the solutions
\begin{eqnarray}
\label{eq:System_with_Unpolarized_target2}
N(\tau) & = & N(0)\ \displaystyle{e^{\,-\,n\,\nu\,\sigma^\mathrm{\ out}_+\,\tau}} \ = \ N_0\ \displaystyle{e^{\,-\,n\,\nu\,\sigma^\mathrm{\ out}_+\,\tau}} \,,\nonumber\\[2ex]
J(\tau) & = & J(0)\ \displaystyle{e^{\,-\,n\,\nu\,\left(\,\sigma^\mathrm{\ out}_+ \, + \,2\,\sigma_{+-} \,\right)\,\tau}} \,,
\end{eqnarray}
and we see that $N(\tau)$ will decrease exponentially, provided that $\sigma^\mathrm{\ out}_+ \neq 0$, and $J(\tau)$ which is zero initially will always remain zero ({\it i.e.}\ if $J(0) = 0$ then $J(\tau) = 0$ for all $\tau$).  So there will be no polarization buildup.  Also in the case when the beam is initially polarized ($J(0) \neq 0$) its polarization will decrease exponentially to zero, remembering the cross-sections are positive quantities.   \ $\Box$

We later show that when the internal target is not polarized $\sigma^\mathrm{\,out}_+ \, = \, \sigma^\mathrm{\,out}_-$ and $\sigma_{+-}\, = \, \sigma_{-+}$.  Thus there will be no polarization buildup by spin filtering if the internal target is unpolarized.
\\
\\
{\bf Lemma 2}
\\
\\
{\it If $\sigma^\mathrm{\,out}_+ \, = \, \sigma^\mathrm{\,out}_- \, = \,0$ there will be no loss of beam intensity $N(\tau)= \mbox{Constant} = N_0$, but there may still be polarization buildup.}
\\
\\
{\bf \emph{Proof:}}
\\
\\
When $\sigma^\mathrm{\,out}_+ \,=\, \sigma^\mathrm{\,out}_- \,=\, 0 $ (this happens when the maximum scattering angle for the process is less than the ring acceptance angle) the polarization evolution equations reduce to
\begin{eqnarray}
\label{eq:Meyers_No_Loss_System}
\frac{\mathrm{d}}{\mathrm{d}\,\tau}\left[\begin{array}{c} N_+ \\[2ex] N_-\end{array}\right] & = & - \,n\,\nu \ \left[\begin{array}{ccc} \sigma_{+-} & -\,\sigma_{-+}\\[2ex] - \,\sigma_{+-} & \sigma_{-+}\end{array}\right] \ \left[\begin{array}{c} N_+ \\[2ex] N_-\end{array}\right] \,,
\end{eqnarray}
{\it i.e.}\ 
\begin{eqnarray}
\label{eq:Meyers_No_Loss_System3}
\frac{\mathrm{d}\,N_+}{\mathrm{d}\,\tau} & = & - \,n\,\nu \ \left(\,\sigma_{+-}\,N_+ \ - \ \sigma_{-+}\,N_-\,\right) \,,\nonumber \\[2ex]
\frac{\mathrm{d}\,N_-}{\mathrm{d}\,\tau} & = & - \,n\,\nu \ \left(\,-\,\sigma_{+-}\,N_+ \ + \ \sigma_{-+}\,N_-\,\right) \,,
\end{eqnarray}
and adding these gives
\begin{eqnarray*}
\frac{\mathrm{d}\,N}{\mathrm{d}\,\tau} & = & \frac{\mathrm{d}}{\mathrm{d}\,\tau}\,\left(\,N_+ \ + \ N_-\,\right) \ = \ \frac{\mathrm{d}\,N_+}{\mathrm{d}\,\tau} \ + \ \frac{\mathrm{d}\,N_-}{\mathrm{d}\,\tau} \,,\\[2ex]
& = & - \,n\,\nu \ \left(\,\sigma_{+-}\,N_+ \ - \ \sigma_{-+}\,N_- \ - \ \sigma_{+-}\,N_+ \ + \ \sigma_{-+}\,N_-\,\right) \ = \ 0\,.
\end{eqnarray*}
\begin{figure}
\centering
\input{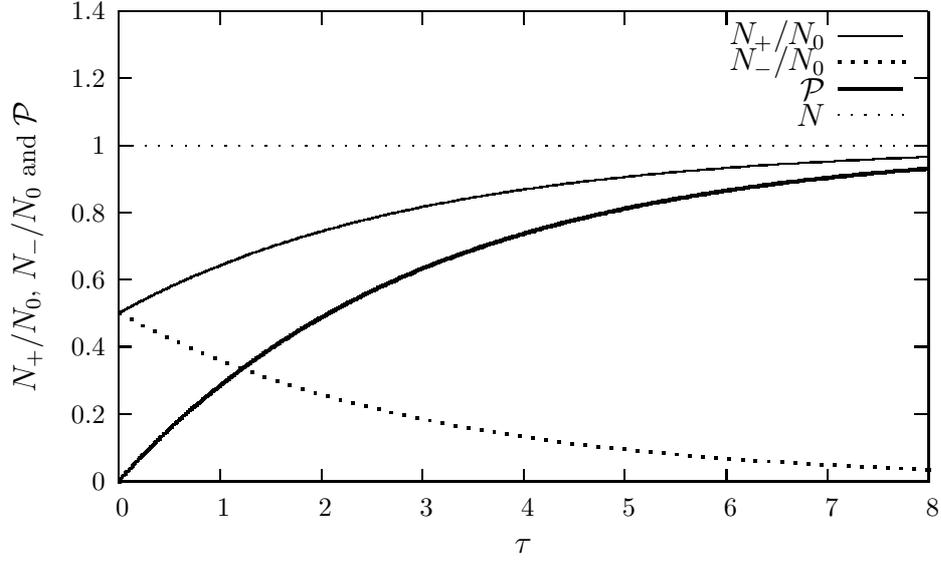}
\caption{\small{\it{A schematic graph of the system treated in Lemma 2 and Corollary 1, showing that the number of particles in the \lq spin up' state ($N_+$) increases with time while the number of particles in the \lq spin down' state ($N_-$) decreases with time.  Hence the beam polarization $\mathcal{P} = (N_+ - N_-)/(N_+ + N_-)$ increases with time.  One sees that the relations $\mathrm{d}N_+/\mathrm{d}\tau =-\,\mathrm{d}N_-/\mathrm{d}\tau$ and $N = N_+ + N_- = N_0$ are satisfied throughout.  The graph just shows general trends, therefore we do not define the units of the time axis.
%The scaling factors for the N's cancel in the polarization hence dont effect the polarization curve.  See notes 1-2-08 on eps graph of this. 
}}
}
\label{fig:N+N-P_Plot_no_loss}
\end{figure}
So we have $\mathrm{d}\,/\,\mathrm{d}\,\tau \ \left[\,N_+(\tau) \ + \ N_-(\tau)\,\right] = 0$ which implies $N_+(\tau) \ + \ N_-(\tau) = \mbox{constant} = N(0)$.  Thus there will be no loss of particles, as expected.  Subtracting eqs.~(\ref{eq:Meyers_No_Loss_System3}) from each other gives
\begin{eqnarray*}
\frac{\mathrm{d}\,J}{\mathrm{d}\,\tau} & = & \frac{\mathrm{d}}{\mathrm{d}\,\tau}\,\left(\,N_+ \ - \ N_-\,\right) \ \,=\, \ \frac{\mathrm{d}\,N_+}{\mathrm{d}\,\tau} \ - \ \frac{\mathrm{d}\,N_-}{\mathrm{d}\,\tau} \,,\\[2ex]
& = & - \,2\,n\,\nu \ \left(\,\sigma_{+-}\,N_+ \ - \ \sigma_{-+}\,N_-\,\right)\,,
\end{eqnarray*}
which leads to a non-zero $J(\tau)$ ({\it i.e.}\ non-zero polarization) provided that $\sigma_{+-} \,\neq\, \sigma_{-+}$. \ $\Box$

The system without scattering out of the ring described in eq.~(\ref{eq:Meyers_No_Loss_System}) is very similar to the system which describes the Sokolov-Ternov effect presented in section~\ref{sec:The_Sokolov-Ternov_Effect}.  The systems describing these two physical processes should be similar as they are both governed solely by spin-flip transitions.  In spin filtering the spin-flip transitions are induced by scattering off the polarized internal target while in the Sokolov-Ternov effect the spin-flip transitions are induced by spontaneous synchrotron radiation emission of photons while the charged particles of the beam are being bent in the magnetic field of the ring.  In fact these systems are identical except for the interpretations of the matrix entries, and that because there is no target in the Sokolov-Ternov effect, the system of equations describing it does not depend on a target areal density $n$.  
%While electrons and positrons circulating in a storage ring acquire a useful polarization in a relatively short time due to the Sokolov-Ternov effect \cite{Lee:1997,Leader:2005}, an unrealistically large time would be required for protons or antiprotons in a storage ring at currently achievable energies, due to their high mass \cite{Krisch:1986nt}.
  The solution of the system presented in eq.~(\ref{eq:Meyers_No_Loss_System}) is identical to the solution of the Sokolov-Ternov system presented in detail in section~\ref{sec:The_Sokolov-Ternov_Effect}. 
\\
\\
{\bf Corollary 1}
\\
\\
{\it When there is no scattering out of the beam, {\it i.e.}\ $\sigma^\mathrm{\,out}_+ \, = \, \sigma^\mathrm{\,out}_- \, = \,0$, the condition
\begin{equation}
\label{eq:No_Loss_Conservation_of_N}
\frac{\mathrm{d}\,N_+}{\mathrm{d}\,\tau} \ = \ - \, \frac{\mathrm{d}\,N_-}{\mathrm{d}\,\tau} \,,
\end{equation}
must be satisfied.}
\\
\\
{\bf \emph{Proof:}}
\\
\\
We have shown that when there is no scattering out of the ring the equations reduce to eqs.~(\ref{eq:Meyers_No_Loss_System3}) which can immediately be seen to satisfy eq.~(\ref{eq:No_Loss_Conservation_of_N}). \ $\Box$
%\\

\pagebreak
\noindent
{\bf Lemma 3}
\\
\\
{\it When there is no spin-flip, {\it i.e.}\ $\sigma_{+-}\, = \, \sigma_{-+}\, = \,0$, the change in one spin state should not depend on the number of particles in the other spin state, {\it i.e.}\ the equations should decouple.}
\\
\\
{\bf \emph{Proof:}}
\\
\\
When there is no spin-flip $\sigma_{+-}\, = \, \sigma_{-+}\, = \,0$ and thus the polarization evolution equations reduce to
\begin{eqnarray}
\label{eq:No_Spin_Flip_System}
\frac{\mathrm{d}\,N_+}{\mathrm{d}\,\tau} & = & -\,n\,\nu\,\sigma^\mathrm{\ out}_+ \,N_+  \,,\nonumber \\[2ex]
 \frac{\mathrm{d}\,N_-}{\mathrm{d}\,\tau} & = & -\,n\,\nu\,\sigma^\mathrm{\ out}_- \,N_- \,,
\end{eqnarray}
which is an uncoupled system of equations as required. \ $\Box$

The solutions of the above equations, for an initially unpolarized beam, are easily found to be
\begin{equation}
\label{eq:Solutions_to_No_Spin_Flip_System}
N_+(\tau) \ = \ \frac{N_0}{2}\ \displaystyle{e^{-\,n\,\nu\,\sigma^\mathrm{\ out}_+\,\tau} }
\hspace{2em} \mbox{and} \hspace{2em}
N_-(\tau) \ = \ \frac{N_0}{2}\ \displaystyle{e^{-\,n\,\nu\,\sigma^\mathrm{\ out}_-\,\tau}} \ .
\end{equation}
One sees that if in addition $\sigma^\mathrm{\,out}_+ \,=\,\sigma^\mathrm{\,out}_-$ in eqs.~(\ref{eq:Solutions_to_No_Spin_Flip_System}) then no polarization buildup occurs, as shown in Lemma 1.  It is claimed by the Budker-J\"ulich groups that the spin-flip transition rates are negligible for antiprotons scattering off polarized electrons in a hydrogen target \cite{Milstein:2005bx,Nikolaev:2006gw}.  As we show in Chapter~\ref{ch:Numerical_results}, the maximum scattering angle of antiprotons scattering off atomic electrons is $0.54 \ \mbox{mrad}$, below the acceptance angle of a typical storage ring, thus there is no scattering out of the beam.  Since there is no scattering out of the beam, and spin-flip transitions are negligible the Budker-J\"ulich groups conclude that polarized electrons in an atomic target are not effective in transferring polarization to an antiproton beam by spin filtering \cite{Milstein:2005bx,Nikolaev:2006gw}.  To force some antiprotons to be scattered out of the beam, and to avoid the problem of loss of beam intensity due to antiprotons annihilating with the protons in an atomic target, we suggest to use an opposing polarized electron beam of sufficient energy to scatter some antiprotons beyond the ring acceptance angle \cite{O'Brien:2007hu}.  Such a system is treated in Chapter~\ref{ch:Numerical_results}.  The rate of polarization buildup using this method is slow due to the low densities of polarized electron beams currently available \cite{O'Brien:2007hu}, but the enhanced cross-sections at low energies suggested in Refs.~\cite{Walcher:2007sj,Arenhovel:2007gi,Milstein:2008tc} may counteract this difficulty.

Differentiating eq.~(\ref{eq:Polarization_definition_J/N}) and rearranging gives
\begin{equation}
\label{eq:Differentiating_polarization_definition}
\frac{\mathrm{d}\,\mathcal{P}}{\mathrm{d}\,\tau} \ = \ \frac{1 - \mathcal{P}^{\,2}}{2}\,\left(\,\frac{1}{N_+}\,\frac{\mathrm{d}\,N_+}{\mathrm{d}\,\tau} \ - \ \frac{1}{N_-}\,\frac{\mathrm{d}\,N_-}{\mathrm{d}\,\tau}\,\right)\,,
\end{equation}
and on substituting in eqs.~(\ref{eq:No_Spin_Flip_System}) one obtains
\begin{equation}
\label{eq:Naive_1994_differential_equation}
\frac{\mathrm{d}\,\mathcal{P}}{\mathrm{d}\,\tau} \ = \ \frac{-\,n\,\nu}{2}\, \left(\,1 \ -\ \mathcal{P}^{\,2}\,\right)\,\left(\,\sigma^\mathrm{\ out}_+ \ - \ \sigma^\mathrm{\ out}_-\,\right)\,,
\end{equation}
which is exactly the important equation (3) from Ref.~\cite{Meyer:1994}, the first theoretical description of the FILTEX results.  Integrating the above equation, and using a result from the next section eq.~(\ref{eq:Loss_cross_section_difference}), leads to
\begin{eqnarray}
\label{eq:Solution_of_Naive_1994_system}
\mathcal{P}(\tau) & = & \tanh \left[\,\frac{-\,n\,\nu}{2}\,\left(\,\sigma^\mathrm{\ out}_+ \ - \ \sigma^\mathrm{\ out}_-\,\right)\tau\,\right]\,, \nonumber \\[2ex]
 & = & \tanh \left(\,-\,n\,\nu\,\mathcal{P}_T\,A_\mathrm{\,out}\,\tau\,\right)\,.
\end{eqnarray}
Which was proposed initially as a model of the rate of polarization buildup in spin filtering \cite{Rathmann:1993xf,Meyer:1994}.  Since then the importance of scattering within the beam has been highlighted and a more complex treatment of spin filtering is required, involving the polarization transfer to and depolarization of particles scattering within ring acceptance.  Therefore it is seen that when spin-flip effects are neglected the theoretical treatment of spin filtering presented in this thesis reduces to the initial naive treatments where polarization transfer and depolarization effects of scattering within ring acceptance were not included.

%\pagebreak

\subsection{$\sigma^\mathrm{\ out}_+$, $\sigma^\mathrm{\ out}_-$, $\sigma_{+-}$, $\sigma_{-+}$ and the spin observables}
\label{subsec:Relating_cross-sections_to_spin_observables}

The spin observables of a spin 1/2 - spin 1/2 scattering process are defined in section~\ref{sec:Spin_observables}.  In spin filtering where the polarization of the recoiled target particle is not important one is interested in the polarization transfer, depolarization and double spin asymmetry spin observables.  These have been calculated for electromagnetic antiproton-proton and antiproton-electron elastic scattering in Chapter~\ref{ch:Specific_helicity_amplitudes_and_spin_observables}.  The spin transfer observable has been calculated for low energy antiproton-positron scattering in Ref.~\cite{Arenhovel:2007gi}.  A large increase of this spin transfer cross-section at very low energies is the basis for the proposal to polarize antiprotons by interaction with a co-moving polarized positron beam presented in Ref.~\cite{Walcher:2007sj}.  It is claimed in Ref.~\cite{Milstein:2008tc} that the polarization transfer cross-section for $e\,p$ or $e^+\,\bar{p}$ (like charges) scattering is enhanced at very low relative velocities, but by much less than that claimed in Refs.~\cite{Walcher:2007sj,Arenhovel:2007gi}.  An experiment has been proposed to test, and distinguish between, these claims \cite{PAX:2007_2}.

The cross-sections $\sigma^\mathrm{\ out}_+$, $\sigma^\mathrm{\ out}_-$, $\sigma_{+-}$ and $\sigma_{-+}$ can be related to the spin observables that have been calculated in Chapters~(\ref{ch:Generic_helicity_amplitudes_and_spin_observables} and \ref{ch:Specific_helicity_amplitudes_and_spin_observables}) by the following relations \cite{Nikolaev:2006gw}:
\begin{eqnarray}
\label{eq:Meyers_to_Ours_Relation1}
\sigma^\mathrm{\ out}_+ & \equiv &  I_\mathrm{\,out} \ +\  \mathcal{P}_T \ A_\mathrm{\,out} \,,\\[2ex]
\label{eq:Meyers_to_Ours_Relation2}
\sigma^\mathrm{\ out}_- & \equiv & I_\mathrm{\,out} \ - \ \mathcal{P}_T \ A_\mathrm{\,out} \,,\\[2ex]
\label{eq:Meyers_to_Ours_Relation3}
\sigma_{+-}    & \equiv & L_\mathrm{\,in} \ + \  \frac{\mathcal{P}_T}{2} \ \left(\,A_\mathrm{\,in} \ - \ K_\mathrm{\,in}\,\right) \,,\\[2ex]
\label{eq:Meyers_to_Ours_Relation4}
\sigma_{-+}    & \equiv & L_\mathrm{\,in} \ - \  \frac{\mathcal{P}_T}{2} \ \left(\,A_\mathrm{\,in} \ -\  K_\mathrm{\,in}\,\right) \,,
\end{eqnarray}
where $\mathcal{P}_T$ is the polarization of the target, and $L_\mathrm{\,in} \, = \, \left(\,I_\mathrm{\,in} - D_\mathrm{\,in}\,\right)\,/\,2$ is a loss of polarization quantity.   These relations involve integration of the spin observables presented in Chapter~\ref{ch:Specific_helicity_amplitudes_and_spin_observables} over the following angular ranges, as seen in Table~\ref{tab:Transverse_and_Longitudinal_polarization}.  The {\bf``in''} subscript refers to particles that are scattered at small angles $\leq \theta_{\mathrm{acc}}$ remaining in the beam, and the {\bf``out''} subscript refers to particles that are scattered out of the beam.  Thus the integrals over scattering angle $\theta$ are labeled {\bf``in''} where the range of integration is $\theta_{\mathrm{min}} \leq \theta \leq \theta_{\mathrm{acc}}$, {\bf``out''} where the range of integration is\footnote{While not occurring in the case of antiprotons which we focus on here, an additional effect must be accounted for in the case of polarization buildup of a proton beam by spin filtering off a hydrogen target \cite{Meyer:2008}, as in the FILTEX experiment.  Because the final state particles are identical and hence indistinguishable, $u$-channel $p\,p$ scattering can contribute; {\it i.e.}\ protons from the hydrogen target can be back scattered into the circulating proton beam.  This happens when the beam protons are back scattered into the angular range $\left(\pi \,-\,\theta_\mathrm{acc}\right) \leq \theta \leq\pi$ in the CM frame.  This effect can be accounted for in the above formalism by changing the angular ranges in the CM frame to $\theta_{\mathrm{min}} \leq \theta \leq \theta_{\mathrm{acc}}\,$ plus $\,\left(\pi -\theta_\mathrm{acc}\right) \leq \theta \leq \pi$ for the {\bf``in''} integrations and $\theta_\mathrm{acc} < \theta < \left(\pi -\theta_\mathrm{acc}\right)$ for the {\bf``out''} integrations.  The physical result of this effect is to lessen the rate of decrease of beam intensity.} $\theta_\mathrm{acc} < \theta \leq \theta_\mathrm{max}$, where $\theta_\mathrm{max}$ is the maximum scattering angle for the process in the given reference frame\footnote{In the CM frame $\theta^\mathrm{\,cm}_\mathrm{max}\,=\,\pi$ corresponding to total backward scattering, but in other frames, for example the LAB frame, this extreme value is not reached for some reactions and $\theta^\mathrm{\,lab}_\mathrm{max}\,<\,\pi$.}, and {\bf``all''} $=$ {\bf``in''} $+$ {\bf``out''} where the range of integration is $\theta_\mathrm{min} \leq \theta \leq \theta_\mathrm{max}$; as seen in Table~\ref{tab:Transverse_and_Longitudinal_polarization}.  $I = \mathrm{d}\sigma\,/\,\mathrm{d}\Omega$ is the spin-averaged differential cross-section and $A$, $K$ and $D$ are the double spin asymmetry, polarization transfer and depolarization spin observables respectively as calculated in Chapter~\ref{ch:Specific_helicity_amplitudes_and_spin_observables}.\\
\\
All cross-sections and spin observables contributing to spin filtering are azimuthally averaged, due to the geometry of the scattering, where the scattering plane can be at any azimuthal angle.  Hence single spin observables, for example the analyzing power, do not contribute to the polarization evolution equations because they vanish when azimuthally averaged.  The cylindrical symmetry of the system also implies that the ring acceptance angle has no azimuthal dependence.

%I think the difference in the signs of the $\mathcal{P}_T$ term comes from the state of the target polarization.  e.g.\ if $\mathcal{P}_T = 0.9$ most target particles are in the $+$ spin state so this contributes negatively to scattering of antiprotons in the $-$ spin state.
\noindent
Note the following linear combinations of the cross-sections
\begin{eqnarray}
I_\mathrm{\,out} & = &  \frac{\sigma^\mathrm{\ out}_+ \ +\  \sigma^\mathrm{\ out}_-}{2} \,,\\[2ex]
\label{eq:Loss_cross_section_difference}
\mathcal{P}_T\,A_\mathrm{\,out} & = & \frac{\sigma^\mathrm{\ out}_+ \ - \ \sigma^\mathrm{\ out}_-}{2} \,,\\[2ex]
L_\mathrm{\,in} & = &  \frac{\sigma_{+-} \ +\  \sigma_{-+}}{2} \,,\\[2ex]
\label{eq:Spin_flip_cross_section_difference}
\mathcal{P}_T \ \left(\,A_\mathrm{\,in} \ - \  K_\mathrm{\,in}\,\right) & = & \sigma_{+-} \ - \ \sigma_{-+} \,.
\end{eqnarray}
Again to ensure consistency and to highlight the physical properties we are trying to describe mathematically we prove a number of short Lemma's on the above relations between the cross-sections and the spin observables.
\\
\\
{\bf Lemma 4}
\\
\\
{\it If the target is unpolarized ($\mathcal{P}_T = 0$) then one has that $\sigma^\mathrm{\ out}_+ = \sigma^\mathrm{\ out}_-$ and $\sigma_{+-} = \sigma_{-+}$ so no polarization buildup will occur.}
\\
\\
{\bf \emph{Proof:}}
\\
\\
Setting $\mathcal{P}_T = 0$ into the eqs.~(\ref{eq:Meyers_to_Ours_Relation1}, \ref{eq:Meyers_to_Ours_Relation2}, \ref{eq:Meyers_to_Ours_Relation3} and \ref{eq:Meyers_to_Ours_Relation4}) one immediately obtains $\sigma^\mathrm{\ out}_+ \, = \,  I_\mathrm{\,out}  \, = \, \sigma^\mathrm{\ out}_-$ and $\sigma_{+-} \, = \, L_\mathrm{\,in}  \, = \,\sigma_{-+}$.  Once this is satisfied it is proved in Lemma 1 that no polarization buildup will occur in this case. \ $\Box$
\\
\\
{\bf Lemma 5}
\\
\\
{\it The spin-flip cross-sections should depend only on spin observables relating to particles scattering within the ring, {\it i.e.}\ only to {\bf ``in''} spin observables which are integrated from $\theta_\mathrm{min}$ to $\theta_\mathrm{acc}$.}
\\
\\
{\bf \emph{Proof:}}
\\
\\
This is immediately satisfied by the relations in eqs.~(\ref{eq:Meyers_to_Ours_Relation3} and \ref{eq:Meyers_to_Ours_Relation4}). \ $\Box$
\\
\\
%The latter of these is what Meyer calls $\sigma_{pol} = \sigma_{+-} \ - \sigma_{-+}$ in the case of no losses (Walcher's idea) and he says we must determine what this is in terms of the spin observables, he says we agree its not $\sigma_0\,K_{i00j}$.  I think the above relation solves his problem, and it should be proportional to the target polarization as it is.
%
%
%\pagebreak
%
\noindent
{\bf Lemma 6}
\\
\\
{\it The cross-section differences $\sigma^\mathrm{\ out}_+ \, - \, \sigma^\mathrm{\ out}_-$ and $\sigma_{+-} \, - \,\sigma_{-+}$ should both be proportional to the target polarization $\mathcal{P}_T$.}
\\
\\
{\bf \emph{Proof:}}
\\
\\
This is immediately satisfied by the relations in eqs.~(\ref{eq:Loss_cross_section_difference} and \ref{eq:Spin_flip_cross_section_difference}). \ $\Box$
\\

While the system of polarization evolution equations involving the variables $N_+(\tau)$ and $N_-(\tau)$ presented in eq.~(\ref{eq:Meyers_System}) is very transparent, one is more interested in the variables $N(\tau) \,=\, N_+(\tau) \,+\, N_-(\tau)$ and $J(\tau) \,=\, N_+(\tau) \,-\, N_-(\tau)$ which immediately lead to $\mathcal{P}(\tau) = J(\tau)\,/\,N(\tau)$.  We can transform the system of two first order ODE's in variables $N_+(\tau)$ and $N_-(\tau)$ presented in eq.~(\ref{eq:Meyers_System}) to the following system of two first order ODE's in variables $N(\tau)$ and $J(\tau)$ \cite{Nikolaev:2006gw,O'Brien:2007hu,Buttimore:2007cj}\,:
\begin{eqnarray}
\label{eq:HomogeneousSystem}
  \frac{\mathrm{d}}{\mathrm{d}\,\tau}
\left[
        \begin{array}{c} N \\[2ex] J \end {array}
\right]
\,  = \, - \, n \, \nu
\left[
\begin{array}{ccc}
         I_\mathrm{\, out} && \mathcal{P}_T \, A_\mathrm{\, out}
\\[2ex]
    \mathcal{P}_T \, \left(\,A_\mathrm{\, all} \, - \, K_\mathrm{\,in}\,\right)
&&
         I_\mathrm{\, all} -  D_\mathrm{\, in}
\end {array}
\right]
\,
\left[
        \begin{array}{c} N \\[2ex] J \end {array}
\right] \ ,
\end{eqnarray}
where we have also transformed from the cross-sections to the spin observables which have already been calculated.  The parameters $n$ and $\nu$ are the target areal density and the beam revolution frequency respectively.

\begin{table}[!h]
\begin{center}
\begin{tabular}{|c|c|} 
\hline
 & \\
Transverse polarization requires & Longitudinal polarization requires \\[2ex] \hline
 & \\
$\displaystyle{
\,
   I_\mathrm{\,out}
\, =
\, 2 \, \pi \!\int_{\theta_\mathrm{acc}}^{\theta_\mathrm{max}}
\!
   \left(\frac{\mathrm{d}\,\sigma}{\mathrm{d}\,\Omega}\right) \sin\theta \, \mathrm{d}\theta
}
$
& 
$\displaystyle{
\,
   I_\mathrm{\,out}
\, =
\, 2 \, \pi \!\int_{\theta_\mathrm{acc}}^{\theta_\mathrm{max}}
\!
   \left(\frac{\mathrm{d}\,\sigma}{\mathrm{d}\,\Omega}\right) \sin\theta \, \mathrm{d}\theta
}
$\\[4ex]
$\displaystyle{
    A_\mathrm{\,out}
   =
 2\,\pi \!\int_{\theta_\mathrm{acc}}^{\theta_\mathrm{max}}
\!
   \left( \frac{A_\mathrm{XX} +  A_\mathrm{YY}}{2} \ 
   \frac{\mathrm{d}\,\sigma}{\mathrm{d}\,\Omega} \right) \sin\theta \, \mathrm{d}\theta 
}
$
&
$\displaystyle{
\,
    A_\mathrm{\,out}
\,
   =
\, 2\,\pi \!\int_{\theta_\mathrm{acc}}^{\theta_\mathrm{max}}
\!
   \left( A_\mathrm{ZZ} \ 
   \frac{\mathrm{d}\,\sigma}{\mathrm{d}\,\Omega} \right) \sin\theta \, \mathrm{d}\theta 
}
$\\[4ex]
$ \displaystyle{
   A_\mathrm{\,all}
 =
 2\,\pi \!\int_{\theta_\mathrm{min}}^{\theta_\mathrm{max}}
\!
  \left( \frac{A_\mathrm{XX}  +  A_\mathrm{YY}}{2} \ 
   \frac{\mathrm{d}\,\sigma}{\mathrm{d}\,\Omega} \right) \sin\theta \, \mathrm{d}\theta
}
$
&
$ \displaystyle{ 
\,  
   A_\mathrm{\,all}
\, =
\, 2\,\pi \!\int_{\theta_\mathrm{min}}^{\theta_\mathrm{max}}
\!
    \left( A_\mathrm{ZZ} \ 
   \frac{\mathrm{d}\,\sigma}{\mathrm{d}\,\Omega} \right) \sin\theta \, \mathrm{d}\theta
}
$\\[4ex]
$\displaystyle{
    K_\mathrm{\,in}
   =
  2\,\pi \!\int_{\theta_\mathrm{min}}^{\theta_\mathrm{acc}}
\!
   \left( \frac{K_\mathrm{XX} +  K_\mathrm{YY}}{2} \ 
   \frac{\mathrm{d}\,\sigma}{\mathrm{d}\,\Omega} \right)  \sin\theta \, \mathrm{d}\theta
}
$
&
$\displaystyle{
\,
    K_\mathrm{\,in}
\,
   =
\,  2\,\pi \!\int_{\theta_\mathrm{min}}^{\theta_\mathrm{acc}}
\!
  \left( K_\mathrm{ZZ}\ 
   \frac{\mathrm{d}\,\sigma}{\mathrm{d}\,\Omega} \right) \sin\theta \, \mathrm{d}\theta
}
$\\[4ex]
$\displaystyle{
    D_\mathrm{\,in}
   =
  2\,\pi \!\int_{\theta_\mathrm{min}}^{\theta_\mathrm{acc}}
\!
   \left( \frac{D_\mathrm{XX}  +  D_\mathrm{YY}}{2}\ 
   \frac{\mathrm{d}\,\sigma}{\mathrm{d}\,\Omega} \right)  \sin\theta \, \mathrm{d}\theta
}
$
&
$\displaystyle{
\,
    D_\mathrm{\,in}
\,
   =
\,  2\,\pi \!\int_{\theta_\mathrm{min}}^{\theta_\mathrm{acc}}
\!
  \left( D_\mathrm{ZZ}\ 
   \frac{\mathrm{d}\,\sigma}{\mathrm{d}\,\Omega} \right)  \sin\theta \, \mathrm{d}\theta
}$
\\[4ex]\hline
\end{tabular}
\end{center}
\caption{\small{\it{The entries in the system of equations for polarization buildup involve angular integration over the spin observables presented in Chapter~\ref{ch:Specific_helicity_amplitudes_and_spin_observables}.  $\mathrm{X}$, $\mathrm{Y}$ and $\mathrm{Z}$ are the coordinate axes where the beam is moving in the positive $\mathrm{Z}$ direction.  The minimum value for $\theta$ ($\theta_\mathrm{min}$) relates to the average transverse electron separation for a pure electron target and to the Bohr radius for an atomic gas target, $\theta_\mathrm{acc}$ is the ring acceptance angle and $\theta_\mathrm{max}$ is the maximum scattering angle for the process.}}}
\label{tab:Transverse_and_Longitudinal_polarization}
\end{table}

The systems presented in eqs.~(\ref{eq:Meyers_System} and \ref{eq:HomogeneousSystem}) are identical and from now on we concentrate on the latter as its solution is more illustrative of the underlying physical phenomena, and the dependence on the target polarization is explicit.  In particular one immediately sees that when the target is unpolarized no beam polarization buildup occurs, as when $\mathcal{P}_T = 0$ the system reduces to two uncoupled separable first order ODE's as in eq.~(\ref{eq:System_with_Unpolarized_target1}) with solutions as presented in eq.~(\ref{eq:System_with_Unpolarized_target2}) showing $\mathcal{P}(\tau) = 0 \ \mbox{for all} \ \tau \ \mbox{if} \ \mathcal{P}_T = 0$.   The parameters in the matrix of coefficients of eq.~(\ref{eq:HomogeneousSystem}) depend on the state of the target polarization, {\it i.e.}\ longitudinal or transverse, as seen in Table~\ref{tab:Transverse_and_Longitudinal_polarization}.

\pagebreak

\section{Solving the polarization evolution equations}
\label{sec:Solving_the_polarization_evolution_equations}

We have shown in the previous section that when circulating at frequency $\nu$, for a time $\tau$, in a ring with a polarized internal target of areal density $n$ and polarization $\mathcal{P}_T$ oriented normal to the ring plane, (or longitudinally with rotators)
\begin{eqnarray}
\label{eq:HomogeneousSystem_again}
  \frac{\mathrm{d}}{\mathrm{d}\,\tau}
\left[
        \begin{array}{c} N \\[2ex] J \end {array}
\right]
\,  = \, - \, n \, \nu
\left[
\begin{array}{ccc}
         I_\mathrm{\, out} && \mathcal{P}_T \, A_\mathrm{\, out}
\\[2ex]
    \mathcal{P}_T \, \left(\,A_\mathrm{\, all} -  K_\mathrm{\,in}\,\right)
&&
         I_\mathrm{\, all} -  D_\mathrm{\, in}
\end {array}
\right]
\,
\left[
        \begin{array}{c} N \\[2ex] J \end {array}
\right] \ ,
\end{eqnarray}
 describes the rate of change of the number of beam particles $N(\tau) = N_+(\tau) + N_-(\tau)$ and their total spin $J(\tau) = N_+(\tau) - N_-(\tau)$ \cite{Nikolaev:2006gw}.  In this section we solve this system of polarization evolution equations.  The eigenvalues of the matrix of coefficients are found to be 
\begin{equation}
\label{eq:Eigenvalues}
\lambda_1 \ = \ - \,n\,\nu\, \left(\,I_\mathrm{\,out} \,+\, L_\mathrm{\,in} \,+\, L_\mathrm{\,d}\,\right) 
\hspace*{1em} \mbox{and} \hspace*{1em}
\lambda_2 \ = \ - \,n\,\nu\, \left(\,I_\mathrm{\,out} \,+\, L_\mathrm{\,in} \,-\, L_\mathrm{\,d}\,\right) \, ,
\end{equation}
where the discriminant $L_\mathrm{\,d}$ of the quadratic equation
 for the eigenvalues is
\begin{equation}
\label{eq:Discriminant}
   L_\mathrm{\,d}
\ =
\ \sqrt{\, P_T^{\,2} \, A_\mathrm{\,out} \left( A_\mathrm{\,all}
\, -
\, K_\mathrm{\,in} \right) \, + \, L_\mathrm{\,in}^{\,2} }  \ \, .
\end{equation}
Note that $I_\mathrm{\,out}$, $L_\mathrm{\,in}$ and $L_\mathrm{\,d}$ are all non-negative.  As a consequence the eigenvalues are non-positive and $\lambda_1 \leq \lambda_2 \leq 0$.  When there is no scattering out of the ring all of the {\bf \lq\lq out''} integrations are zero, and one finds that $L_\mathrm{\,d} \,=\, L_\mathrm{\,in}$ and hence $\lambda_1 \,=\, -\,2\,n\,\nu\,L_\mathrm{\,d}$ and $\lambda_2 \,=\, 0$.
 
Now enforcing the initial conditions $N(0) \,=\, N_0$ the total number of particles in the beam initially, and $J(0) \,=\, 0 \Rightarrow N_+(0) \,=\, N_-(0) \,=\, N_0\,/\,2$ {\it i.e.}\ initially the beam is unpolarized, one obtains the solutions:
\begin{eqnarray}
\label{eq:Homogeneous_N}
N(\tau) & = & \frac{\left[\,e^{\,\lambda_1\,\tau}\, \left(\,L_\mathrm{\,d} - L_\mathrm{\,in}\,\right) \, + \, e^{\,\lambda_2\,\tau}\, \left(L_\mathrm{\,d} + L_\mathrm{\,in}\right)\,\right]\,N_0}{2\, L_\mathrm{\,d}} \, ,\\[2ex]
\label{eq:Homogeneous_J}
J(\tau) & = & \displaystyle{ \frac{\left(\,e^{\,\lambda_1\,\tau}-e^{\,\lambda_2\,\tau}\,\right)\,\left(A_\mathrm{\,all} - K_\mathrm{\,in}\right)\,N_0\,P_T}{2\, L_\mathrm{\,d}}}\,. 
\end{eqnarray}
The time ($\tau$) dependence of the polarization of the beam is given by
\begin{eqnarray}
\label{eq:HomogeneousPolarizationBuildup}
P(\tau) \ \,=\, \ \frac{J(\tau)}{N(\tau)} \ \,=\, \ \frac{ -\,P_T\,\left(\,A_\mathrm{\,all} \, - \, K_\mathrm{\,in}\,\right)}{L_\mathrm{\,in} \, + \, L_\mathrm{\,d} \, \coth\left(L_\mathrm{\,d}\, n\,\nu\,\tau\right)} \, .
\end{eqnarray}
The expression for $P(\tau)$ is proportional to $P_T$ which confirms that if the target polarization is zero there will be no polarization buildup in the beam.
The approximate rate of change of polarization for sufficiently short times, and the limit of the polarization for large times are respectively:
\begin{eqnarray}
\label{eq:Homogeneous_initial_and_max_polarization}
   \frac{\mathrm{d}\,P}{\mathrm{d}\,\tau} & \approx & -\, n \, \nu \, P_T
\,
\left( A_\mathrm{\,all} \, - \, K_\mathrm{\,in} \right)\,, \\[2ex]
\mathcal{P}_\mathrm{max} & = & \displaystyle{\lim_{\tau \to \,\infty} P(\tau) \ = \ -\, P_T
\,
\frac{ A_\mathrm{\,all} \, - \, K_\mathrm{\,in}
}
{   L_\mathrm{\,in} \, + \, L_\mathrm{\,d}
}}
\,.
\end{eqnarray}
In order to compare to earlier treatments of spin filtering notice that in the absence of scattering within the ring, when all {\bf \lq\lq in''} spin observables are zero, one has that $L_\mathrm{\,d} \,=\,\mathcal{P}_T\,A_\mathrm{\,out}$ and eq.~(\ref{eq:HomogeneousPolarizationBuildup}) reduces to eq.~(\ref{eq:Solution_of_Naive_1994_system}) which was the initial treatment of spin filtering proposed in 1993 where scattering within ring acceptance was not included.  Having said that, the general behaviour of equations \ref{eq:Solution_of_Naive_1994_system} and \ref{eq:HomogeneousPolarizationBuildup} are similar. 

For pure electromagnetic scattering the double spin asymmetries equal the polarization transfer spin observables \cite{O'Brien:2006zt}, thus one can simplify the above equations using $A_\mathrm{\,in} \,=\, K_\mathrm{\,in}$, $A_\mathrm{\,out} \,=\, K_\mathrm{\,out}$ and $A_\mathrm{\,all} \,=\, K_\mathrm{\,all}$\,; hence $A_\mathrm{\,all} \,-\, K_\mathrm{\,in} \,=\, K_\mathrm{\,out}$\,.  

At this point one may wish to find expressions for $N_+(\tau)$ and $N_-(\tau)$ which can easily be obtained from eqs.~(\ref{eq:Polarization_definition_J/N}, \ref{eq:Homogeneous_N} and \ref{eq:Homogeneous_J}):
\begin{eqnarray}
N_+(\tau) & = & \frac{N(\tau) \ + \ J(\tau)}{2}  \,, \nonumber \\[2ex]
 & = & \frac{N_0}{4\,L_\mathrm{\,d}}\,\left\{\,e^{\,\lambda_1\,\tau}\,\left[\,L_\mathrm{\,d} \,-\,L_\mathrm{\,in} \,+\,  \left(\,A_\mathrm{\,all} \,-\, K_\mathrm{\,in}\,\right)\,\mathcal{P}_T\,\right]\nonumber \right. \\[2ex]
& & \left. \qquad \qquad + \ e^{\,\lambda_2\,\tau}\,\left[\,L_\mathrm{\,d} \,+\,L_\mathrm{\,in} \,-\,  \left(\,A_\mathrm{\,all} \,-\, K_\mathrm{\,in}\,\right)\,\mathcal{P}_T\,\right]\,\right\} \,,\\[6ex]
N_-(\tau) & = & \frac{N(\tau) \ - \ J(\tau)}{2}  \,, \nonumber\\[2ex]
 & = & \frac{N_0}{4\,L_\mathrm{\,d}}\,\left\{\,e^{\,\lambda_1\,\tau}\,\left[\,L_\mathrm{\,d} \,-\,L_\mathrm{\,in} \,-\,  \left(\,A_\mathrm{\,all} \,-\, K_\mathrm{\,in}\,\right)\,\mathcal{P}_T\,\right] \nonumber \right. \\[2ex]
& & \left. \qquad \qquad + \ e^{\,\lambda_2\,\tau}\,\left[\,L_\mathrm{\,d} \,+\,L_\mathrm{\,in} \,+\,  \left(\,A_\mathrm{\,all} \,-\, K_\mathrm{\,in}\,\right)\,\mathcal{P}_T\,\right]\,\right\}\,,
\end{eqnarray}
which can easily be seen to satisfy $N_+(0) \,=\,N_-(0)\,=\,N_0\,/\,2$, the correct initial conditions.  If the target is unpolarized ($\mathcal{P}_T \,=\,0$) one sees that $N_+(\tau)\,=\,N_-(\tau)$ therefore no polarization buildup will occur, as was required by Lemma 4.
%Is it possible to simplify the above expressions???? Try to simplify them.

\subsection{Beam lifetime and figure of merit}
\label{subsec:Beam_lifetime_and_figure_of_merit}

The beam lifetime $\tau_*$, the time taken for the beam intensity to decrease by a factor of $e \approx 2.718$, {\it i.e.}\ $N\left(\,\tau_*\,\right)=N_0/e$, can be obtained from eq.~(\ref{eq:Homogeneous_N}).  One finds 
\begin{equation}
\label{eq:Beam_lifetime}
\tau_{*} \ \approx \ \displaystyle{\frac{1}{n\,\nu\,I_\mathrm{\,out}}} \  ,
\end{equation}
a more accurate form of which is derived in section~\ref{subsec:Fraction_of_antiprotons_lost_per_revolution}.
% where $I_\mathrm{\,out} + L_\mathrm{\,in} \gg L_\mathrm{\,d}$ when there is scattering out of the ring.

The Figure Of Merit (FOM) provides a measure of the quality of the polarized beam, and is given by
\begin{equation}
\mathrm{FOM}(\tau) \ = \ \mathcal{P}^{\,2}(\tau)\ N(\tau) \ = 
 \ \frac{J^{\,2}(\tau)}{N(\tau)} \,.
\end{equation}
The figure of merit for the above case is
\begin{eqnarray}
\label{eq:Homogeneous_FOM}
\hspace*{-3.5em}\mathrm{FOM}(\tau) &\! =\! & \frac{\left(A_\mathrm{\,all} - K_\mathrm{\,in}\right)^{\,2} N_0\,\mathcal{P}_T^{\,2}}{2\,L_\mathrm{\,d}} \left[ \,\frac{\displaystyle{\left(\,e^{\,\lambda_1\,\tau} - e^{\,\lambda_2\,\tau}\,\right)^{\,2}}}{e^{\,\lambda_1\,\tau}\, \left(\,L_\mathrm{\,d} - L_\mathrm{\,in}\,\right) \, + \, e^{\,\lambda_2\,\tau}\, \left(L_\mathrm{\,d} + L_\mathrm{\,in}\right)}\,\right] .
\end{eqnarray}
\begin{figure}
\centering
\input{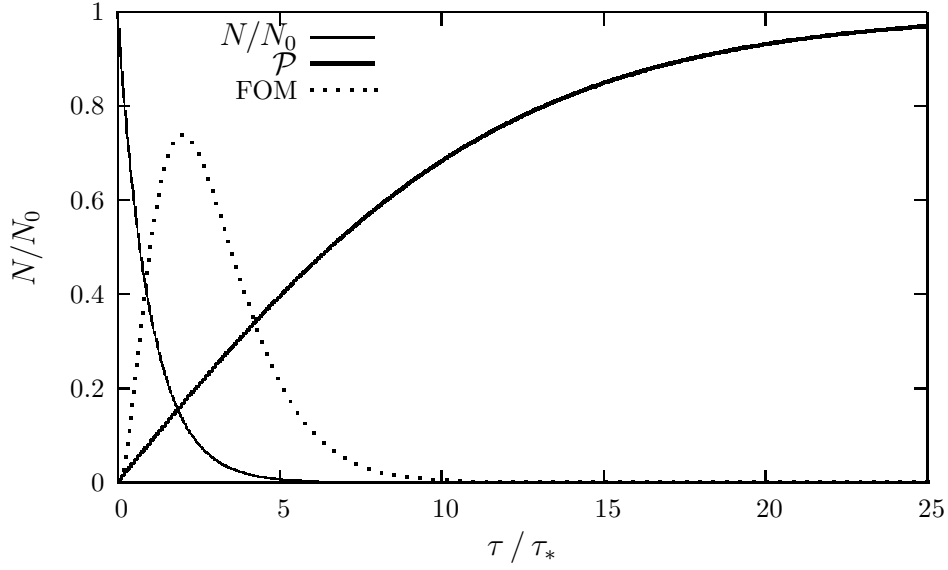} %x-range only to 25
\caption{\small{\it{A schematic graph showing the behaviour of the beam intensity $N$ and beam polarization $\mathcal{P}$ as time (scaled by the beam lifetime $\tau_*$) increases.  The behaviour of the figure of merit $\mathrm{FOM}$ is also shown on the graph, being blown up to clearly show it has a maximum at twice the beam lifetime.  
%The graph just shows general trends and is not in exact numerical correspondence to the equations, which will be presented in Chapter~\ref{ch:Numerical_results}.
}}
}
\label{fig:FOM_Plot_standard}
\end{figure}
\!\!Maximizing the figure of merit gives the optimum polarization buildup time, taking into account the trade-off between decreasing beam intensity and increasing beam polarization.  Solving $\mathrm{d}\,\mathrm{FOM}\,/\,\mathrm{d}\,\tau \, = \, 0$ yields 
\begin{equation}
\label{eq:Optimum_polarization_buildup_time}
\tau_\mathrm{optimum} \ \approx \ \frac{2}{n\,\nu\,I_\mathrm{\,out}} \ \approx \ 2\,\tau_* \ , 
\end{equation}
approximately twice the beam lifetime.  Thus the optimum time for polarization buildup is twice the lifetime of the beam, as also found in Ref.~\cite{Rathmann:2004pm}.

The behaviours of the beam intensity, beam polarization and the figure of merit as time increases are shown in Figure~\ref{fig:FOM_Plot_standard}.  Note the characteristic trade-off of spin filtering: as the beam polarization increases the beam intensity greatly decreases.\\
\noindent
In Chapter \ref{ch:Various_scenarios_of_spin_filtering} we address possible ways of circumventing this drawback, for instance by continuously inputting particles into the beam.

\subsection{Pure electromagnetic scattering}
\label{subsec:Pure_electromagnetic_scattering}

For pure electromagnetic scattering $\, A_{ij} = K_{ij}$ where $i,j \in \{\mathrm{\,X,Y,Z\,}\}$, so the system simplifies to
\begin{eqnarray}
\label{eq:System_for_pure_EM_scattering}
  \frac{\,\mathrm{d}}{\mathrm{d}\,\tau}
\left[
        \begin{array}{c} N \\[2ex] J \end {array}
\right]
\  = \ - \, n \, \nu
\left[
\begin{array}{ccc}
         I_\mathrm{\, out} && \mathcal{P}_T \, K_\mathrm{\, out}
\\[2ex]
    \mathcal{P}_T \, K_\mathrm{\, out}
&&
         I_\mathrm{\, all} -  D_\mathrm{\, in}
\end {array}
\right]
\,
\left[
        \begin{array}{c} N \\[2ex] J \end {array}
\right] \,,
\end{eqnarray}
the coefficient matrix of which is symmetric, and the solutions become
\begin{equation}
\label{eq:Ptau_for_pure_EM_scattering}
   \mathcal{P}(\tau) \ = \ \frac{J(\tau)}{N(\tau)} \ = 
\ 
\frac{ -\,K_\mathrm{\,out}\, \mathcal{P}_T}
{ L_\mathrm{\,in} \, + \, L_\mathrm{\,d} \, \coth\left(L_\mathrm{\,d}\, n\,\nu\,\tau\right)}\,,
\end{equation}
\begin{equation}
\label{eq:Ld_for_pure_EM_scattering}
   L_\mathrm{\,d}
\ =
\ \sqrt{\, \mathcal{P}_T^{\,2} \, K_\mathrm{\,out}^{\,2} \, + \, L_\mathrm{\,in}^{\,2} } \ .
\end{equation}
The approximate rate of change of polarization for sufficiently short times, and the limit of the polarization for large times simplify to respectively:
\begin{equation}
\label{eq:Pmax_for_pure_EM_scattering}
   \frac{\mathrm{d}\,\mathcal{P}}{\mathrm{d}\tau} \  \approx \  -\, n \, \nu \, \mathcal{P}_T
\,
 K_\mathrm{\,out}
\hspace*{3em}
\displaystyle{\mathcal{P}_\mathrm{max} \ = \ 
   \lim_{\tau \to \infty} \mathcal{P}(\tau) \ = 
\
\frac{ -\,K_\mathrm{\,out}\, \mathcal{P}_T
}
{   L_\mathrm{\,in} \, + \, L_\mathrm{\,d}
}} \ .
\end{equation}
We now show that the maximum polarization achieved cannot exceed one, {\it i.e.}\ $|\mathcal{P}_\mathrm{max}| \,\le\, 1$.  From the definition of $L_\mathrm{\,d}$ for the case of pure electromagnetic scattering we have $L_\mathrm{\,d}^{\,2}\, =\, \mathcal{P}_T^{\,2} \, K_\mathrm{\,out}^2 \, + \, L_\mathrm{\,in}^{\,2}\Rightarrow L_\mathrm{\,d} \ge |\mathcal{P}_T\,K_\mathrm{\,out}|$ which can be used to obtain
\begin{equation}
\label{eq:Pmax_inequality_for_pure_EM_scattering}
\displaystyle{|\mathcal{P}_\mathrm{max}| \ \le \
\frac{ L_\mathrm{\,d}
}
{   L_\mathrm{\,in} \, + \, L_\mathrm{\,d}
}
\ \le \ 
\frac{ L_\mathrm{\,d}
}
{   L_\mathrm{\,d}
} \ = \ 1}\,,
\end{equation}
since $L_\mathrm{\,in}$ is a non-negative quantity. $\ \Box$  

Note the upper bound $|\mathcal{P}_\mathrm{max}| = 1$ only happens for $L_\mathrm{\,in} = 0$, {\it i.e.}\ no polarization is lost.

\pagebreak

\subsection{Fraction of antiprotons lost per revolution}
\label{subsec:Fraction_of_antiprotons_lost_per_revolution}

We now calculate the fraction of the antiproton beam that is lost per revolution.  Define $\Delta \tau$ to be the time taken for one revolution, thus $\nu \,\Delta \tau = 1$, and $\Delta N$ to be the change in beam intensity during this time.  Manipulating the $\mathrm{d}\,N\,/\,\mathrm{d}\,\tau$ equation gives us
\begin{eqnarray}
\frac{\Delta N}{\Delta \tau} & = & -\,n\,\nu\,I_\mathrm{\,out}\,N  \ -\  n\,\nu\,\mathcal{P}_T\,A_\mathrm{\,out}\,J \,, \nonumber \\[1ex]
%\therefore \ 
%\Delta N & = & -\,n\,\nu\,I_\mathrm{\,out}\,N\,\Delta \tau - n\,\nu\,\mathcal{P}_T\,A_\mathrm{\,out}\,J\,\Delta \tau \nonumber \\[1ex]
%\therefore \ 
\frac{\Delta N}{N} & = & -\,n\,\nu\,I_\mathrm{\,out}\,\Delta \tau \ - \ n\,\nu\,\mathcal{P}_T\,A_\mathrm{\,out}\,\frac{J}{N}\,\Delta \tau \nonumber  \,,\\[1ex]
& = & -\,n\,I_\mathrm{\,out} \ -\  n\,\mathcal{P}_T\,A_\mathrm{\,out}\,\frac{J}{N} 
%\ = \  -\,n\,I_\mathrm{\,out} \ -\  n\,\mathcal{P}_T\,A_\mathrm{\,out}\,\mathcal{P}_{\bar{p}}
 \,, \nonumber \\[2ex]
& = &-\, n \left(\,I_\mathrm{\,out} \ + \ \mathcal{P}_{\bar{p}}\,\mathcal{P}_T\,A_\mathrm{\,out}\,\right) \,.
\end{eqnarray}
Since $\Delta N$ is the change in beam intensity during time $\Delta \tau$, and the beam intensity decreases with time, the quantity $\Delta N$ is negative.  But the fraction of the antiprotons lost per revolution is $n \left(\,I_\mathrm{\,out} \,+\, \mathcal{P}_{\bar{p}}\,\mathcal{P}_T\,A_\mathrm{\,out}\,\right)$ which is positive, and thus the fraction of antiprotons remaining in the beam per revolution is $1 \,-\, n \left(\,I_\mathrm{\,out} \,+\, \mathcal{P}_{\bar{p}}\,\mathcal{P}_T\,A_\mathrm{\,out}\,\right)$.  The fraction of antiprotons lost per revolution is not constant in time, it depends on the polarization of the antiproton beam $\mathcal{P}_{\bar{p}}$\,, which is zero initially.  %$\mathcal{P}_{\bar{p}}$ and $\mathcal{P}_T$ have opposite signs, as our equations tell us, then since $A_\mathrm{\,out} \geq 0$ the complete second term $\mathcal{P}_{\bar{p}}\,\mathcal{P}_T\,A_\mathrm{\,out} \leq 0$
The fraction of antiprotons lost per revolution decreases as the polarization of the antiproton beam increases.  This makes physical sense, since as the antiproton beam polarization increases there will be fewer particles in the spin state that is scattered out more often.

This can also be used to obtain the beam lifetime.  Since the frequency $\nu$ is the number of revolutions per second, the fraction of particles lost from the beam per second is $n\,\nu \left(\,I_\mathrm{\,out} \,+\, \mathcal{P}_{\bar{p}}\,\mathcal{P}_T\,A_\mathrm{\,out}\,\right)$, hence the fraction of particles lost from the beam in $\tau$ seconds is $n\,\nu \tau \left(\,I_\mathrm{\,out} \,+\, \mathcal{P}_{\bar{p}}\,\mathcal{P}_T\,A_\mathrm{\,out}\,\right)$\,.  The beam lifetime is the time taken for the beam intensity to decrease by a factor of $1\,/\,e$, {\it i.e.}\ the time taken for a fraction $1-1\,/\,e = (e-1)\,/\,e$ of the beam particles to be lost due to scattering out of the ring.  Thus we can calculate the beam lifetime ($\tau_*$) by solving
\begin{equation}
 n\,\nu\,\tau_* \left(\,I_\mathrm{\,out} \,+\, \mathcal{P}_{\bar{p}}\,\mathcal{P}_T\,A_\mathrm{\,out}\,\right) \ = \ \frac{e \,-\, 1}{e}\,,
\end{equation}
leading to the beam lifetime
\begin{eqnarray}
\tau_*  & = &  \frac{e - 1}{n\,\nu\,e\,\left(\,I_\mathrm{\,out} \,+\, \mathcal{P}_{\bar{p}}\,\mathcal{P}_T\,A_\mathrm{\,out}\,\right)} \nonumber \,,\\[2ex]
& = &  \frac{1}{n\,\nu\,\left(\,I_\mathrm{\,out} \,+\, \mathcal{P}_{\bar{p}}\,\mathcal{P}_T\,A_\mathrm{\,out}\,\right)} \ - \ \frac{1}{n\,\nu\,e\,\left(\,I_\mathrm{\,out} \,+\, \mathcal{P}_{\bar{p}}\,\mathcal{P}_T\,A_\mathrm{\,out}\,\right)}\,,
\end{eqnarray}
the first term of which is $e \approx 2.718$ time larger than the second.  Note that since $I_\mathrm{\,out} \,>\, \mathcal{P}_{\bar{p}}\,\mathcal{P}_T\,A_\mathrm{\,out}$, this expression for the beam lifetime limits to the one calculated by the other method above are equal to leading order.  Of importance here is the fact that the beam lifetime is not constant, it increases as the beam polarization increases; which makes physical sense as there will be fewer particles in the spin state that is scattered out more often.

\chapter{Various scenarios of spin filtering}
\label{ch:Various_scenarios_of_spin_filtering}

\vspace*{5ex}
\begin{minipage}{4.5cm}
\end{minipage}
\hfill
\begin{minipage}{12cm}
\begin{quote}
\emph{\lq\lq Let no one ignorant of Mathematics enter here.\rq\rq}
%\begin{right}
\flushright{Plato, inscription over the entrance to the Academy.}
%\end{right}
\end{quote}
\end{minipage}
\vspace{8ex}

A major problem with spin filtering is that as the beam polarization increases the beam intensity decreases, since particles are being continuously scattered out of the beam.  While one may obtain a polarized antiproton beam its intensity may be too low to be of use in any experiment.  In this chapter we investigate the possibility of continuously inputting unpolarized particles into the beam to counteract this loss of beam intensity.  We present a thorough investigation of spin filtering under various alternate scenarios of interest to any practical project to produce a high intensity polarized antiproton beam.  These scenarios are: 1) spin filtering while the beam is being accumulated, {\it i.e.}\ unpolarized particles are continuously being fed into the beam at a constant rate, 2) unpolarized particles are continuously being fed into the beam at a linearly increasing rate, {\it i.e.}\ the particle input rate is ramped up, 3) the particle input rate is equal to the rate at which particles are being lost due to scattering beyond the ring acceptance angle, the beam intensity remaining constant, 4) increasing the initial polarization of a stored beam by spin filtering, and finally 5) the input of particles into the beam is stopped after a certain amount of time, but spin filtering continues.  The five sections of this chapter each treat one of the scenarios of spin filtering listed above, in that order.

\section{Accumulation of antiprotons in the ring}
\label{sec:Accumulation_of_antiprotons_in_the_ring}

In the discussion so far we have only considered polarizing an antiproton beam when the beam is already accumulated in the storage ring.  The $\mathcal{P}\mathcal{A}\mathcal{X}$ Collaboration plans to obtain their antiproton beam by collecting the produced antiprotons from high energy interactions of protons on targets of light nuclei, such as Beryllium.  The antiprotons will be continuously fed into the storage ring at a fixed rate and accumulated, hence increasing the beam intensity, allowing for a greater luminosity in an experiment.  The $\mathcal{P}\mathcal{A}\mathcal{X}$ Collaboration estimates the production rate of antiprotons as being $10^7$ per second \cite{Barone:2005pu}.  Since $10^{11}$ antiprotons are required in the storage ring, antiprotons will be fed into the storage ring at a rate of $10^7$ per second for $10^4$ seconds \cite{Barone:2005pu}.

We now consider a system where spin filtering occurs as the antiprotons are being fed into the ring.  The original system of equations must be amended to account for this constant accumulation.  The effect will be to add a term $\beta$ to the $\mathrm{d}\,N(\tau)\,/\,\mathrm{d}\,\tau$ equation, where $\beta$ is the constant rate at which antiprotons are fed into the ring; while the $\mathrm{d}\,J(\tau)\,/\,\mathrm{d}\,\tau$ equation remains unchanged.  The initial conditions are $N(0) = N_0$\,, which will be set to zero in section~\ref{subsec:No_initial_beam}, and $J(0) = 0$.  The new system of differential equations is
\begin{eqnarray}
\label{eq:InhomogeneousSystem_N}
\frac{\mathrm{d}\,N(\tau)}{\mathrm{d}\,\tau} & = & \, - \, n \, \nu \ \left[\, I_\mathrm{\, out} \ N(\tau) \ + \  \mathcal{P}_T \, A_\mathrm{\, out} \ J(\tau)\, \right] \ + \  \beta  \, , \\[1ex]
\label{eq:InhomogeneousSystem_J}
\frac{\mathrm{d}\,J(\tau)}{\mathrm{d}\,\tau} & = &\, - \, n \, \nu \  \left[\, \mathcal{P}_T \, \left(\,A_\mathrm{\, all} -  K_\mathrm{\,in}\,\right)\ N(\tau) \  + \ \left(\,I_\mathrm{\, all} -  D_\mathrm{\, in}\,\right)\ J(\tau)\, \right] \,.
\end{eqnarray}
By differentiating eq.~(\ref{eq:InhomogeneousSystem_J}) with respect to $\tau$ and substituting in eq.~(\ref{eq:InhomogeneousSystem_N}) one obtains an inhomogeneous second order linear differential equation with constant coefficients for $J(\tau)$:
\begin{equation}
\label{eq:Inhomogeneous_Second_Order_ODE}
\frac{\mathrm{d}^{\,2}\,J(\tau)}{\mathrm{d}\,\tau^2} \ - \ \left(\,\lambda_1 + \lambda_2\,\right)\,\frac{\mathrm{d}\,J(\tau)}{\mathrm{d}\,\tau} \ +\  \lambda_1\,\lambda_2\,J(\tau) \ = \  - \,n\, \nu \,\mathcal{P}_T\,\left(\,A_\mathrm{\,all} - K_\mathrm{\,in}\,\right) \, \beta \,,
\end{equation}
the solution of which is
\begin{eqnarray}
\label{eq:Inhomogeneous_J}
J(\tau) & = & \frac{\mathcal{P}_T\,\left(\,A_\mathrm{\,all} - K_\mathrm{\,in}\,\right)}{2\,L_\mathrm{d}\,\lambda_1\,\lambda_2}\ \left[\,\lambda_2\,\left(\,\lambda_1\,N_0 \,+\, \beta \,\right)\,e^{\,\lambda_1\,\tau} \ - \ \lambda_1\,\left(\,\lambda_2\,N_0 \,+\, \beta \,\right)\,e^{\,\lambda_2\,\tau}  \right. \nonumber \\[1ex]
& & \qquad \qquad \qquad \qquad \qquad \qquad \left. \ + \ \beta\,\left(\,\lambda_1 \,-\, \lambda_2\,\right) 
\,\right] \,.
\end{eqnarray}
Differentiating eq.~(\ref{eq:Inhomogeneous_J}) with respect to $\tau$ and substituting into eq.~(\ref{eq:InhomogeneousSystem_J}) gives an expression for $N(\tau)$\,:
\begin{eqnarray}
\label{eq:Inhomogeneous_N}
N(\tau)  & = & \frac{1}{2\,L_\mathrm{d}\,\lambda_1\,\lambda_2} \left[\,
\lambda_2\,\left(\,\lambda_1\,N_0 + \beta \,\right) \,\left(\,L_\mathrm{d} - L_\mathrm{\,in}\,\right)\,e^{\,\lambda_1\,\tau}
 \right. \\[2ex]
& & \left. + \ \lambda_1\,\left(\,\lambda_2\,N_0 + \beta\,\right) \,\left(\,L_\mathrm{\,in} + L_\mathrm{d}\,\right)\,e^{\,\lambda_2\,\tau} \, + \, \beta\,\left(\,I_\mathrm{\,all} - D_\mathrm{\,in}\,\right)\,\left(\,\lambda_2 - \lambda_1\,\right)\,\right]\nonumber \,.
\end{eqnarray}
As a consistency check one can see that these solutions for $J(\tau)$ and $N(\tau)$ satisfy the initial conditions $J(0) = 0$ and $N(0) = N_0$, and in the particular case when $\beta = 0$ the above expressions reduce to the solution of the homogeneous system presented in section~\ref{sec:Solving_the_polarization_evolution_equations}.
Dividing $J(\tau)$ by $N(\tau)$ we obtain an expression for the polarization $\mathcal{P}(\tau)$ as a function of time,
\begin{equation}
\label{eq:Inhomogeneous_Polarization_Buildup}
\mathcal{P}(\tau) \ = \ \frac{-\,\left(\,A_\mathrm{\,all} - K_\mathrm{\,in}\,\right)\,\mathcal{P}_T}{L_\mathrm{\,in} + L_\mathrm{\,d}\,\displaystyle{\left[\,\frac{2}{1 - \frac{\lambda_2\,\left[\,e^{\,\lambda_1\,\tau}\,\left(\,\lambda_1\,N_0 + \beta\,\right)\, - \, \beta\,\right]
}{\lambda_1\,\left[\,e^{\,\lambda_2\,\tau}\,\left(\,\lambda_2\,N_0 + \beta\,\right)\, - \, \beta\,\right]}}\,-\,1\,\right]
}}\,.
\end{equation}
When the particle input rate is zero ({\it i.e.}\ $\beta = 0$) the above equation simplifies to 
\begin{equation}
\mathcal{P}(\tau) \, = \, \frac{-\,\left(\,A_\mathrm{\,all} - K_\mathrm{\,in}\,\right)\,\mathcal{P}_T}{L_\mathrm{\,in} + L_\mathrm{\,d}\,\left[\,\displaystyle{\frac{2}{1 - e^{\,\left(\,\lambda_1 - \lambda_2\,\right)\,\tau}}} - 1 \,\right]}
\, = \, 
\frac{ -\,\left(\,A_\mathrm{\,all} \, - \, K_\mathrm{\,in}\,\right)\,\mathcal{P}_T}{L_\mathrm{\,in} \, + \, L_\mathrm{\,d} \, \coth\left(L_\mathrm{\,d}\, n\,\nu\,\tau\right)} \, ,
\end{equation}
which is the solution of the homogeneous case presented in eq.~(\ref{eq:HomogeneousPolarizationBuildup}).

Using a Taylor Series expansion we find that the approximate initial rate of polarization buildup for each of these cases ($N_0 \neq 0$ with $\beta \neq 0$ and $N_0 = 0$ with $\beta \neq 0$) is the same as in the homogeneous case ($N_0 \neq 0$ with $\beta = 0$):
\begin{equation}
   \frac{\mathrm{d}\,\mathcal{P}}{\mathrm{d}\tau} \, \, \approx \, -\, n \, \nu \, \mathcal{P}_T
\,
\left( A_\mathrm{\,all} \, - \, K_\mathrm{\,in} \right)\,.
\end{equation}
The maximum polarization achievable is the limit as time approaches infinity:
\begin{equation}
\mathcal{P}_\mathrm{max} \, = \,  \lim_{\tau \to \,\infty} \mathcal{P}(\tau) \, = \, \frac{-\,\mathcal{P}_T\,\left(\,A_\mathrm{\,all} - K_\mathrm{\,in}\,\right)}{I_\mathrm{\,all} - D_\mathrm{\,in}} \ ,
\end{equation}
which is independent of both $N_0$ and $\beta$, however note that in taking this limit we used the fact that $\beta \neq 0$.  If $\beta$ was equal to zero then the maximum polarization achievable would equal that from the homogeneous case; as can be easily seen from eq.~(\ref{eq:Inhomogeneous_Polarization_Buildup}) remembering that $\lambda_1 \leq \lambda_2 \leq 0$.  Thus for the complete case there are just two values of the maximum polarization, one for $\beta = 0$ and one for all $\beta \neq 0$.
The figure of merit for this inhomogeneous case is:
\begin{eqnarray}
\hspace*{-2em}\mathrm{FOM}(\tau) & = & \mathcal{P}^{\,2}(\tau)\,N(\tau) \ = \ \frac{J^{\,2}(\tau)}{N(\tau)} \ = \ \frac{\left(\,A_\mathrm{\,all} - K_\mathrm{\,in}\,\right)^{\,2}\,\mathcal{P}_T^{\,2}}{2\,L_\mathrm{\,d}\,\lambda_1\,\lambda_2} \times \nonumber\\[2ex]
& & \hspace*{-8em}   \frac{\left[\,
c_1\,e^{\,\lambda_1\,\tau} - c_2\,e^{\,\lambda_2\,\tau} + \beta\,\left(\,\lambda_1 - \lambda_2\,\right)\,\right]^{\,2}}{c_1\,\left(\,L_\mathrm{\,d} - L_\mathrm{\,in}\,\right)\,e^{\,\lambda_1\,\tau} +  c_2\,\left(\,L_\mathrm{\,in} + L_\mathrm{\,d}\,\right)\,e^{\,\lambda_2\,\tau} + \beta\,\left(\,I_\mathrm{\,all} - D_\mathrm{\,in}\,\right)\,\left(\,\lambda_2 - \lambda_1\,\right)} \, ,
\end{eqnarray}
where for convenience we have defined the two constants $c_1 = \lambda_2\,\left(\,\lambda_1\,N_0 + \beta\,\right)$ and $c_2 = \lambda_1\,\left(\,\lambda_2\,N_0 + \beta\,\right)$.  Note the FOM will not have a maximum in finite time if the accumulation rate $\beta$ is high enough to make the beam intensity a constant or increase with time.  If this happens the FOM will increase monotonically.

\subsection{No initial beam}
\label{subsec:No_initial_beam}

Of interest is the particular case when $N_0 = 0$, {\it i.e.}\ there are no particles in the beam initially.  In this case the above solutions simplify to
\begin{eqnarray}
\label{eq:Inhomogeneous_J_with_N0_0}
J(\tau) & = & \frac{\beta\,\mathcal{P}_T\,\left(\,A_\mathrm{\,all} - K_\mathrm{\,in}\,\right)}{2\,L_\mathrm{d}\,\lambda_1\,\lambda_2}\ \left[\, \lambda_2\,\left(\,e^{\,\lambda_1\,\tau} \,-\, 1\,\right) \,+ \, \lambda_1\, \left(\,1\,-\,e^{\,\lambda_2\,\tau}\,\right)\,\right]  \, ,\\[4ex]
\label{eq:Inhomogeneous_N_with_N0_0}
N(\tau)  & = & \frac{\beta}{2\,L_\mathrm{d}\,\lambda_1\,\lambda_2} \left[\,
\lambda_2\,\left(\,L_\mathrm{d} - L_\mathrm{\,in}\,\right)\,e^{\,\lambda_1\,\tau}
\,+ \, \lambda_1\,\left(\,L_\mathrm{\,in} + L_\mathrm{d}\,\right)\,e^{\,\lambda_2\,\tau} \right. \\[2ex]
& & \left. \qquad \qquad \qquad \qquad \qquad \qquad + \ \left(\,I_\mathrm{\,all} - D_\mathrm{\,in}\,\right)\,\left(\,\lambda_2 - \lambda_1\,\right)\,\right] \, ,\nonumber \\[4ex]
\label{eq:Polarization_Inhomogeneous_N_equals_0}
\mathcal{P}(\tau) &  = &  \frac{-\,\left(\,A_\mathrm{\,all} - K_\mathrm{\,in}\,\right)\,\mathcal{P}_T}{L_\mathrm{\,in} + \displaystyle{L_\mathrm{\,d}\, \left[\,\frac{2}{1 - \frac{\left(\,1 - e^{\,\lambda_1\,\tau}\,\right)\,\lambda_2}{\left(\,1 - e^{\,\lambda_2\,\tau}\,\right)\,\lambda_1}}\,-\,1\,\right] 
}}\,,\\[4ex]
\label{eq:Inhomogeneous_FOM_with_N_equals_0}
\mathrm{FOM}(\tau) &  = & \frac{\left(\,A_\mathrm{\,all} - K_\mathrm{\,in}\,\right)^{\,2}\,\mathcal{P}_T^{\,2}\,\beta}{2\,L_\mathrm{\,d}\,\lambda_1\,\lambda_2} \times  \\[2ex]
& & \hspace*{-0.5em} \frac{\left[\,
\lambda_2\,\left(\,e^{\,\lambda_1\,\tau}\,-\,1\,\right)  \ + \ \lambda_1\,\left(\,1\,-\,e^{\,\lambda_2\,\tau}\,\right) \,\right]^{\,2}}{\lambda_2\,\left(\,L_\mathrm{\,d} - L_\mathrm{\,in}\,\right)\,e^{\,\lambda_1\,\tau} +  \lambda_1\,\left(\,L_\mathrm{\,in} + L_\mathrm{\,d}\,\right)\,e^{\,\lambda_2\,\tau} + \left(\,I_\mathrm{\,all} - D_\mathrm{\,in}\,\right)\,\left(\,\lambda_2 - \lambda_1\,\right)} \nonumber \,.
\end{eqnarray}
Interestingly the $\beta$ dependence of $\mathcal{P}(\tau)$ vanishes in this case, {\it i.e.}\ the polarization buildup rate is independent of the rate at which antiprotons are fed into the ring, if there are no particles in the beam initially.  But we have used the fact that $\beta \neq 0$ to obtain the above result.   We should note the obvious physical fact that if $N_0 = 0$ and $\beta = 0$ {\it i.e.}\ there are no particles in the beam initially and no particles are fed into the beam, then there will never be any particles in the beam; so measuring the beam polarization is meaningless.  Notice that in this case where $N_0=0$, the figure of merit is proportional to the particle input rate $\beta$.

\subsubsection{Summary}
\label{subsec:Summary}

The results obtained thus far for the polarization buildup in various scenarios can be summarized as
\begin{eqnarray}
\mathcal{P}(\tau) \ = \ \left\{
\begin{array}{ll}
0 & \mbox{for} \ \beta = 0 \ \ \& \ \ N_0 = 0\\[5ex]
\displaystyle{\frac{ -\,\left(\,A_\mathrm{\,all} \, - \, K_\mathrm{\,in}\,\right)\,\mathcal{P}_T}{L_\mathrm{\,in} \, + \, L_\mathrm{\,d} \, \coth\left(L_\mathrm{\,d}\, n\,\nu\,\tau\right)}}  & \mbox{for} \ \beta = 0 \ \ \& \ \ N_0 \neq 0 \nonumber\\[5ex]
\displaystyle{\frac{-\,\left(\,A_\mathrm{\,all} - K_\mathrm{\,in}\,\right)\,\mathcal{P}_T}{L_\mathrm{\,in} + L_\mathrm{\,d}\,\displaystyle{\left[\,\frac{2}{1 - \frac{\left(\,1 - e^{\,\lambda_1\,\tau}\,\right)\,\lambda_2}{\left(\,1 - e^{\,\lambda_2\,\tau}\,\right)\,\lambda_1}}\,-\,1\,\right]
}}} &  \mbox{for} \ \beta \neq 0 \ \ \& \ \ N_0 = 0 \nonumber \\[13ex]
\displaystyle{
\frac{-\,\left(\,A_\mathrm{\,all} - K_\mathrm{\,in}\,\right)\,\mathcal{P}_T}{L_\mathrm{\,in} + L_\mathrm{\,d}\,\displaystyle{\left[\,\frac{2}{1 - \frac{\lambda_2\,\left[\,e^{\,\lambda_1\,\tau}\,\left(\,\lambda_1\,N_0 + \beta\,\right)\, - \, \beta\,\right]
}{\lambda_1\,\left[\,e^{\,\lambda_2\,\tau}\,\left(\,\lambda_2\,N_0 + \beta\,\right)\, - \, \beta\,\right]}}\,-\,1\,\right]
}}} &  \mbox{for} \ \beta \neq 0 \ \ \& \ \ N_0 \neq 0 \nonumber 
\end {array}
\right.\nonumber
\end{eqnarray}
where, as usual, $\beta$ is the constant rate at which particles are fed into the beam and $N_0$ is the number of particles in the beam initially.

%\pagebreak

\section{Constant beam intensity}
\label{sec:Constant_beam_intensity}

In this case the accumulation rate is set specifically so that extra particles are fed into the beam at such a rate so that the beam intensity is kept constant, {\it i.e.}\ fed in at such a rate to balance the rate at which particles are scattered out of the beam.  The system of equations is much simpler in this case.   Here $N(\tau) = N_0$ is a constant, hence $\mathrm{d}\,N(\tau)/\mathrm{d}\,\tau =0$, and the $J(\tau)$ equation becomes a first order linear ODE with constant coefficients
\begin{equation}
\frac{\mathrm{d}\,J(\tau)}{\mathrm{d}\,\tau} \, + \,  n\,\nu \, \left(\,I_\mathrm{\,all} - D_\mathrm{\,in}\,\right)\,J(\tau) \, = \,  - \, n\,\nu \, \left(\,A_\mathrm{\,all} - K_\mathrm{\,in}\,\right)\,\mathcal{P}_T \ N_0 \,,
\end{equation}
and imposing the initial conditions $N(0) = N_0$ and $J(0) = 0$ one obtains the solution
\begin{eqnarray}
J(\tau) & = & \frac{-\,\left(\,A_\mathrm{\,all} - K_\mathrm{\,in}\,\right)\,\mathcal{P}_T \ N_0}{\left(\,I_\mathrm{\,all} - D_\mathrm{\,in}\,\right)} \ \left[\,1\,-\,e^{\,-\, n\,\nu \, \left(\,I_\mathrm{\,all} \ - \ D_\mathrm{\,in}\,\right)\, \tau}\ \right] \,.
\end{eqnarray}
Now the polarization as a function of time can be presented
\begin{equation}
\label{eq:Polarization_in_N_Constant_case}
\mathcal{P}(\tau) \, = \, \frac{J(\tau)}{N(\tau)} \, = \, \frac{J(\tau)}{N_0} \, = \ \frac{-\,\left(\,A_\mathrm{\,all} - K_\mathrm{\,in}\,\right)\,\mathcal{P}_T}{\left(\,I_\mathrm{\,all} - D_\mathrm{\,in}\,\right)} \ \left[\,1\,-\,e^{\,-\, n\, \nu \,  \left(\,I_\mathrm{\,all}\ - \ D_\mathrm{\,in}\,\right)\, \tau}\ \right] \,.
\end{equation}
To find the maximum polarization achievable, {\it i.e.}\ the limit as time tends to infinity, we note that $I_\mathrm{\,all} > D_\mathrm{\,in}$ thus $-\, n\,\nu \, \left(\,I_\mathrm{\,all} - D_\mathrm{\,in}\,\right) < 0$ and hence one obtains
\begin{equation}
\mathcal{P}_\mathrm{max} \ = \ \lim_{\tau \to \,\infty} \mathcal{P}(\tau) \ = \ \frac{-\,\left(\,A_\mathrm{\,all} - K_\mathrm{\,in}\,\right)\,\mathcal{P}_T}{I_\mathrm{\,all} - D_\mathrm{\,in}} \,,
\end{equation}
which is the same as in the inhomogeneous case of section~\ref{sec:Accumulation_of_antiprotons_in_the_ring} when $\beta \neq 0$.

The initial rate of polarization buildup can be obtained by expanding $\mathcal{P}(\tau)$ as a Taylor expansion in $n\,\nu\,\tau$.  Assuming $n\,\nu\,\tau$ is small we neglect terms of second or higher order giving  
\begin{equation}
\frac{\mathrm{d}\,\mathcal{P}}{\mathrm{d}\,\tau} \  \approx \   -\,n\,\nu\,\left(\,A_\mathrm{\,all} - K_\mathrm{\,in}\,\right)\,\mathcal{P}_T \, ,
\end{equation}
as it was in the homogeneous case presented in eq.~(\ref{eq:Homogeneous_initial_and_max_polarization}).  The figure of merit in this case is easily obtained
\begin{eqnarray}
\mathrm{FOM}(\tau) 
& = & \frac{\left(\,A_\mathrm{\,all} - K_\mathrm{\,in}\,\right)^{\,2}\,\mathcal{P}_T^{\,2}\,N_0}{\left(\,I_\mathrm{\,all} - D_\mathrm{\,in}\,\right)^{\,2}} \ \left[\,1 \, - \, e^{\,-\, n\,\nu \,\left(\,I_\mathrm{\,all}\ - \ D_\mathrm{\,in}\,\right)\, \tau}\ \right]^{\,2} ,
\end{eqnarray}
and increases monotonically with time.

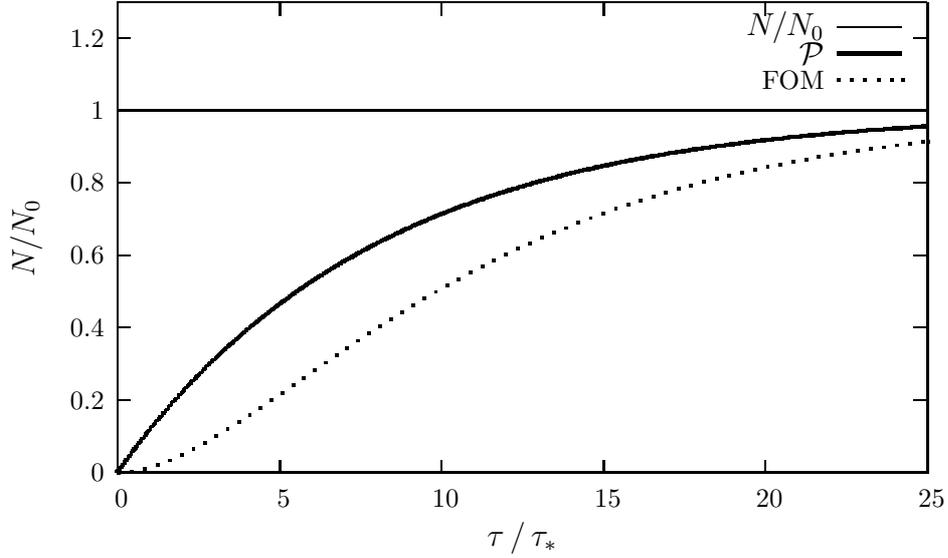
\begin{figure}
\centering
\input{FOM_Plot_Constant_N2.tex} %x-range to 25 
\caption{\small{\it{A schematic graph showing the behaviour of the beam intensity $N$, which is constant in this case, and beam polarization $\mathcal{P}$ as time (scaled by the beam lifetime $\tau_*$) increases.    The figure of merit is also shown on the graph, with the same scales as the other functions.  The graph just shows general trends and is not in exact numerical correspondence to the equations, which will be presented in Chapter~\ref{ch:Numerical_results}.  The beam lifetime is defined as earlier, in the absence of particle input to the beam.
}}
}
\label{fig:FOM_Plot_Constant_N}
\end{figure}

The behaviour of the beam intensity, beam polarization and figure of merit as functions of time are plotted in Figure~\ref{fig:FOM_Plot_Constant_N}.  One notes that in this case the figure of merit increases monotonically and hence does not have a maximum in finite time.  Therefore in this scenario it would be best to continue the spin filtering process for as long as possible before extracting the polarized antiproton beam.

\subsection{Approximating the critical input rate}
\label{subsec:Approximating_the_critical_input_rate}

The accumulation rate needed to keep the beam intensity constant is important, as this critical rate divides the solution of the system into two physically distinct cases.  Smaller accumulation rates than this critical value cause the beam intensity to decrease, hence the FOM will have a maximum in finite time.  Larger values than the critical value cause the beam intensity to increase continuously, hence the FOM will increase monotonically.  We can see from eq.~(\ref{eq:Homogeneous_N}) that $N(\tau)$ does not decrease linearly with time $\tau$.  So the accumulation rate needed to keep the beam intensity constant, say $f(\tau)$, will not be linear in $\tau$.  We now derive the function  $f(\tau)$ and obtain a linear approximation to it, which can be used in the inhomogeneous case treated in section~\ref{sec:Accumulation_of_antiprotons_in_the_ring}.  We must solve $N(\tau) = N_{hom}(\tau) + f(\tau) = N_0$, {\it i.e.}\
\begin{equation}
N(\tau)  \, = \,  \frac{\left[\,e^{\,\lambda_1\,\tau}\, \left(\,L_\mathrm{\,d} - L_\mathrm{\,in}\,\right) \, + \, e^{\,\lambda_2\,\tau}\, \left(L_\mathrm{\,d} + L_\mathrm{\,in}\right)\,\right]\,N_0}{2\, L_\mathrm{\,d}} + f(\tau) \, = \, N_0 \, ,
\end{equation}
which leads to
\begin{eqnarray}
f(\tau) 
 & = & \frac{N_0}{2\, L_\mathrm{\,d}}\ \left[\,2\, L_\mathrm{\,d} - e^{\,\lambda_1\,\tau}\, \left(\,L_\mathrm{\,d} - L_\mathrm{\,in}\,\right) - e^{\,\lambda_2\,\tau}\, \left(L_\mathrm{\,d} + L_\mathrm{\,in}\right)\,\right] \,.
\end{eqnarray}
A linear approximation $f_L(\tau)$ to $f(\tau)$ can be found by Taylor expanding the exponentials to order $\tau$ (which is valid since $n\,\nu\,\tau$ is small), to obtain
\begin{equation}
f_L(\tau)  \ = \ n\,\nu\,N_0\, I_\mathrm{\,out}\,\tau \, ,
\end{equation}
which is in the linear form $+\,\beta_c\,\tau$ where $\beta_c = n\,\nu\,N_0\, I_\mathrm{\,out}$ is the critical value of $\beta$ which when added to the $\mathrm{d}\,N(\tau)\,/\,\mathrm{d}\,\tau$ differential equation  in section~\ref{sec:Accumulation_of_antiprotons_in_the_ring} approximately makes the beam intensity constant.

\section{The input rate is ramped up}
\label{sec:The_input_rate_is_ramped_up}

In this section we investigate a scenario where unpolarized particles are input into the beam at a linearly increasing rate, {\it i.e.}\ the input rate is ramped up.  This is accounted for by the following system of polarization  evolution equations
\begin{eqnarray}
\label{eq:ODE_for_N}
\hspace*{-2em} \frac{\mathrm{d}\,N(\tau)}{\mathrm{d}\,\tau} & = & \, - \, n \, \nu \ \left[\, I_{{\mathrm{\,out}}} \ N(\tau) \ + \ \mathcal{P}_T \, A_{{\mathrm{\,out}}} \ J(\tau)\, \right] \ + \  \beta \, \tau \,,\\[1ex]
\label{eq:ODE_for_J}
\hspace*{-2em} \frac{\mathrm{d}\,J(\tau)}{\mathrm{d}\,\tau} & = &\, - \, n \, \nu \  \left[\, \mathcal{P}_T \, \left(\,A_{\mathrm{\, all}} -  K_{\mathrm{\,in}}\,\right)\ N(\tau) \  + \ \left(\,I_{\mathrm{\, all}} -  D_{\mathrm{\, in}}\,\right)\ J(\tau)\, \right] ,
\end{eqnarray}
where $\beta\,\tau$ is the rate at which particles are fed in, the input ramped up at a rate proportional to the time elapsed.  The initial conditions are $N(0) = N_0$ which we may later set to zero, and $J(0) = 0$.  By differentiating eq.~(\ref{eq:ODE_for_J}) with respect to $\tau$ and substituting in eq.~(\ref{eq:ODE_for_N}) one obtains a second order linear inhomogeneous differential equation for $J(\tau)$:
\begin{eqnarray}
\label{eq:Inhomogeoeous_J_Differential_Equation}
\hspace*{-2em}\frac{\mathrm{d}^{\,2}\,J(\tau)}{\mathrm{d}\,\tau^2}  \ - \ \left(\,\lambda_1 + \lambda_2\,\right)\,\frac{\mathrm{d}\,J(\tau)}{\mathrm{d}\,\tau} \ + \ \lambda_1\,\lambda_2\,J(\tau) \ = \ - \,n\, \nu \,\mathcal{P}_T \,\left(\,A_{\mathrm{\,all}} - K_{\mathrm{\,in}}\,\right) \, \beta \, \tau
\,
\end{eqnarray}
the solution of which is
\begin{equation}
\label{eq:Inhomogeneous_J_solution}
J(\tau) \ = \ F_{\lambda_2\,\lambda_1}\,e^{\,\lambda_1\,\tau} \ +\  F_{\lambda_1\,\lambda_2}\,e^{\,\lambda_2\,\tau} \ +\ \beta \,\left(\,A_1\,\tau + A_2 \,\right)\,.
\end{equation}
Where for convenience we have defined the constants
\begin{eqnarray}
A_1 & \equiv & \frac{-\,n\,\nu\,\mathcal{P}_T \,\left(\,A_{\mathrm{\,all}} - K_{\mathrm{\,in}}\,\right)}{\lambda_1\,\lambda_2} \,,\\[2ex]
%
%\hspace*{1em} \mbox{and} \hspace*{1em}
%
A_2 & \equiv & \frac{2\,n^{\,2}\,\nu^{\,2}\,\mathcal{P}_T \,\left(\,A_{\mathrm{\,all}} - K_{\mathrm{\,in}}\,\right)\left(\,L_{\mathrm{\,in}} + I_{\mathrm{\,out}}\,\right)}{\lambda_1^{\,2}\,\lambda_2^{\,2}} \,,
\end{eqnarray}
\begin{equation}
F_{\lambda_2\,\lambda_1} \ \equiv \ \frac{n\,\nu\,\left(\,A_\mathrm{\,all} - K_\mathrm{\,in}\,\right)\,N_0\,\mathcal{P}_T \ +\  \beta \, \left(\,A_1 - \lambda_2\,A_2\,\right)}{\lambda_2 - \lambda_1} \,,
\end{equation}
obtained by imposing the initial conditions $J(0) = 0$ and $N(0) = N_0$ thus $\mathrm{d}\,J (0) / \mathrm{d}\,\tau = -\,n \, \nu \,\mathcal{P}_T \,\left(\,A_{\mathrm{\,all}} - K_{\mathrm{\,in}}\,\right)\,N_0$.  The function $F_{\lambda_1\,\lambda_2}$ is $F_{\lambda_2\,\lambda_1}$ with $\lambda_1$ and $\lambda_2$ interchanged.  
Differentiating eq.~(\ref{eq:Inhomogeneous_J_solution}) with respect to $\tau$ and substituting into eq.~(\ref{eq:ODE_for_J}) gives an expression for $N(\tau)$:
\begin{eqnarray}
\label{eq:Inhomogeneous_N_Solution}
N(\tau) & = & \frac{-1}{\left(\,A_{\mathrm{\,all}} - K_{\mathrm{\,in}}\,\right)\,\mathcal{P}_T}\, \left\{\,
\phantom{\frac{1}{2}} \hspace*{-0.9em}
F_{\lambda_2\,\lambda_1}\,e^{\,\lambda_1\,\tau}\left(\,L_{\mathrm{\,in}} \,-\, L_{\mathrm{\,d}}\,\right) \,+\, F_{\lambda_1\,\lambda_2}\,e^{\,\lambda_2\,\tau}\left(\,L_{\mathrm{\,in}} \,+\, L_{\mathrm{\,d}}\,\right)\right. \nonumber \\[1ex]
&   &  \left. \qquad \qquad \qquad \quad \, + \, \beta \, \left[\,\frac{A_1}{n\,\nu} \,+\, \left(\,I_{\mathrm{\,out}}\,+\, 2\,L_{\mathrm{\,in}}\,\right)\,\left(\,A_1\,\tau + A_2\,\right)\,\right]
\,\right\}.
\end{eqnarray}
As a consistency check it can be seen that the inhomogeneous solutions for $J(\tau)$ and $N(\tau)$ satisfy the initial conditions, and that when $\beta = 0$ they reduce to the solutions of the homogeneous system eq.~(\ref{eq:HomogeneousSystem}) presented in section~\ref{sec:Solving_the_polarization_evolution_equations}.

Dividing $J(\tau)$ by $N(\tau)$ we obtain an expression for the polarization as a function of time ($\tau$),
\begin{equation}
{\mathcal{P}}(\tau) \ = \  \frac{-\,\mathcal{P}_T\,\left(\,A_{\mathrm{\,all}} - K_{\mathrm{\,in}}\,\right)}{L_{\mathrm{\,in}} + L_{\mathrm{\,d}}\,\displaystyle{\left[\,\frac{2}{1 - \frac{e^{\,\lambda_1\,\tau}\,F_{\lambda_2\,\lambda_1}\,\left(\,\lambda_2 - \lambda_1\,\right)\, - \, \beta\,\left[\,A_1\,\left(\,1 - \lambda_2\,\tau\,\right) - \lambda_2\,A_2\,\right]
}{e^{\,\lambda_2\,\tau}\,F_{\lambda_1\,\lambda_2}\,\left(\,\lambda_1 - \lambda_2\,\right)\, - \, \beta\,\left[\,A_1\,\left(\,1 - \lambda_1\,\tau\,\right) - \lambda_1\,A_2\,\right]}}\,-\,1\,\right]
}}\,.
\end{equation}
When $\beta = 0$ the above equation simplifies to 
\begin{equation}
{\mathcal{P}}(\tau) \ = \ 
\frac{-\, \mathcal{P}_T\,\left(\, A_{\mathrm{\,all}} \, - \, K_{\mathrm{\,in}}\,\right)}{L_{\mathrm{\,in}} \, + \, L_{\mathrm{\,d}} \, \coth\left(L_{\mathrm{\,d}}\, n\,\nu\,\tau\right)}\,,
\end{equation}
which is the solution of the homogeneous case eq.~(\ref{eq:HomogeneousSystem}) presented in section~\ref{sec:Solving_the_polarization_evolution_equations}.

Of interest is the case when $N(0) = N_0 = 0$, {\it i.e.}\ there are no particles in the beam initially.  To obtain this result we set $N_0 = 0$ in the above equation to obtain
\begin{equation}
{\mathcal{P}}(\tau) \ = \ \frac{-\,\mathcal{P}_T\,\left(\,A_{\mathrm{\,all}} - K_{\mathrm{\,in}}\,\right)}{L_{\mathrm{\,in}} + L_{\mathrm{\,d}}\,\displaystyle{\left[\,\frac{2}{1 - \frac{\left(\,e^{\,\lambda_1\,\tau} - 1\,\right)\,\lambda_2\,A_2 - A_1\,\left(\,e^{\,\lambda_1\,\tau} + \lambda_2\,\tau -1\,\right)}{\left(\,e^{\,\lambda_2\,\tau}-1\,\right)\,\lambda_1\,A_2 - A_1\,\left(\,e^{\,\lambda_2\,\tau} + \lambda_1\,\tau -1\,\right)}}\,-\,1\,\right]
}}\,,
\end{equation}
where for $\beta \neq 0$ the $\beta$ dependence vanishes.  Using a Taylor Series expansion we obtain the approximate initial rate of polarization buildup 
\begin{equation}
   \frac{{\mathrm{d}}\,{\mathcal{P}}}{{\mathrm{d}}\tau} \, \, \approx \, -\, n \, \nu \, \mathcal{P}_T
\,
\left( A_{\mathrm{\,all}} \, - \, K_{\mathrm{\,in}} \right)\,,
\end{equation}
identical to that of the homogeneous case presented in section~\ref{sec:Solving_the_polarization_evolution_equations}.  The maximum polarization achievable is the limit as time approaches infinity:
\begin{equation}
{\mathcal{P}}_{\mathrm{max}} \ = \ \lim_{\tau \to \infty} {\mathcal{P}}(\tau) \ =\  \frac{-\,\mathcal{P}_T\,\left(\,A_{\mathrm{\,all}} - K_{\mathrm{\,in}}\,\right)}{I_{\mathrm{\,all}} - D_{\mathrm{\,in}}} \ =  \frac{-\,\mathcal{P}_T\,\left(\,A_{\mathrm{\,all}} - K_{\mathrm{\,in}}\,\right)}{I_{\mathrm{\,out}} + 2\,L_{\mathrm{\,in}}}\,.
\end{equation}
The above expression is only valid for $\beta \neq 0$, the $\beta = 0$ expression is presented in section~\ref{sec:Solving_the_polarization_evolution_equations}.

For this inhomogeneous case the figure of merit is:
\begin{eqnarray}
\mathrm{FOM}(\tau) & = & {\mathcal{P}}^{\,2}(\tau)\,N(\tau) \ =  \frac{J^{\,2}(\tau)}{N(\tau)} \ = \  \\[2ex]
& & \hspace*{-6em} \frac{- \,\mathcal{P}_T\,\left(\,A_{\mathrm{\,all}} - K_{\mathrm{\,in}}\,\right)\,\left[\,F_{\lambda_2\,\lambda_1}\,e^{\,\lambda_1\,\tau} \ + \  F_{\lambda_1\,\lambda_2}\,e^{\,\lambda_2\,\tau} \ + \ \beta \,\left(\,A_1\,\tau \,+\, A_2 \,\right)\,\right]^{\,2}}{F_{\lambda_2\,\lambda_1}\,e^{\,\lambda_1\,\tau}\left(L_{\mathrm{\,in}} + L_{\mathrm{\,d}}\right)- F_{\lambda_1\,\lambda_2}\,e^{\,\lambda_2\,\tau}\left(L_{\mathrm{\,in}} + L_{\mathrm{\,d}}\right) +  \beta  \left[\,\frac{A_1}{n\,\nu} + \left(I_{\mathrm{\,all}} - D_{\mathrm{\,in}}\right)\left(A_1\,\tau + A_2\right)\,\right]} \nonumber \,.
\end{eqnarray}
If the particle accumulation rate $\beta\,\tau$ is high enough to make the beam intensity constant or increase with time the figure of merit will be a monotonically increasing function of time, {\it i.e.}\ it will not have a maximum in finite time.

%\pagebreak

\section{Stored beam with initial polarization}
\label{sec:Stored_beam_with_initial_polarization}

We now solve the homogeneous system where the initial polarization is not zero.  This will be used if two methods of polarizing antiprotons are combined, {\it i.e.}\ if antiprotons were produced with a small polarization by some other method and one wanted to increase that polarization by spin filtering in a storage ring, where the luminosity could also be increased.    In this section the beam has been stored and there is no further input of particles into the beam.

The system of differential equations, the eigenvalues and eigenvectors are the same as section~\ref{sec:Solving_the_polarization_evolution_equations}, but one of the initial conditions is different.  The new initial conditions are $N(0) = N_0 > 0$ the total number of particles in the beam initially, and $J(0) = J_0 \neq 0 \Rightarrow N_+(0) \neq N_-(0)$ {\it i.e.}\ initially the beam is polarized.  Note that since the number of particles in the beam in one particular spin state must not be greater than the total number of particles in the beam the bound $|J_0| \leq N_0$ is respected.  A negative value for $J_0$ simply implies that the antiproton beam is initially polarized in the opposite direction to the polarization direction of the target.  
Enforcing these initial conditions leads to the solutions
\begin{equation}
\label{eq:Initial_Polarization_N}
 N(\tau) \ = \  \frac{\left(\,J_0\,\mathcal{P}_T\,A_\mathrm{\,out} \,-\,N_0\,L_\mathrm{\, in}\,\right)\, \left(\,e^{\,\lambda_1\,\tau}\, - \, e^{\,\lambda_2\,\tau}\,\right)\, +\, N_0 \,L_\mathrm{\, d}\, \left(\,e^{\,\lambda_1\,\tau} \,+\, e^{\,\lambda_2\,\tau}\,\right)}{2\,L_\mathrm{\, d}} \,,
\end{equation}
\vspace*{1ex}
\begin{equation}
\label{eq:Initial_Polarization_J}
\hspace*{-0.5em} J(\tau) \ = \ \frac{\left[\,N_0 \,\mathcal{P}_T \left(\,A_\mathrm{\,all} - K_\mathrm{\,in}\,\right) + J_0\,L_\mathrm{\, in} \,\right] \left(\,e^{\,\lambda_1\,\tau} - e^{\,\lambda_2\,\tau}\,\right) \,+\, J_0\,L_\mathrm{\, d} \left(\,e^{\,\lambda_1\,\tau} + e^{\,\lambda_2\,\tau}\,\right)}{2\,L_\mathrm{\, d}} \,,
\end{equation}
which reduce to the solutions of the original homogeneous system presented in eqs.~(\ref{eq:Homogeneous_N} and \ref{eq:Homogeneous_J}) when $J_0 \rightarrow 0$.  The beam lifetime is the same to leading approximation as in the homogeneous case when $J(0) = 0$.  Dividing $J(\tau)$ by $N(\tau)$ provides the time dependence of the polarization of the beam
\begin{eqnarray}
\hspace*{-2em} \mathcal{P}(\tau)  & = &   \frac{L_\mathrm{\, d}\,J_0 - \,\tanh\left(L_\mathrm{\, d}\,n\,\nu\,\tau\,\right)\,\left[\,L_\mathrm{\, in}\,J_0 \,+\, N_0\,\mathcal{P}_T\,\left(\,A_\mathrm{\,all} - K_\mathrm{\,in}\,\right)\,\right]}{L_\mathrm{\, d}\,N_0 \,+\, \tanh\left(L_\mathrm{\, d}\,n\,\nu\,\tau\,\right)\,\left[\,L_\mathrm{\, in}\,N_0 \,-\, J_0\,\mathcal{P}_T\,A_\mathrm{\,out}\,\right]} \,.
\end{eqnarray}
Denoting the initial polarization $\mathcal{P}(0) = J_0\,/\,N_0 \equiv \mathcal{P}_0$ the above can be written as
\begin{equation}
\label{eq:Polarization_buildup_with_initial_polarization}
\mathcal{P}(\tau)  \ = \  \frac{L_\mathrm{\, d}\,\mathcal{P}_0 \,-\, \tanh\left(L_\mathrm{\, d}\,n\,\nu\,\tau\,\right)\,\left[\,L_\mathrm{\, in}\,\mathcal{P}_0 \,+\, \mathcal{P}_T\,\left(\,A_\mathrm{\,all} - K_\mathrm{\,in}\,\right)\,\right]}{L_\mathrm{\, d} \,+\, \tanh\left(L_\mathrm{\, d}\,n\,\nu\,\tau\,\right)\,\left[\,L_\mathrm{\, in}\,-\, \mathcal{P}_0\,\mathcal{P}_T\,A_\mathrm{\,out}\,\right]}\,.
\end{equation}
The approximate rate of change of polarization for sufficiently short times is found by Taylor expanding to first order in $\tau$
\begin{equation}
   \frac{\mathrm{d}\,\mathcal{P}}{\mathrm{d}\tau} \, \ \approx \  n\,\nu\,\left\{\,\left[\,A_\mathrm{\,out}\,\mathcal{P}_0^{\,2} \,-\, \left(\,A_\mathrm{\,all} - K_\mathrm{\,in}\,\right)\,\right]\,\mathcal{P}_T\, -\, 2\,\mathcal{P}_0\,L_\mathrm{\, in}\,\right\}\,.
\end{equation}
The limit as time goes to infinity of $\mathcal{P}(\tau)$ in eq.~(\ref{eq:Polarization_buildup_with_initial_polarization}) is
\begin{equation}
 \lim_{\tau \to \,\infty} \ \mathcal{P}(\tau) \ = 
\ 
\frac{ \mathcal{P}_0\,\left(\,L_\mathrm{\,d} \, - \, L_\mathrm{\,in}\,\right)\,-\, \mathcal{P}_T\,\left(\,A_\mathrm{\,all} \, - \, K_\mathrm{\,in}\,\right)
}
{   \left(\,L_\mathrm{\,in} \, + \, L_\mathrm{\,d}\,\right) - A_\mathrm{\,out}\,\mathcal{P}_0\ \mathcal{P}_T
} \, ,
\end{equation}
which of course agrees with the earlier maximum polarization if $J_0 = 0$ ({\it i.e.}\ $\mathcal{P}_0 = 0$).  
The figure of merit for this case is:
\begin{eqnarray}
\mathrm{FOM}(\tau) & = & \mathcal{P}^{\,2}(\tau)\,N(\tau) \ = \ \frac{J^{\,2}(\tau)}{N(\tau)} \ = \\[2ex]
& & \hspace*{-6em}  \displaystyle{ \frac{N_0\,\left\{\,\left[\,\mathcal{P}_T\,\left(\,A_\mathrm{\,all} - K_\mathrm{\,in}\,\right) + \mathcal{P}_0\,L_\mathrm{\, in} \,\right] \left(\,e^{\,\lambda_1\,\tau} - e^{\,\lambda_2\,\tau}\,\right) + \mathcal{P}_0\,L_\mathrm{\, d} \left(\,e^{\,\lambda_1\,\tau} + e^{\,\lambda_2\,\tau}\,\right)\,\right\}^{\,2}}
{2\,L_\mathrm{\,d}\,\left\{\,\left(\,\mathcal{P}_0\,\mathcal{P}_T\,A_\mathrm{\,out} \,-\,L_\mathrm{\, in}\,\right)\, \left(\,e^{\,\lambda_1\,\tau}\, - \, e^{\,\lambda_2\,\tau}\,\right)\, +\,L_\mathrm{\, d}\, \left(\,e^{\,\lambda_1\,\tau} \,+\, e^{\,\lambda_2\,\tau}\,\right)\,\right\}}} \,.\nonumber 
\end{eqnarray}
Note that in terms of the number of particles in each spin state the initial conditions for an initially polarized beam, $N(0) \,=\,N_0$ and $J(0)\,=\,J_0$, are
\begin{equation}
N_+(0) \ = \ \frac{N_0}{2}\,\left(\,1\,+\,\mathcal{P}_0\,\right)
\hspace*{2em} \hbox{and} \hspace*{2em}
N_-(0) \ = \ \frac{N_0}{2}\,\left(\,1\,-\,\mathcal{P}_0\,\right)\,.
\end{equation}

\subsection{An unpolarized target}
\label{subsec:An_unpolarized_target}

A special case of this system deserves comment.  Given that the beam is initially polarized what happens if the target is unpolarized?  One would imagine that the beam polarization should decrease and eventually reach zero.  We now analyze the equations of section~\ref{sec:Stored_beam_with_initial_polarization} when the target is unpolarized ({\it i.e.}\ $\mathcal{P}_T = 0$) and use the fact that $L_\mathrm{\,d} = \sqrt{\, \mathcal{P}_T^{\,2} \, A_\mathrm{\,out} \left( A_\mathrm{\,all}\, - \, K_\mathrm{\,in} \right) \, + \, L_\mathrm{\,in}^{\,2} } = L_\mathrm{\,in}$ when $\mathcal{P}_T = 0$ to obtain the beam polarization as a function of time
\begin{eqnarray}
\mathcal{P}(\tau)  & = & 
\mathcal{P}_0\ \displaystyle{e^{\,\left(\,\lambda_1 \,-\, \lambda_2\,\right)\ \tau}} \, ,
\end{eqnarray}
which is an exponentially decreasing function of $\tau$ for $\lambda_1 - \lambda_2 < 0$.  The beam polarization will not decrease in the special case of $\lambda_1 - \lambda_2 = 0$, but this only happens when $L_\mathrm{\,in} = 0$, {\it i.e.}\ when there is no depolarization.  The special case of $\lambda_1 - \lambda_2 = 0$, which does not lead to polarization buildup as seen from eq.~(\ref{eq:Initial_Polarization_J}), would be avoided by any experimental effort, thus is omitted from the rest of the discussion.  

The limit of beam polarization for large times when $\mathcal{P}_T = 0$ is
\begin{equation}
   \lim_{\tau \to \,\infty} \ \mathcal{P}(\tau) \ = 
\ 0\,.
\end{equation}
Thus, as expected, if the beam is initially polarized and the target unpolarized then the beam polarization will decrease with time and eventually the beam polarization will reduce to zero.  Thus a beam cannot gain polarization from an unpolarized target by spin filtering.

The figure of merit in this case simplifies to $\mathrm{FOM}(\tau) = N_0\,\mathcal{P}_0^{\,2}\,e^{\,\left(\,2\,\lambda_1 - \lambda_2\,\right)\,\tau}$ which is a monotonically decreasing function of $\tau$.  One can derive a polarization half-life in this case, the time taken for the polarization to decrease by a factor of $2$, by solving $\mathcal{P}(\tau) = \mathcal{P}_0\,/\,2$ to obtain
\begin{equation}
\tau_{\, \frac{1}{2}} \ = \ \frac{\ln 2}{\lambda_2 \,-\, \lambda_1} \ = \ \frac{\ln 2}{2\,n\,\nu\,L_\mathrm{\,d}} \,.
\end{equation}
This scenario occurs in an electron cooler, a device used to focus the beam in many storage rings\footnote{Electron cooling is described in section~\ref{sec:Beam_Cooling}.}.  The beam passes through a co-moving beam of unpolarized electrons with low transverse momentum, in order to dampen the transverse momentum of the antiprotons in the stored beam.  But the low electron areal densities in cooler beams, where typically $n \, \approx \, 10^{-18}~\mbox{fm}^{-2} \, = \, 10^{-19}~\mbox{mb}^{-1} $, causes the polarization half-life to be very large.  Thus our work shows that this depolarization effect is negligible, in agreement with Ref.~\cite{Anderson:1986gu}.  Since electron cooling is a necessary part of the spin filtering process of polarization buildup it is very important that the depolarization caused by electron cooling is negligible.

It has recently been suggested that the positron-antiproton polarization transfer observable is very much enhanced at low energies \cite{Arenhovel:2007gi,Milstein:2008tc}.  This enhancement is the basis of the recent proposal by Th.~Walcher \emph{et al.}\ to polarize an antiproton beam by repeated interaction with a co-moving polarized positron beam in a storage ring \cite{Walcher:2007sj}.  All of the antiprotons remain within the beam in this scenario and one avoids the problem of the antiprotons annihilating with protons in an atomic gas target.  This large enhancement of the polarization transfer observable at low energies is due to the unlike charges of the positron and antiproton, and does not occur for the like charges case of antiproton-electron scattering.  Hence this does not affect the conclusion that depolarization of an antiproton beam in an electron cooler is negligible.

\subsection{A critical value for the target polarization}
\label{subsec:A_critical_value_for_the_target_polarization}

The beam polarization will also decrease for low values of the target polarization.  In fact there is a critical value of the target polarization $\mathcal{P}_T$ which keeps the beam polarization constant.  If the target polarization is above this critical value the polarization of the beam will increase, and if the target polarization is below this critical value the beam polarization will decrease.  The critical value is obtained by solving
\begin{equation}
\mathcal{P}(\tau) \ = \  \frac{L_\mathrm{\, d}\,\mathcal{P}_0 \, - \,\tanh\left(L_\mathrm{\, d}\,n\,\nu\,\tau\,\right)\,\left[\,L_\mathrm{\, in}\,\mathcal{P}_0 \, + \, \mathcal{P}_T\,\left(\,A_\mathrm{\,all} - K_\mathrm{\,in}\,\right)\,\right]}{L_\mathrm{\, d} \, + \, \tanh\left(L_\mathrm{\, d}\,n\,\nu\,\tau\,\right)\,\left[\,L_\mathrm{\, in} \, - \, \mathcal{P}_0\,\mathcal{P}_T\,A_\mathrm{\,out}\,\right]} \ = \ \mathcal{P}_0 \, ,
\end{equation}
for $\mathcal{P}_T$, where the time dependence will cancel leading to
\begin{equation}
\mathcal{P}_T^\mathrm{\,critical} \ = \ \frac{2\,\mathcal{P}_0\,L_\mathrm{\, in}}{\mathcal{P}_0^{\,2}\,A_\mathrm{\,out} \ - \ \left(\,A_\mathrm{\,all} \,-\, K_\mathrm{\,in}\,\right)}\,.
\end{equation}
For target polarizations below this critical value the maximum beam polarization occurs at time $\tau = 0$, and for target polarizations above this critical value the maximum beam polarization occurs at large times $\tau \rightarrow \infty$.

\section{Particles fed in for a limited time}
\label{sec:Particles_fed_in_for_a_limited_time}

The Heaviside step function could be used in the system of equations to explain the case of particles input into a beam for a certain amount of time after which the input is turned off and no more particles are fed into the beam, but spin filtering continues\footnote{This section makes extensive use of Laplace transform methods in solving differential equations.  All results needed are presented in Appendix~\ref{Appendix:Laplace_transform_methods}.}.  This scenario is under consideration by the $\mathcal{P}\mathcal{A}\mathcal{X}$ Collaboration \cite{Barone:2005pu}, but there has been no theoretical treatment of it to date.  The Heaviside function is a piecewise continuous function which is zero in one region and one everywhere else, it is used in many mathematical modeling problems to describe an external effect turned on or off after a certain duration of time.  In our case this external effect is the input of particles into the beam.  The Heaviside function is defined as  
\begin{eqnarray}
\label{eq:Heaviside_function_definition}
 H\left(\,\tau \,-\, \tau_c\,\right) \ = \ \left\{ 
\begin{array}{ll} 
0  & \hspace*{2em} \mbox{if} \ \tau < \tau_c\\[2ex]  
1 & \hspace*{2em} \mbox{if} \ \tau \geq \tau_c
\end {array}
\right.
\end{eqnarray}
and is used to describe an external effect turned on at time $\tau_c$, but we require an external effect on initially and turned off at time $\tau_c$, thus we need 
\begin{eqnarray}
\label{eq:Reverse_Heaviside_function_definition}
H\left( \tau \right) \,-\, H \left(\,\tau \,-\, \tau_c\,\right) \ = \ \left\{ 
\begin{array}{ll} 
1  & \hspace*{2em} \mbox{if} \ 0 \leq \tau < \tau_c \\[2ex] 0 & \hspace*{2em} \mbox{if} \ \tau \geq \tau_c
\end {array}
\right.
\end{eqnarray}
Note that in the special case when $\tau_c = 0$, $\left[\,H\left( \tau \right) - H\left(\tau\right)\,\right]  = 0$, as this is in the second region.  This describes a physical situation where particles are being fed in for zero seconds, which is the same as saying no particles are fed in, so the second order ODE should be homogeneous in this case, which it is.  The extra term to add to the $\mathrm{d}\,N\,/\,\mathrm{d}\,\tau$ equation to account for particles being fed in at a constant rate $\beta$ per second for $\tau_c$ seconds after which the input is switched off is $\beta\,\left[\,H\left( \tau \right) - H \left(\,\tau - \tau_c\,\right)\,\right]$.

The equations are now broken into two pieces, {\it i.e.}\ discontinuous, but piecewise continuous.  The solutions will be broken into two regions $0\leq \tau < \tau_c$ and $\tau \geq \tau_c$\,, where the solutions in the region $0 \leq \tau < \tau_c$ should equal those in the inhomogeneous case presented in section~\ref{sec:Accumulation_of_antiprotons_in_the_ring}.  The initial conditions will be $N(0) = 0$ and $J(0) = 0$, thus $J'(0)= 0$.  The Heaviside function above is included in the second order ODE for $J(\tau)$ to obtain:
\begin{eqnarray}
& & \hspace*{-2em}\frac{\mathrm{d}^{\,2}\,J(\tau)}{\mathrm{d}\,\tau^2}  \ - \  \left(\,\lambda_1 \,+\, \lambda_2\,\right)\,\frac{\mathrm{d}\,J(\tau)}{\mathrm{d}\,\tau} \ + \ \lambda_1 \,\lambda_2\, J(\tau)  \\[1ex]
&  & \qquad \qquad \quad = \ - \,n\, \nu \,\mathcal{P}_T\,\left(\,A_\mathrm{\,all} - K_\mathrm{\,in}\,\right) \, \beta\,\left[\,H\left( \tau \right) \,-\, H\left(\,\tau \,-\, \tau_c\,\right)\,\right] \, , \nonumber
\end{eqnarray}
which can be solved by Laplace Transform methods, as described in Appendix~\ref{Appendix:Laplace_transform_methods}, to obtain
\begin{eqnarray}
\label{eq:Heaviside_J_solution}
  \displaystyle{\frac{J(\tau)}{C_1}}  =  \left\{  \begin{array}{ll}
\lambda_1\, \left(\,1\,-\,e^{\,\lambda_2\,\tau}\,\right) \ + \  \lambda_2\,\left(\,e^{\,\lambda_1\,\tau} \,-\, 1\,\right)
&  \ \,\mbox{if} \ 0 \leq \tau < \tau_c \\[5ex] 
\lambda_1\, \left(\,e^{\,\lambda_2\,\left(\,\tau \,-\, \tau_c\,\right)}\,-\, e^{\,\lambda_2\,\tau}\,\right) \ + \ \lambda_2\,\left(\,e^{\,\lambda_1\,\tau} \,-\, e^{\,\lambda_1\,\left(\,\tau \,-\, \tau_c\,\right)}\,\right)
&  \ \,\mbox{if} \ \tau \geq \tau_c
\end {array}
\right.
\end{eqnarray}
where for convenience we have defined the constant factor
\begin{equation}
C_1 \ \equiv \ \frac{\beta\,\mathcal{P}_T \left(\,A_\mathrm{\,all} - K_\mathrm{\,in}\,\right)}{2\,L_\mathrm{d}\,\lambda_1\,\lambda_2} \,.
\end{equation}
One sees from eq.~(\ref{eq:Heaviside_J_solution}) that $J(\tau) = 0 \ \mbox{for all} \ \tau$ when $\tau_c = 0$, which is physically reasonable as there are never any particles in the beam if $\tau_c = 0$.  Also the $\mathcal{P}_T$ factor in $C_1$ indicates that $J(\tau)$ will always be zero if $\mathcal{P}_T = 0$ ({\it i.e.}\ if the target is unpolarized).  It can also be seen that the complete solution in the region $\tau \geq \tau_c$ is the combination of the solution in the region $0 \leq \tau < \tau_c$ and an additional part dependent on $\tau_c \,$; which is
\begin{equation}
C_1\,\left[\,\lambda_1\, \left(\,e^{\,\lambda_2\,\left(\,\tau \,-\, \tau_c\,\right)} \,-\, 1\,\right) \ + \  \lambda_2\,\left(\,1\,-\,e^{\,\lambda_1\,\left(\,\tau \,-\, \tau_c\,\right)}\,\right)\,\right] \, , \nonumber
\end{equation}
and immediately one sees that when $\tau = \tau_c$ this additional part vanishes.  So when $\tau = \tau_c$\,, {\it i.e.}\ at the boundary between the two regions, the two solutions match.  Therefore the solution for $J(\tau)$ is continuous as expected.  The expression for $J(\tau)$ in the first region $0 \leq \tau < \tau_c$  of eq.~(\ref{eq:Heaviside_J_solution}) is equal to the solution of the inhomogeneous system presented in eq.~(\ref{eq:Inhomogeneous_J_with_N0_0}).

A similar analysis as that done for $J(\tau)$ reveals the second order ODE for $N(\tau)$
\begin{eqnarray}
& & \hspace*{-3em} \frac{\mathrm{d}^{\,2}\,N(\tau)}{\mathrm{d}\,\tau^2}  \ - \  \left(\,\lambda_1 \,+\, \lambda_2\,\right)\,\frac{\mathrm{d}\,N(\tau)}{\mathrm{d}\,\tau} \ + \ \lambda_1 \,\lambda_2\, N(\tau) \nonumber \\[1ex]
&  & \qquad \qquad \quad = \ n\, \nu \,\left(\,I_\mathrm{\,all} - D_\mathrm{\,in}\,\right) \, \beta \,\left[\,H\left( \tau \right) \,-\, H\left(\,\tau \,-\, \tau_c\,\right)\,\right]\,.
\end{eqnarray}
The initial conditions are $N(0) = 0$ and $N'(0) = \beta\, \left[\,1 - H\left(\,-\tau_c\,\right)\,\right]$, the latter of which deserves comment.  The rate $N'(0)$ should be $\beta$ when $\tau_c > 0$ and $0$ when $\tau_c = 0$, corresponding to the case of particles being fed in for zero seconds.  Note that $H\left(\,-\tau_c\,\right) = 1 \ \mbox{when} \ \tau_c = 0$ and $H\left(\,-\tau_c\,\right) = 0 \ \mbox{when} \ \tau_c > 0$, and note physically that $\tau_c$\,, the duration for which particles are fed into the beam, cannot be negative.  The Heaviside function in this initial condition forces the solution for $N(\tau)$ to be split into three regions.  On solving by Laplace Transform methods one obtains
\begin{eqnarray}
\label{eq:Heaviside_beam_intensity_solution}
\frac{N(\tau)}{C_2} =  \left\{  \begin{array}{ll} 
   \lambda_1 \,\left[\, 1\,-\, \left(\,\displaystyle{\frac{L_\mathrm{\,in} \,+\, L_\mathrm{d}}{I_\mathrm{\,all} - D_\mathrm{\,in}}}\,\right)\,e^{\,\lambda_2\,\tau} \,\right] \\[4ex]
\qquad \qquad + \ \lambda_2 \,\left[\,\left(\,\displaystyle{\frac{L_\mathrm{\,in} \,-\, L_\mathrm{d}}{I_\mathrm{\,all} - D_\mathrm{\,in}}}\,\right)\,e^{\,\lambda_1\,\tau}  \,-\, 1\,\right]
&  \mbox{if} \ 0 \leq \tau < \tau_c \\[9ex] 
\lambda_1 \,\left[\, e^{\,\lambda_2\,\left(\,\tau \,-\, \tau_c\,\right)}- \left(\,\displaystyle{\frac{L_\mathrm{\,in} + L_\mathrm{d}}{I_\mathrm{\,all} - D_\mathrm{\,in}}}\,\right)\,e^{\,\lambda_2\,\tau} \,\right] \\[4ex]
\qquad \qquad + \ \lambda_2 \,\left[\,\left(\,\displaystyle{\frac{L_\mathrm{\,in} \,-\, L_\mathrm{d}}{I_\mathrm{\,all} - D_\mathrm{\,in}}}\,\right)\,e^{\,\lambda_1\,\tau} \,-\, e^{\,\lambda_1\,\left(\,\tau \,-\, \tau_c\,\right)}\,\right]
&  \mbox{if} \ \tau \geq \tau_c > 0 \\[7ex]
0 & \mbox{if} \ \tau_c = 0 
\end{array}
\right.
\end{eqnarray}
where again for convenience we have defined a constant factor
\begin{equation}
C_2 \ \equiv \ \frac{-\,\beta\,\left(\,I_\mathrm{\,all} - D_\mathrm{\,in}\,\right)}{2\,L_\mathrm{d}\,\lambda_1\,\lambda_2} \,.
\end{equation}
Again one sees that the complete solution in the region $\tau \geq \tau_c > 0$ is the combination of the solution in the region $0 \leq \tau < \tau_c$ plus an additional part dependent on $\tau_c \,$; which is
\begin{equation}
C_2\,\left[\,\lambda_1\, \left(\,e^{\,\lambda_2\,\left(\,\tau \,-\, \tau_c\,\right)} \,-\, 1\,\right) \ + \  \lambda_2\,\left(\,1\,-\,e^{\,\lambda_1\,\left(\,\tau \,-\, \tau_c\,\right)}\,\right)\,\right] \,. \nonumber
\end{equation}
Immediately we see that when $\tau = \tau_c$ this additional part vanishes, thus the solution for $N(\tau)$ is continuous.  The solution in the first region $0 \leq \tau < \tau_c$ is equal to the solution from our inhomogeneous case presented in eq.~(\ref{eq:Inhomogeneous_N_with_N0_0}), and it satisfies the initial condition $N(0) = 0$.

We now present results for the polarization $\mathcal{P}(\tau) = J(\tau)\,/\,N(\tau)$ as a function of time, in both regions.  The polarization is undefined when there are no particles in the beam, thus we need not treat the case $\tau_c = 0$.  As expected in the $0 \leq \tau < \tau_c$ region $\mathcal{P}(\tau)$ equals the solution of our inhomogeneous case presented in eq.~(\ref{eq:Polarization_Inhomogeneous_N_equals_0}), and in the region $\tau \geq \tau_c > 0$ one finds
{\small
\begin{eqnarray}
\hspace*{-1.5em}  
& \hspace*{-1.5em}  & \mathcal{P}(\tau) \ =  \\[2ex]
& \hspace*{-1.5em} & \displaystyle{\frac{\mathcal{P}_T \left(\,A_\mathrm{\,all} - K_\mathrm{\,in}\,\right)\,\left[\,\lambda_1\,e^{\,\lambda_2\,\tau}\,\left(\,e^{\,-\,\lambda_2\,\tau_c}\,-\, 1\,\right) \ + \  \lambda_2\,e^{\,\lambda_1\,\tau}\,\left(\,1 \,-\, e^{\,-\,\lambda_1\,\tau_c}\,\right)\,\right]
}{ \lambda_1 \,e^{\,\lambda_2\,\tau} \left[\, \left(\,I_\mathrm{\,all} - D_\mathrm{\,in}\,\right) e^{-\lambda_2\,\tau_c} - \left(\,L_\mathrm{\,in} + L_\mathrm{d}\,\right) \,\right] + \lambda_2 \,e^{\,\lambda_1\,\tau} \left[\,\left(\,L_\mathrm{\,in} - L_\mathrm{d}\,\right)  - \left(\,I_\mathrm{\,all} - D_\mathrm{\,in}\,\right) e^{-\lambda_1\,\tau_c}\,\right]} \nonumber \,.
}
\end{eqnarray}
}
The approximate initial rate of polarization buildup and the maximum polarization achievable will both reside in the $0 \leq \tau < \tau_c$ region, and thus will be identical to those presented in section~\ref{sec:Accumulation_of_antiprotons_in_the_ring}.  This is because the maximum polarization achievable occurs when the input rate is never switched off, {\it i.e.}\ in the $0 \leq \tau < \tau_c$ region.

%\pagebreak

\chapter{Numerical results}
\label{ch:Numerical_results}

\vspace*{5ex}
\begin{minipage}{6cm}
\end{minipage}
\hfill
\begin{minipage}{10cm}
\begin{quote}
\emph{\lq\lq The whole point of physics is to work out a number, with decimal points etc.! Otherwise you haven't done anything.\rq\rq}
%\begin{right}
\flushright{Richard Feynman}
%\end{right}
\end{quote}
\end{minipage}
\vspace{8ex}

As an application of the theoretical work presented throughout the thesis we now investigate a possible method to produce a high intensity polarized antiproton beam by spin filtering off an opposing polarized electron beam.  It is also outlined how this work can be applied to polarizing antiprotons by spin filtering off a polarized hydrogen target.  Firstly a description of the electron cooling technique to refocus the beam after scattering off the target each revolution in order to maintain high beam density is presented in section~\ref{sec:Beam_Cooling}.  Then the various experimental input parameters, such as revolution frequency, target areal density, target polarization and the effective acceptance angle, needed to obtain realistic numerical estimates from our mathematical formalism are each described in section~\ref{sec:Input_parameters}.  The benefits of using a lepton target are described in section~\ref{sec:Spin_filtering_off_a_polarized_electron_beam}, before analyzing the case of spin filtering off an opposing polarized electron beam.  Finally spin filtering off a polarized hydrogen target is discussed in section~\ref{sec:Spin_filtering_off_a_polarized_hydrogen_target}, in the three cases of hydrogen with only electrons polarized, hydrogen with only protons polarized and finally hydrogen with both electrons and protons polarized.  In section~\ref{subsec:Electromagnetic_and_hadronic_scattering} it is shown that electromagnetic effects dominate hadronic effects in $\bar{p}\,p$ scattering in the region of low momentum transfer of interest in spin filtering. 

%\subsection{Coulomb screening}
%\label{subsec:Coulomb_screening}
%
%The electromagnetic cross-sections have a photon pole, where the cross-sections rise to infinity as the scattering angle tends to zero.  Fortunately there is a minimum scattering angle $\theta_\mathrm{min}$ scattering below which is prevented by Coulomb screening. 

\section{Beam cooling}
\label{sec:Beam_Cooling}

After interaction with the internal target in the storage ring many of the beam particles do not move exactly along the beam axis, {\it i.e.}\ they have acquired a small deflection angle.  This causes the beam to spread transversely and this process is called beam \lq\lq heating", in analogy to the random motion of atoms in a hot thermodynamic gas.  In order to maintain a well ordered beam, where all particles move as collinearly as possible and also to increase the beam transverse density, a method to counteract this beam spread is required.  Fortunately such a method exists and has been utilized successfully in many experiments world wide over the past three decades.  

The method is called {\it Electron Cooling} and was invented in 1966 by G.~I.~Budker at the Institute for Nuclear Physics (INP) laboratory (later renamed the Budker Institute of Nuclear Physics (BINP) in his honour) in Novosibirsk \cite{Budker:1967sd}, as a way to increase the luminosity of $p\,p$ and $\bar{p}\,p$ colliders.  Electron cooling was first tested in 1974 with $68 \ \mbox{MeV}$ protons at the NAP-M storage ring of INP Novosibirsk \cite{Budker:1975xp}, and is currently operational at over a dozen storage rings world wide.

The terminology of \lq\lq cooling" is analogous to thermodynamic cooling, as the random transverse motion of the (anti)protons is dampened by cooling the beam.  A \lq\lq hot" beam has many particles with large transverse motion, whereas a \lq\lq cool" (or \lq\lq cold") beam has low transverse motion, {\it i.e.}\ all particles move collinear to the beam axis.

An electron cooler is a device inserted into the storage ring, where the antiproton beam passes through a co-moving cold electron beam, and on multiple Coulomb scattering with the electrons the transverse motion of the antiprotons is reduced, {\it i.e.}\ the antiproton beam phase-space density is increased.  One immediately thinks of an analogy with the temperatures of mixed gases: Gas A with a high temperature is mixed with gas B having a low temperature, after some time the combined mixture tends to a uniform temperature which is midway between the initial temperatures of the individual gases.  If the gases could be separated afterwards, one could say that gas B has reduced the temperature of gas A.  Fortunately the electrons can be injected into, and extracted out of, an antiproton beam easily by magnets which deflect charged particles at different angles depending on the mass of the charged particles.  Given that the mass of the antiprotons is approximately 1800 times that of the electrons, the electrons can easily be completely removed from the mixture.  The velocity of the electrons in a cooler is carefully set to equal the average velocity of the antiprotons, to maximize the interaction time.  The antiprotons undergo Coulomb scattering in the electron \lq\lq gas" and lose transverse energy, which is transfered to the co-moving electrons until some thermal equilibrium is attained.  
%One could say that hot antiprotons can be cooled by cold electrons just as hot coffee is cooled by milk.  
The electrons get \lq\lq heated up" but are discarded after each pass and new cold electrons are injected continuously.

Electron cooling is conventionally used on low to medium energy (anti)protons.  Many laboratories are now investigating high energy electron cooling \cite{CERNcourier:09_2007} and it has been shown to work on $8.9 \ \mbox{GeV}$ antiprotons in Fermilab \cite{Nagaitsev:2005jr}.  GSI Darmstadt are investigating electron cooling the antiproton beam in the HESR at up to about $8 \ \mbox{GeV}$ \cite{Barone:2001sp}.  It is expected that in the near future high energy electron cooling will be commonplace in many laboratories.

In section \ref{sec:Stored_beam_with_initial_polarization} of this thesis we have proven that the depolarization of an antiproton beam due to electron cooling is negligible because of the low areal density of electrons in a cooler beam.  Therefore the beam can be refocused after interaction with the target each revolution, without losing significant beam polarization.

\section{Input parameters}
\label{sec:Input_parameters}

As shown in the system of polarization evolution equations of Chapter~\ref{ch:Polarization_buildup_by_spin_filtering}, to give the highest possible antiproton polarization after a given filter time, the maximum antiproton revolution frequency, maximum target areal density and maximum target polarization are required.  We now investigate each of these in turn before computing the numerical quantities.  

\subsection{Maximum revolution frequency}
\label{subsec:Maximum_revolution_frequency}

An upper bound on the velocity of the antiprotons is the speed of light $c = 3 \ \times \ 10^8 \ \mbox{m\,s}^{-1}$.  The revolution frequency ($\nu$) is simply the reciprocal of the time taken for one revolution, which in turn is the circumference of the storage ring ($L$) divided by the velocity of the antiprotons ($v = \beta\,c$), {\it i.e.}\
\begin{equation}
\label{eq:Revolution_frequency}
\nu \ = \ \frac{\beta\,c}{L} \,.
\end{equation}
Obviously one can maximize the revolution frequency by using a very high energy beam in an extremely small circumference storage ring, but this is limited by the power of the magnets to bend high energy antiprotons around such a small circumference.  A plausible example is treated in \cite{Walcher:2007sj} with an antiproton polarizer ring of circumference $L = 75 \ \mbox{m}$ and a beam velocity of $\beta = v\,/\,c = 0.5$ giving a revolution frequency of $2\ \mbox{MHz}$.  The $\mathcal{P}\mathcal{A}\mathcal{X}$ Collaboration proposes an Antiproton Polarizer Ring (APR) of circumference $L = 86.5 \ \mbox{m}$ with antiprotons of kinetic energy $250 \ \mbox{MeV}$ which, using the relation $p =  \sqrt{T \left(\,T + 2 \, M\,\right)}$, corresponds to a momentum of $729.13 \ \mbox{MeV}/c$ and a velocity of $\beta = 0.6136$, thus they propose a revolution frequency of $\nu = 2.12662 \ \mbox{MHz}$ \cite{Barone:2005pu}.
%  An extreme case provides an upper-bound on the maximum revolution frequency achievable in a storage ring: assume a ring of circumference $L = 30 \ \mbox{m}$ with antiprotons traveling at very close to the speed of light $\beta = v\,/\,c = 0.95$, this gives a revolution frequency of $9.5 \ \mbox{MHz}$. 
%  Many different scenarios are under investigation so we stress that in this thesis we are not confining our analysis to the parameters of the $\mathcal{P}\mathcal{A}\mathcal{X}$ proposal.
The High Energy Storage Ring (HESR) of the proposed Facility for Antiproton and Ion Research (FAIR) at GSI Darmstadt provides a high energy example.  Antiprotons with momentum $15 \ \mbox{GeV}/c$ will be stored in a $574 \ \mbox{metre}$ circumference ring, giving a velocity of $\beta = 0.998$ and a revolution frequency of $521628 \ \mbox{Hz}$ \cite{Barone:2005pu}. 

Having more than one target in the ring, or having the electron beam overlap with the antiproton beam at more than one point in the ring, has exactly the same effect as increasing the revolution frequency,  {\it i.e.}\ using two targets has the same effect as doubling the revolution frequency, using three targets has the same effect as tripling the revolution frequency etc.  This is obvious since we are using the revolution frequency as a measure of the number of times the beam passes through the target.  
%So to maximize the polarization buildup time in the numerical calculations we could use an effective $\nu = 10$ and use two or three targets if necessary to achieve this value.  
Having $R$ targets in the ring will increase the rate of polarization buildup by a factor of $R$.  The ring is limited by space, and such targets would have to be purposefully built, so realistically $R = 1$ is most likely for $\mathcal{P}\mathcal{A}\mathcal{X}$, anything above $R = 5$ would be very challenging.

The effect of multiple targets in the storage ring could be included in our system of polarization evolution equations simply by multiplying the entire coefficient matrix by a parameter $R$, where $R$ is the number of targets in the ring.  This would carry directly into each of the solutions presented in this thesis under the substitution $n\,\nu \rightarrow n\,\nu\,R$.  Note one must set $R = 1$ to compare to other work in the field which assume only one target in the ring.

One could choose $\nu = 5 \ \mbox{MHz}$ as a best case scenario available in the near future.  If necessary two interaction regions of the opposing electron beam with the antiproton beam could be used to achieve this effective revolution frequency. 

\subsection{Maximum target areal density}
\label{subsec:Maximum_target_areal_density}

As shown in Chapter \ref{ch:Polarization_buildup_by_spin_filtering} the rate of polarization buildup is highly dependent on the target areal density.  One requires as high as possible a target areal density to achieve the highest rate of polarization buildup.  In this section we review the maximum areal densities of different types of target currently available.

The FILTEX polarized hydrogen target, developed in the early 1990's, had an areal density of $6 \ \times \ 10^{13} \ \mbox{atoms per} \ \mbox{cm}^2$ \cite{Rathmann:1993xf}.  Since then some advances have been made in polarized atomic gas targets, in particular by the HERMES Collaboration.  The HERMES Collaboration has produced and used polarized hydrogen and deuterium targets with densities of up to $10^{14} \ \mbox{atoms per} \ \mbox{cm}^2$ \cite{Airapetian:2004yf}.  It is expected that this maximum areal density of polarized atomic gas targets could be increased by a factor of 100 in the near future.  For a recent review of polarized gas targets see Ref.~\cite{Steffens:2003}.  The PANDA Collaboration, also at the Facility for Antiproton and Ion Research (FAIR) at GSI Darmstadt, aim to produce a hydrogen pellet target of areal density up to $10^{16} \ \mbox{atoms per} \ \mbox{cm}^2$ \cite{Brinkmann:2005we}, similar to the target in operation at the WASA detector in COSY J\"ulich.
%  For a realistic maximum value of the polarized hydrogen gas areal density available in the near future we use $10^{15} \ \mbox{atoms per} \ \mbox{cm}^2$ in our numerical calculations.  

The maximum areal densities of polarized electron beams is many orders of magnitude lower than that of atomic targets, because of the electromagnetic repulsion felt by the like-charge electrons in the beam.  This effect is absent for atoms which are electrically neutral.  Typical areal densities of polarized electron beams produced thus far is $10^{8} \ \mbox{electrons per} \ \mbox{cm}^2$.  Given the enormous research and development effort that is currently been afforded to the International Linear Collider (ILC) project, which will use electron and positron beams, the maximum electron beam areal densities can be expected to be increased in the near future \cite{MoortgatPick:2005cw,Brau:2007zza}.  Perhaps polarized electron beams with areal densities of up to $10^{9} \ \mbox{electrons per} \ \mbox{cm}^2$, or even $10^{10} \ \mbox{electrons per} \ \mbox{cm}^2$ may be available in the next decade.  It has been claimed that electrons stored in a Penning trap may soon reach areal densities of $10^{12} \ \mbox{electrons per} \ \mbox{cm}^2$ \cite{Rathmann:2004pm}.

%Electron beam areal densities of greater than $10^{10} \ \mbox{electrons per} \ \mbox{cm}^2$ are greatly effected by space-charge effects, and are considered unachievable for the foreseeable future.   

For a realistic maximum value of the polarized electron beam areal density available in the near future we use $10^{12} \ \mbox{electrons per} \ \mbox{cm}^2$ in our numerical calculations.

\subsection{Maximum target polarization}
\label{subsec:Maximum_target_polarization}

Beams of electrons and positrons with polarizations of up to $\mathcal{P}_T = 0.9$ have been produced and utilized in many laboratories \cite{Alley:1995ia,Aulenbacher:2004yg}.  Polarized internal targets of atomic hydrogen and deuterium with polarizations of nuclei, electrons or both of up to $\mathcal{P}_\mathrm{T} = 0.9$ have also been constructed, in particular the HERMES polarized hydrogen and deuterium targets which have been operated both longitudinally and transversely polarized \cite{Airapetian:2004yf,Steffens:2003}.  There is an uncertainty of about $3\%$ to $5\%$ in the measurement of these polarizations, thus polarizations of above $0.95$ are impractical to produce.  As a target polarization that should be state of the art in the coming decade before $\mathcal{P}\mathcal{A}\mathcal{X}$ will be realized we pick a target polarization of $\mathcal{P}_\mathrm{T} = 0.9$ in the following numerical calculations.

\subsection{Effective acceptance angle}
\label{subsec:Effective_acceptance_angle}

The acceptance angle $\theta_\mathrm{acc}$ introduced earlier, is an idealistic simplification.  It assumes all beam particles are moving exactly along the infinitely narrow beam axis, hence neglecting the following two effects:
\\[2ex]
{\bf (a)} In reality the beam has a finite extent, {\it i.e.}\ a r.m.s.\ radius of about $8 \ \mbox{mm}$ at the target \cite{Barone:2005pu}.  So some particles are moving collinearly with the beam but at a distance of a few millimetres from the beam axis.  
\\[2ex]
{\bf (b)} Not all particles move collinearly, many particles have a slight angle of motion with respect to the beam axis due to scattering off the target.
\\[2ex]
Some of these particles that are scattered at or even slightly less than the idealistic acceptance angle will be lost.  Hence there is an effective ring acceptance angle $\theta_\mathrm{acc}^\mathrm{\,effective}$ which is always less than the idealistic ring acceptance angle $\theta_\mathrm{acc}^\mathrm{\,naive}$\,, {\it i.e.}\ $\theta_\mathrm{acc}^\mathrm{\,effective} < \theta_\mathrm{acc}^\mathrm{\,naive}$\,.  The effect of this is to lessen the region of integration for the {\bf \lq\lq in''} spin observables thus reducing them, and increase the region of integration for the {\bf \lq\lq out''} spin observables thus increasing them.  This effect also reduces the beam lifetime.
% and we must introduce an effective beam lifetime which is less than the naive beam lifetime introduced earlier. 

These effects have been investigated by the $\mathcal{P}\mathcal{A}\mathcal{X}$ Collaboration for the polarized proton beam scattering off an internal hydrogen target at COSY J\"ulich by comparing the theoretically calculated idealistic loss cross-section to the experimentally measured beam lifetime \cite{Stancari:2007}; the results were as follows.

They calculate the ratio of the effective spin-averaged loss cross-section $\sigma^\mathrm{\,out}_\mathrm{effective}$ to the naive spin-averaged loss cross-section $\sigma^\mathrm{\,out}_\mathrm{naive}$, the former of which is proportional to $(\theta_\mathrm{acc}^\mathrm{\,effective})^{-2}$ and the latter to $(\theta_\mathrm{acc}^\mathrm{\,naive})^{-2}$, because scattering is dominated by the Coulomb interaction.  One has
\begin{equation}
\label{eq:Stancari_F_ratio1}
 F \ = \ \frac{\sigma^\mathrm{\,out}_\mathrm{effective}}{\sigma^\mathrm{\,out}_\mathrm{naive}} \ = \ \frac{(\theta_\mathrm{acc}^\mathrm{\,naive})^{\,2}}{(\theta_\mathrm{acc}^\mathrm{\,effective})^{\,2}}\,,
\end{equation}
%
% the second = above could be an \approx ????
%
therefore the ratio between the effective and naive acceptance angles is
\begin{equation}
\label{eq:Stancari_F_ratio2}
\frac{\theta_\mathrm{acc}^\mathrm{\,effective}}{\theta_\mathrm{acc}^\mathrm{\,naive}} \ = \ \frac{1}{\sqrt{F}}\,.
\end{equation}
For the COSY polarized proton beam scattering off an internal hydrogen target it was found that $F = 1.14$ \cite{Stancari:2007}, {\it i.e.}\ $1/\sqrt{F} = 0.94$ and one has
\begin{equation}
\label{eq:Stancari_F_ratio_final}
\theta_\mathrm{acc}^\mathrm{\,effective} \ = \ \frac{1}{\sqrt{F}} \ \theta_\mathrm{acc}^\mathrm{\,naive} \ = \ 0.94 \ \theta_\mathrm{acc}^\mathrm{\,naive} \,.
\end{equation}
In the numerical calculations that follow we shall take these results and use an effective acceptance angle that is $0.94$ times the idealistic acceptance angle of the storage ring.

%The effective beam lifetime can be related to the naive beam lifetime as follows 
%%
%\begin{equation}
%\tau_\mathrm{\,lifetime}^\mathrm{\,effective} \ = \ 0.83 \ \tau_\mathrm{\,lifetime}^\mathrm{\,naive}
%\end{equation}
% 

\subsection{Minimum scattering angle}
\label{subsec:Minimum_scattering_angle}

From quantum mechanics one has the relation, $|{\bf q}|\,b\,=\,\hbar$, between the modulus of the three-momentum transfer ($|{\bf q}|$) in an interaction and the impact parameter ($b$).  Since $\hbar$ is set to one throughout this thesis one can relate the modulus of the minimum three-momentum transfer of an interaction to the maximum impact parameter by:
\begin{equation}
\label{eq:q_min}
|{\bf q}_\mathrm{min}| \ = \ \frac{1}{b_\mathrm{max}} \,,
\end{equation}
and using $-\,t\,=\,|{\bf q}|^{\,2}$ one has that the minimum squared momentum transfer is
\begin{equation}
\label{eq:t_min}
-\,t_\mathrm{min} \ = \ \frac{1}{b^{\,2}_\mathrm{max}} \,.
\end{equation}
This minimum value for $|\,t\,|$ ensures that there is no singularity from the $1/t$ dependence of many of the spin observables presented in Chapter~\ref{ch:Specific_helicity_amplitudes_and_spin_observables}. 

There are two cases of interest to us:
\\
\\
{\bf 1) A polarized hydrogen target}
\\
\\
For the case of a polarized hydrogen target the maximum impact parameter is given by the Bohr radius $a_B$.  For an antiproton scattering off a hydrogen atom at impact parameters greater $a_B$ the Coulomb fields of the atomic electron and proton screen each other, hence the antiproton sees the atom as electrically neutral and does not interact with it.  Hence we have the relation
\begin{equation}
\label{eq:t_min_HydrogenTarget}
-\,t_\mathrm{min} \ = \ \frac{1}{a^{\,2}_B}  \ = \ \alpha^{\,2} \,m^2 \  = \ 0.000013912 \ (\mbox{MeV}/c)^{\,2} \,,
\end{equation}
where $\alpha$ is the fine structure constant and $m$ the mass of the electron, and the relation $a_B\,=\,\left(\alpha\,m\right)^{-1}$ has been used.

One can convert this to the minimum laboratory frame scattering angle $\theta_\mathrm{min}$ using eq.~(\ref{eq:t_to_LAB_angle}) to obtain \cite{Jackson:1999}
\begin{equation}
\label{eq:Theta_min_to_Bohr_radius}
\theta_\mathrm{min} \ = \ \frac{1}{p_\mathrm{lab}\,a_B}\,,
\end{equation}
where $p_\mathrm{lab}$ is the laboratory frame antiproton momentum.  Using the relations $p\,=\,\sqrt{T\,\left(\,T\,+\,2\,M\,\right)}$ and $a_B\,=\,\left(\alpha\,m\right)^{-1}$, where $T$ is the Laboratory frame kinetic energy of the antiprotons, one can rewrite this as \cite{Nikolaev:2006gw}
\begin{equation}
\label{eq:Theta_min}
\theta_\mathrm{min} \ = \ \frac{\alpha\,m}{\sqrt{T\,\left(T\,+\,2\,M\,\right)}}\,,
\end{equation}
where $M$ is the mass of the (anti)proton.  For FILTEX kinetic energies of $T\,=\,23 \ \mbox{MeV}$ one finds that $\theta_\mathrm{min}\,\approx\,0.02 \ \mbox{mrad}$, far below both the acceptance angle of any storage ring and the maximum angle antiprotons are scattered off stationary electrons, hence verifying eq.~(\ref{eq:Kolyas_angle_inequalities}).  For scattering angle less than this, corresponding to impact parameters greater than the Bohr radius, the Coulomb fields of the atomic electron and proton of the hydrogen atom screen each other, hence antiprotons do not interact with the hydrogen atom.
\\
\\
{\bf 2) A polarized electron beam}
\\
\\
For an electron beam of areal density $n$ particles per femtometre squared the average distance between electrons is $1/\sqrt{n} \ \mbox{fm}$.  If an antiproton passes exactly equidistant from two electrons in the beam it will feel no force as the Coulomb fields of the two electrons will cancel each other.  The maximum impact parameter is one-half of the average electron separation, {\it i.e.}\ $b_\mathrm{max} \,=\,1/(2\,\sqrt{n}\,)$.  Therefore
\begin{equation}
\label{eq:t_min_ElectronBeam}
-\,t_\mathrm{min} \ = \ \frac{1}{b^{\,2}_\mathrm{max}} \ = \ 4\,n \,,
\end{equation}
where one must convert the areal density $n$ into units of $(\mbox{MeV}/c)^{\,2}$ using the conversion factor $1 \ \mbox{fm}^{-2} \,=\, 38937.9323 \ (\mbox{MeV}/c)^{\,2}$ \cite{Amsler:2008zz}.

By Taylor expanding the left hand side of eq.~(\ref{eq:Ring_frame_theta_to_t_Taylor_expansion}) with respect to the ring frame\footnote{The ring frame is described in detail in section~\ref{subsec:An_opposing_electron_beam}.} scattering angle $\theta^{\,r}$ one obtains the approximate relation, valid for small $\theta^{\,r}$ and small $|\,t\,|$
\begin{equation}
\label{eq:Approx_Theta_to_t_ring_frame}
\theta^{\,r} \ \approx \ \frac{\sqrt{-\,t}}{p_{\bar p}^{\,r}}\,,
\end{equation}
where $p_{\bar p}^{\,r}$ is the antiproton momentum in the ring frame.  Therefore the minimum ring frame scattering angle in this case is
\begin{equation}
\label{eq:Theta_ring_min}
\theta^{\,r}_\mathrm{min} \ \approx \ \frac{\sqrt{-\,t_\mathrm{min}}}{p_{\bar p}^{\,r}} \ = \ \frac{\sqrt{4\,n}}{p_1} \,.
\end{equation}
Results for the minimum squared momentum transfer and minimum ring frame scattering angle for various values of the electron beam areal density are presented in Table~\ref{table:t_min_and_Theta_min_for_various_n}.
\begin{table}[!h]
\begin{center}
\begin{tabular}{|c|c|c|c|}
\hline
$$             & $$   & $$    & $$ \\[-1ex]
         $n \ [\mbox{cm}^{-2}]$      & $n \ [\mbox{fm}^{-2}]$   &  $t_\mathrm{min} \ [(\mbox{MeV}/c)^{\,2}]$    & $\theta^{\,r}_\mathrm{min} \ [\mbox{mrad}]$ \\[2ex]
\hline
\hline
 & & & \\[-2ex]
$10^{8}\ $  & $10^{-18}$ & $-\,1.5575 \times 10^{-13}$ & $2.6 \times 10^{-11}$ \\[0.5ex]  \hline
 & & & \\[-2ex]
$10^{9}\ $  & $10^{-17}$ & $-\,1.5575 \times 10^{-12}$ & $8.3\times 10^{-11}$ \\[0.5ex] \hline
 & & & \\[-2ex]
$10^{10}$ & $10^{-16}$ & $-\,1.5575 \times 10^{-11}$ & $2.6 \times 10^{-10}$ \\[0.5ex]  \hline 
 & & & \\[-2ex]
$10^{11}$ & $10^{-15}$ & $-\,1.5575 \times 10^{-10}$ & $8.3 \times 10^{-10}$ \\[0.5ex] \hline
 & & & \\[-2ex]
$10^{12}$ & $10^{-14}$ & $\!\!-\,1.5575 \times 10^{-9}$ & $\!\!2.6 \times 10^{-9}$ \\[0.5ex]
\hline
\end{tabular}
\end{center}
\caption{\small{\it{Results for the minimum squared momentum transfer $t_\mathrm{min}$ and minimum ring frame scattering angle $\theta^{\,r}_\mathrm{min}$ for various values of the electron beam areal density $n$.  The ring frame antiproton momentum is fixed at $p_1 = 15 \ \mbox{GeV}/c$.  The conversion factor $1 \ \mbox{cm}^{-2} \,=\, 10^{-26} \ \mbox{fm}^{-2}$ has been used.}}}
\label{table:t_min_and_Theta_min_for_various_n}
\end{table}

\section{Spin filtering off a polarized electron beam}
\label{sec:Spin_filtering_off_a_polarized_electron_beam}

\subsection{The advantages of using a lepton target}
\label{subsec:The_advantages_of_using_a_lepton_target}

There are many advantages of using a polarized lepton target (or beam) over a polarized internal atomic target (hydrogen or deuterium) for spin filtering:

\begin{enumerate}

\item There is no loss of beam intensity due to annihilation of the antiprotons with protons as there is in the nuclear targets.

\item The polarization observables for antiproton - electron scattering are calculable in perturbative QED (as presented in this thesis), whereas for an atomic target currently less known hadronic polarization observables contribute.

\item Residual gas does not build up over time in the storage ring as would happen if an atomic target was used.

%\item The antiproton beam could be accelerated first and then polarized, or polarized while accelerating, without worrying about decreasing hadronic spin transfer cross sections at high energy.
%
%****Note I could omit this last point as I'm not as sure about it, and anyway three advantages are enough.  ****

\end{enumerate}

The first of these is by far the most important, and has caused many groups to investigate methods to polarize antiprotons by spin filtering off lepton beams and targets.  We now investigate various scenarios of spin filtering off pure lepton beams and targets.

\subsection{Antiprotons scattering off stationary electrons}
\label{subsec:Antiprotons_scattering_off_stationary_electrons}

\begin{figure}
\hspace*{-0.7em}
     \includegraphics{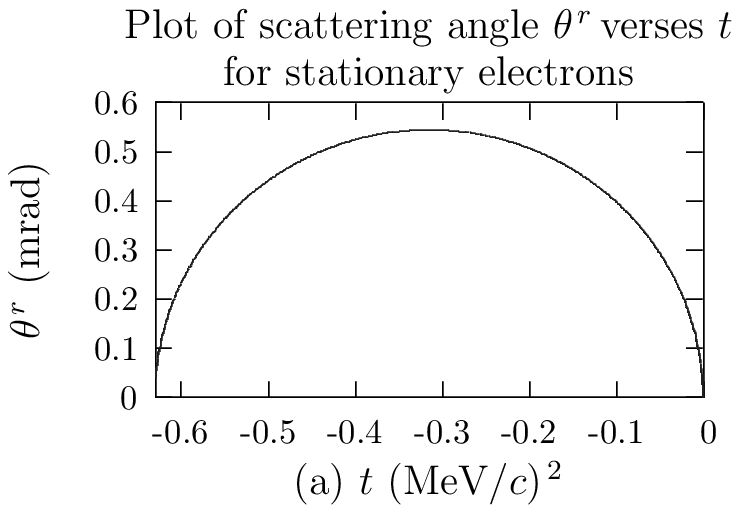}
\hspace*{-0.5em}
     \includegraphics{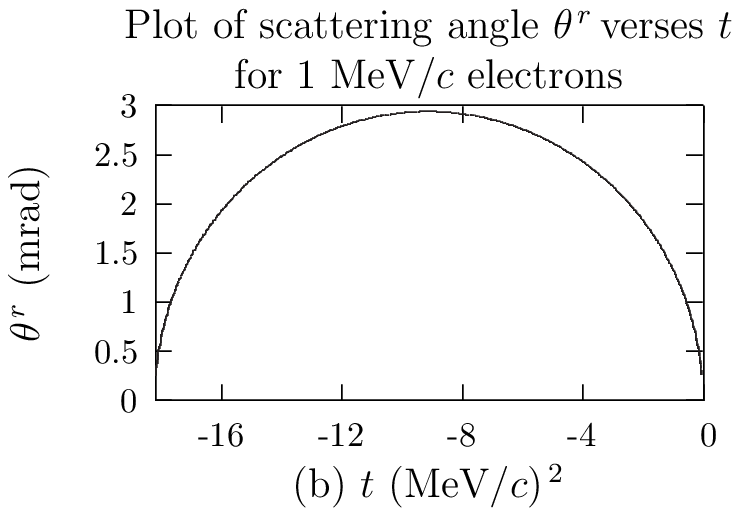}
\\[3ex]
\hspace*{-0.7em}
     \includegraphics{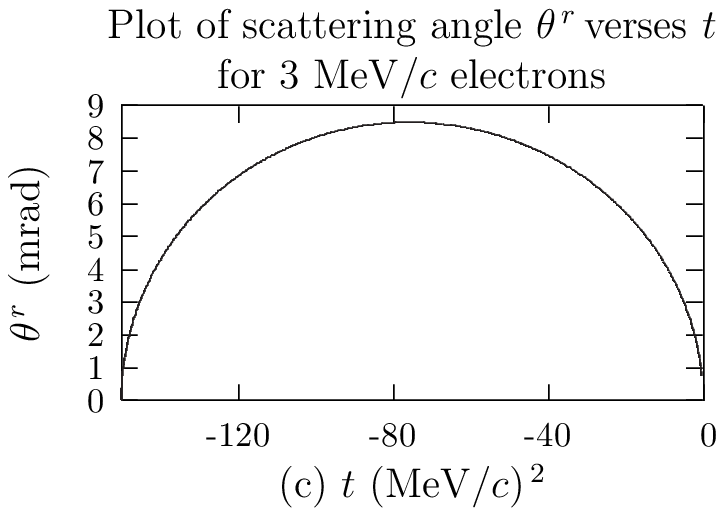}
\hspace*{-0.5em}
     \includegraphics{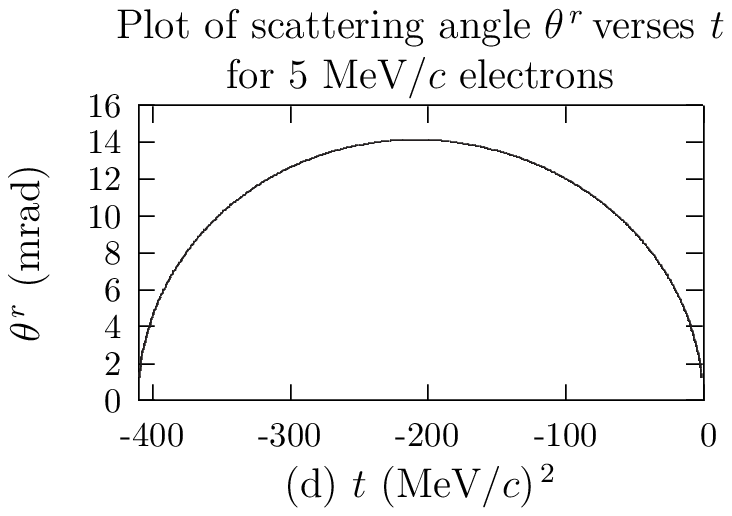}
\\[3ex]
\hspace*{-0.7em}
     \includegraphics{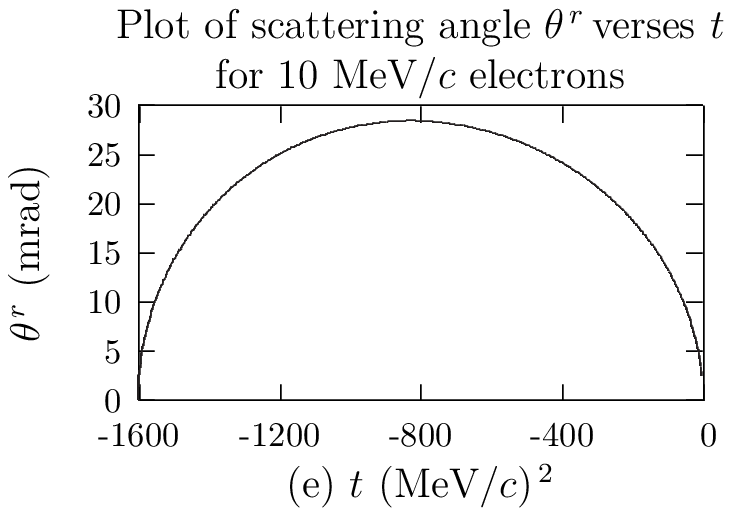}
\hspace*{-0.5em}
     \includegraphics{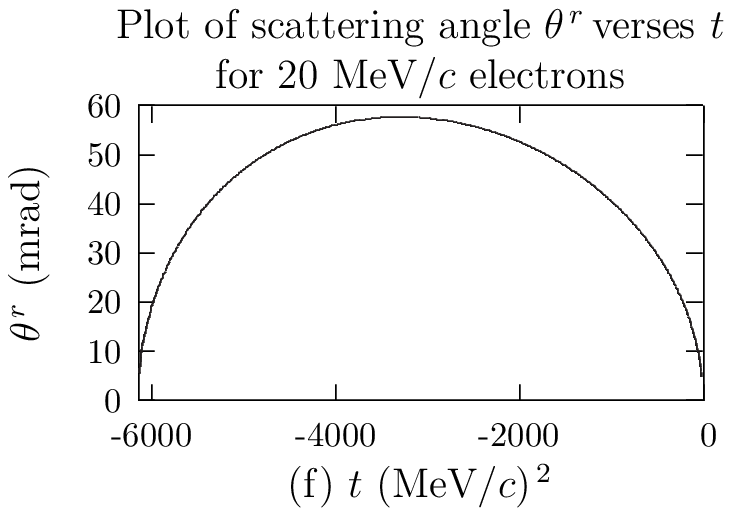}
\caption{\small{\it{(a) Ring frame antiproton scattering angle $\theta^{\,r}$\! versus squared momentum transfer $t$ for antiprotons of momentum $729 \ \mbox{MeV}/c$ scattering off electrons in an atomic target, where one confirms that the maximum antiproton scattering angle is $0.54 \ \mbox{mrad}$.  (b)--(f) With a colliding electron beam the maximum ring frame antiproton scattering angle increases as the electron beam momentum increases in a direction opposite to that of the antiproton beam.  The plots show the relationship between $\theta^{\,r}$\! and $t$ given in eq.~(\ref{eq:Ring_frame_theta_to_t}).
}}}
\label{fig:Theta_versus_t_plots}
\end{figure}
\begin{figure}[!h]
\hspace*{-0.7em}
     \includegraphics{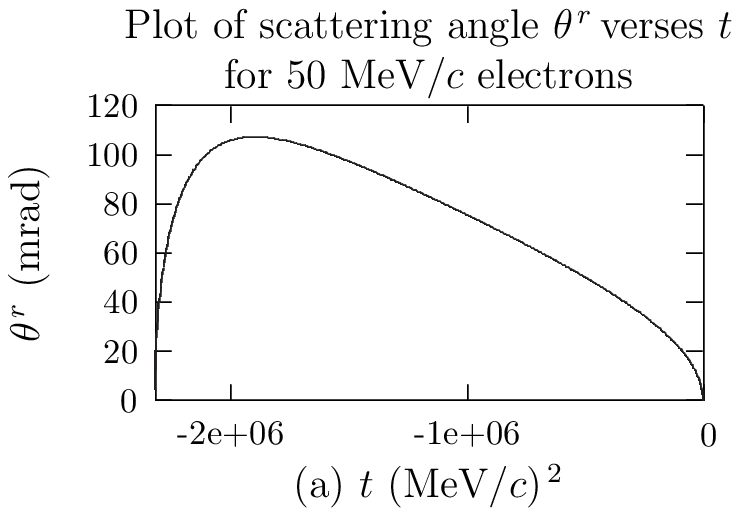}
\hspace*{-0.5em}
     \includegraphics{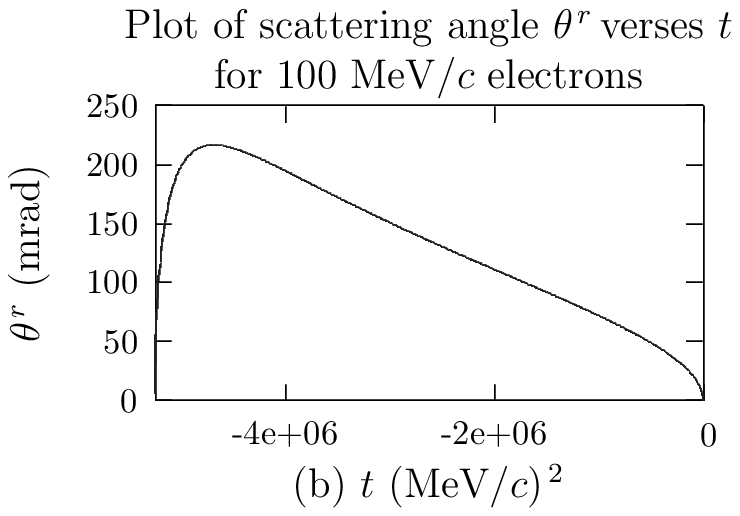}
\\[3ex]
\hspace*{-0.7em}
     \includegraphics{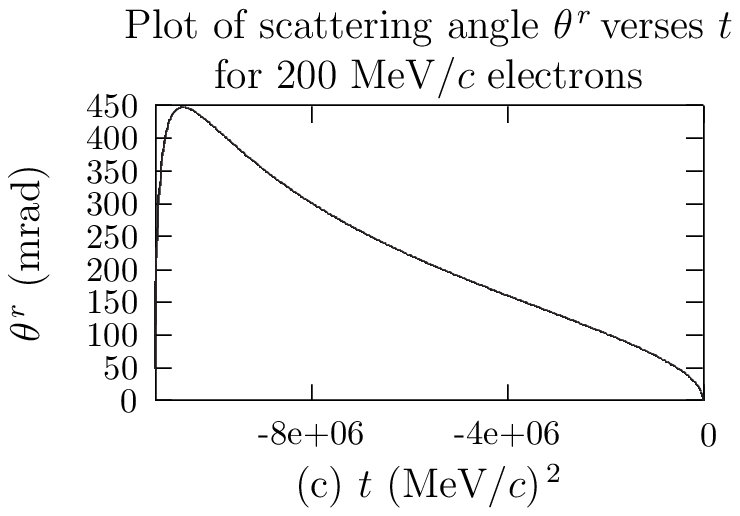}
\hspace*{-0.5em}
     \includegraphics{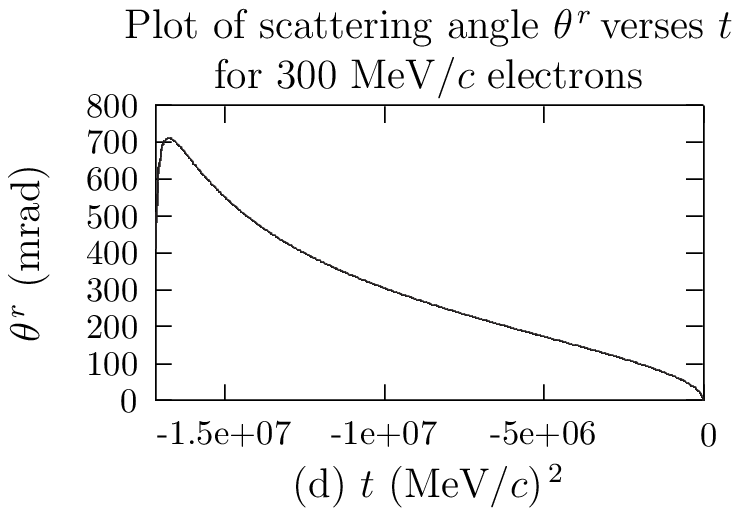}
\caption{\small{\it{Ring frame antiproton scattering angle $\theta^{\,r}$\! versus squared momentum transfer $t$ for antiprotons of momentum $15 \ \mbox{GeV}/c$ scattering off an opposing electron beam of increasing energy.  The plots become more skewed towards higher $|\,t\,|$ for higher opposing electron beam
momentum.  The plots show the relationship between $\theta^{\,r}$\! and $t$ given in eq.~(\ref{eq:Ring_frame_theta_to_t}).}}}
\label{fig:Theta_versus_t_plots_p2_15000}
\end{figure}

The maximum laboratory frame scattering angle for antiprotons scattering off stationary electrons is $m\,/\,M = 0.54\ \mbox{mrad}$ as shown in eq.~(\ref{eq:Theta_max_for_stationary_electrons}) and Figure.~\ref{fig:Theta_versus_t_plots} (a).  This is below the acceptance angle of any storage ring \cite{Barone:2005pu,Rathmann:2004pm} so that all scattering off atomic electrons will be within the ring.  Therefore stationary electrons can only contribute to the polarization buildup of the antiproton beam by selective spin-flip.  The Budker and J\"ulich groups claim that spin-flip effects while scattering within the ring are small for $\bar{p}\,e^- \rightarrow \bar{p}\,e^-$ scattering; hence stationary electrons, and very low energy electrons such as in an atomic target, are not effective in transferring polarization to an antiproton beam \cite{Milstein:2005bx,Nikolaev:2006gw}.  We propose the use of an opposing polarized electron beam of sufficient energy to increase the scattering angles of the antiprotons beyond acceptance as seen in Figure~\ref{fig:Theta_versus_t_plots} (b)--(f).  This is the subject of the remainder of the chapter.

Note another solution to this problem would be to use a polarized muon target.  Stationary muons, having much more mass than electrons ($m_\mu \approx 200\,m_e$), would provide a maximum laboratory frame antiproton scattering angle of $\theta^{\,\mu}_\mathrm{max} \ \approx \ m_\mu \,/\,M \ = \ 113 \ \mbox{mrad}$.  This is far beyond the ring acceptance angles under consideration at $\mathcal{PAX}$, hence allows selective scattering out of the beam to contribute to polarization buildup.  Another positive aspect of muons is that they can be produced automatically highly polarized, through the decays of charged pions.  Charged pions decay into muons and (anti)neutrinos: $\pi^+ \rightarrow \mu^+ \, \nu_{\mu}$ and $\pi^- \rightarrow \mu^- \, \bar{\nu}_{\mu}$\,.  Since neutrino's have only one possible polarization state,  angular momentum conservation forces the produced muons to be polarized.  Polarized muon beams have been used in many experiments, among them the seminal EMC and SMC experiments which ushered in a new era of interest in spin physics and in particular in the spin structure of nucleons.  But it is feared that the density of such polarized muon beams will be too low with today's technologies to provide a reliable method of polarizing antiprotons \cite{Parkhomchuk:1999wj}.

At higher energies the plots of the relationship between $\theta^{\,r}$\! and $t$ become skewed towards higher $|\,t\,|$ as shown in Figure~\ref{fig:Theta_versus_t_plots_p2_15000} above for HESR energies, where the antiprotons have ring frame momentum $15 \ \mbox{GeV}/c$.

%\subsection{Ring frame kinematics}
%\label{subsec:Ring_frame_kinematics}
\subsection{An opposing electron beam}
\label{subsec:An_opposing_electron_beam}

A crucial variable for spin filtering is the acceptance angle, as it defines which particles are scattered out of the beam and which are scattered at small angles remaining in the beam.  This angle is with respect to the beam axis, so for a stationary target in a storage ring the acceptance angle is the scattering angle in the LAB frame.   We are now investigating the use of a colliding electron beam instead of a stationary target.  Here the acceptance angle does not correspond to either the scattering angle in the Centre-of-Mass frame or the LAB frame.  We want to use a frame in which the antiproton scattering angle with respect to the beam axis still corresponds to the angle featuring in our equations.  Note the scattering angle of the electrons is irrelevant here, and its use will be avoided by conservation of four-momentum.  A \emph{ring frame} for elastic antiproton electron scattering 
\begin{equation*}
\bar{p}\left(\,P_1\,,\,M\,\right) \ + \ e^-\left(\,P_2\,,\,m\right) \ \rightarrow \ \bar{p}\left(\,P_3\,,\,M\,\right) \ + \ e^-\left(\,P_4\,,\,m\right)
\end{equation*}
is defined as follows\footnote{If one of the particles is initially at rest the ring frame equals the laboratory frame.}:
%
%\hspace*{-3em}
\begin{table}[ht]
\begin{center}
\begin{tabular}{|ccc|ccc|}\hline
& & & & &\\[-1ex]
$ P_1 \!\!$&$  = $&$ \!\!\left(\,E_1,\,0,\,0,\,p_1\,\right) $& $ P_3 \!\!$&$ = $&$ \!\!\left(\,E_{\,3},\,p_3\,\sin\theta^{\,r},\,0,\,p_3\,\cos\theta^{\,r}\,\right)$\\[1ex]
$ P_2 \!\!$&$ \!\!\!=\!\!\!$&$ \!\!\left(E_{\,2},0,\,0,-p_{\,2}\,\right) $&$ P_4 \!\!$&$ \!\!\!\!\!\!\!\!\! = \!\!\!\!\!\!\!\!\! $&$ \!\!\left(E_1 \!-E_{\,3} + E_{\,2},-p_3\,\sin\theta^{\,r},0,\,p_1 - p_{\,2} - p_3\, \cos\theta^{\,r}\,\right)$\\[1ex]\hline
\end{tabular}
\end{center}
\end{table}

\vspace*{-3ex}
\noindent
%Since all particles are external, they are on-shell.  
Looking at $E^2 = p^{\,2} + m^2$ for the final state electron gives
\begin{equation*}
\left(\,E_1 + E_{\,2} - E_{\,3}\,\right)^{\,2} \ = \ p_3^{\,2}\, \sin^2\theta^{\,r} \ + \ \left(\,p_1 - p_{\,2} - p_3\, \cos\theta^{\,r} \,\right)^{\,2} \ + \ m^2 \,. 
\end{equation*}
After substituting out the unknowns $E_{\,3}$ and $p_3$ this gives\footnote{Many of the expressions in this chapter can be re-written in terms of the relativistic velocity $\beta_i = p_i\,/E_i$ of each particle, where $i \in \{1,2\}$.  In particular $p_1\, E_{\,2} + p_{\,2}\, E_1 = E_1\,E_2 \left(\beta_1 + \beta_2\right)$ and $E_1\,E_2 - p_1\,p_2 = E_1\,E_2 \left(1 - \beta_1\,\beta_2\right)$.} an equation for the ring frame antiproton scattering angle $\theta^{\,r}$ in terms of the Mandelstam variable $t = \left(\,p_3 \,-\, p_1 \,\right)^{\,2}$,
\begin{equation}
\label{eq:Ring_frame_theta_to_t}
\cos\theta^{\,r} \ = \ \frac{t  \ -\  \left[\,2\, M^{\,2} \, + \, \frac{E_1 \left(\,t\,-\,2\, E_1^{\,2}\,\right)\, p_{\,2} \ -\ E_1\left(\,2\, E_1\, E_{\,2} \,+\, t\,\right)\, p_1}{p_1\,E_{\,2} \ +\  p_{\,2}\,E_1}\right]}
{2 \, p_1\, \sqrt{\left[\frac{\left(\,2\, E_1\, E_{\,2} \,+\, t\,\right) \, p_1 \ +\ 
\left(\,2\, E_1^{\,2} \,-\, t \,\right)\, p_{\,2}}{2\,\left(\,p_1\, E_{\,2} \ +\  p_{\,2}\, E_1\,\right)}\right]^{\,2}  -\, M^{\,2}\ }} \,.
\end{equation} 
A Taylor expansion to $\mathcal{O}(t)$ of the above provides some clarity of the behaviour at small $|\,t\,|$:
\begin{equation}
\label{eq:Ring_frame_theta_to_t_Taylor_expansion}
\cos\theta^{\,r} \ \approx \ 1 \ + \ \frac{t}{2\,p^{\,2}_1} \,.
\end{equation}
Equation~(\ref{eq:Ring_frame_theta_to_t}), the ring frame analogy of the much simpler Centre-of-Mass frame relation eq.~(\ref{eq:t_to_CM_theta}), is used to graph $\theta^{\,r}$ versus $t$ in Figure~\ref{fig:Theta_versus_t_plots}.  There is no backward antiproton scattering, {\it i.e.}\ $\theta^{\,r}_\mathrm{max} \leq \pi/2$, for opposing electron beam momentum of
\begin{equation}
\label{eq:Upperbound_for_E_momentum}
p_2 \ < \ \frac{M\,p_1}{E_1 \,+\, p_1 \,+\, M} \,,
\end{equation}
%
%See my notes and email to Nigel dated 13-02-07 for work on this, also note Nigel's reply to this email.  The above inequality is strict because of the $p_e \approx E_e$ approximation. 
using the $p_2 \approx E_2$ approximation to make the inequality strict.  In this region the maximum ring frame antiproton scattering angle is given, via eq.~(\ref{eq:Ring_frame_theta_to_t}), by \cite{Milstein:2005bx}
\begin{equation}
\label{eq:SinTheta_max}
 \sin\theta^{\,r}_\mathrm{max}  \ =  \ \frac{p_1\, E_{\,2} \ + \ p_{\,2}\, E_1}{M \left(\,p_1 \ - \ p_{\,2}\,\right)} \,,
\end{equation}
which, using $\sin\theta \approx \theta$ for small angles, limits to the correct formula
\begin{equation}
\label{eq:Theta_max_for_stationary_electrons}
\theta^{\,r}_\mathrm{max} \ \approx \  \sin\theta^{\,r}_\mathrm{max}  \ = \  \frac{m}{M}  \ = \ 0.54 \ \mbox{mrad} \,,
\end{equation}
for stationary electrons ({\it i.e.}\ $p_{\,2} = 0$ and hence $E_{\,2} = m$).  Notice this maximum scattering angle for stationary electrons is independent of the antiproton momentum.  One can raise $\theta^{\,r}_\mathrm{max}$, and hence scatter more antiprotons out of the beam, simply by increasing the electron momentum $p_{\,2}$.  Raising it sufficiently beyond the acceptance angle will increase $K_\mathrm{\,out}$ and the rate of buildup of polarization.

Equation~(\ref{eq:SinTheta_max}) can be used to derive a relation for the electron momentum needed to scatter antiprotons beyond the ring acceptance angle $\theta^{\,r}_\mathrm{acc}$, hence allowing selective scattering out of the beam to contribute to polarization buildup.  This happens for electron momentum $p_{\,2} > p^\mathrm{out}$ where on assuming $\theta^{\,r}_\mathrm{acc}$ is small and $p_{\,2} \approx E_{\,2}$, {\it i.e.}\ that the electron mass is small compared to its momentum, one obtains
\begin{equation}
\label{eq:E_momentum_to_scatter_Pbar_out}
p^\mathrm{out} \ \approx \ \frac{M\,p_1\,\theta^{\,r}_\mathrm{acc}}{E_1 \ + \ p_1} \,.
\end{equation}

\begin{table}[!h]
\begin{center}
\begin{tabular}{|c||c|c|c|c|c|c|}
\hline
 & & & & & &  \\*[-2ex]
 $\theta^{\,r}_\mathrm{acc}$ \ [mrad] \ \ \,$=$ & $1$ & $2$ & $5$ & $10$ & $20$ & $50$\\[0.5ex]
\hline
 & & & & & &  \\*[-2ex]
  $p^\mathrm{out}$ \ [MeV$/c$] $=$ & $0.468678$ & $0.937356$ & $2.34339$ & $4.68678$ & $9.37356$ & $23.4339$\\[0.5ex]
\hline
\end{tabular}
\end{center}
\caption{\small{\it{This table, from eq.~(\ref{eq:E_momentum_to_scatter_Pbar_out}), shows the opposing electron momentum needed to scatter antiprotons of momentum $15 \ \mbox{GeV}/c$ out of the ring for various ring acceptance angles.  The upper-bound for the electron momentum, given this antiproton momentum, is found from eq.~(\ref{eq:Upperbound_for_E_momentum}) to be $454.478 \ \mbox{MeV}/c$.  Typical storage rings have acceptance angles in the range of $1 \ \mbox{mrad}$ to $50 \ \mbox{mrad}$.}}}
\label{table:Pout_for_various_THETAacc}
\end{table}
%The table entries can be verified by the graphs in figure {fig:Theta_versus_t_plots}
%
\noindent
Note that typical storage rings have acceptance angles in the range of $1 \ \mbox{mrad}$ to $50 \ \mbox{mrad}$.

The Lorentz invariant $\lambda$ in the ring frame is 
\begin{equation}
\label{eq:Ring_frame_invariant_lambda}
\lambda \ = \ 4\, k_\mathrm{cm}^{\,2}\,s \ = \ 4\, \left(\,p_1\,E_{\,2} \ + \ p_{\,2}\,E_1\,\right)^{\,2}\,,
\end{equation}
and the Mandelstam $s$ variable is
\begin{equation}
\label{eq:Ring_frame_Mandelstam_s}
s \ = \ M^{\,2} \ + \ m^2 \ + \ 2\,E_1\,E_2 \ + \ 2\,p_1\,p_2 \,.
\end{equation}
The maximum squared momentum transfer $|\,t\,|$, corresponding to total backward scattering, is given in the ring frame by
\begin{equation}
\label{eq:Ring_frame_t_4}
t_4 \ = \ -\,4\,k^{\,2}_\mathrm{cm} \ = \ -\,\frac{\lambda}{s}\ = \ \frac{-\,4\,\left(\,p_1\,E_{\,2} \ + \ p_{\,2}\,E_1\,\right)^{\,2}}{ M^{\,2} \, + \, m^2 \, + \, 2\,E_1\,E_2 \, + \, 2\,p_1\,p_2}\ ,
\end{equation}
using the ring frame expressions of eqs.~(\ref{eq:Ring_frame_invariant_lambda} and \ref{eq:Ring_frame_Mandelstam_s}).

\begin{figure}[p]
\centering
\includegraphics[height=6.5cm]{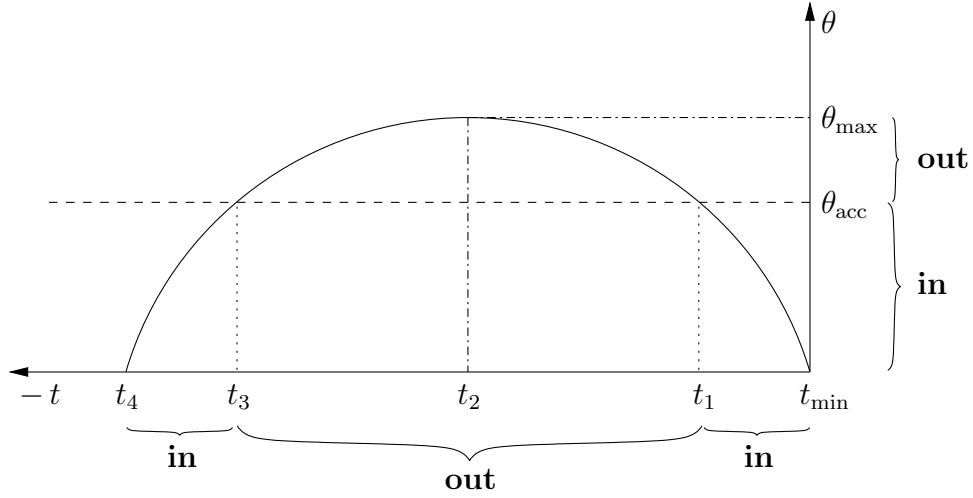}
\vspace*{-2ex}
\caption{\small{\it{Plot of scattering angle $\theta$ versus squared momentum transfer $t$, in the region of opposing electron beam momentum defined by eq.~(\ref{eq:Upperbound_for_E_momentum}).  The acceptance angle $\theta_\mathrm{acc}$ is plotted as a horizontal dashed line showing the region where particles are scattered {\bf \lq\lq out"} of the ring ($t_1$ to $t_3$) and the regions where particles are scattered at small angles remaining {\bf \lq\lq in"} the ring ($t_\mathrm{min}$ to $t_1$ and $t_3$ to $t_4$).  Note $t_\mathrm{min}$ corresponds to the minimum scattering angle $\theta_\mathrm{min}$, scattering below which is prevented by Coulomb screening.  The squared momentum transfer $t_2$ corresponds to the maximum scattering angle $\theta_\mathrm{max}$\,.}}}
\label{fig:Acceptance_Angle_Plot4}
\end{figure}

The spin observables needed for spin filtering presented in section~\ref{sec:Observables_needed_for_spin_filtering} can be converted into the ring frame by using eq.~(\ref{eq:Ring_frame_invariant_lambda}).  Figure~\ref{fig:Acceptance_Angle_Plot4} shows how the regions of scattering angle defined by the acceptance angle $\theta^{\,r}_\mathrm{acc}$ can be converted into regions of squared momentum transfer $t$.  Hence the angular regions of integration of the spin observables, presented in Table~\ref{tab:Transverse_and_Longitudinal_polarization} can be presented as regions of integration over squared momentum transfer $t$ in Table~\ref{tab:Transverse_and_Longitudinal_polarization_using_t}.
\psfrag{theta}{$\theta$}
\psfrag{Acc}{$\theta_\mathrm{acc}$}
\psfrag{Tmax}{$\theta_\mathrm{max}$}
\psfrag{t0}{$t_\mathrm{min}$}
\psfrag{t1}{$t_1$}
\psfrag{t2}{$t_2$}
\psfrag{t3}{$t_3$}
\psfrag{t4}{$t_4$}
\psfrag{t5}{$-\,t$}
\psfrag{in}{{\bf in}}
\psfrag{out}{{\bf out}}
%Consider shading the "in" regions of the above diagram, do it in xfig.
%
\begin{table}[p]
\begin{center}
\begin{tabular}{|c|c|} 
\hline
 & \\[-1.5ex]
Transverse polarization requires & Longitudinal polarization requires \\[1ex] \hline
 & \\[-1.5ex]
$\displaystyle{
\,
   I_\mathrm{\,out}
\, =
\, 2\,\pi \int_{\,t_3}^{\,t_1}
\,
   \frac{\mathrm{d}\,\sigma}{\mathrm{d}\,t} \ \mathrm{d}\,t
}
$
& 
$\displaystyle{
\,
   I_\mathrm{\,out}
\, =
\, 2\,\pi \! \int_{\,t_3}^{\,t_1}
\,
   \frac{\mathrm{d}\,\sigma}{\mathrm{d}\,t} \ \mathrm{d}\,t
}
$\\[3ex]
$\displaystyle{
    A_\mathrm{\,out}
   \,=\,2\,\pi \! \int_{\,t_3}^{\,t_1}
   \left( \frac{A_\mathrm{XX}  \,+\,  A_\mathrm{YY}}{2} \right)
 \frac{\mathrm{d}\,\sigma}{\mathrm{d}\,t} \ \mathrm{d}\,t
}
$
&
$\displaystyle{
\,
    A_\mathrm{\,out}
\,
   =
\, 2\,\pi \! \int_{\,t_3}^{\,t_1}
   A_\mathrm{ZZ}\,\,
\frac{\mathrm{d}\,\sigma}{\mathrm{d}\,t} \ \mathrm{d}\,t
}
$\\[3ex]
$ \displaystyle{
   A_\mathrm{\,all}
 \,=\,
 2\,\pi \! \int_{\,t_4}^{\,t_\mathrm{min}}
 \left( \frac{A_\mathrm{XX}  \,+\,  A_\mathrm{YY}}{2} \right)
 \frac{\mathrm{d}\,\sigma}{\mathrm{d}\,t} \ \mathrm{d}\,t
}
$
&
$ \displaystyle{ 
\,  
   A_\mathrm{\,all}
\, =
\, 2\,\pi \! \int_{\,t_4}^{\,t_\mathrm{min}}
    A_\mathrm{ZZ}\,\,
 \frac{\mathrm{d}\,\sigma}{\mathrm{d}\,t} \ \mathrm{d}\,t
}
$\\[3ex]
$\displaystyle{
    K_\mathrm{\,in}
  \, =\,
  2\,\pi \! \int_{\,t_1}^{\,t_\mathrm{min}}
   \left( \frac{K_\mathrm{XX}  \,+\,  K_\mathrm{YY}}{2} \right)
 \frac{\mathrm{d}\,\sigma}{\mathrm{d}\,t} \ \mathrm{d}\,t 
}
$
&
$
\displaystyle{
\,
    K_\mathrm{\,in}
\,
  = 
\,  2\,\pi \! \int_{\,t_1}^{\,t_\mathrm{min}}
  K_\mathrm{ZZ}\,\,
 \frac{\mathrm{d}\,\sigma}{\mathrm{d}\,t} \ \mathrm{d}\,t
}
$
\\[3ex]
$\displaystyle{
\ \ \ \ \ \ \ \
\ \ + \ 2\,\pi \! \int_{\,t_4}^{\,t_3}
   \left( \frac{K_\mathrm{XX}  \,+\,  K_\mathrm{YY}}{2} \right)
 \frac{\mathrm{d}\,\sigma}{\mathrm{d}\,t} \ \mathrm{d}\,t 
}
$
&
$\displaystyle{
\ \ \ \ \ \ \ \ +
\  2\,\pi \! \int_{\,t_4}^{\,t_3}
  K_\mathrm{ZZ}\,\,
 \frac{\mathrm{d}\,\sigma}{\mathrm{d}\,t} \ \mathrm{d}\,t 
}
$
\\[3ex]
$\displaystyle{
    D_\mathrm{\,in}
   \,=\,  2\,\pi \! \int_{\,t_1}^{\,t_\mathrm{min}}
 \left( \frac{D_\mathrm{XX}  \,+\,  D_\mathrm{YY}}{2} \right)
 \frac{\mathrm{d}\,\sigma}{\mathrm{d}\,t} \ \mathrm{d}\,t 
}
$
&
$\displaystyle{
\,
    D_\mathrm{\,in}
\,
   =
\, 2\,\pi \! \int_{\,t_1}^{\,t_\mathrm{min}}
  D_\mathrm{ZZ}\,\,
 \frac{\mathrm{d}\,\sigma}{\mathrm{d}\,t} \ \mathrm{d}\,t 
}$
\\[3ex]
$\displaystyle{
\ \ \ \ \ \ \  + \ 2\,\pi \! \int_{\,t_4}^{\,t_3}
 \left( \frac{D_\mathrm{XX}  \,+\,  D_\mathrm{YY}}{2} \right)
 \frac{\mathrm{d}\,\sigma}{\mathrm{d}\,t} \ \mathrm{d}\,t 
}
$
&
$\displaystyle{
\ \ \ \ \ \ \ +
\  2\,\pi \! \int_{\,t_4}^{\,t_3}
  D_\mathrm{ZZ}\,\,
 \frac{\mathrm{d}\,\sigma}{\mathrm{d}\,t} \ \mathrm{d}\,t 
}
$
\\[3ex]\hline
\end{tabular}
\end{center}
\vspace*{-3ex}
\caption{\small{\it{The entries in the system of polarization evolution equations, for both longitudinal and transverse polarization, integrated with respect to squared momentum transfer $t$.  Table~\ref{tab:Transverse_and_Longitudinal_polarization} shows similar expressions but integrated with respect to scattering angle $\theta$.}}} 
\label{tab:Transverse_and_Longitudinal_polarization_using_t}  
\end{table}
Using the expressions of section~\ref{subsec:Antiproton_electron_scattering} and Table~\ref{tab:Transverse_and_Longitudinal_polarization_using_t}, one has that for longitudinal\footnote{Since for antiproton-electron scattering the longitudinal spin-transfer observable from section~\ref{subsec:Antiproton_electron_scattering} is greater than the transverse spin-transfer observable we do the numerical calculation for the longitudinal case.  It has been shown at RHIC that one can rotate the polarization of the beam from longitudinal to transverse, or vice versa, without any loss of polarization.  Since the stable spin direction in a storage ring is transverse, it is likely that the beam will circulate in the ring with transverse polarization but be rotated to longitudinal directly before the target and back to transverse directly after the target.  This is how RHIC operates \cite{Bai:2007zza}.  The antiproton beam will eventually be utilized in a transversely polarized state in order to measure $A_{TT}$ and hence obtain information on the transversity distribution of quarks in the nucleon.} polarization:
\begin{eqnarray}
\label{eq:Kout_Pbar_E}
K_\mathrm{\,out} & = & \displaystyle{\frac{8\, \pi^{\,2} \, \alpha^2\, \mu_p}{\lambda }\left(\,s \,-\, m^2 \,-\, M^{\,2}\,\right)} \, \int_{\,t_3}^{\,t_1} \frac{1}{t} \ \mathrm{d}\,t \,, \nonumber \\[2ex]
& = & \displaystyle{\frac{-\,8\, \pi^{\,2} \, \alpha^2\, \mu_p}{\lambda }\left(\,s \,-\, m^2 \,-\, M^{\,2}\,\right)} \, \ln \frac{t_3}{t_1} \,.
\end{eqnarray}
%
% All logarithms are now positive
%
\begin{eqnarray}
\label{eq:Lin_Pbar_E}
L_\mathrm{\,in} & = & \displaystyle{\frac{-\,16\,\pi^{\,2}\,M^{\,2}\,\alpha^2\,\left(\,s \,+\, m^2 \,-\, M^{\,2}\,\right)^2}{\lambda^2}} \ \left(\ \int_{\,t_1}^{\,t_\mathrm{min}} \frac{1}{t} \ \mathrm{d}\,t \ +\  \int_{\,t_4}^{\,t_3} \frac{1}{t} \ \mathrm{d}\,t \ \right) \,, \nonumber \\[2ex]
 & = & \displaystyle{\frac{16\,\pi^{\,2}\, M^{\,2}\,\alpha^2\,\left(\,s \,+\, m^2 \,-\, M^{\,2}\,\right)^2}{\lambda^2}} \ \left(\ \ln \frac{t_1}{t_\mathrm{min}} \ +\  \ln \frac{t_4}{t_3}\ \right) \,, \nonumber \\[2ex]
 & = & \displaystyle{\frac{16\,\pi^{\,2}\, M^{\,2}\,\alpha^2\,\left(\,s \,+\, m^2 \,-\, M^{\,2}\,\right)^2}{\lambda^2}} \ \ln \frac{t_1\ t_4}{t_\mathrm{min}\ t_3}  \,.
\end{eqnarray}
Now $L_\mathrm{\,d}$ follows directly from the above expressions via eq.~(\ref{eq:Ld_for_pure_EM_scattering}).  The natural logarithms are expressed so that they are all positive, {\it i.e.}\ their arguments are greater than one, by using 
\begin{equation*}
\ln \left(\frac{a}{b}\right) \ = \ \ln \left(\frac{b}{a}\right)^{-1} \ = \ -\,\ln \left(\frac{b}{a}\right) \,.
\end{equation*} 
To present numerical results for the case of section~\ref{sec:Constant_beam_intensity} where particles are fed into the beam at such a rate that the beam intensity remains constant, we also need $I_\mathrm{out}$\,:
\begin{eqnarray}
\label{eq:Iout_Pbar_E}
I_\mathrm{\,out} & = & \frac{8\,\pi^{\,2}\,\alpha^{\,2}}{\lambda}\ \left(\,s\,-\,m^2\,-\,M^{\,2} \,\right)^{\,2} \, \int_{\,t_3}^{\,t_1}  \frac{1}{t^{\,2}} \ \mathrm{d}\,t \,, \nonumber \\[2ex]
& = & \frac{8\,\pi^{\,2}\,\alpha^{\,2}}{\lambda}\ \left(\,s\,-\,m^2\,-\,M^{\,2} \,\right)^{\,2} \,\left(\ \frac{1}{t_3} \ -\  \frac{1}{t_1}\ \right) \,.
\end{eqnarray}
%
%See notes 18-3-2008 for the above calculation.
%
Since all antiproton-electron scattering is purely electromagnetic we now have all expressions needed to present the polarization buildup as a function of time from eq.~(\ref{eq:Ptau_for_pure_EM_scattering}).

One can obtain the values of $t_1$ and $t_3$ for various acceptance angles by solving eq.~(\ref{eq:Ring_frame_theta_to_t}).  Table~\ref{table:t1_and_t3_for_various_Theta_acc} presents results for values of $p_1 = 15 \ \mbox{GeV}/c$ and $p_2 = 50 \ \mbox{MeV}/c$, where one sees from eq.~(\ref{eq:SinTheta_max}) that $\theta^{\,r}_\mathrm{max} = 107.248 \ \mbox{mrad}$.  The maximum electron beam momentum for this value of antiproton momentum, in order for the antiproton scattering angle to be less than $\pi/2$, is found from eq.~(\ref{eq:Upperbound_for_E_momentum}) to be $p_2 = 454.478 \ \mbox{MeV}/c$.
%As a check of this I can put this into my formula above and I get two identical values for $t$, confirming that it is the maximum.  This is done in the Mathematica notebook Local-Frame-Max-Angle.nb.
%
\begin{table}[!h]
\begin{center}
\begin{tabular}{|c||c|c|c|c|}
\hline
$$             & $$      & $$ & $$      & $$\\[-2ex]
$\theta^{\,r}_\mathrm{acc} \ [\mbox{mrad}]$   &         $t_1 \ [(\mbox{MeV}/c)^2]$    & $t_3 \ [(\mbox{MeV}/c)^2]$  &  $\displaystyle{\ln \frac{t_3}{t_1}}$ & $\displaystyle{\ln \frac{t_1\,t_4}{t_\mathrm{min}\,t_3}}$\\[2ex]
\hline
\hline
 & & & & \\[-2ex]
$1$  & $-\,224.988$ & $-\,2.32222 \times 10^6$ & $9.24199$ & $25.6962$\\[0.5ex] \hline
 & & & & \\[-2ex]
$2$  & $-\,899.809$ & $-\,2.32219 \times 10^6$ & $7.85584$ & $27.0824$\\[0.5ex] \hline
 & & & & \\[-2ex]
$5$  & $-\,5617.55$ & $-\,2.32195 \times 10^6$ & $6.02427$ & $28.9140$\\[0.5ex] \hline
 & & & & \\[-2ex]
$10$ & $-\,22381.3$ & $-\,2.32111 \times 10^6$ & $4.64158$ & $30.2966$\\[0.5ex] \hline 
 & & & & \\[-2ex]
$20$ & $-\,88135.2$ & $-\,2.31770 \times 10^6$ & $3.26946$ & $31.6688$\\[0.5ex]\hline
 & & & & \\[-2ex]
$50$ & $-\,498035$  & $-\,2.29123 \times 10^6$ & $1.52617$ & $33.4120$\\[0.5ex]
%$40$             & $$   & $$   & $$    & $$    & $$    & $$\\
\hline
\end{tabular}
\end{center}
\caption{\small{\it{Values of $t_1$ and $t_3$, in units of $(\mbox{MeV}/c)^2$, and the natural logarithms involving them appearing in the integrated spin observables, for various acceptance angles, in units of $\mbox{mrad}$, obtained by solving eq.~(\ref{eq:Ring_frame_theta_to_t}).  Results are for values of $p_1 = 15 \ \mbox{GeV}/c$ and $p_2 = 50 \ \mbox{MeV}/c$, where one sees from eq.~(\ref{eq:SinTheta_max}) that $\theta^{\,r}_\mathrm{max} = 107.248\ \mbox{mrad}$ and from eq.~(\ref{eq:Ring_frame_t_4}) that $t_4 = \ -\,2.32223 \times 10^6 \ (\mbox{MeV}/c)^{\,2}$.  Here we take the most optimistic $t_\mathrm{min}=-\,1.5575 \times 10^{-9} \ (\mbox{MeV}/c)^2$ corresponding to an electron beam areal density of $10^{12} \, \mbox{cm}^2$ as discussed in Table~\ref{table:t_min_and_Theta_min_for_various_n}.}}}
\label{table:t1_and_t3_for_various_Theta_acc}
\end{table}

Raising the opposing electron beam momentum has the same physical effect, that more antiprotons are scattered out of the beam, as lowering the acceptance angle.  The acceptance angle is a fixed parameter of a storage ring whereas the momentum of the opposing electron beam can easily be altered.  Therefore it makes sense to pick a typical ring acceptance angle and investigate what effect changing the opposing electron beam momentum has on the polarization buildup time.  This is done in Table~\ref{table:t1_and_t3_for_various_p2} which follows.

As correctly emphasized in Ref.~\cite{Walcher:2007sj} the important practical parameters for a method to polarize antiprotons are:
\\[1.5ex]
- the polarization buildup time,
\\
- the degree of polarization achieved after this time,
\\
- the number of antiprotons available after this time, and
\\
- the phase space of the polarized antiprotons.
\\[1.5ex]
The latter is a measure of how focused the final polarized antiproton beam is, and given that the beam can be focused by electron cooling we do not worry about this parameter here.  In section~\ref{subsec:Beam_lifetime_and_figure_of_merit} it has been shown that the figure of merit $\mathrm{FOM}\left(\tau\right) = \mathcal{P}^{\,2}\left(\tau\right)\,N\left(\tau\right)$ has a maximum at twice the beam lifetime.  Of interest then is the polarization achieved after two beam lifetimes, {\it i.e.}\ the polarization achieved when the beam intensity has decreased by a factor of $e^2 \approx 7.389$.  The later provides a combined measure of the second and third points listed above.

\begin{figure}[!h]
\hspace*{4.5em}
\rotatebox{-90}{\includegraphics[height=13.45cm]{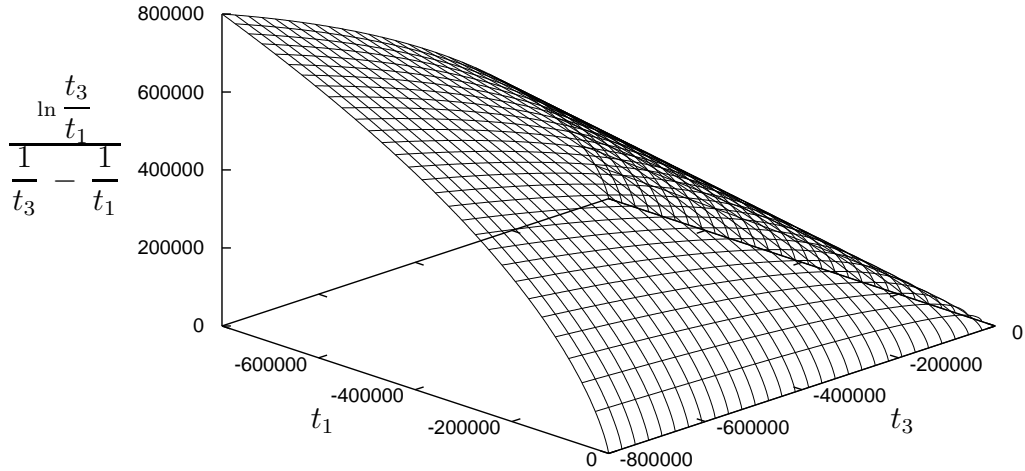}}
\caption{\small{\it{A surface plot of the dependence on $t_1$ and $t_3$ in $\mathcal{P}\left(\,\tau_\mathrm{optimum}\,\right)$.  The units of each axis are $(\mbox{MeV}/c)^{\,2}$.}}}
\label{fig:3D_Plot}
\end{figure}

Equations~(\ref{eq:Optimum_polarization_buildup_time} and \ref{eq:Pmax_for_pure_EM_scattering}) can be used to provide an estimate of the polarization achieved after two beam lifetimes:
\begin{equation}
\label{eq:P_of_optimum_time1}
\mathcal{P}\left(\tau_\mathrm{optimum}\right) \ = \ \mathcal{P}\left(\,2\,\tau_*\,\right) \ \approx \ \frac{-\,2\,\mathcal{P}_T\,K_\mathrm{\,out}}{I_\mathrm{\,out}} \,.
\end{equation}
It is interesting to notice that this optimum polarization is not dependent on the electron beam areal density $n$ or the antiproton beam revolution frequency $\nu$, although the time taken to achieve this polarization is strongly dependent on both of these parameters, as shown in eq.~(\ref{eq:Optimum_polarization_buildup_time}).  In fact to this approximation, which is valid at times of twice the beam lifetime, the optimum polarization achieved is dependent only on energy and the ring acceptance angle, {\it i.e.}\ $\theta_\mathrm{acc}$ or $t_1$ and $t_3$\,.  Using the expressions for $K_\mathrm{\,out}$ and $I_\mathrm{\,out}$ presented in eqs.~(\ref{eq:Kout_Pbar_E} and \ref{eq:Iout_Pbar_E}) one can write
\begin{equation}
\label{eq:P_of_optimum_time2}
\mathcal{P}\left(\tau_\mathrm{optimum}\right) \ = \ \mathcal{P}\left(\,2\,\tau_*\,\right) \ \approx \ \frac{2\,\mu_p\,\mathcal{P}_T}{s\,-\,m^2\,-\,M^{\,2}}\ \frac{\displaystyle{\ln\frac{t_3}{t_1}}}{\displaystyle{\frac{1}{t_3} \ - \ \frac{1}{t_1}}} \ ,
\end{equation}
which we now maximize with respect to acceptance angle and energy.  One finds that $\mathcal{P}\left(\tau_\mathrm{optimum}\right)$ is maximal for large $t_1 \approx t_3$, {\it i.e.}\ for large $\theta_\mathrm{acc} \approx \theta_\mathrm{max}$\,, as shown in Figure~\ref{fig:3D_Plot}, and also for high energies.  Note these equalities cannot be strict otherwise $\mathcal{P}\left(\tau_\mathrm{optimum}\right)$ is undefined, but this is physically reasonable as if $\theta_\mathrm{acc} = \theta_\mathrm{max}$ there is no scattering out of the ring and hence the beam lifetime $\tau_*$ is undefined.
\psfrag{t1}{$t_1$}
\psfrag{t3}{$t_3$}
\psfrag{FXY}{$\hspace*{-3em}\frac{\ln \displaystyle{\frac{t_3}{t_1}}}{\displaystyle{\frac{1}{t_3} \,-\, \frac{1}{t_1}}}$}
%
%Results for the first three parameters above are now presented for the case of spin filtering of a stored antiproton beam.  

Since $\mathcal{P}\left(\,\tau_\mathrm{optimum}\,\right)$ is maximal at high energies we perform the following numerical calculations at the HESR energies.  Thus the antiproton beam will have ring frame momentum $15 \ \mbox{GeV}/c$.  Stored antiprotons, having already been accelerated to these energies, could be polarized directly in the HESR ring.  This eliminates the need for a purpose built low energy Antiproton Polarizing Ring, providing another advantage of this method over spin filtering off a polarized atomic gas target.  This will also avoid the problem of having to accelerate a polarized antiproton beam past depolarizing resonances in the storage ring.

%\pagebreak

\subsubsection{Numerical results: Initial treatment}

Taking the discussion of section~\ref{subsec:Effective_acceptance_angle} into account we fix the effective acceptance angle at $\theta^{\,r}_\mathrm{acc} = 50 \ \mbox{mrad}$, the highest acceptance angle under consideration by the $\mathcal{PAX}$ Collaboration \cite{Barone:2005pu}.  We shall investigate spin filtering with HESR parameters of antiproton momentum of $15 \ \mbox{GeV}$ and revolution frequency of $521628 \ \mbox{Hz}$.  For this ring acceptance angle and antiproton momentum one finds from eqs.~(\ref{eq:Upperbound_for_E_momentum} and \ref{eq:E_momentum_to_scatter_Pbar_out}) that the minimum and maximum opposing electron beam momentum, in order to provide scattering out of the beam but no backward scattering, are $23.4339 \ \mbox{MeV}/c$ and $454.478\ \mbox{MeV}/c$ respectively.
\begin{table}[!h]
\begin{center}
\begin{tabular}{|c||c|c|c|c|c|}
\hline
$$             & $$    & $$ & $$   & $$    & $$\\[-2ex]
           $p_2 \ [\mbox{MeV}/c]$    &         $t_1 \ [(\mbox{MeV}/c)^2]$    & $t_3 \ [(\mbox{MeV}/c)^2]$  & $t_4 \ [(\mbox{MeV}/c)^2]$ & $\displaystyle{\ln \frac{t_3}{t_1}}$ & $\displaystyle{\ln \frac{t_1\,t_4}{t_\mathrm{min}\,t_3}}$\\[2ex]\hline
\hline
 & & & & & \\[-2ex]
$50$  & $-\,498035$ & $-\,2.291 \times 10^6$ & $-\,2.322 \times 10^6$ & $1.52617$ & $33.4120$\\[0.5ex] \hline
 & & & & & \\[-2ex]
$100$ & $-\,520975$ & $-\,5.230 \times 10^6$ & $-\,5.238 \times 10^6$ & $2.30641$ & $33.4452$\\[0.5ex] \hline
 & & & & & \\[-2ex]
$200$ & $-\,539229$ & $-\,1.119 \times 10^7$ & $-\,1.119 \times 10^7$ & $3.03262$ & $33.4782$\\[0.5ex]\hline
 & & & & & \\[-2ex]
$300$ & $-\,546436$ & $-\,1.718 \times 10^7$ & $-\,1.718 \times 10^7$ & $3.44796$ & $33.4914$\\[0.5ex] \hline
 & & & & & \\[-2ex]
$400$ & $-\,550267$ & $-\,2.317 \times 10^7$ & $-\,2.317 \times 10^7$ & $3.74039$ & $33.4983$\\[0.5ex] \hline
 & & & & & \\[-2ex]
$454$ & $-\,551672$ & $-\,2.641 \times 10^7$ & $-\,2.641 \times 10^7$ & $3.86869$ & $33.5009$\\[0.5ex]
\hline
\end{tabular}
\end{center}
\caption{\small{\it{Values of $t_1$, $t_3$ and $t_4$, in units of $(\mbox{MeV}/c)^2$, and the natural logarithms involving them appearing in the integrated spin observables, for various opposing electron beam momenta, in units of $\mbox{MeV}/c$, obtained by solving eq.~(\ref{eq:Ring_frame_theta_to_t}).  Results are for fixed values of $\theta^{\,r}_\mathrm{acc} = 50 \ \mbox{mrad}$ and $p_1 = 15 \ \mbox{GeV}/c$, where one sees from eq.~(\ref{eq:E_momentum_to_scatter_Pbar_out}) that $p^\mathrm{\,out} = 23.4339 \ \mbox{MeV}/c$.  Again we take the most optimistic $t_\mathrm{min}=-\,1.5575 \times 10^{-9} \ (\mbox{MeV}/c)^2$ corresponding to an electron beam areal density of $10^{12} \, \mbox{cm}^2$ as discussed in Table~\ref{table:t_min_and_Theta_min_for_various_n}.}}}
\label{table:t1_and_t3_for_various_p2}
\end{table}

Using the values\footnote{At these high energies $t_3 \approx t_4$, due to the skewness of the relationship between $\theta$ and $t$ in the ring frame, as seen in Figure~\ref{fig:Theta_versus_t_plots_p2_15000}.  Hence the major contribution to $\ln(t_1\,t_4/t_\mathrm{min}\,t_3)$ comes from $\ln(t_1/t_\mathrm{min})$.} of $\ln(t_3/t_1)$ and $\ln(t_1\,t_4/t_\mathrm{min}\,t_3)$ presented in Table~\ref{table:t1_and_t3_for_various_p2} one can obtain numerical values of $K_\mathrm{\,out}$, $L_\mathrm{\,in}$, $L_\mathrm{\,d}$ and $I_\mathrm{\,out}$ using eqs.~(\ref{eq:Kout_Pbar_E} to \ref{eq:Iout_Pbar_E}).  Hence using the results of section~\ref{sec:Solving_the_polarization_evolution_equations} one can obtain numerical values of the maximum polarization achievable for various opposing electron beam momenta, as presented in Table~\ref{table:Polarization_Todays_Technology}.  The beam lifetime, $\tau_*$, is the time taken for the beam intensity to decrease by a factor of $e=2.718$.  The figure of merit has a maximum at twice the beam lifetime, as described in section~\ref{subsec:Beam_lifetime_and_figure_of_merit}, hence we are principally interested in the polarization achieved after this time.
%
% time taken to reach $20\%$ polarization $\tau_{20\%}$ and the fraction of particles left in the beam after this time $N\!\left(\tau_{20\%}\right)$
%
%\pagebreak 

\begin{table}[!h]
\begin{center}
\begin{tabular}{|c||c|c|c|c|}
\hline
$$             & $$   & $$    & $$ & $$  \\[-2ex]
           $p_2 \ [\mbox{MeV}/c]$    &$I_\mathrm{\,out} \ [\mbox{mb}]$  &  $K_\mathrm{\,out} \ [\mbox{mb}]$    & $L_\mathrm{\,in} \ [\mbox{mb}]$  & $L_\mathrm{\,d} \ [\mbox{mb}]$\\[1ex]
\hline
\hline
 & &  & &  \\[-2ex]
$50$  & \ $0.00257272$ \  & \ $-\,0.00232374$ \ & \ $0.01068$ \     & \ $0.0108828$ \  \\[0.5ex]\hline
 & &  & &  \\[-2ex]
$100$ & $0.00282946$ & $-\,0.0017559$  & $0.00267275$  & $0.00310499$  \\[0.5ex]\hline
 & &  & &  \\[-2ex]
$200$ & $0.00288982$ & $-\,0.00115439$ & $0.000668854$ & $0.00123563$  \\[0.5ex]\hline
 & &  & & \\[-2ex]
$300$ & $0.00290078$ & $-\,0.000874996$ & $0.000297386$ & $0.000841777$ \\[0.5ex]\hline
 & &  & &  \\[-2ex]
$400$ & $0.00290458$ & $-\,0.000711905$ & $0.000167314$ & $0.0006622$ \\[0.5ex]\hline
 & &  & &  \\[-2ex]
$454$ & $0.00290566$ & $-\,0.000648743$ & $0.00012989$ & $0.000598142$ \\[0.5ex]
\hline
\end{tabular}
\end{center}
\caption{\small{\it{The integrated spin observables, in units of millibarns, for various opposing electron beam momenta, in units of $\mbox{MeV}/c$.  Results are for fixed values of $\theta^{\,r}_\mathrm{acc} = 50 \ \mbox{mrad}$, $p_1 = 15 \ \mbox{GeV}/c$ and electron beam polarization $\mathcal{P}_T = 0.9$.}}}
\label{table:Integrated_spin_observables}
\end{table}
%
%\vfill
%
\begin{table}[!h]
\begin{center}
\begin{tabular}{|c||c|c|c|}
\hline
 & & & \\[-2ex]
        $p_2 \ [\mbox{MeV}/c]$   & $\mathcal{P}_\mathrm{max}$  & $\tau_* \ [\mbox{seconds}]$ & $\mathcal{P}\left(\,2\,\tau_*\,\right)$ \\[1ex]\hline
\hline
 & & & \\[-2ex]
$50$  & \ $0.096990$ \ & \ $7.45156 \times 10^{11}$ \ & \ $0.096990$ \ \\[0.5ex] \hline
 & & & \\[-2ex]
$100$ & $0.273516$ & $6.77541 \times 10^{11}$ & $0.269873$ \\[0.5ex] \hline
 & & & \\[-2ex]
$200$ & $0.545529$ & $6.63390 \times 10^{11}$ & $0.424074$ \\[0.5ex] \hline
 & & & \\[-2ex]
$300$ & $0.691294$ & $6.60883 \times 10^{11}$ & $0.412933$ \\[0.5ex] \hline
 & & & \\[-2ex]
$400$ & $0.772397$ & $6.60019 \times 10^{11}$ & $0.372750$ \\[0.5ex] \hline
 & & & \\[-2ex]
$454$ & $0.801983$ & $6.59772 \times 10^{11}$ & $0.350905$ \\[0.5ex]
\hline
\end{tabular}
\end{center}
\caption{\small{\it{The maximum polarization achievable, the beam lifetime and the optimum polarization achieved after two beam lifetimes for a stored antiproton beam, for various opposing electron beam momenta, in units of $\mbox{MeV}/c$.  Results are for fixed values of $\theta^{\,r}_\mathrm{acc} = 50 \ \mbox{mrad}$, $p_1 = 15 \ \mbox{GeV}/c$, electron beam polarization $\mathcal{P}_T=0.9$, electron beam areal density $n=10^{12} \ \mbox{cm}^{-2} = 10^{-15} \ \mbox{mb}^{-1}$ and antiproton revolution frequency $521628 \ \mbox{Hz}$.  The values for the integrated spin observables are taken from Table~\ref{table:Integrated_spin_observables}, and inserted into eq.~(\ref{eq:Pmax_for_pure_EM_scattering}).}}}
\label{table:Polarization_Todays_Technology}
\end{table}

\pagebreak

%The negative sign of both the maximum polarization achievable and the optimum polarization achieved after two beam lifetimes simply means the antiproton beam polarization obtained is in the opposite direction to that of the polarized electron beam.  This is a result of the positive sign of each of the integrated spin observables presented in Figure~\ref{table:Integrated_spin_observables} and the overall minus sign of eqs.~(\ref{eq:Ptau_for_pure_EM_scattering} and \ref{eq:Pmax_for_pure_EM_scattering}).
%
%\noindent
%
The positive values for the induced antiproton polarization indicate that it is orientated in the same direction as the polarization of the electron beam.  This is a result of the negative sign of $K_\mathrm{\,out}$, the positive signs of the other integrated spin observables presented in Table~\ref{table:Integrated_spin_observables} and the overall minus sign of eqs.~(\ref{eq:Ptau_for_pure_EM_scattering} and \ref{eq:Pmax_for_pure_EM_scattering}). 

The maximum polarization achievable, satisfying the relation of eq.~(\ref{eq:Pmax_for_pure_EM_scattering}), increases with increasing opposing electron beam momentum as a consequence of $L_\mathrm{\,in}$ and $L_\mathrm{\,d}$ decreasing faster than $K_\mathrm{\,out}$ with increasing opposing electron beam momentum. 

While the polarization achieved after two beam lifetimes in the above investigations are very high one must note that the time taken to reach these polarizations is impractically long.  In practice one requires the beam lifetime to be of the order of a few hours, as opposed to the $10^{11} \ \mbox{seconds} \approx 3171 \ \mbox{years}$ in the above cases!

One notes that the closer $t_1$ is to $t_3$, {\it i.e.}\ the closer $\theta_\mathrm{acc}$ is to $\theta_\mathrm{max}$, the lower the rate that particles are scattered out of the beam and hence the longer the beam lifetime.  Since spin filtering must continue for two beam lifetimes to achieve the optimum polarization the beam lifetime in practice must not be greater than a few hours.  Therefore one should maximize $\mathcal{P}\left(\,2\,\tau_*\,\right)$ in eq.~(\ref{eq:P_of_optimum_time2}) subject to the constraint that $\tau_* < 5 \ \mbox{hours}$, which will allow for the optimum polarization to be achieved in less than $10$ hours.  This constraint can be presented in terms of $t_1$ and $t_3$ as
\begin{eqnarray}
\label{eq:Constraint}
\tau_* \ = \ \frac{2}{n\,\nu\,I_\mathrm{\,out}} & \leq & 5 \ \mbox{hours} \ = \ 18000 \ \mbox{seconds} \ \ \Rightarrow \ \ I_\mathrm{\,out} \ \geq \ \frac{1}{9000\,n\,\nu} \ , \nonumber \\[2ex]
\therefore \ \ \ \ \frac{1}{t_3}\ -\ \frac{1}{t_1} & \geq & \frac{\lambda}{72000\,\pi^2\,\alpha^2\,n\,\nu\left(\,s\,-\,m^2\,-\,M^{\,2}\,\right)^{\,2}} \ > \ 0 \,.
\end{eqnarray}  
The minimum value for this difference is essentially energy independent, but highly dependent on the electron beam areal density $n$ and the antiproton beam revolution frequency $\nu$.  We shall present the numerics for two cases: (1) for $n=10^{12} \ \mbox{cm}^{-2}$ the best electron beam areal density that will be available now or in the near future, and (2) for an ideal case of $n=10^{20} \ \mbox{cm}^{-2}$ assuming great advances in electron beam areal densities.  The latter simply provides a verification that the method of antiproton polarization buildup by spin filtering off an opposing polarized electron beam works in principle.  One finds the minimum value for the difference of reciprocals of $t$, as presented in eq.~(\ref{eq:Constraint}), to be $5.28527 \times 10^{\,6} \ (\mbox{MeV}/c)^{-2}$ for case (1) and $0.0528527 \ (\mbox{MeV}/c)^{-2}$ for case (2).

Unfortunately with today's electron beam areal densities and antiproton beam revolution frequencies the rate at which antiprotons are scattered out of the ring will be very low, even with the lowest possible ring acceptance angles of about $1 \ \mbox{mrad}$.  An extreme case provides a lower bound on the antiproton beam lifetime with today's technologies: $15 \ \mbox{GeV}/c$ antiprotons scattering off a $454 \ \mbox{MeV}/c$ opposing electron beam in a ring with lowest possible acceptance angle of $1 \ \mbox{mrad}$ and antiproton beam revolution frequency $521628 \ \mbox{Hz}$, which still gives a beam lifetime of $2.63468 \times 10^8 \ \mbox{seconds} \approx 8.35 \ \mbox{years}$.  Hence with today's electron beam areal densities and antiproton beam revolution frequencies the beam lifetime will always be of the order of years instead of the required hours.  Thus the constraint presented in eq.~(\ref{eq:Constraint}) is too strict with current parameters, and one cannot maximize the polarization achieved after two beam lifetimes subject to it.

We now investigate the scenario where unpolarized antiprotons are continuously fed into the beam at such a rate to cancel the rate that antiprotons are being scattered out of the beam, the beam intensity remaining constant, as presented in section~\ref{sec:Constant_beam_intensity}.  The time taken to reach $8\%$ polarization without any loss of beam intensity is presented for various opposing electron beam momenta.  Inserting the integrated spin observables presented in Table~\ref{table:Integrated_spin_observables} into eq.~(\ref{eq:Polarization_in_N_Constant_case}), where $A_\mathrm{\,all} \,-\, K_\mathrm{\,in} \,=\, K_\mathrm{\,out}$ in the pure electromagnetic case of interest here and $I_\mathrm{\,all} \,-\, D_\mathrm{\,in} \,=\, I_\mathrm{\,out} \,+\, 2\,L_\mathrm{\,in}$\,, gives the results presented in Table~\ref{table:time_to_reach_polarization_of_0.08_Constant_N} below.
\begin{table}[!h]
\begin{center}
\begin{tabular}{|c||c|c|}
\hline
 & & \\[-2ex]
$p_2 \ [\mbox{MeV}/c]$  & $\mathcal{P}_\mathrm{max}$  & $\tau_{8\%} \ [\mbox{seconds}]$ \\[1ex]
\hline
\hline
 & & \\[-2ex]
$50$  & \ $0.087385$ \ & \ $1.97920 \times 10^{11}$\ \\[0.5ex] \hline
 & & \\[-2ex]
$100$ & $0.193311$ & $1.25265 \times 10^{11}$\\[0.5ex] \hline
 & & \\[-2ex]
$200$ & $0.245759$ & $1.78586 \times 10^{11}$\\[0.5ex] \hline
 & & \\[-2ex]
$300$ & $0.225285$ & $2.40580 \times 10^{11}$\\[0.5ex] \hline
 & & \\[-2ex]
$400$ & $0.197800$ & $3.06730 \times 10^{11}$\\[0.5ex] \hline
 & & \\[-2ex]
$454$ & $0.184451$ & $3.44399 \times 10^{11}$\\[0.5ex]
\hline
\end{tabular}
\end{center}
\caption{\small{\it{The maximum polarization achievable and the time taken, in units of seconds, to reach $8\%$ polarization in the scenario where unpolarized particles are fed into the beam at the same rate particles are being scattered out of the beam, the beam intensity remaining constant $N(\tau) = N_0$, for various opposing electron beam momenta, in units of $\mbox{MeV}/c$.  Results are for fixed values of $\theta^{\,r}_\mathrm{acc} = 50 \ \mbox{mrad}$, $p_1 = 15 \ \mbox{GeV}/c$, electron beam polarization $\mathcal{P}_T=0.9$, electron beam areal density $n=10^{12} \ \mbox{cm}^{-2} = 10^{-15} \ \mbox{mb}^{-1}$ and antiproton revolution frequency $521628 \ \mbox{Hz}$.  The values for the integrated spin observables are taken from Table~\ref{table:Integrated_spin_observables}, and inserted into eq.~(\ref{eq:Polarization_in_N_Constant_case}).}}}
\label{table:time_to_reach_polarization_of_0.08_Constant_N}
\end{table}

Due to the long times taken to reach significant polarization it is apparent that neither of these methods of obtaining a high intensity polarized antiproton beam are practical at present.  The key parameters limiting the rate of polarization buildup are the areal density of the opposing polarized electron beam, and the revolution frequency of the antiproton beam in the storage ring.  The antiproton beam revolution frequency and the areal density of polarized electron beams would have to increase by many orders of magnitude in order for these methods to provide significant polarization in a few hours, which would be required.  However, considering the immense research and development that will take place in the near future on electron and positron beams at the International Linear Collider (ILC), advances in electron beam areal densities can be expected in the coming years \cite{MoortgatPick:2005cw,Brau:2007zza}.

\subsubsection{Numerical results: An ideal case}

To conclude this treatment let us now investigate an ideal case, assuming fanciful values for the key parameters that are not currently achievable, but may be achievable in the future, to show that in principle the method works.  Let us assume a ring frame acceptance angle of $50 \ \mbox{mrad}$ and a very high electron beam areal density of $10^{20} \ \mbox{cm}^{-2}$ which gives $t_\mathrm{min}=-\,1.5575 \times 10^{-1} \ (\mbox{MeV}/c)^{\,2}$.  We shall investigate spin filtering with HESR parameters of antiproton momentum of $15 \ \mbox{GeV}$ and revolution frequency of $521628 \ \mbox{Hz}$.  For this ring acceptance angle and antiproton momentum one finds from eqs.~(\ref{eq:Upperbound_for_E_momentum} and \ref{eq:E_momentum_to_scatter_Pbar_out}) that the minimum and maximum opposing electron beam momentum, in order to provide scattering out of the beam but no backward scattering, are $23.4339 \ \mbox{MeV}/c$ and $454.478\ \mbox{MeV}/c$ respectively.  Since $t_\mathrm{min}$ is much larger in this case compared to the previous treatment one expects $L_\mathrm{\,in}$ to be much smaller in this treatment.  As a consequence in this case both the maximum polarization achievable and the optimum polarization achieved after two beam lifetime should be larger, and the beam lifetime should be much shorter.  A similar analysis to that presented above gives the results presented in the following tables.

\begin{table}[!h]
\begin{center}
\begin{tabular}{|c||c|c|c|c|c|}
\hline
$$             & $$    & $$ & $$   & $$    & $$\\[-2ex]
           $p_2 \ [\mbox{MeV}/c]$    &         $t_1 \ [(\mbox{MeV}/c)^2]$    & $t_3 \ [(\mbox{MeV}/c)^2]$  & $t_4 \ [(\mbox{MeV}/c)^2]$ & $\displaystyle{\ln \frac{t_3}{t_1}}$ & $\displaystyle{\ln \frac{t_1\,t_4}{t_\mathrm{min}\,t_3}}$\\[2ex]
\hline
\hline
 & & & & & \\[-2ex]
$50$  & $-\,498035$ & $-\,2.291 \times 10^6$ & $-\,2.322 \times 10^6$ & $1.52617$ & $14.9914$ \\[0.5ex] \hline
 & & & & & \\[-2ex]
$100$  & $-\,520975$ & $-\,5.230 \times 10^6$ & $-\, 5.238 \times 10^6$ & $2.30641$ & $15.0246$\\[0.5ex] \hline
 & & & & & \\[-2ex]
$200$  & $-\,539229$ & $-\,1.119 \times 10^7$ & $-\,1.119 \times 10^7$ & $3.03262$ & $15.0575$\\[0.5ex]\hline
 & & & & & \\[-2ex]
$300$ & $-\,546436$ & $-\,1.718 \times 10^7$ & $-\,1.718 \times 10^7$ & $3.44796$ & $15.0707$\\[0.5ex] \hline
 & & & & & \\[-2ex]
$400$ & $-\,550267$ & $-\,2.317 \times 10^7$ & $-\,2.317 \times 10^7$ & $3.74039$ & $15.0777$\\[0.5ex] \hline
 & & & & & \\[-2ex]
$454$ & $-\,551672$ & $-\,2.641 \times 10^7$ & $-\,2.641 \times 10^7$ & $3.86869$ & $15.0802$\\[0.5ex]
\hline
\end{tabular}
\end{center}
\caption{\small{\it{Values of $t_1$, $t_3$ and $t_4$, in units of $(\mbox{MeV}/c)^2$, and the natural logarithms involving them appearing in the integrated spin observables, for various opposing electron beam momenta, in units of $\mbox{MeV}/c$, obtained by solving eq.~(\ref{eq:Ring_frame_theta_to_t}).  Results are for idealistic fixed values of $\theta^{\,r}_\mathrm{acc} = 50 \ \mbox{mrad}$ and $p_1 = 15 \ \mbox{GeV}/c$.  Here we take the very far fetched $t_\mathrm{min}=-\,1.5575 \times 10^{-1} \ (\mbox{MeV}/c)^2$ corresponding to an electron beam areal density of $10^{20} \, \mbox{cm}^2$.}}}
\label{table:t1_and_t3_for_various_p2_3}
\end{table}
\begin{table}[!h]
\begin{center}
\begin{tabular}{|c||c|c|c|c|}
\hline
$$             & $$   & $$    & $$ & $$  \\[-2ex]
           $p_2 \ [\mbox{MeV}/c]$    &$I_\mathrm{\,out} \ [\mbox{mb}]$  &  $K_\mathrm{\,out} \ [\mbox{mb}]$    & $L_\mathrm{\,in} \ [\mbox{mb}]$  & $L_\mathrm{\,d} \ [\mbox{mb}]$\\[1ex]
\hline
\hline
 & &  & &  \\[-2ex]
$50$  & \ $0.00257272$ \ & $-\,0.00232374$ & $0.00479191$ & \ $0.00522841$ \  \\[0.5ex]\hline
 & &  & &  \\[-2ex]
$100$ & $0.00282946$ & $-\,0.0017559$ & $0.00120068$ & $0.00198469$  \\[0.5ex]\hline
 & &  & &  \\[-2ex]
$200$ & $0.00288982$ & $-\,0.00115439$ & $0.000300831$ & $0.00108163$  \\[0.5ex]\hline
 & &  & & \\[-2ex]
$300$ & $0.00290078$ & $-\,0.000874996$ & $0.00013382$ & $0.000798785$ \\[0.5ex]\hline
 & &  & &  \\[-2ex]
$400$ & $0.00290458$ & $-\,0.000711905$ & $0.0000753084$ & $0.000645125$ \\[0.5ex]\hline
 & &  & &  \\[-2ex]
$454$ & $0.00290566$ & $-\,0.000648743$ & $0.000058469$ & $0.000586789$ \\[0.5ex]
\hline
\end{tabular}
\end{center}
\caption{\small{\it{The integrated spin observables, in units of millibarns, for various opposing electron beam momenta, in units of $\mbox{MeV}/c$.  Results are for fixed values of $\theta^{\,r}_\mathrm{acc} = 50 \ \mbox{mrad}$, $p_1 = 15 \ \mbox{GeV}/c$ and electron beam polarization $\mathcal{P}_T = 0.9$.}}}
\label{table:Integrated_spin_observables_Ideal_Case}
\end{table}

\begin{table}[!h]
\begin{center}
\begin{tabular}{|c||c|c|c|}
\hline
 & & & \\[-2ex]
        $p_2 \ [\mbox{MeV}/c]$   &  $\mathcal{P}_\mathrm{max}$ & $\tau_* \ [\mbox{seconds}]$ & $\mathcal{P}\left(\,2\,\tau_*\,\right)$ \\[1ex]
% &$\tau_{20\%} \ [\mbox{seconds}]$ &  $N\!\left(\tau_{20\%}\right)$\\[1ex]
\hline
\hline
 & & & \\[-2ex]
$50$  & \ $0.208713$ \ & $7452$ & \ $0.208648$\ 
% & $340.366$ & $$
\\[0.5ex] \hline
 & & & \\[-2ex]
$100$ & $0.496115$ & $6775$ & $0.459284$
% & $380.003$ & $$
\\[0.5ex] \hline
 & & & \\[-2ex]
$200$ & $0.751524$ & $6634$ & $0.517906$
% & $495.257$ & $$
\\[0.5ex] \hline
 & & & \\[-2ex]
$300$ & $0.844405$ & $6609$ & $0.455736$
% & $614.839$ & $$
\\[0.5ex] \hline
 & & & \\[-2ex]
$400$ & $0.889346$ & $6600$ & $0.395042$
% & $708.286$ & $$
\\[0.5ex] \hline
 & & & \\[-2ex]
$454$ & $0.904861$ & $6598$ & $0.367339$
\\[0.5ex]
\hline
\end{tabular}
\end{center}
\caption{\small{\it{The maximum polarization achievable, the beam lifetime $\tau_*$ and the polarization achieved after spin filtering for two beam lifetimes
% and the time $\tau_{20\%}$ taken, in units of seconds, to reach $20\%$ polarization of a stored antiproton beam and the fraction of particles left in the beam after this time $N\!\left(\tau_{20\%}\right)$
for various opposing electron beam momenta, in units of $\mbox{MeV}/c$.  Results are for idealistic fixed values of $\theta^{\,r}_\mathrm{acc} = 50 \ \mbox{mrad}$, $p_1 = 15 \ \mbox{GeV}/c$, electron beam polarization $\mathcal{P}_T=0.9$, electron beam areal density $n=10^{20} \ \mbox{cm}^{-2} = 10^{-7} \ \mbox{mb}^{-1}$ and antiproton revolution frequency $521628 \ \mbox{Hz}$.  The values for the integrated spin observables are taken from Table~\ref{table:Integrated_spin_observables_Ideal_Case}, and inserted into eq.~(\ref{eq:Ptau_for_pure_EM_scattering}).}}}
\label{table:Polarization_Ideal_Case}
\end{table}

%\pagebreak

One emphasizes the best case presented above as achieving an antiproton beam polarization of $51.8\%$ after just $1.84 \ \mbox{hours}$ when the beam intensity has decreased by a factor of $e^2 = 7.389$, by spin filtering off an opposing electron beam of momentum $200 \ \mbox{MeV}/c$.

Since the beam intensity has decreased significantly by the time the polarization reaches a high percentage, we now redo the analysis for the scenario where unpolarized particles are fed into the beam at the same rate particles are being scattered out of the beam, the beam intensity remaining constant $N(\tau) = N_0$, as presented in section~\ref{sec:Constant_beam_intensity}.  The time taken to reach $15\%$ without any loss of beam intensity is presented for various opposing electron beam momenta.  Inserting the integrated spin observables presented in Table~\ref{table:Integrated_spin_observables_Ideal_Case} into eq.~(\ref{eq:Polarization_in_N_Constant_case}), where $A_\mathrm{\,all} \,-\, K_\mathrm{\,in} \,=\, K_\mathrm{\,out}$ in the pure electromagnetic case of interest here and $I_\mathrm{\,all} \,-\, D_\mathrm{\,in} \,=\, I_\mathrm{\,out} \,+\, 2\,L_\mathrm{\,in}$\,, gives the results presented in Table~\ref{table:time_to_reach_polarization_of_0.15_Constant_N}.
\begin{table}[!h]
\begin{center}
\begin{tabular}{|c||c|c|}
\hline
 & & \\[-2ex]
$p_2 \ [\mbox{MeV}/c]$  & $\mathcal{P}_\mathrm{max}$  & $\tau_{15\%} \ [\mbox{seconds}]$ \\[1ex]
\hline
\hline
 & & \\[-2ex]
$50$  & \ $0.172036$ \  & $3241$\\[0.5ex] \hline
 & & \\[-2ex]
$100$ & $0.302115$ & $2515$\\[0.5ex] \hline
 & & \\[-2ex]
$200$ & $0.297568$ & $3851$\\[0.5ex] \hline
 & & \\[-2ex]
$300$ & $0.248546$ & $5597$\\[0.5ex] \hline
 & & \\[-2ex]
$400$ & $0.209713$ & $7882$\\[0.5ex] \hline
 & & \\[-2ex]
$454$ & $0.193168$ & $9504$\\[0.5ex]
\hline
\end{tabular}
\end{center}
\caption{\small{\it{The maximum polarization achievable and the time taken, in units of seconds, to reach $15\%$ polarization in the scenario where unpolarized particles are fed into the beam at the same rate particles are being scattered out of the beam, the beam intensity remaining constant $N(\tau) = N_0$, for various opposing electron beam momenta, in units of $\mbox{MeV}/c$.  Results are for fixed values of $\theta^{\,r}_\mathrm{acc} = 50 \ \mbox{mrad}$, $p_1 = 15 \ \mbox{GeV}/c$, electron beam polarization $\mathcal{P}_T=0.9$, electron beam areal density $n=10^{20} \ \mbox{cm}^{-2} = 10^{-7} \ \mbox{mb}^{-1}$ and antiproton revolution frequency $521628 \ \mbox{Hz}$.  The values for the integrated spin observables are taken from Table~\ref{table:Integrated_spin_observables_Ideal_Case}, and inserted into eq.~(\ref{eq:Polarization_in_N_Constant_case}).}}}
\label{table:time_to_reach_polarization_of_0.15_Constant_N}
\end{table}

%Interestingly note that the maximum polarization achievable in this case is very small.  This is because the unpolarized particles that are continuously fed into the ring dilute the polarization of the stored beam.  In this scenario, for pure electromagnetic scattering, the polarization buildup is proportional to $-\,\mathcal{P}_T\,K_\mathrm{\,out}\,/\,\left(\,I_\mathrm{\,out} + 2\,L_\mathrm{\,in}\,\right)$, and the $I_\mathrm{\,out}$ in the denominator is very large compared to the other spin observables.  This large $I_\mathrm{\,out}$ factor is not present in the denominator of the scenario of spin filtering a stored beam (without any external particle input), hence the maximum polarizations achievable are greater in that case.

Notice that in the best case presented in Table~\ref{table:time_to_reach_polarization_of_0.15_Constant_N}, for an opposing electron beam of momentum $100 \ \mbox{MeV}/c$, the antiproton beam polarization builds up to $15 \%$ in only $2515 \ \mbox{seconds} \,\approx\, 42 \ \mbox{minutes}$, while maintaining constant beam intensity.

While this idealistic treatment might be far from today's technologies, requiring an increase of eight orders of magnitude in the product $n\,\nu$, it highlights that the method of polarizing an antiproton beam by spin filtering off an opposing polarized electron beam works very well in principle.

\pagebreak

\subsection{A co-moving lepton beam}
\label{subsec:A_co-moving_lepton_beam}
\noindent
It has recently been proposed by Th.~Walcher {\it et al.}~\cite{Walcher:2007sj,Aulenbacher:2008zz} that a polarized positron beam, with very low relative momentum to an antiproton beam, could transfer polarization to the antiproton beam.  This method of spin filtering is entirely based on selective spin flip while scattering within the ring, as a co-moving positron beam cannot scatter antiprotons out of the ring.  Coulomb forces dominate the spin transfer observables at such low energies, and the antiproton-positron interaction is chosen because of the attraction of unlike charges.  A positron beam can easily be polarized by the Sokolov-Ternov effect as described in sections~(\ref{subsec:Spontaneous_synchrotron_radiation_emission} and \ref{sec:The_Sokolov-Ternov_Effect}).  The basis for this proposal is a dramatic enhancement of the polarization transfer cross-section in the reaction $\bar{p}\,e^{+\,\uparrow} \rightarrow \bar{p}^{\,\uparrow}\,e^+$ as calculated in Ref.~\cite{Arenhovel:2007gi}.  Such a dramatic enhancement, of over nine orders of magnitude, has been called into question by many groups, who point out that, if true, multiple scattering effects should not be neglected.  In particular Ref.~\cite{Milstein:2008tc} claims that there is an enhancement at low energies but by many orders of magnitude less than that claimed in Ref.~\cite{Arenhovel:2007gi}, and that the polarization transfer cross-section is still far too low to make the Walcher {\it et al.}\ proposal practical at present.  It also remains to be seen if the depolarization observable also gets enhanced greatly at such low relative velocities.  An experiment has been proposed to test these claims \cite{PAX:2007_2}.  As encountered earlier the relatively low areal densities of polarized positron beams is a crucial disadvantage of this method.  Any advances in the technology of obtaining high intensity positron beams would greatly benefit this proposal.

An interesting application of a great increase in the areal densities of polarized electron beams is that an electron cooler could utilize a high intensity polarized electron beam.  This would allow for an antiproton beam (or a beam of any other particles for that matter) to be both cooled and polarized after repeated interaction with the electron cooler.  This would represent a major advance in accelerator physics and hadron storage rings.

%\pagebreak

\section{Spin filtering off a polarized hydrogen target}
\label{sec:Spin_filtering_off_a_polarized_hydrogen_target}

\subsection{Electromagnetic and hadronic scattering}
\label{subsec:Electromagnetic_and_hadronic_scattering}

The critical squared momentum transfer $t_c$\,, below which electromagnetic effects dominate and above which hadronic effects dominate in $\bar{p}\,p$ scattering, is now derived.  

For both hadronic and electromagnetic scattering, at low $|\,t\,|$, the non-spin-flip amplitudes $\phi_1 +\phi_3 \equiv \phi_+$ dominate the spin-averaged differential cross-section.  The leading $t$ imaginary part of the hadronic amplitude is given by the Optical Theorem:
\begin{equation}
\label{eq:Optical_Theorem}
\left. \mathcal{I}m\left\{\,\phi^h_+(s,t)\,\right\}\right|_{t=0}\  = \ \frac{2\,k_\mathrm{cm}\,\sqrt{s}\,\sigma^{\,\bar{p}\,p}_\mathrm{tot}}{8\,\pi} \,,
\end{equation}
and the leading $t$ part of the $\bar{p}\,p$ electromagnetic amplitude is given by
\begin{equation}
\label{eq:Electromagnetic_amplitude}
\frac{-\,\alpha\,\left(\,s\,-\,2\,M^{\,2}\,\right)\,F^{\,2}_1\!\left(t\right)}{t} \ e^{\,\delta \,i}  \,,
\end{equation}
where $F_1 \approx 1$ for small $|\,t\,|$ and $e^{\,\delta \,i} \approx 1$ as $\delta$ is small \cite{Buttimore:1998rj}.  The Coulomb phase shift, $e^{\,\delta \,i}$, accounts for the small correction to the single-photon exchange amplitude coming from multi-photon exchange \cite{Bethe:1958}.  The hadronic amplitude is
\begin{equation}
\label{eq:Hadronic_amplitude}
\frac{2\,k_\mathrm{cm}\,\sqrt{s}\,\sigma^{\,\bar{p}\,p}_\mathrm{tot}}{8\,\pi}\, \left(\,i \,+\,\rho\,\right)\,e^{\,b\,t}  \,,
\end{equation}
where $\rho = \rho\!\left(s,t\right) = \mathcal{R}e \{\phi^h_+\}\,/\,\mathcal{I}m \{\phi^h_+\}$ the ratio of real to imaginary parts of the hadronic non-flip amplitude.  For small $|\,t\,|$ one has that $e^{\,b\,t} \approx 1$ \cite{Buttimore:1978ry,Kopeliovich:1974ee}.  Thus to leading order in small $|\,t\,|$ the spin-averaged $\bar{p}\,p$ cross-section is 
\begin{equation}
\label{eq:EM_equals_Hadronic1}
\frac{\mathrm{d}\,\sigma}{\mathrm{d}\,t} \ \propto \ \left|\, \frac{\alpha\,\left(\,s\,-\,2\,M^{\,2}\,\right)}{t} \ + \ \frac{2\,k_\mathrm{cm}\,\sqrt{s}\,\sigma^{\,\bar{p}\,p}_\mathrm{tot}}{8\,\pi}\,\left(\,i\,+\,\rho\,\right)\,\right|^{\,2}  \,.
\end{equation}
The electromagnetic and hadronic amplitudes are of equal size when
\begin{equation}
\label{eq:EM_equals_Hadronic2}
\left|\,\frac{-\,\alpha\,\left(\,s\,-\,2\,M^{\,2}\,\right)}{t_c}\,\right| \ = \ \left|\,\frac{2\,k_\mathrm{cm}\,\sqrt{s}\,\sigma^{\,\bar{p}\,p}_\mathrm{tot}}{8\,\pi}\,\left(\,i\,+\,\rho\,\right)\,\right|  \,,
\end{equation}
{\it i.e.}\ when
\begin{equation}
\label{eq:EM_equals_Hadronic3}
 t_c \ = \ -\,\frac{8\,\pi\,\alpha\,\left(\,s\,-\,2\,M^{\,2}\,\right)}{2\,k_\mathrm{cm}\,\sqrt{s}\,\sigma^{\,\bar{p}\,p}_\mathrm{tot}\,\sqrt{1\,+\,\rho^2}}  \ = \ -\,\frac{8\,\pi\,\alpha\,E_\mathrm{lab}}{p_\mathrm{lab}\,\sigma^{\,\bar{p}\,p}_\mathrm{tot}\,\sqrt{1\,+\,\rho^2}}\,,
\end{equation}
where the relations $k_\mathrm{cm}\,\sqrt{s} = p_\mathrm{lab}\,M$ and $E_\mathrm{lab} = \left(s - 2\,M^{\,2}\,\right)/(2M)$ have been used.  Finally using the relativistic laboratory velocity $\beta_\mathrm{lab} = p_\mathrm{lab}\,/\,E_\mathrm{lab}$ gives \cite{Buttimore:1978ry,Kopeliovich:1974ee}
\begin{equation}
\label{eq:t_c_derived_final}
 t_c \ = \ -\,\frac{8\,\pi\,\alpha}{\beta_\mathrm{lab}\,\sigma^{\,\bar{p}\,p}_\mathrm{tot}\,\sqrt{1\,+\,\rho^2}}  \,,
\end{equation}
the critical squared momentum transfer below which electromagnetic effects dominate and above which hadronic effects dominate $\Box$.  

\noindent
At high energies $\beta_\mathrm{lab} \approx 1$ and $\rho^2 \approx 0$ \cite{Ashford:1985mm,Klempt:2002ap}, giving the often used result
\begin{equation}
t_c \ = \ -\,\frac{8\,\pi\,\alpha}{\sigma^{\,\bar{p}\,p}_\mathrm{tot}}  \,.
\end{equation}
This region of momentum transfer is referred to in the literature as the Coulomb-Nuclear-Interference (CNI) region.

In analogy to $t_c$ above, we now derive an expression for the critical antiproton laboratory frame scattering angle below which the electromagnetic interaction dominates the hadronic interaction in $\bar{p}\,p$ scattering.  It is then shown that the scattering angles of importance in spin filtering are below this critical angle, and hence the electromagnetic $\bar{p}\,p$ cross-sections calculated in Chapters~\ref{ch:Generic_helicity_amplitudes_and_spin_observables} and \ref{ch:Specific_helicity_amplitudes_and_spin_observables} provide a good approximation to the total $\bar{p}\,p$ interaction in this region.

Using eqs.~(\ref{eq:CM_to_LAB_momenta} and \ref{eq:t_to_CM_theta}) and the relation between the scattering angles in the Centre-of-Mass and LAB frames
\begin{equation}
\label{eq:CM_to_LAB_angles}
\frac{\sin\theta_\mathrm{cm}}{\sin\theta_\mathrm{lab}} \ = \ \frac{p_\mathrm{lab}}{k_\mathrm{cm}}\,,
\end{equation}
and Taylor expanding $\sin^2\theta_\mathrm{lab}$ for small $\theta_\mathrm{lab}$ gives
\begin{equation}
\label{eq:t_to_LAB_angle}
-\,t \ \approx \ p^{\,2}_\mathrm{lab}\,\theta^{\,2}_\mathrm{lab} \,.
\end{equation}
So the critical scattering angle below which the electromagnetic interaction dominates the hadronic interaction is
\begin{equation}
\label{eq:t_c_to_LAB_angle}
\theta^{\,c}_\mathrm{lab} \ \approx \ \frac{\sqrt{-\,t_c}}{p_\mathrm{lab}}\,,
\end{equation}
and inserting the expression for $t_c$ in eq.~(\ref{eq:t_c_defined}) one obtains that electromagnetic effects dominate hadronic effects for laboratory scattering angles 
\begin{equation}
\label{eq:CNI_LAB_angle}
\theta \ < \ \theta^{\,c}_\mathrm{lab} \ \approx \ \sqrt{\frac{8\,\pi\,\alpha}{T\,\left(\,T + 2\,M\,\right)\,\sigma^{\,{\bar p}\,p}_\mathrm{tot}\,\beta_\mathrm{lab}\,\sqrt{1\,+\,\rho^2}}} \ \approx \ 152 \ \mbox{mrad} \,,
\end{equation}
%
% NB Numerical calculation done in notes 4-4-2008, page 2.
%
where $M$ and $T$ are the antiproton mass and kinetic energy in the Laboratory frame respectively \cite{Nikolaev:2006gw}, and the numerical result is for FILTEX kinetic energies of $T = 23 \ \mbox{MeV}$ where $\sigma^{\,{\bar p}\,p}_\mathrm{tot} \approx 325 \ \mbox{mb}$ \cite{Amsler:2008zz,Klempt:2002ap} and $\rho \approx 0.1$ \cite{Klempt:2002ap,Bruckner:1985zi}.  This gives the strong inequality \cite{Nikolaev:2006gw}
\begin{equation}
\label{eq:Kolyas_angle_inequalities}
\theta_\mathrm{min} \ \ll \ \theta_e \ \ll \ \theta_\mathrm{acc} \ \ll \ \theta^{\,c}_\mathrm{lab} \,,
\end{equation}
where $\theta_e = 0.54 \ \mbox{mrad}$ is the maximum angle antiprotons are scattered by stationary (or atomic) electrons as shown in eq.~(\ref{eq:Theta_max_for_stationary_electrons}) and Figure~\ref{fig:Theta_versus_t_plots} (a).  Since spin filtering utilizes angles below $152 \ \mbox{mrad}$, electromagnetic (QED) effects provide a good approximation to the total $\bar{p}\,p$ interaction in this region. 

In conventional particle physics experiments particles must be scattered out of the beam pipe into detectors for measurements to be made, hence no direct experimental observations can be made for scattering at angles below the ring acceptance angle.  Thus there is very little experimental data on this region of very low angle scattering, which is of interest only in storage rings.  In particular hadronic antiproton-proton amplitudes are completely unknown in this kinematical region.  The LHC very forward detector TOTEM hopes to obtain some data on low angle proton-proton scattering in the near future \cite{Avati:2003qj}.

One can derive an expression for the antiproton momentum in the laboratory frame that makes the squared momentum transfer for total backward scattering equal to the critical squared momentum transfer below which electromagnetic effects dominate hadronic effects.  Solving the equation $t_c = -\,4\,k^{\,2}_\mathrm{cm} = -\,4\,M^{\,2}\,p^{\,2}_\mathrm{lab}\,/s$ one obtains $p_\mathrm{lab} \approx 31.6 \ \mbox{MeV}/c$, below which all scattering is electromagnetically dominated.  
%Low energies are also preferred for spin filtering off an atomic gas target, in particular energies below the pion production threshold, to avoid a large loss of beam intensity due to inelastic collisions.

\subsection{Polarization states of a hydrogen target}
\label{subsec:Polarization_states_of_a_hydrogen_target}

Unpolarized hydrogen atoms in a strong magnetic field equally populate each of four hyperfine states:
\begin{eqnarray*}
|\uparrow_{p}\,\downarrow_{e}\,\rangle \hspace*{3em} |\downarrow_{p}\,\downarrow_{e}\,\rangle \hspace*{3em} |\downarrow_{p}\,\uparrow_{e}\,\rangle \hspace*{3em} |\uparrow_{p}\,\uparrow_{e}\,\rangle 
\end{eqnarray*}
It is explained in section~\ref{subsec:Polarizing_hydrogen_gas} how these hyperfine states and pairs of hyperfine states can be isolated to give polarized hydrogen.  We are particularly interested in three types of polarized hydrogen target, with all atoms in the hyperfine states as follows
\begin{eqnarray}
\label{eq:Hydrogen_polarization_state1}
|\uparrow_{p}\,\uparrow_{e}\,\rangle \ + \ |\uparrow_{p}\,\downarrow_{e}\,\rangle \hspace*{1em} \implies \hspace*{1em} \mathcal{P}_p \ = \ 1 \hspace*{1em} \mbox{and} \hspace*{1em} \mathcal{P}_e \ = \ 0 \\[2ex]
\label{eq:Hydrogen_polarization_state2}
|\uparrow_{p}\,\uparrow_{e}\,\rangle \ + \ |\downarrow_{p}\,\uparrow_{e}\,\rangle \hspace*{1em} \implies \hspace*{1em} \mathcal{P}_p \ = \ 0 \hspace*{1em} \mbox{and} \hspace*{1em} \mathcal{P}_e \ = \ 1 \\[2ex]
\label{eq:Hydrogen_polarization_state3}
|\uparrow_{p}\,\uparrow_{e}\,\rangle \hspace*{1em} \implies \hspace*{1em} \mathcal{P}_p \ = \ 1 \hspace*{1em} \mbox{and} \hspace*{1em} \mathcal{P}_e \ = \ 1 
\end{eqnarray}
where we denote the polarization of the electrons in the hydrogen by $\mathcal{P}_e$ and the polarization of the protons in the hydrogen by $\mathcal{P}_p$\,.  In practice the atoms are not perfectly isolated in certain hyperfine states, thus the electron and proton polarizations in polarized hydrogen are less than one.  The HERMES Collaboration have utilized polarized hydrogen targets with $\mathcal{P}_e \,=\,0.9$ and/or $\mathcal{P}_p\,=\,0.9$ \cite{Airapetian:2004yf}.  Spin filtering off a polarized hydrogen target in each of the polarization states presented in eqs.~(\ref{eq:Hydrogen_polarization_state1}--\ref{eq:Hydrogen_polarization_state3}) can be treated similarly to the treatment presented in section~\ref{sec:Spin_filtering_off_a_polarized_electron_beam} using the spin observables for antiproton-proton and antiproton-electron scattering presented in sections~(\ref{subsec:Antiproton_proton_scattering} and \ref{subsec:Antiproton_electron_scattering}) respectively.

In the second case where the electrons in the hydrogen target are polarized but the protons are unpolarized one has that $\sigma^\mathrm{\,out}_+ \,=\, \sigma^\mathrm{\,out}_-$, {\it i.e.}\ particles in both spin states are scattered out of the beam at equal rates, since only the protons are massive enough to scatter the antiprotons beyond the ring acceptance angle.  Thus while there is scattering out of the ring, it is not spin-dependent and does not lead to a buildup of beam polarization.  Therefore only selective spin-flip in electromagnetic antiproton-electron elastic scattering can contribute to polarization buildup in this case, yet one also has the negative effects of antiprotons being scattered out of the beam and annihilating with the protons in the hydrogen target; decreasing the beam intensity.  The only advantage this case has over using a pure lepton target, considering it has the disadvantages listed above, is that the areal densities of electrons in atomic targets is greater than those achievable in pure lepton targets to date. 

Some experimental tests must now be carried out to decide which of the three possible states of a polarized atomic target would be most effective in polarizing an antiproton beam by spin filtering \cite{PAX:2006a}.  The Antiproton Decelerator (AD) storage ring at CERN is the only source of antiprotons in the required energy range.  Consequently spin filtering studies of antiprotons scattering off a polarized hydrogen target at the AD ring are planned in the near future \cite{Nass:2007zz,PAX:2005,Lenisa:2007zz}.  In particular these experiments should determine which of the antiproton-proton spin-dependent cross-sections 
\begin{equation}
\label{eq:Delta_sigma_L_and_T}
\Delta \sigma^{\,\bar{p}\,p}_L \ =\  \sigma^{\,\bar{p}\,p}_{\rightleftarrows} \ -\  \sigma^{\,\bar{p}\,p}_{\rightrightarrows} 
\hspace*{2em} \mbox{and} \hspace*{2em} 
\Delta \sigma^{\,\bar{p}\,p}_T \ =\  \sigma^{\,\bar{p}\,p}_{\uparrow\downarrow} \ -\  \sigma^{\,\bar{p}\,p}_{\uparrow\uparrow} \ ,
\end{equation}
for longitudinal and transverse polarizations respectively is largest, and at what laboratory frame antiproton kinetic energies in the AD range $50 -200 \ \mbox{MeV}$ they are maximal.  This should provide essential information on the optimal parameters of a proposed Antiproton Polarizing Ring (APR) to produce a high intensity polarized antiproton beam at the FAIR facility at GSI Darmstadt.

It is well known that there is a significant spin-dependence of the proton-proton total cross-section.  This was the basis of the FILTEX experiment, and the resulting polarization buildup of the proton beam was a consequence of this spin-dependence.  At FILTEX laboratory frame kinetic energies of $23 \ \mbox{MeV}$ one has that $\frac{1}{2}\left(\,\sigma^{\,p\,p}_{\uparrow\downarrow} \,-\, \sigma^{\,p\,p}_{\uparrow\uparrow}\,\right) \,=\,122 \ \mbox{mb}$ \cite{Horowitz:1994}.  Any measured polarization buildup of the AD antiproton beam during the proposed spin filtering experiments will provide direct evidence of, and the polarization buildup rate will provide a measure of, a spin-dependence in the antiproton-proton total cross-section.

Measurements of the spin-dependent antiproton-proton cross-sections at the AD ring will also provide the first experimental results to test and distinguish between the current models \cite{Dmitriev:2007ms,Klempt:2002ap,Haidenbauer:1984dz,Mull:1994gz,Haidenbauer:2007zza} of the hadronic antiproton-proton interaction in the non-perturbative regime.

We have pointed out that a pure lepton target has many advantages over an atomic target, but has the disadvantage of lower target densities because of the electromagnetic repulsion of the leptons in a pure lepton target which is less severe in the electrically neutral atomic targets.  Consequently it is now appropriate to outline some possible, albeit far fetched, solutions to this problem: 1) Positronium, an electrically neutral electron-positron bound state could in future be used as a high density polarized pure lepton target.  Being electrically neutral it should allow for similar densities as atomic targets.  2) Similarly muonium, an electrically neutral electron-antimuon bound state could possibly be used as above.  3) A polarized muon target/beam would allow for spin-dependent scattering out of the ring, as discussed in section~\ref{subsec:Antiprotons_scattering_off_stationary_electrons}.  Because muons have approximately 200 times the mass of electrons using a muon target will enhance the polarization transfer cross-sections $K_\mathrm{XX}$ and $K_\mathrm{YY}$ as seen in eq.~(\ref{eq:Pbar_e_spin_transfer_observables}).  4) A polarized tau lepton target or beam would be even more favourable to our needs, because of the very large mass of the tau lepton ($m_\tau \approx 3477\,m_e \approx 17\,m_\mu$).  We stress that these ideas are far fetched, and very far from today's technologies, in particular because of the very low mean lifetimes of the particles discussed.  However they may be practical in future, and in the mean time they might prompt other solutions to the problem of low target areal densities.

%\pagebreak

\chapter{Conclusions}
\label{ch:Conclusions}

\vspace*{5ex}
\begin{minipage}{6cm}
\end{minipage}
\hfill
\begin{minipage}{10cm}
\begin{quote}
\emph{\lq\lq I was born not knowing and have had only a little time to change that here and there.\rq\rq}
\flushright{Richard Feynman}
\end{quote}
\end{minipage}
\vspace{8ex}

There has been much recent research into possible methods of polarizing an antiproton beam, the most promising being spin filtering, the theoretical understanding of which is currently incomplete.  The method of polarization buildup by spin filtering requires many of the beam particles to remain within the beam after repeated interaction with an internal target in a storage ring.   Hence small scattering angles, where it is shown that electromagnetic effects dominate hadronic effects, are important.  

All spin-averaged and spin-dependent cross-sections and spin observables for elastic spin 1/2 - spin 1/2 scattering, for both point-like particles and non-point-like particles with internal structure defined by electromagnetic form factors, have been presented to first order in QED.  Particular attention is paid to spin transfer and depolarization cross-sections in antiproton-proton, antiproton-electron and positron-electron scattering, in the low $|\,t\,|$ region of momentum transfer.  Of the spin-averaged formula derived we highlight that a generalization of the Rosenbluth formula has been presented in a new compact Lorentz invariant form.  It is a two-fold generalization in that the masses of both particles are included and both particles are taken to have internal structure determined by electromagnetic form factors.  While these results are eventually applied to spin filtering later in the thesis they are not limited to this application.  The complete set of spin 1/2 - spin 1/2 helicity amplitudes and spin observables should prove useful to many other areas in particle physics.  

The complete set of spin 0 - spin 1 electromagnetic helicity amplitudes have also been presented to first order in QED.  These are useful in describing the spin-dependent scattering of deuterons off carbon nuclei for example.

A thorough mathematical treatment of spin filtering has also been presented, identifying the two key physical processes involved: (a) selective scattering out of the ring and (b) selective spin flip while remaining in the ring.  The dynamical properties of the physical system have also been highlighted and analyzed.  Sets of differential equations which describe the buildup of polarization by spin filtering have been presented and solved in many different scenarios of interest.  These scenarios are: 1) spin filtering of a stored beam, 2) spin filtering while the beam is being accumulated, {\it i.e.}\ unpolarized particles are continuously being fed into the beam at a constant rate, 3) unpolarized particles are continuously being fed into the beam at a linearly increasing rate, {\it i.e.}\ the particle input rate is ramped up, 4) the input rate is equal to the rate at which particles are being lost due to scattering beyond the ring acceptance angle, the beam intensity remaining constant, 5) increasing the initial polarization of a stored beam by spin filtering, 6) the input of particles into the beam is stopped after a certain amount of time, but spin filtering continues.

The depolarization of a polarized beam on interaction with an unpolarized target or beam, as in the important case of electron cooling, has also been investigated and shown to be negligible.

There are advantages of using a lepton target instead of an atomic gas target for spin filtering, principal amongst them that antiprotons will not annihilate with the target as they do with the protons in the atomic targets, leading to a loss of beam intensity.  Since electrons in an atomic target are not massive enough to scatter antiprotons beyond the acceptance angle of any storage ring we have proposed using an opposing polarized electron beam, of momentum large enough to provide scattering of antiprotons beyond ring acceptance, as a possible method to polarize antiprotons by spin filtering.  This is presented as a practical application of the theoretical work presented throughout the thesis.  The areal density of the polarized electron beam is identified as the key parameter limiting the rate of antiproton polarization buildup in this proposal.  After analyzing this proposal it is concluded that the areal densities of electron beams currently available would have to increase significantly in order for this method of polarizing an antiproton beam to be practical. 

While this thesis is devoted to investigating spin-dependent antiproton interactions and the theoretical background to spin filtering in light of a possible method to polarize antiprotons, we emphasize that much of the work presented is applicable to many other areas of particle physics.  

Possible extensions to this work would include redoing the analysis for the case of antiprotons repeatedly interacting with a polarized deuterium target.  Polarized deuterium targets have successfully been utilized by the HERMES experiment, and would be available for spin filtering studies in future.  In the kinematical regime of interest in spin filtering antiproton-deuterium scattering consists of antiproton-electron and antiproton-deuteron scattering.  The helicity amplitudes and spin observables for electromagnetic antiproton-electron scattering have been presented in this thesis, and those for antiproton (spin 1/2) - deuteron (spin 1) scattering could be derived in analogy to those for spin 1/2 - spin 1/2 and spin 0 - spin 1 scattering presented here. 

We hope that the treatment of spin filtering presented in this thesis will clarify some of the confusion in the theoretical literature, and perhaps play some part in the eventual achievement of a high intensity polarized antiproton beam.  Measurements obtained using such a beam should lead to a better understanding of the spin structure of the protons and neutrons.  Given that the proton is the nucleus of the hydrogen atom, the most abundant element in the Universe, one cannot overstate the importance of a better understanding of its internal structure.

\pagebreak

\appendix

% Replace the line referring to Appendices in the .toc file to \vskip 7pt\noindent {\bf{\Large Appendices}} immediately before I compile my thesis the final time.  This removes the page number from word ``Appendices'' in TOC. 

%\appendixpage
%\addappheadtotoc
%\noappendicestocpagenum

\addtocontents{toc}{\vspace*{3ex}}
\addcontentsline{toc}{chapter}{{\large Appendices}}

\chapter{Dirac algebra}
\label{Appendix:Dirac_algebra}

\renewcommand{\theequation}{\thechapter .\arabic{equation}}
%The above command labels equations (A.1) instead of (A.0.1) could use - instead of . to give (A-1), but I think (A.1) is better.

The $4 \times 4$ Dirac gamma matrices, in the Dirac-Pauli representation, are
\begin{eqnarray}
\label{eq:Dirac_Gamma_Matrices}
 \gamma^0 \ = \ 
 \left[ \begin{array}{cc}
	{\bf I} & {\bf 0}\\ {\bf 0} & -{\bf I}
 \end{array} \right]
 \hspace*{1.5em}
  \gamma^5 \ = \ 
 \left[ \begin{array}{cc}
	{\bf 0} & {\bf I}\\ {\bf I} & {\bf 0}
 \end{array} \right]
 \hspace*{1.5em}
  \gamma^j \ = \ 
 \left[\begin{array}{cc}
	{\bf 0} & {\bf \sigma}_j\\ -{\bf \sigma}_j & {\bf 0}
 \end{array}\right] \,,
\end{eqnarray}
for  $j \in \left\{1,2,3\right\}$, where 
\begin{eqnarray}
\label{eq:Pauli_spin_matrices}
 {\bf \sigma}_1 \ = \ 
 \left[\begin{array}{cc}
	0 & 1\\ 1 & 0
 \end{array}\right]
 \hspace*{1.5em}
  {\bf \sigma}_2 \ = \ 
 \left[\begin{array}{cc}
	0 & -i\\ i & 0
 \end{array}\right]
  \hspace*{1.5em}
  {\bf \sigma}_3 \ = \ 
 \left[\begin{array}{cc}
	1 & 0\\ 0 & -1
 \end{array}\right] \,,
\end{eqnarray}
\begin{eqnarray*}
  {\bf I} \ = \ 
 \left[\begin{array}{cc}
	1 & 0\\ 0 & 1
 \end{array}\right]
 \hspace*{2em}
   {\bf 0} \ = \ 
 \left[\begin{array}{cc}
	0 & 0\\ 0 & 0
 \end{array}\right] \,,\nonumber
\end{eqnarray*}
are the $2 \times 2$ Pauli spin matrices, with $i \,=\,\sqrt{-1}$, and the $2 \times 2$ identity and zero matrices respectively.  Feynman slash notation $\slashed{p} \,=\,\gamma^\mu\,p_\mu$, and the usual convention of Greek characters representing four dimensional space-time indices $\left\{0,1,2,3\right\}$ and Latin characters representing three dimensional space indices $\left\{1,2,3\right\}$ are used throughout the thesis.  The {\bf Dirac equation} is 
\begin{equation}
\label{eq:Dirac_equation}
\left(\, i \,\gamma^{\,\mu} \, \partial_\mu \,-\, m \,\right)\psi(x) \ =\ 0\,,
\end{equation}
where the plane wave solutions are
\begin{equation}
\label{eq:Solutions_to_Dirac_equation}
\psi(x) \ =\ u(p)\,e^{-ip \cdot x}
\hspace*{2em} \hbox{and} \hspace*{2em}
\psi(x) \ =\ v(k)\,e^{+ik \cdot x}\,,
\end{equation}
for particles and antiparticles respectively, where $u$ are the spinors for particles and $v$ are the anti-spinors for antiparticles.  Applying the operator $-\,\left(\, i \gamma^\mu \partial_\mu + m \,\right)$ to the left of both sides of the Dirac equation yields the {\bf Klein-Gordon equation}:
\begin{equation}
\label{eq:Klein_Gordon_equation}
\left(\,\partial^{\,\mu} \,\partial_\mu \,+\, m^2\,\right)\psi(x) \ = \ 0\,.
\end{equation}
The hermitian conjugates of spinors $u = u(\,\vec{p}, \lambda\,)$ are as follows
\begin{eqnarray}
\label{eq:Spinor_algebra_relations}
u^\dagger & = & \bar{u}\,\gamma^0 \nonumber \\[1ex]
\left(\gamma^0\right)^\dagger & = & \gamma^0 \nonumber\\[1ex]
\gamma_5^\dagger & = & \gamma_5 \nonumber\\[1ex]
\bar{u}^\dagger & = & \left(\gamma^0\right)^\dagger  u \ =\  \gamma^0 \,u \nonumber\\[1ex]
\gamma_\mu^\dagger & = & \gamma^0 \,\gamma_\mu \,\gamma^0 \\[1ex]
\gamma^0 & = & \gamma_0 \nonumber\\[1ex]
\left( \gamma_0 \right)^2 & = &  \gamma^0 \,\gamma_0 \ =\  {\bf I}_{\,4 \times 4} \,.\nonumber
\end{eqnarray}
The completeness relations for the sum of spinors are
\begin{eqnarray}
\label{eq:Completeness_relations}
 \sum_{\lambda} u(p, \lambda)\, \bar{u}(p, \lambda) & = & \slashed{p} \,+\, m \,,\\[2ex]
\label{eq:Completeness_relations1}
 \sum_{\lambda'} v(k, \lambda')\, \bar{v}(k, \lambda') & = & \slashed{k} \,-\, m \,,\\[2ex]
\label{eq:Completeness_relations2}
 u(p, \lambda)\, \bar{u}(p, \lambda) & = & \frac{1}{2}\left(\slashed{p} \,+\, m \right)\left[1\,+\,\gamma_5\,\slashed{S}(p,\lambda)\right]  \,,\\[2ex]
\label{eq:Completeness_relations3}
  v(k, \lambda')\, \bar{v}(k, \lambda') & = & \frac{1}{2}\left(\slashed{k} \,-\, m \right)\left[1\,+\,\gamma_5\,\slashed{S}(k,\lambda')\right] \,,
\end{eqnarray}
where the polarization four-vector $S^\mu(p,\lambda)$ of a particle is orthogonal to its momentum four vector, {\it i.e.}\ $S^\mu(p,\lambda) \,p_\mu \,=\, 0$, and is normalized such that $S^\mu S_\mu \,=\, - 1$.  The spinors satisfy the Dirac equation as follows:
\begin{eqnarray}
\label{eq:Spinor_Dirac_equation}
\left(\, \slashed{p} - m \,\right)u(p) & = & 0 \,,\nonumber\\[1ex]
\left(\, \slashed{k} + m \,\right)v(k) & = & 0 \,,
\end{eqnarray}
and conjugating these equations yields
\begin{eqnarray}
\label{eq:Conjugate_spinor_Dirac_equation}
\bar{u}(p)\left(\, \slashed{p} - m \,\right) & = & 0 \,,\nonumber\\[1ex]
\bar{v}(k)\left(\, \slashed{k} + m \,\right) & = & 0 \,.
\end{eqnarray}
Gamma matrix calculations can be simplified using: 
\begin{eqnarray}
\label{eq:Gamma_matrix_products}
\gamma^\mu \,\gamma_\mu & = & 4 \,,\nonumber \\[2ex]
\gamma^\mu \,\gamma^\nu \,\gamma_\mu & = & -2\,\gamma^\nu \,,\\[2ex]
\gamma^\mu \,\gamma^\nu \,\gamma^\rho \,\gamma_\mu & = & 4\,\eta^{\,\nu \rho} \,,\nonumber \\[2ex]
\gamma^\mu \,\gamma^\nu \,\gamma^\rho \,\gamma^\sigma \,\gamma_\mu & = & -\,2\,\gamma^\sigma \,\gamma^\rho \,\gamma^\nu \nonumber\,.
\end{eqnarray}
Traces of products of gamma matrices are as follows
\begin{eqnarray}
\label{eq:Gamma_matrix_trace_theorems}
\mbox{Tr}\left[\,\mbox{odd} \ \# \ \mbox{of}\  \gamma \ \mbox{matrices}\,\right] & = & 0 \,,\nonumber \\[2ex]
\mbox{Tr}\left[\,\gamma^\mu \,\gamma^\nu\,\right] & = & 4 \,\eta^{\mu\,\nu} \,,\nonumber \\[2ex]
\mbox{Tr}\left[\,\gamma^\mu \,\gamma^\nu \,\gamma^\rho \,\gamma^\sigma\,\right] & = & 4\,\left( \,\eta^{\mu\,\nu}\,\eta^{\rho\,\sigma} \,-\, \eta^{\mu\,\rho}\,\eta^{\nu\,\sigma}\, + \,\eta^{\mu\,\sigma}\,\eta^{\nu\,\rho} \,\right) \,,\nonumber\\[2ex]
\mbox{Tr}\left[\,\gamma^5\,\right] \ = \ \mbox{Tr}\left[\,\gamma^5\,\gamma^\mu\,\right] & = & \mbox{Tr}\left[\,\gamma^\mu \,\gamma^\nu\,\gamma^5\,\right] \ = \ 0 \,,\\[2ex]
\mbox{Tr}\left[\,\gamma^\mu \,\gamma^\nu \,\gamma^\rho \,\gamma^\sigma\,\gamma^5\,\right] & = & -\,4\,i\,\epsilon^{\,\mu\,\nu\,\rho\,\sigma} \nonumber\,.
\end{eqnarray}
% See wikipedia or any standard text book for a proof of each of these.  Know these proofs for my Viva.
%
The totally antisymmetric permutation tensor, also known as the Levi-Civita symbol, is defined as
\begin{equation}
\epsilon^{\,\mu \,\nu \,\rho \,\sigma} \ = \ \left\{ 
\begin{array}{c} 
\,+\,1 \ \ \ \ \ \mbox{if} \ \mu \,\nu \,\rho \,\sigma \ \mbox{is an even permutation of} \ 0123 \\[1ex]
-\,1 \ \ \ \ \ \mbox{if} \ \mu \,\nu \,\rho \,\sigma \ \mbox{is an odd permutation of}  \ 0123 \\[1ex]
\hspace*{-12.3em} 0 \hspace*{1.7em} \mbox{otherwise}
\end{array} \right.
\end{equation}
%
% This definition agrees with the Mathematica version, but is minus what we said in our first Prague paper.  But any of the calculation in the thesis come from Mathematica where the above version is used, so stick to that.
%
and it satisfies the following contraction identities:
\begin{eqnarray}
\epsilon^{\,\mu \,\nu \,\rho \,\sigma}\, \epsilon_{\,\mu \,\omega \,\lambda \,\tau} & = & \delta^{\,\nu}_{\,\tau} \,\delta^{\,\rho}_{\,\lambda} \,\delta^{\,\sigma}_{\,\omega} + \delta^{\,\nu}_{\,\omega} \,\delta^{\,\rho}_{\,\tau} \,\delta^{\,\sigma}_{\,\lambda} + \delta^{\,\nu}_{\,\lambda} \,\delta^{\,\rho}_{\,\omega} \,\delta^{\,\sigma}_{\,\tau} - \delta^{\,\nu}_{\,\omega} \,\delta^{\,\rho}_{\,\lambda} \,\delta^{\,\sigma}_{\,\tau} - \delta^{\,\nu}_{\,\tau} \,\delta^{\,\rho}_{\,\omega} \,\delta^{\,\sigma}_{\,\lambda} - \delta^{\,\nu}_{\,\lambda} \,\delta^{\,\rho}_{\,\tau}\, \delta^{\,\sigma}_{\,\omega} \,,\nonumber \\[2ex]
\epsilon^{\,\mu \,\nu \,\rho \,\sigma}\, \epsilon_{\,\mu \,\nu \,\lambda \,\tau} & = & -\,2\,(\,\delta^{\,\rho}_{\,\lambda} \,\delta^{\,\sigma}_{\,\tau} \,-\, \delta^{\,\rho}_{\,\tau} \,\delta^{\,\sigma}_{\,\lambda}\,) \,,\nonumber \\[2ex]
\epsilon^{\,\mu \,\nu \,\rho \,\sigma} \,\epsilon_{\,\mu \,\nu \,\rho \,\tau}    & = & -\,6\,\delta^{\,\sigma}_{\,\tau} \,,\\[2ex]
\epsilon^{\,\mu \,\nu \,\rho \,\sigma} \,\epsilon_{\,\mu \,\nu \,\rho \,\sigma}   & = & -\,24  \,,\nonumber 
\end{eqnarray}
%
% I have checked these in a calculation on 9-1-08, and they all agree with Peskin and Schroeder. Checked all of this once and for all, they are all correct now.
%
where the Kronecker delta is defined, not using the Einstein summation convention, as
\begin{equation}
\delta^{\,\mu}_{\,\nu} \ = \ \left\{ \begin{array}{c} 1 \ \ \ \ \mbox{if}\  \mu \,=\,\nu \\[2ex]
0 \ \ \ \ \mbox{if} \ \mu \,\neq \,\nu  \end{array} \right.
\end{equation}
and when using the Einstein summation convention is used to sum over repeated indices $\mu \in \left\{\,0,1,2,3\,\right\}$ one has $\delta^{\,\mu}_{\,\mu} \,=\, \delta^0_0 + \delta^1_1 +\delta^2_2 +\delta^3_3 \,=\, 4$.

\pagebreak

\chapter{Relations between Mandelstam variables}
\label{Appendix:Relations_between_Mandelstam_variables}

\begin{figure}[!h]
\centering
\includegraphics[width=5cm]{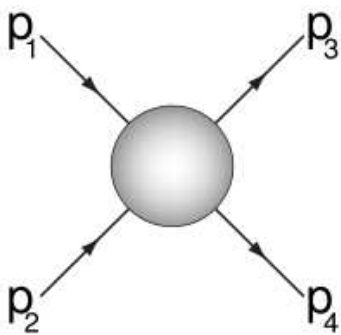}
\caption{\small{\it{The Mandelstam variables are defined from the general two particle to two particles scattering process, where $p_1$ and $p_2$ are the momenta of the incoming particles and $p_3$ and $p_4$ are the momenta of the outgoing particles.}}}
\label{fig:Mandelstam_Variables}
\end{figure}
\noindent
The Mandelstam variables \cite{Mandelstam:1958xc} are defined as follows
\begin{eqnarray}
\label{eq:Mandelstam_Variables_Definition}
s & = & (\,p_1 + p_2\,)^{\,2} \ =\  (\,p_3 + p_4\,)^{\,2} \,,\nonumber \\[1ex]
t & = & (\,p_4 - p_2\,)^{\,2} \ =\  (\,p_1 - p_3\,)^{\,2} \,,\\[1ex]
u & = & (\,p_3 - p_2\,)^{\,2} \ =\  (\,p_1 - p_4\,)^{\,2} \,,\nonumber
\end{eqnarray}
where we have used the law of conservation of four momentum
\begin{equation}
\label{eq:Conservation_of_four_momentum}
p_1 \,+\, p_2 \ = \ p_3 \,+\, p_4\,.
\end{equation}
Initial and final state particles, which in elastic scattering are the same, are on-shell thus $p_1 \cdot p_1 \,=\, p_3 \cdot p_3 \,=\, M^{\,2}$ and $p_2 \cdot p_2 \,=\, p_4 \cdot p_4 \,=\, m^2$, where $M$ and $m$ are the masses of the two particles in the elastic process.  One can square the above relations to obtain
\begin{eqnarray}
s & = & m^2 + M^{\,2} + 2\,p_1 \cdot p_2 \ =\ m^2 + M^{\,2} + 2\,p_3 \cdot p_4  \,,\nonumber\\[1ex]
t & = & 2\,m^2 - 2\,p_2 \cdot p_4   \ =\ 2\,M^{\,2} - 2\,p_1 \cdot p_3 \,,\\[1ex]
u & = & m^2 + M^{\,2} - 2\,p_2 \cdot p_3\ =\ m^2 + M^{\,2} - 2\,p_1 \cdot p_4\,,\nonumber
\end{eqnarray}
hence one sees that
\begin{equation}
p_1 \cdot p_2 \ = \ p_3 \cdot p_4
\hspace*{2em} \hbox{and} \hspace*{2em}
p_1 \cdot p_4 \ = \ p_2 \cdot p_3 \,,
\end{equation}
and adding, using conservation of four momentum, gives the defining relation for Mandelstam variables
\begin{equation}
\label{eq:Elastic_sum_of_Mandelstam_variables}
 s \,+\, t \,+\, u \ = \ 2\,m^2 \,+\, 2\,M^{\,2} \,.
\end{equation}
The above relation is the special case for elastic scattering of the general relation
\begin{equation}
\label{eq:Sum_of_Mandelstam_variables}
 s \,+\, t \,+\, u \ =\ \sum_{i=1}^4 m_i^2 \,.
\end{equation}
Rearranging the $t$ equation gives
\begin{equation}
p_1 \cdot p_3 \ = \ \left(\,M^{\,2} \,-\, \frac{t}{2}\,\right)
\hspace*{2em} \hbox{and} \hspace*{2em}
p_2 \cdot p_4 \ = \ \left(\,m^2 \,-\, \frac{t}{2}\,\right)\,.
\end{equation}
Similarly for $s$ and $u$
\begin{eqnarray}
p_1 \cdot p_4 & = & p_2  \cdot p_3 \ = \  \frac{1}{2} \left(\,m^2 + M^{\,2} - u \,\right) \,,\\[1ex]
p_1 \cdot p_2 & = & p_3  \cdot p_4 \ = \ -\frac{1}{2} \left(\,m^2 + M^{\,2} - s \,\right)\,.
\end{eqnarray} 
For convenience define
\begin{equation*}
R^{\,\mu}  \ = \ p_1^{\,\mu} + p_3^{\,\mu}
\hspace*{2em} \hbox{and} \hspace*{2em}
r^{\,\nu}  \ = \ p_2^{\,\nu} + p_4^{\,\nu} \ ,
\end{equation*}
thus we have
\begin{eqnarray}
R^{\,2} & = & R^{\,\mu} R_{\,\mu} \ = \ p_1^{\,2} + 2\,p_1 \cdot p_3 + p_3^{\,2} \ = \ 2\,M^{\,2} + 2\,M^{\,2} - t \ = \ 4\,M^{\,2} - t \,,\nonumber\\[1ex]
r^2 & = & r^\nu r_\nu \ = \ \cdot \cdot \cdot \cdot \ = \ \left(\,4\,m^2 - t \,\right)\,.
\end{eqnarray}
Now
\begin{eqnarray}
R \cdot r & =  & R^{\,\mu} \,r_\mu \ = \ \left(\,p_1^{\,\mu} + p_3^{\,\mu} \,\right)\left(\,{p_2}_{\,\mu} + {p_4}_{\,\mu} \,\right) \ = \  p_1 \cdot p_2 + p_1 \cdot p_4 + p_2 \cdot p_3 + p_3 \cdot p_4 \,,\nonumber\\[2ex]
& = &  2 \left[-\frac{1}{2} \left(\, m^2 + M^{\,2} -s \,\right) \right] + 2 \left[\frac{1}{2} \left( \,m^2\, + M^{\,2} - u \,\right) \right] \ = \  s-u\,,
\end{eqnarray}
similarly
\begin{equation}
\label{eq:r_dot_p}
r \cdot p_1 \ = \ r \cdot p_3 \ = \ R \cdot p_2 \ = \ R \cdot p_4 \ = \ \frac{1}{2} \left(\, s-u \,\right)\,.
\end{equation}

\chapter{Derivation of the Gordon decomposition identities}
\label{Appendix:Derivation_of_the_Gordon_decomposition_identities}

The Dirac gamma matrix structure of the most general proton electromagnetic current can greatly be simplified using the Gordon Decomposition Identity \cite{Gordon:1928,Bjorken:1964}, which we now derive.

The commutation and anticommutation relations
\begin{equation}
\label{eq:Gamma_matrix_commutation_relations}
\frac{i}{2}\left[\,\gamma^\mu,\gamma^\nu \,\right] \ =\ \sigma^{\mu\nu}  \hspace*{1cm} \mbox{and} \hspace*{1cm} \left\{\,\gamma^\mu,\gamma^\nu \,\right\} \ =\  2\,\eta^{\mu\nu}\,,
\end{equation}
where 
\begin{equation}
\label{eq:Definition_of_commutation_and_anticommutation_relations}
\left[\,\gamma^\mu,\gamma^\nu \,\right] \ =\ \gamma^\mu \,\gamma^\nu \,-\, \gamma^\nu \gamma^\mu 
\hspace*{1cm} \mbox{and} \hspace*{1cm} 
\left\{\,\gamma^\mu,\gamma^\nu \,\right\} \ =\  \gamma^\mu \,\gamma^\nu \,+\, \gamma^\nu \gamma^\mu \,,
\end{equation}
and $\eta^{\mu\nu} \,=\,\mbox{diag}(+1,-1,-1,-1)$ the Minkowski metric tensor, can be used to write
\begin{equation*}
\left[\,\gamma^{\,\mu},\gamma^{\,\nu} \,\right] \,=\, \gamma^{\,\mu} \,\gamma^{\,\nu} \,-\, \gamma^{\,\nu} \gamma^{\,\mu} \ =\  \gamma^{\,\mu} \,\gamma^{\,\nu} \,-\, \left( 2\,\eta^{\,\mu\,\nu} \,-\, \gamma^{\,\mu} \,\gamma^{\,\nu} \right) \ =\  2\,\gamma^{\,\mu} \,\gamma^{\,\nu} \,-\, 2\,\eta^{\,\mu\,\nu} \,.
\end{equation*}
Hence
\begin{equation*}
i\,\sigma^{\,\mu\,\nu} \ =\  - \left(\,\gamma^{\,\mu} \,\gamma^{\,\nu} \,-\, \eta^{\,\mu\,\nu}\,\right) \ =\   \eta^{\,\mu\,\nu} \,-\, \gamma^{\,\mu} \,\gamma^{\,\nu}\,,
\end{equation*}
but equivalently
\begin{equation*}
i\,\sigma^{\,\mu\,\nu} \ =\   \eta^{\,\mu\,\nu} \,-\, \left(2\,\eta^{\,\mu\,\nu} \,-\, \gamma^{\,\nu} \,\gamma^{\,\mu}\,\right) \ =\  \gamma^{\,\nu} \,\gamma^{\,\mu} \,-\, \eta^{\,\mu\,\nu}\,.
\end{equation*}
Using the above relations one can compute
\begin{eqnarray}
\label{eq:Sigma_matrix_manipulation}
\bar{u}(p')\,i\,\sigma^{\,\mu\,\nu}\left(\,p'_{\,\nu} \,-\, p_{\,\nu}\,\right)\,u(p) & = & 
\bar{u}(p')\,\left[\,i\,\sigma^{\,\mu\,\nu}\,p'_{\,\nu} \ -\  i\,\sigma^{\,\mu\,\nu}\,p_{\,\nu}\,\right] u(p)\,, \nonumber \\[1ex]
& = & \bar{u}(p')\left[\left( \gamma^{\,\nu} \,\gamma^{\,\mu} \,-\, \eta^{\,\mu\,\nu}\right)p'_{\,\nu} \ -\  \left(\eta^{\,\mu\,\nu} \,-\, \gamma^{\,\mu} \,\gamma^{\,\nu}\,\right)p_{\,\nu} \right]u(p)\,,\nonumber\\[1ex]
& = & \bar{u}(p')\left[\gamma^{\,\nu} p'_{\,\nu}\gamma^{\,\mu} \ - \ p'^{\,\mu} - p^{\,\mu} \ +\  \gamma^{\,\mu} \gamma^{\,\nu} p_{\,\nu} \right]u(p)\,,\nonumber\\[1ex]
& = & \bar{u}(p')\left[ \slashed{p}' \,\gamma^{\,\mu} \ -\  \left(\,p \,+\, p' \,\right)^{\,\mu} \ +\  \gamma^{\,\mu}\, \slashed{p} \right]u(p) \,,
\end{eqnarray}
where the Feynman slash notation $\slashed{p} \,=\,\gamma^\mu\,p_\mu$ has been used.  Now simplify by using the Dirac equation and its conjugate
\begin{eqnarray}
\left(\, \slashed{p} \,-\, M \,\right)u(p) & = & 0 \hspace{0.5cm} \implies \hspace{0.5cm} \slashed{p} \,\, u(p) \ = \  M \,u(p)\,, \nonumber\\[1ex]
\bar{u}(p')\left(\, \slashed{p}' \,-\, M \,\right) & = & 0 \hspace{0.5cm} \implies\hspace{0.5cm}  \bar{u}(p')\,\,\slashed{p}' \ =\  \bar{u}(p')\,M\,,
\end{eqnarray}
where $M$ is the mass of the particle, to obtain
\begin{eqnarray*}
\bar{u}(p')\,i\,\sigma^{\,\mu\,\nu}\left(\,p'{\,_\nu} \,-\, p_{\,\nu}\,\right)\,u(p) & = & \bar{u}(p')\left[ \,M \, \gamma^{\,\mu} \ -\  \left(\,p \,+\, p' \,\right)^{\,\mu} \ +\ \gamma^{\,\mu}\,M\, \right]u(p)\,,\\[1ex]
& = & \bar{u}(p')\left[ \,2\,M \, \gamma^{\,\mu} \ -\  \left(\,p \,+\, p'\,\right)^{\,\mu} \, \right]u(p)\,.
\end{eqnarray*}
Rearranging gives the general form of the {\bf Gordon Decomposition identity}:
\begin{eqnarray}
\label{eq:Gordon_decomposition_identity1}
\bar{u}(p')\,\gamma^{\,\mu}\,u(p) \ = \ \bar{u}(p')\,\left[\frac{\left(\,p \,+\, p' \,\right)^{\,\mu}}{2\,M} \ +\  \frac{i\,\sigma^{\,\mu\,\nu}\left(\,p'\,-\, p\,\right)_\nu}{2\,M}\,\right]u(p)  \,.
\end{eqnarray}
Another Gordon decomposition identity, which we now derive, can be used to simplify the gamma matrix structure of the antiproton current using the $v$ anti-spinors.  Equation~(\ref{eq:Sigma_matrix_manipulation}) for anti-spinors gives
\begin{equation*}
\bar{v}(p')\,i\,\sigma^{\,\mu\,\nu}\left(\,p'_{\,\nu} \,-\, p_{\,\nu}\,\right)\,v(p) \ = \ \bar{v}(p')\left[ \slashed{p}' \,\gamma^{\,\mu} \ -\  \left(\,p \,+\, p' \,\right)^{\,\mu} \ +\  \gamma^{\,\mu}\, \slashed{p} \right]v(p) \,.
\end{equation*}
But now we must use the Dirac equation for antiparticles and its conjugate
\begin{eqnarray}
\left(\, \slashed{p} \,+\, M \,\right)v(p) & = & 0 \hspace{0.5cm} \implies \hspace{0.5cm} \slashed{p} \,\, v(p) \ = \  -\,M \,v(p)\,, \nonumber\\[1ex]
\bar{v}(p')\left(\, \slashed{p}' \,+\, M \,\right) & = & 0 \hspace{0.5cm} \implies\hspace{0.5cm}  \bar{v}(p')\,\,\slashed{p}' \ =\  \bar{v}(p')\,\left(\,-\,M\,\right) \,,
\end{eqnarray}
on the above equation to obtain
\begin{eqnarray*}
\bar{v}(p')\,i\,\sigma^{\,\mu\,\nu}\left(\,p'{\,_\nu} \,-\, p_{\,\nu}\,\right)\,v(p) & = & \bar{v}(p')\left[ \,-\,M \, \gamma^{\,\mu} \ -\  \left(\,p \,+\, p' \,\right)^{\,\mu} \ +\ \gamma^{\,\mu}\,\left(\,-\,M\,\right)\, \right]v(p)\,,\\[1ex]
& = & \bar{v}(p')\left[ \,-\,2\,M \, \gamma^{\,\mu} \ -\  \left(\,p \,+\, p'\,\right)^{\,\mu} \, \right]v(p)\,.
\end{eqnarray*}
Now rearranging gives the {\bf Gordon Decomposition identity for anti-spinors}:
\begin{eqnarray}
\label{eq:Gordon_decomposition_identity_for_antiparticles}
\bar{v}(p')\,\gamma^{\,\mu}\,v(p) \ = \ \bar{v}(p')\,\left[\,-\,\frac{\left(\,p \,+\, p' \,\right)^{\,\mu}}{2\,M} \ -\  \frac{i\,\sigma^{\,\mu\,\nu}\left(\,p'\,-\, p\,\right)_\nu}{2\,M}\,\right]v(p)  \,.
\end{eqnarray}

\chapter{Feynman rules for QED}
\label{Appendix:Feynman_rules_for_QED}

The Lagrangian for Quantum Electrodynamics (QED) is
\begin{equation}
\mathcal{L} \ = \ \bar{\psi} \left(\,i\,\slashed{\partial}\,-\,m\,\right) \psi \ - \ \frac{1}{4}\,\left( F_{\mu\,\nu}\right)^{\,2} \ - \ e\,\bar{\psi}\,\gamma^{\,\mu}\,\psi \,A_{\,\mu}\,.
\end{equation}
Momentum is conserved at each vertex, fermion loops receive an additional factor of $(-1)$ and undetermined loop momenta are integrated over by: 
\begin{equation}
\int  \frac{\mathrm{d}^{\,4}\,p}{(2\,\pi)^{\,4}} \,.
\end{equation}
We work in the Feynman gauge where the gauge parameter is set to $\xi \,=\,1$.

\begin{table}[!h]
\begin{tabular}{rcl}
Fermion propagator: & \hspace*{3em} \ \includegraphics[width=2.6cm]{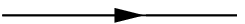} \hspace*{3em}  &  $=\ \displaystyle{\frac{i \left(\,\slashed{p} \,+\,m\,\right)}{p^{\,2} \,-\,m^2 \,+\,i\,\epsilon}}$\\[4ex]
Photon propagator: & \ \includegraphics[width=2.6cm]{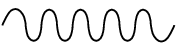} &  $=\ \displaystyle{\frac{-\,i\,\eta_{\,\mu\,\nu}}{p^{\,2} \,+\,i\,\epsilon}}$\\[4ex]
QED vertex: & 
\begin{minipage}{6em} \centering \includegraphics{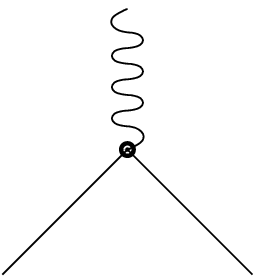} \end{minipage}
 &  $=\ i\,Q\,e\,\gamma^{\,\mu}$ \\[10ex]
External fermions: & 
\begin{minipage}{6em} \centering \includegraphics{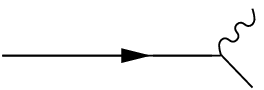} \end{minipage}  &  $=\ u(p,\,\lambda)$ \ \ \ (initial) \\[3ex]
                   & \begin{minipage}{6em} \centering \includegraphics{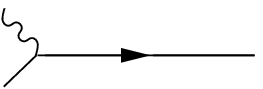} \end{minipage} & $=\ \bar{u}(p,\,\lambda)$ \ \ \ (final) \\[3ex]
External antifermions: &  \begin{minipage}{6em} \centering \includegraphics{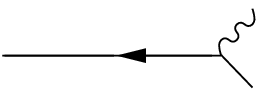} \end{minipage} & $=\ \bar{v}(p,\,\lambda)$ \ \ \ (initial) \\[3ex]
                 &  \begin{minipage}{6em} \centering \includegraphics{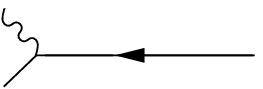} \end{minipage} &  $=\ v(p,\,\lambda)$ \ \ \ (final) \\[3ex]
External photons: &  \begin{minipage}{6em} \centering \includegraphics[width=2.6cm]{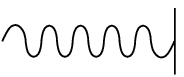} \end{minipage} &   $=\ \epsilon_{\,\mu}(p)$ \ \ \ \ \ (initial)\\[3ex]
                & \begin{minipage}{6em} \centering \includegraphics[width=2.6cm]{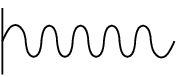} \end{minipage} &  $=\ \epsilon^*_{\,\mu}(p)$ \ \ \ \ \ (final)
\end{tabular}
\label{tab:Feynman_Rules}
\end{table}
The charge factor $Q\,=\,-1$ for an electron or antiproton, and $Q\,=\,+1$ for a positron or proton.  Time increases from left to right in all Feynman diagrams throughout the thesis.

\chapter{Sample Mathematica code}
\label{Appendix:Sample_Mathematica_code}

The {\tt Mathematica} code to derive the generic depolarization equation from eq.~(\ref{eq:Generic_calculation_with_epsilons}) is included in this appendix.  The {\tt Mathematica} add on package for High Energy Physics {\tt Tracer.m} \cite{Jamin:1991dp} is used, which defines {\tt G5}$\ \equiv \gamma^5$ and uses {\tt Spur[]} for Trace.  All other QED calculations in the thesis are done analogously.  Only input commands are included.  All parameters are as defined and used throughout the thesis.
\\
\\
{\tt
(* Import the file Tracer.m *)\\[2ex]
<< Tracer/Tracer.m;\\
VectorDimension[4];\\
Spur[1];\\[2ex]
(* Spin projection operators *)\\[2ex]
SP1 = G[le, U] + G[le, G5]**G[le, S1];\\
SP3 = G[le, U] + G[le, G5]**G[le, S3];\\[2ex]
(* Spin trace *)\\[2ex]
SpinTrace = G[le, p1 + M U]**SP1**(H G[le, \{nu\}] + F (p1.\{nu\}\\
 + p3.\{nu\}))**G[le, p3 + M U]**SP3**(H G[le, \{mu\}]\\
 + F (p1.\{mu\} + p3.\{mu\}));\\
SpinTrace2 = FullSimplify[SpinTrace/.le -> 1];\\[2ex]
(* Spinless trace *)\\[2ex]
SpinlessTrace = G[le2, p4 + m U]**(h G[le2, \{mu\}] + f (p2.\{mu\}\\
 + p4.\{mu\}))**G[le2, p2 + m U]**(h G[le2,\{nu\}] + f (p2.\{nu\} + p4.\{nu\}));\\
NoSpur[1];\\
Spur[2];\\
SpinlessTrace2 = FullSimplify[SpinlessTrace/.le2 -> 2];\\[2ex]
(* Contract all Lorentz indices *)\\[2ex]
msq1 = Simplify[ContractEpsGamma[SpinTrace2 SpinlessTrace2]];\\
msq2 = FullSimplify[msq1];\\[2ex]
(* Take the depolarization observable away from the spin-averaged *)\\[2ex]
SpinAveraged = Simplify[msq2/.\{S1 -> 0, S3 -> 0\}];\\
SpinTerm1 = SpinAveraged - msq2;\\[2ex]
(* Now make algebraic simplifications *)\\[2ex]
SpinTerm2 =  Simplify[SpinTerm1/.\{p2.p4 -> m\^\ \!\!\!2 - t/2, p1.p3 -> M\^\ \!\!\!2\\
- t/2, p4.p4 -> m\^\ \!\!\!2, p3.p3 -> M\^\ \!\!\!2, p2.p2 -> m\^\ \!\!\!2, p1.p1 -> M\^\ \!\!\!2,\\
p3.p4 -> s/2 - M\^\ \!\!\!2/2 - m\^\ \!\!\!2/2, p1.p2 -> s/2 - M\^\ \!\!\!2/2 - m\^\ \!\!\!2/2,\\
 p3.p2 -> \ M\^\ \!\!\!2/2 + m\^\ \!\!\!2/2 - u/2, p1.p4 -> \ M\^\ \!\!\!2/2 + \ m\^\ \!\!\!2/2 - u/2,\\
 S1.p1 -> 0, S3.p3 -> 0\}];\\[2ex]
SpinTerm3 = Simplify[SpinTerm2/.\{p3 - p1  -> q, p2 - p4 -> q,\\
u -> 2m\^\ \!\!\!2 + 2M\^\ \!\!\!2 - s - t\}];\\[2ex]
SpinTerm4 = Simplify[SpinTerm3/.\{p3.S1 -> q.S1, p1.S3 -> - q.S3\}];\\[2ex]
SpinTerm5 =  FullSimplify[SpinTerm4 /.\{F -> - F\_2/2M, f -> - f\_2/2m\}];\\[2ex]
GenericDepolarisationEquation = FullSimplify[SpinTerm5 /.\{H -> F\_1\\
+ F\_2, h -> f\_1 + f\_2, p1.S3 - p3.S3 -> - q.S3, - p1.S3 + p3.S3 ->\\
 q.S3, p2.S1 - p4.S1 -> q.S1\}]\\
}

\noindent
The output is now the generic depolarization equation, for the electromagnetic scattering of two non-identical spin 1/2 particles with internal structure defined by form factors, used in the thesis.  All helicity amplitudes and spin observables for elastic spin 1/2 - spin 1/2 scattering to first order in QED can be obtained from eq.~(\ref{eq:Generic_calculation_with_epsilons}) in a similar fashion.

\chapter{Laplace transform methods}
\label{Appendix:Laplace_transform_methods}

Laplace transform methods are used to solve the differential equations involving the Heaviside step function in section~\ref{sec:Particles_fed_in_for_a_limited_time}.  All Laplace transform results needed are presented here.

The Laplace transform $\mathcal{L}\left\{\,f(\tau)\,\right\} \ = \ F(s)$ of a function $f(\tau)$ is defined as
\begin{equation}
\mathcal{L}\left\{\,f(\tau)\,\right\} \ = \ F(s) \ = \ \int_0^{\,\infty} e^{\,-s\,\tau}\,f(\tau)\,\mathrm{d}\,\tau \,.
\end{equation}
It is a linear operator, which can be proven as follows:

\begin{eqnarray*}
\mathcal{L}\left\{\,c_1\,f_1(\tau) + c_2\,f_2(\tau)\,\right\} & = & \int_0^{\,\infty} e^{\,-s\,\tau}\,\left[\,c_1\,f_1(\tau) \ + \ c_2\,f_2(\tau)\,\right]\,\mathrm{d}\,\tau \,,\\[3ex]
& = & c_1\int_0^{\,\infty} e^{\,-s\,\tau}\,f_1(\tau)\,\mathrm{d}\,\tau \ + \ c_2\int_0^{\,\infty} e^{\,-s\,\tau}\,f_2(\tau)\,\mathrm{d}\,\tau  \,,\\[3ex]
& = & c_1\,\mathcal{L}\left\{\,f_1(\tau)\,\right\} \ + \ c_2\,\mathcal{L}\left\{\,f_2(\tau)\,\right\} \,,
\end{eqnarray*}
where $c_1$ and $c_2$ are constants and $f_1(\tau)$ and $f_2(\tau)$ are arbitrary functions.

Differential equations involving the Heaviside step function can be solved by taking the Laplace transform of the entire equation, applying the relations which follow, solving for $\mathcal{L}\{\,f(\tau)\,\}$ and then taking the inverse Laplace transform of what is left to obtain the solution $f(\tau)$ to the differential equation.

\pagebreak

The Laplace transforms used in solving the differential equations in section~\ref{sec:Particles_fed_in_for_a_limited_time} are:

\begin{eqnarray}
\mathcal{L}\left\{\,f'(\tau)\,\right\} & = & s\,\mathcal{L}\left\{\,f(\tau)\,\right\} \ -\  f(0) \,,\\[2ex]
\mathcal{L}\left\{\,f''(\tau)\,\right\} & = & s^{\,2}\,\mathcal{L}\left\{\,f(\tau)\,\right\} \ -\  s \, f(0) \ -\  f'(0) \,,\\[2ex]
\mathcal{L}\left\{\,H \left(\,\tau \,-\, \tau_c\,\right)\,\right\} & = & \frac{e^{\,-\,\tau_c\,s}}{s} \,,\\[2ex]
\mathcal{L}\left\{\,H\left(\,\tau \,\right)\,\right\} & = & \frac{1}{s} \ = \ \mathcal{L}\left\{\,1\,\right\}\,,
\end{eqnarray}
where $f(\tau)$ is an arbitrary function, $'$ denotes first derivative and $''$ denotes second derivative.  One obtains $f(0)$ and $f'(0)$ from the initial conditions of the differential equation.  The Heaviside step function, $H\left(\,\tau \,-\, \tau_c\,\right)$, is  defined as  
\begin{eqnarray}
\label{eq:Heaviside_function_definition2}
 H\left(\,\tau \,-\, \tau_c\,\right) \ = \ \left\{ 
\begin{array}{ll} 
0  & \hspace*{2em} \mbox{if} \ \tau < \tau_c \\[2ex]  
1 & \hspace*{2em} \mbox{if} \ \tau \geq \tau_c
\end{array}
\right.
\end{eqnarray}
The Inverse Laplace Transforms go from right to left in the above list, for example
\begin{equation}
\mathcal{L}^{-1}\left\{\,\frac{e^{\,-\,\tau_c\,s}}{s}\,\right\} \ = \ H \left(\,\tau - \tau_c\,\right)\,.
\end{equation}
The Laplace inverse is defined, as always, such that 
\begin{equation}
\mathcal{L}\left\{\mathcal{L}^{-1}\left\{f(\tau)\right\}\right\} \ = \ f(\tau) \ = \ \mathcal{L}^{-1}\left\{\mathcal{L}\left\{f(\tau)\right\}\right\} \,.
\end{equation}

\end{document}

%% file: dedication.tex
\vspace*{4.7ex}

\begin{center}
{\Large To my parents, Ann and Dan, for everything.}
\end{center}

%% file: theabstract.tex
\onehalfspacing

{\normalsize

There has been much recent research into possible methods of polarizing an antiproton beam, the most promising being spin filtering, the theoretical understanding of which is currently incomplete.  The method of polarization buildup by spin filtering requires many of the beam particles to remain within the beam after repeated interaction with an internal target in a storage ring.   Hence small scattering angles, where we show that electromagnetic effects dominate hadronic effects, are important.  All spin-averaged and spin-dependent electromagnetic cross-sections and spin observables for elastic spin 1/2 - spin 1/2 scattering, for both point-like particles and non-point-like particles with internal structure defined by electromagnetic form factors, are derived to first order in QED.  Particular attention is paid to spin transfer and depolarization cross-sections in antiproton-proton, antiproton-electron and positron-electron scattering, in the low $|\,t\,|$ region of momentum transfer.  A thorough mathematical treatment of spin filtering is then presented, identifying the key physical processes involved and highlighting the dynamical properties of the physical system.  We present and solve sets of differential equations which describe the buildup of polarization by spin filtering in many different scenarios of interest.  The advantages of using a lepton target are outlined, and finally a proposal to polarize antiprotons by spin filtering off an opposing polarized electron beam is investigated.

}

%% file: summary.tex
Immense efforts, both theoretical and experimental, have been afforded to gaining a better understanding of the spin structure of the nucleon since the startling results from the EMC experiment at CERN in 1988 that the intrinsic valence quarks contribute only a small fraction of the proton's spin.  Yet to this day almost nothing is known about the transversity distribution of quarks in the nucleon, the last remaining leading twist piece of the QCD description of the partonic structure of the nucleon in the collinear limit.  A high intensity polarized antiproton beam would be required to best analyze the transversity distribution function, via Drell-Yan lepton pair production in the scattering of polarized antiprotons off polarized protons.  Unfortunately no high intensity polarized antiproton beam has been achieved to date.

Hence there has been much recent research into possible methods of polarizing an antiproton beam, instigated by the recent proposal of the $\mathcal{PAX}$ (Polarized Antiproton eXperiments) Collaboration at GSI, Darmstadt. The most promising method under consideration is spin filtering, the theoretical understanding of which is currently incomplete.  The method of polarization buildup by spin filtering requires many of the beam particles to remain within the beam after repeated interaction with an internal target in a storage ring.   Hence small scattering angles, where we show that electromagnetic effects dominate hadronic effects, are important.  The theoretical background to this effort is investigated in this thesis.

We derive fully relativistic expressions for all spin-averaged and spin-dependent electromagnetic cross-sections and spin observables for elastic spin 1/2 - spin 1/2 scattering, for both point-like particles and non-point-like particles with internal structure defined by electromagnetic form factors, to first order in QED.  Particular attention is paid to spin transfer and depolarization cross-sections in antiproton-proton, antiproton-electron and positron-electron scattering, in the low $|\,t\,|$ region of momentum transfer.
% of interest in storage rings.  
Of the spin-averaged formula derived we highlight that a generalization of the Rosenbluth formula is presented in a new compact Lorentz invariant form.  It is a two-fold generalization in that the masses of both particles are included and both particles are taken to have internal structure determined by electromagnetic form factors.  While these results are eventually applied to spin filtering later in the thesis they are not limited to this application.  The complete set of spin 1/2 - spin 1/2 helicity amplitudes and spin observables should prove useful to many other areas in particle physics.  

The complete set of spin 0 - spin 1 electromagnetic helicity amplitudes are also presented to first order in QED.  These are useful in describing the spin-dependent scattering of deuterons off carbon nuclei for example. 

A thorough mathematical treatment of spin filtering is then presented, identifying the two key physical processes involved: (a) selective scattering out of the ring and (b) selective spin flip while remaining in the ring.  The dynamical properties of the physical system under investigation are highlighted.  Sets of differential equations are presented and solved which describe the buildup of polarization by spin filtering in many different scenarios of interest.  These scenarios are: 1) spin filtering of a stored beam, 2) spin filtering while the beam is being accumulated, {\it i.e.}\ unpolarized particles are continuously being fed into the beam at a constant rate, 3) unpolarized particles are continuously being fed into the beam at a linearly increasing rate, {\it i.e.}\ the particle input rate is ramped up, 4) the input rate is equal to the rate at which particles are being lost due to scattering beyond the ring acceptance angle, the beam intensity remaining constant, 5) increasing the initial polarization of a stored beam by spin filtering, 6) the input of particles into the beam is stopped after a certain amount of time, but spin filtering continues.

The depolarization of a polarized beam on interaction with an unpolarized target or beam, as in the important case of electron cooling, is also investigated and shown to be negligible.

We show that there are advantages of using a lepton target instead of an atomic gas target for spin filtering, principal amongst them that antiprotons will not annihilate with the target as they do with the protons in the atomic targets, leading to a loss of beam intensity.  After showing that electrons in an atomic target are not massive enough to scatter antiprotons beyond the acceptance angle of any storage ring we propose using an opposing polarized electron beam, of momentum large enough to provide scattering of antiprotons beyond ring acceptance, as a possible method to polarize antiprotons by spin filtering.  This is presented as a practical application of the theoretical work presented throughout the thesis.  The areal density of the polarized electron beam is identified as the key parameter limiting the rate of antiproton polarization buildup in this proposal.

%% file: acknowledgments.tex
%\section*{Acknowledgments}

% Dedication: To my parents, Ann and Dan, for everything.
% Dedication: To my parents, Ann and Dan, for, quite simply, everything.

%\onehalfspacing{

%This thesis would never have been completed without the help and support of a number of people, whom I express my gratitude to here.

Firstly I thank my supervisor Dr.~Nigel Buttimore, for his constant support and encouragement.  His enthusiasm for the field of research made the process much more enjoyable than it otherwise would have been.

I express a special thank you to my parents, for always being there for me, for instilling in me an insatiable thirst for knowledge and for helping me see this thesis through to completion.  It is with great pleasure that I dedicate this thesis to them.  To my brother Dennis, sister Edel and Laura Gibson for constantly lifting my spirits and for always being around when I need them. 
%for understanding my passion for such an abstract field of study.

I thank my office mates in G29 and WR 2.3 for all the good times, and fellow Ph.D.\ students David Henry, Sinead Keegan, Derek Kitson, Stephan Meissner and Richie Morrin for mutual encouragement to keep us all motivated.  I would also like to thank everyone who offered to proof-read this thesis, perhaps they did not know what they were getting themselves into!

I thank the Irish Research Council for Science, Engineering and Technology (IRCSET) for a Postgraduate Scholarship, without which I would never have been able to embark on this research.  The funding provided offered me the opportunity to attend many International conferences and interact with some of the leading scientists in this field, which proved invaluable in progressing my research.  In particular advice from Professor Mauro Anselmino, Professor Elliot Leader, Professor Hans-Otto Meyer, Dr.~Frank Rathmann and Dr.~Werner Vogelsang proved very useful.

I am very grateful to Professor Peter Hogan for all of his encouragement and advice during my Undergraduate studies.  In particular for the many conversations we had, and the insight he provided, when I was deciding on the field of research for my Ph.D.

I wish to thank the International research institutions DESY Zeuthen, ECT* Trento and the University of Crete, where I spent extended periods.  I learned much about particle physics, and the process of physical research, at these institutions.  The beautiful Alpine surroundings at ECT* Trento provided great inspiration, and is where much of this thesis began to take shape.

Thanks to all my team-mates on the Connaught, UCD, TCD and Brookfield tennis teams, for all the laughs and good times on and off the court over the years.

Finally I thank the staff of the School of Mathematics, in particular the administrative staff Helen and Karen for doing a fantastic job and making all of our lives much easier.

%}

%% file: N+N-Ratio_V_Polarization_Plot.tex
% GNUPLOT: LaTeX picture
\setlength{\unitlength}{0.240900pt}
\ifx\plotpoint\undefined\newsavebox{\plotpoint}\fi
\sbox{\plotpoint}{\rule[-0.200pt]{0.400pt}{0.400pt}}%
\begin{picture}(1500,900)(0,0)
\font\gnuplot=cmr10 at 10pt
\gnuplot
\sbox{\plotpoint}{\rule[-0.200pt]{0.400pt}{0.400pt}}%
\put(181.0,123.0){\rule[-0.200pt]{303.052pt}{0.400pt}}
\put(181.0,123.0){\rule[-0.200pt]{4.818pt}{0.400pt}}
\put(161,123){\makebox(0,0)[r]{-1}}
\put(1419.0,123.0){\rule[-0.200pt]{4.818pt}{0.400pt}}
\put(181.0,188.0){\rule[-0.200pt]{303.052pt}{0.400pt}}
\put(181.0,188.0){\rule[-0.200pt]{4.818pt}{0.400pt}}
\put(161,188){\makebox(0,0)[r]{-0.8}}
\put(1419.0,188.0){\rule[-0.200pt]{4.818pt}{0.400pt}}
\put(181.0,254.0){\rule[-0.200pt]{303.052pt}{0.400pt}}
\put(181.0,254.0){\rule[-0.200pt]{4.818pt}{0.400pt}}
\put(161,254){\makebox(0,0)[r]{-0.6}}
\put(1419.0,254.0){\rule[-0.200pt]{4.818pt}{0.400pt}}
\put(181.0,319.0){\rule[-0.200pt]{303.052pt}{0.400pt}}
\put(181.0,319.0){\rule[-0.200pt]{4.818pt}{0.400pt}}
\put(161,319){\makebox(0,0)[r]{-0.4}}
\put(1419.0,319.0){\rule[-0.200pt]{4.818pt}{0.400pt}}
\put(181.0,385.0){\rule[-0.200pt]{303.052pt}{0.400pt}}
\put(181.0,385.0){\rule[-0.200pt]{4.818pt}{0.400pt}}
\put(161,385){\makebox(0,0)[r]{-0.2}}
\put(1419.0,385.0){\rule[-0.200pt]{4.818pt}{0.400pt}}
\put(181.0,450.0){\rule[-0.200pt]{303.052pt}{0.400pt}}
\put(181.0,450.0){\rule[-0.200pt]{4.818pt}{0.400pt}}
\put(161,450){\makebox(0,0)[r]{ 0}}
\put(1419.0,450.0){\rule[-0.200pt]{4.818pt}{0.400pt}}
\put(181.0,515.0){\rule[-0.200pt]{303.052pt}{0.400pt}}
\put(181.0,515.0){\rule[-0.200pt]{4.818pt}{0.400pt}}
\put(161,515){\makebox(0,0)[r]{ 0.2}}
\put(1419.0,515.0){\rule[-0.200pt]{4.818pt}{0.400pt}}
\put(181.0,581.0){\rule[-0.200pt]{303.052pt}{0.400pt}}
\put(181.0,581.0){\rule[-0.200pt]{4.818pt}{0.400pt}}
\put(161,581){\makebox(0,0)[r]{ 0.4}}
\put(1419.0,581.0){\rule[-0.200pt]{4.818pt}{0.400pt}}
\put(181.0,646.0){\rule[-0.200pt]{303.052pt}{0.400pt}}
\put(181.0,646.0){\rule[-0.200pt]{4.818pt}{0.400pt}}
\put(161,646){\makebox(0,0)[r]{ 0.6}}
\put(1419.0,646.0){\rule[-0.200pt]{4.818pt}{0.400pt}}
\put(181.0,712.0){\rule[-0.200pt]{303.052pt}{0.400pt}}
\put(181.0,712.0){\rule[-0.200pt]{4.818pt}{0.400pt}}
\put(161,712){\makebox(0,0)[r]{ 0.8}}
\put(1419.0,712.0){\rule[-0.200pt]{4.818pt}{0.400pt}}
\put(181.0,777.0){\rule[-0.200pt]{303.052pt}{0.400pt}}
\put(181.0,777.0){\rule[-0.200pt]{4.818pt}{0.400pt}}
\put(161,777){\makebox(0,0)[r]{ 1}}
\put(1419.0,777.0){\rule[-0.200pt]{4.818pt}{0.400pt}}
\put(181.0,123.0){\rule[-0.200pt]{0.400pt}{157.549pt}}
\put(181.0,123.0){\rule[-0.200pt]{0.400pt}{4.818pt}}
\put(181,82){\makebox(0,0){ 0.001}}
\put(181.0,757.0){\rule[-0.200pt]{0.400pt}{4.818pt}}
\put(244.0,123.0){\rule[-0.200pt]{0.400pt}{2.409pt}}
\put(244.0,767.0){\rule[-0.200pt]{0.400pt}{2.409pt}}
\put(328.0,123.0){\rule[-0.200pt]{0.400pt}{2.409pt}}
\put(328.0,767.0){\rule[-0.200pt]{0.400pt}{2.409pt}}
\put(370.0,123.0){\rule[-0.200pt]{0.400pt}{2.409pt}}
\put(370.0,767.0){\rule[-0.200pt]{0.400pt}{2.409pt}}
\put(391.0,123.0){\rule[-0.200pt]{0.400pt}{157.549pt}}
\put(391.0,123.0){\rule[-0.200pt]{0.400pt}{4.818pt}}
\put(391,82){\makebox(0,0){ 0.01}}
\put(391.0,757.0){\rule[-0.200pt]{0.400pt}{4.818pt}}
\put(454.0,123.0){\rule[-0.200pt]{0.400pt}{2.409pt}}
\put(454.0,767.0){\rule[-0.200pt]{0.400pt}{2.409pt}}
\put(537.0,123.0){\rule[-0.200pt]{0.400pt}{2.409pt}}
\put(537.0,767.0){\rule[-0.200pt]{0.400pt}{2.409pt}}
\put(580.0,123.0){\rule[-0.200pt]{0.400pt}{2.409pt}}
\put(580.0,767.0){\rule[-0.200pt]{0.400pt}{2.409pt}}
\put(600.0,123.0){\rule[-0.200pt]{0.400pt}{157.549pt}}
\put(600.0,123.0){\rule[-0.200pt]{0.400pt}{4.818pt}}
\put(600,82){\makebox(0,0){ 0.1}}
\put(600.0,757.0){\rule[-0.200pt]{0.400pt}{4.818pt}}
\put(663.0,123.0){\rule[-0.200pt]{0.400pt}{2.409pt}}
\put(663.0,767.0){\rule[-0.200pt]{0.400pt}{2.409pt}}
\put(747.0,123.0){\rule[-0.200pt]{0.400pt}{2.409pt}}
\put(747.0,767.0){\rule[-0.200pt]{0.400pt}{2.409pt}}
\put(790.0,123.0){\rule[-0.200pt]{0.400pt}{2.409pt}}
\put(790.0,767.0){\rule[-0.200pt]{0.400pt}{2.409pt}}
\put(810.0,123.0){\rule[-0.200pt]{0.400pt}{157.549pt}}
\put(810.0,123.0){\rule[-0.200pt]{0.400pt}{4.818pt}}
\put(810,82){\makebox(0,0){ 1}}
\put(810.0,757.0){\rule[-0.200pt]{0.400pt}{4.818pt}}
\put(873.0,123.0){\rule[-0.200pt]{0.400pt}{2.409pt}}
\put(873.0,767.0){\rule[-0.200pt]{0.400pt}{2.409pt}}
\put(957.0,123.0){\rule[-0.200pt]{0.400pt}{2.409pt}}
\put(957.0,767.0){\rule[-0.200pt]{0.400pt}{2.409pt}}
\put(999.0,123.0){\rule[-0.200pt]{0.400pt}{2.409pt}}
\put(999.0,767.0){\rule[-0.200pt]{0.400pt}{2.409pt}}
\put(1020.0,123.0){\rule[-0.200pt]{0.400pt}{157.549pt}}
\put(1020.0,123.0){\rule[-0.200pt]{0.400pt}{4.818pt}}
\put(1020,82){\makebox(0,0){ 10}}
\put(1020.0,757.0){\rule[-0.200pt]{0.400pt}{4.818pt}}
\put(1083.0,123.0){\rule[-0.200pt]{0.400pt}{2.409pt}}
\put(1083.0,767.0){\rule[-0.200pt]{0.400pt}{2.409pt}}
\put(1166.0,123.0){\rule[-0.200pt]{0.400pt}{2.409pt}}
\put(1166.0,767.0){\rule[-0.200pt]{0.400pt}{2.409pt}}
\put(1209.0,123.0){\rule[-0.200pt]{0.400pt}{2.409pt}}
\put(1209.0,767.0){\rule[-0.200pt]{0.400pt}{2.409pt}}
\put(1229.0,123.0){\rule[-0.200pt]{0.400pt}{157.549pt}}
\put(1229.0,123.0){\rule[-0.200pt]{0.400pt}{4.818pt}}
\put(1229,82){\makebox(0,0){ 100}}
\put(1229.0,757.0){\rule[-0.200pt]{0.400pt}{4.818pt}}
\put(1292.0,123.0){\rule[-0.200pt]{0.400pt}{2.409pt}}
\put(1292.0,767.0){\rule[-0.200pt]{0.400pt}{2.409pt}}
\put(1376.0,123.0){\rule[-0.200pt]{0.400pt}{2.409pt}}
\put(1376.0,767.0){\rule[-0.200pt]{0.400pt}{2.409pt}}
\put(1419.0,123.0){\rule[-0.200pt]{0.400pt}{2.409pt}}
\put(1419.0,767.0){\rule[-0.200pt]{0.400pt}{2.409pt}}
\put(1439.0,123.0){\rule[-0.200pt]{0.400pt}{157.549pt}}
\put(1439.0,123.0){\rule[-0.200pt]{0.400pt}{4.818pt}}
\put(1439,82){\makebox(0,0){ 1000}}
\put(1439.0,757.0){\rule[-0.200pt]{0.400pt}{4.818pt}}
\put(181.0,123.0){\rule[-0.200pt]{303.052pt}{0.400pt}}
\put(1439.0,123.0){\rule[-0.200pt]{0.400pt}{157.549pt}}
\put(181.0,777.0){\rule[-0.200pt]{303.052pt}{0.400pt}}
\put(40,450){\makebox(0,0){\rotatebox{90}{Polarization ($\mathcal{P}$)}\hspace*{1em}}}
\put(810,21){\makebox(0,0){\shortstack{\\[0.5ex]$N_+\,/\,N_-$ ratio}}}
\put(810,839){\makebox(0,0){Polarization versus $N_+\,/\,N_-$ ratio}}
\put(181.0,123.0){\rule[-0.200pt]{0.400pt}{157.549pt}}
\sbox{\plotpoint}{\rule[-0.600pt]{1.200pt}{1.200pt}}%
\put(181,124){\usebox{\plotpoint}}
\put(257,122.01){\rule{0.482pt}{1.200pt}}
\multiput(257.00,121.51)(1.000,1.000){2}{\rule{0.241pt}{1.200pt}}
\put(181.0,124.0){\rule[-0.600pt]{18.308pt}{1.200pt}}
\put(302,123.01){\rule{0.723pt}{1.200pt}}
\multiput(302.00,122.51)(1.500,1.000){2}{\rule{0.361pt}{1.200pt}}
\put(259.0,125.0){\rule[-0.600pt]{10.359pt}{1.200pt}}
\put(332,124.01){\rule{0.723pt}{1.200pt}}
\multiput(332.00,123.51)(1.500,1.000){2}{\rule{0.361pt}{1.200pt}}
\put(305.0,126.0){\rule[-0.600pt]{6.504pt}{1.200pt}}
\put(355,125.01){\rule{0.482pt}{1.200pt}}
\multiput(355.00,124.51)(1.000,1.000){2}{\rule{0.241pt}{1.200pt}}
\put(335.0,127.0){\rule[-0.600pt]{4.818pt}{1.200pt}}
\put(375,126.01){\rule{0.723pt}{1.200pt}}
\multiput(375.00,125.51)(1.500,1.000){2}{\rule{0.361pt}{1.200pt}}
\put(357.0,128.0){\rule[-0.600pt]{4.336pt}{1.200pt}}
\put(390,127.01){\rule{0.723pt}{1.200pt}}
\multiput(390.00,126.51)(1.500,1.000){2}{\rule{0.361pt}{1.200pt}}
\put(378.0,129.0){\rule[-0.600pt]{2.891pt}{1.200pt}}
\put(403,128.01){\rule{0.482pt}{1.200pt}}
\multiput(403.00,127.51)(1.000,1.000){2}{\rule{0.241pt}{1.200pt}}
\put(393.0,130.0){\rule[-0.600pt]{2.409pt}{1.200pt}}
\put(415,129.01){\rule{0.723pt}{1.200pt}}
\multiput(415.00,128.51)(1.500,1.000){2}{\rule{0.361pt}{1.200pt}}
\put(405.0,131.0){\rule[-0.600pt]{2.409pt}{1.200pt}}
\put(426,130.01){\rule{0.482pt}{1.200pt}}
\multiput(426.00,129.51)(1.000,1.000){2}{\rule{0.241pt}{1.200pt}}
\put(418.0,132.0){\rule[-0.600pt]{1.927pt}{1.200pt}}
\put(433,131.01){\rule{0.723pt}{1.200pt}}
\multiput(433.00,130.51)(1.500,1.000){2}{\rule{0.361pt}{1.200pt}}
\put(428.0,133.0){\rule[-0.600pt]{1.204pt}{1.200pt}}
\put(443,132.01){\rule{0.723pt}{1.200pt}}
\multiput(443.00,131.51)(1.500,1.000){2}{\rule{0.361pt}{1.200pt}}
\put(436.0,134.0){\rule[-0.600pt]{1.686pt}{1.200pt}}
\put(451,133.01){\rule{0.482pt}{1.200pt}}
\multiput(451.00,132.51)(1.000,1.000){2}{\rule{0.241pt}{1.200pt}}
\put(446.0,135.0){\rule[-0.600pt]{1.204pt}{1.200pt}}
\put(458,134.01){\rule{0.723pt}{1.200pt}}
\multiput(458.00,133.51)(1.500,1.000){2}{\rule{0.361pt}{1.200pt}}
\put(453.0,136.0){\rule[-0.600pt]{1.204pt}{1.200pt}}
\put(463,135.01){\rule{0.723pt}{1.200pt}}
\multiput(463.00,134.51)(1.500,1.000){2}{\rule{0.361pt}{1.200pt}}
\put(461.0,137.0){\usebox{\plotpoint}}
\put(471,136.01){\rule{0.482pt}{1.200pt}}
\multiput(471.00,135.51)(1.000,1.000){2}{\rule{0.241pt}{1.200pt}}
\put(466.0,138.0){\rule[-0.600pt]{1.204pt}{1.200pt}}
\put(476,137.01){\rule{0.482pt}{1.200pt}}
\multiput(476.00,136.51)(1.000,1.000){2}{\rule{0.241pt}{1.200pt}}
\put(473.0,139.0){\usebox{\plotpoint}}
\put(481,138.01){\rule{0.723pt}{1.200pt}}
\multiput(481.00,137.51)(1.500,1.000){2}{\rule{0.361pt}{1.200pt}}
\put(478.0,140.0){\usebox{\plotpoint}}
\put(486,139.01){\rule{0.723pt}{1.200pt}}
\multiput(486.00,138.51)(1.500,1.000){2}{\rule{0.361pt}{1.200pt}}
\put(484.0,141.0){\usebox{\plotpoint}}
\put(491,140.01){\rule{0.723pt}{1.200pt}}
\multiput(491.00,139.51)(1.500,1.000){2}{\rule{0.361pt}{1.200pt}}
\put(489.0,142.0){\usebox{\plotpoint}}
\put(496,141.01){\rule{0.723pt}{1.200pt}}
\multiput(496.00,140.51)(1.500,1.000){2}{\rule{0.361pt}{1.200pt}}
\put(494.0,143.0){\usebox{\plotpoint}}
\put(501,142.01){\rule{0.723pt}{1.200pt}}
\multiput(501.00,141.51)(1.500,1.000){2}{\rule{0.361pt}{1.200pt}}
\put(499.0,144.0){\usebox{\plotpoint}}
\put(506,143.01){\rule{0.723pt}{1.200pt}}
\multiput(506.00,142.51)(1.500,1.000){2}{\rule{0.361pt}{1.200pt}}
\put(509,144.01){\rule{0.482pt}{1.200pt}}
\multiput(509.00,143.51)(1.000,1.000){2}{\rule{0.241pt}{1.200pt}}
\put(504.0,145.0){\usebox{\plotpoint}}
\put(514,145.01){\rule{0.482pt}{1.200pt}}
\multiput(514.00,144.51)(1.000,1.000){2}{\rule{0.241pt}{1.200pt}}
\put(516,146.01){\rule{0.723pt}{1.200pt}}
\multiput(516.00,145.51)(1.500,1.000){2}{\rule{0.361pt}{1.200pt}}
\put(511.0,147.0){\usebox{\plotpoint}}
\put(521,147.01){\rule{0.723pt}{1.200pt}}
\multiput(521.00,146.51)(1.500,1.000){2}{\rule{0.361pt}{1.200pt}}
\put(524,148.01){\rule{0.482pt}{1.200pt}}
\multiput(524.00,147.51)(1.000,1.000){2}{\rule{0.241pt}{1.200pt}}
\put(526,149.01){\rule{0.723pt}{1.200pt}}
\multiput(526.00,148.51)(1.500,1.000){2}{\rule{0.361pt}{1.200pt}}
\put(519.0,149.0){\usebox{\plotpoint}}
\put(531,150.01){\rule{0.723pt}{1.200pt}}
\multiput(531.00,149.51)(1.500,1.000){2}{\rule{0.361pt}{1.200pt}}
\put(534,151.01){\rule{0.482pt}{1.200pt}}
\multiput(534.00,150.51)(1.000,1.000){2}{\rule{0.241pt}{1.200pt}}
\put(536,152.01){\rule{0.723pt}{1.200pt}}
\multiput(536.00,151.51)(1.500,1.000){2}{\rule{0.361pt}{1.200pt}}
\put(539,153.01){\rule{0.723pt}{1.200pt}}
\multiput(539.00,152.51)(1.500,1.000){2}{\rule{0.361pt}{1.200pt}}
\put(529.0,152.0){\usebox{\plotpoint}}
\put(544,154.01){\rule{0.723pt}{1.200pt}}
\multiput(544.00,153.51)(1.500,1.000){2}{\rule{0.361pt}{1.200pt}}
\put(547,155.01){\rule{0.482pt}{1.200pt}}
\multiput(547.00,154.51)(1.000,1.000){2}{\rule{0.241pt}{1.200pt}}
\put(549,156.01){\rule{0.723pt}{1.200pt}}
\multiput(549.00,155.51)(1.500,1.000){2}{\rule{0.361pt}{1.200pt}}
\put(552,157.01){\rule{0.482pt}{1.200pt}}
\multiput(552.00,156.51)(1.000,1.000){2}{\rule{0.241pt}{1.200pt}}
\put(554,158.01){\rule{0.723pt}{1.200pt}}
\multiput(554.00,157.51)(1.500,1.000){2}{\rule{0.361pt}{1.200pt}}
\put(557,159.01){\rule{0.482pt}{1.200pt}}
\multiput(557.00,158.51)(1.000,1.000){2}{\rule{0.241pt}{1.200pt}}
\put(559,160.01){\rule{0.723pt}{1.200pt}}
\multiput(559.00,159.51)(1.500,1.000){2}{\rule{0.361pt}{1.200pt}}
\put(562,161.01){\rule{0.482pt}{1.200pt}}
\multiput(562.00,160.51)(1.000,1.000){2}{\rule{0.241pt}{1.200pt}}
\put(564,162.01){\rule{0.723pt}{1.200pt}}
\multiput(564.00,161.51)(1.500,1.000){2}{\rule{0.361pt}{1.200pt}}
\put(567,163.01){\rule{0.482pt}{1.200pt}}
\multiput(567.00,162.51)(1.000,1.000){2}{\rule{0.241pt}{1.200pt}}
\put(569,164.51){\rule{0.723pt}{1.200pt}}
\multiput(569.00,163.51)(1.500,2.000){2}{\rule{0.361pt}{1.200pt}}
\put(572,166.01){\rule{0.482pt}{1.200pt}}
\multiput(572.00,165.51)(1.000,1.000){2}{\rule{0.241pt}{1.200pt}}
\put(574,167.01){\rule{0.723pt}{1.200pt}}
\multiput(574.00,166.51)(1.500,1.000){2}{\rule{0.361pt}{1.200pt}}
\put(577,168.01){\rule{0.482pt}{1.200pt}}
\multiput(577.00,167.51)(1.000,1.000){2}{\rule{0.241pt}{1.200pt}}
\put(579,169.01){\rule{0.723pt}{1.200pt}}
\multiput(579.00,168.51)(1.500,1.000){2}{\rule{0.361pt}{1.200pt}}
\put(582,170.51){\rule{0.482pt}{1.200pt}}
\multiput(582.00,169.51)(1.000,2.000){2}{\rule{0.241pt}{1.200pt}}
\put(584,172.01){\rule{0.723pt}{1.200pt}}
\multiput(584.00,171.51)(1.500,1.000){2}{\rule{0.361pt}{1.200pt}}
\put(587,173.01){\rule{0.482pt}{1.200pt}}
\multiput(587.00,172.51)(1.000,1.000){2}{\rule{0.241pt}{1.200pt}}
\put(589,174.51){\rule{0.723pt}{1.200pt}}
\multiput(589.00,173.51)(1.500,2.000){2}{\rule{0.361pt}{1.200pt}}
\put(592,176.01){\rule{0.482pt}{1.200pt}}
\multiput(592.00,175.51)(1.000,1.000){2}{\rule{0.241pt}{1.200pt}}
\put(594,177.01){\rule{0.723pt}{1.200pt}}
\multiput(594.00,176.51)(1.500,1.000){2}{\rule{0.361pt}{1.200pt}}
\put(597,178.51){\rule{0.482pt}{1.200pt}}
\multiput(597.00,177.51)(1.000,2.000){2}{\rule{0.241pt}{1.200pt}}
\put(599,180.01){\rule{0.723pt}{1.200pt}}
\multiput(599.00,179.51)(1.500,1.000){2}{\rule{0.361pt}{1.200pt}}
\put(602,181.51){\rule{0.723pt}{1.200pt}}
\multiput(602.00,180.51)(1.500,2.000){2}{\rule{0.361pt}{1.200pt}}
\put(605,183.51){\rule{0.482pt}{1.200pt}}
\multiput(605.00,182.51)(1.000,2.000){2}{\rule{0.241pt}{1.200pt}}
\put(607,185.01){\rule{0.723pt}{1.200pt}}
\multiput(607.00,184.51)(1.500,1.000){2}{\rule{0.361pt}{1.200pt}}
\put(610,186.51){\rule{0.482pt}{1.200pt}}
\multiput(610.00,185.51)(1.000,2.000){2}{\rule{0.241pt}{1.200pt}}
\put(612,188.01){\rule{0.723pt}{1.200pt}}
\multiput(612.00,187.51)(1.500,1.000){2}{\rule{0.361pt}{1.200pt}}
\put(615,189.51){\rule{0.482pt}{1.200pt}}
\multiput(615.00,188.51)(1.000,2.000){2}{\rule{0.241pt}{1.200pt}}
\put(617,191.51){\rule{0.723pt}{1.200pt}}
\multiput(617.00,190.51)(1.500,2.000){2}{\rule{0.361pt}{1.200pt}}
\put(620,193.51){\rule{0.482pt}{1.200pt}}
\multiput(620.00,192.51)(1.000,2.000){2}{\rule{0.241pt}{1.200pt}}
\put(622,195.51){\rule{0.723pt}{1.200pt}}
\multiput(622.00,194.51)(1.500,2.000){2}{\rule{0.361pt}{1.200pt}}
\put(625,197.01){\rule{0.482pt}{1.200pt}}
\multiput(625.00,196.51)(1.000,1.000){2}{\rule{0.241pt}{1.200pt}}
\put(627,198.51){\rule{0.723pt}{1.200pt}}
\multiput(627.00,197.51)(1.500,2.000){2}{\rule{0.361pt}{1.200pt}}
\put(630,200.51){\rule{0.482pt}{1.200pt}}
\multiput(630.00,199.51)(1.000,2.000){2}{\rule{0.241pt}{1.200pt}}
\put(632,202.51){\rule{0.723pt}{1.200pt}}
\multiput(632.00,201.51)(1.500,2.000){2}{\rule{0.361pt}{1.200pt}}
\put(635,204.51){\rule{0.482pt}{1.200pt}}
\multiput(635.00,203.51)(1.000,2.000){2}{\rule{0.241pt}{1.200pt}}
\put(637,206.51){\rule{0.723pt}{1.200pt}}
\multiput(637.00,205.51)(1.500,2.000){2}{\rule{0.361pt}{1.200pt}}
\put(638.51,210){\rule{1.200pt}{0.723pt}}
\multiput(637.51,210.00)(2.000,1.500){2}{\rule{1.200pt}{0.361pt}}
\put(642,211.51){\rule{0.723pt}{1.200pt}}
\multiput(642.00,210.51)(1.500,2.000){2}{\rule{0.361pt}{1.200pt}}
\put(645,213.51){\rule{0.482pt}{1.200pt}}
\multiput(645.00,212.51)(1.000,2.000){2}{\rule{0.241pt}{1.200pt}}
\put(647,215.51){\rule{0.723pt}{1.200pt}}
\multiput(647.00,214.51)(1.500,2.000){2}{\rule{0.361pt}{1.200pt}}
\put(650,217.51){\rule{0.482pt}{1.200pt}}
\multiput(650.00,216.51)(1.000,2.000){2}{\rule{0.241pt}{1.200pt}}
\put(652,220.01){\rule{0.723pt}{1.200pt}}
\multiput(652.00,218.51)(1.500,3.000){2}{\rule{0.361pt}{1.200pt}}
\put(655,222.51){\rule{0.482pt}{1.200pt}}
\multiput(655.00,221.51)(1.000,2.000){2}{\rule{0.241pt}{1.200pt}}
\put(657,225.01){\rule{0.723pt}{1.200pt}}
\multiput(657.00,223.51)(1.500,3.000){2}{\rule{0.361pt}{1.200pt}}
\put(660,227.51){\rule{0.723pt}{1.200pt}}
\multiput(660.00,226.51)(1.500,2.000){2}{\rule{0.361pt}{1.200pt}}
\put(661.51,231){\rule{1.200pt}{0.723pt}}
\multiput(660.51,231.00)(2.000,1.500){2}{\rule{1.200pt}{0.361pt}}
\put(665,232.51){\rule{0.723pt}{1.200pt}}
\multiput(665.00,231.51)(1.500,2.000){2}{\rule{0.361pt}{1.200pt}}
\put(666.51,236){\rule{1.200pt}{0.723pt}}
\multiput(665.51,236.00)(2.000,1.500){2}{\rule{1.200pt}{0.361pt}}
\put(670,237.51){\rule{0.723pt}{1.200pt}}
\multiput(670.00,236.51)(1.500,2.000){2}{\rule{0.361pt}{1.200pt}}
\put(671.51,241){\rule{1.200pt}{0.723pt}}
\multiput(670.51,241.00)(2.000,1.500){2}{\rule{1.200pt}{0.361pt}}
\put(675,243.01){\rule{0.723pt}{1.200pt}}
\multiput(675.00,241.51)(1.500,3.000){2}{\rule{0.361pt}{1.200pt}}
\put(676.51,247){\rule{1.200pt}{0.723pt}}
\multiput(675.51,247.00)(2.000,1.500){2}{\rule{1.200pt}{0.361pt}}
\put(680,249.01){\rule{0.723pt}{1.200pt}}
\multiput(680.00,247.51)(1.500,3.000){2}{\rule{0.361pt}{1.200pt}}
\put(683,251.51){\rule{0.482pt}{1.200pt}}
\multiput(683.00,250.51)(1.000,2.000){2}{\rule{0.241pt}{1.200pt}}
\put(685,254.01){\rule{0.723pt}{1.200pt}}
\multiput(685.00,252.51)(1.500,3.000){2}{\rule{0.361pt}{1.200pt}}
\put(686.51,258){\rule{1.200pt}{0.723pt}}
\multiput(685.51,258.00)(2.000,1.500){2}{\rule{1.200pt}{0.361pt}}
\put(690,260.01){\rule{0.723pt}{1.200pt}}
\multiput(690.00,258.51)(1.500,3.000){2}{\rule{0.361pt}{1.200pt}}
\put(691.51,264){\rule{1.200pt}{0.964pt}}
\multiput(690.51,264.00)(2.000,2.000){2}{\rule{1.200pt}{0.482pt}}
\put(695,267.01){\rule{0.723pt}{1.200pt}}
\multiput(695.00,265.51)(1.500,3.000){2}{\rule{0.361pt}{1.200pt}}
\put(696.51,271){\rule{1.200pt}{0.723pt}}
\multiput(695.51,271.00)(2.000,1.500){2}{\rule{1.200pt}{0.361pt}}
\put(700,273.01){\rule{0.723pt}{1.200pt}}
\multiput(700.00,271.51)(1.500,3.000){2}{\rule{0.361pt}{1.200pt}}
\put(701.51,277){\rule{1.200pt}{0.723pt}}
\multiput(700.51,277.00)(2.000,1.500){2}{\rule{1.200pt}{0.361pt}}
\put(704.01,280){\rule{1.200pt}{0.964pt}}
\multiput(702.51,280.00)(3.000,2.000){2}{\rule{1.200pt}{0.482pt}}
\put(706.51,284){\rule{1.200pt}{0.723pt}}
\multiput(705.51,284.00)(2.000,1.500){2}{\rule{1.200pt}{0.361pt}}
\put(709.01,287){\rule{1.200pt}{0.964pt}}
\multiput(707.51,287.00)(3.000,2.000){2}{\rule{1.200pt}{0.482pt}}
\put(711.51,291){\rule{1.200pt}{0.723pt}}
\multiput(710.51,291.00)(2.000,1.500){2}{\rule{1.200pt}{0.361pt}}
\put(714.01,294){\rule{1.200pt}{0.964pt}}
\multiput(712.51,294.00)(3.000,2.000){2}{\rule{1.200pt}{0.482pt}}
\put(718,297.01){\rule{0.723pt}{1.200pt}}
\multiput(718.00,295.51)(1.500,3.000){2}{\rule{0.361pt}{1.200pt}}
\put(719.51,301){\rule{1.200pt}{0.964pt}}
\multiput(718.51,301.00)(2.000,2.000){2}{\rule{1.200pt}{0.482pt}}
\put(723,304.01){\rule{0.723pt}{1.200pt}}
\multiput(723.00,302.51)(1.500,3.000){2}{\rule{0.361pt}{1.200pt}}
\put(724.51,308){\rule{1.200pt}{0.964pt}}
\multiput(723.51,308.00)(2.000,2.000){2}{\rule{1.200pt}{0.482pt}}
\put(727.01,312){\rule{1.200pt}{0.964pt}}
\multiput(725.51,312.00)(3.000,2.000){2}{\rule{1.200pt}{0.482pt}}
\put(729.51,316){\rule{1.200pt}{0.964pt}}
\multiput(728.51,316.00)(2.000,2.000){2}{\rule{1.200pt}{0.482pt}}
\put(733,319.01){\rule{0.723pt}{1.200pt}}
\multiput(733.00,317.51)(1.500,3.000){2}{\rule{0.361pt}{1.200pt}}
\put(734.51,323){\rule{1.200pt}{0.964pt}}
\multiput(733.51,323.00)(2.000,2.000){2}{\rule{1.200pt}{0.482pt}}
\put(737.01,327){\rule{1.200pt}{0.964pt}}
\multiput(735.51,327.00)(3.000,2.000){2}{\rule{1.200pt}{0.482pt}}
\put(739.51,331){\rule{1.200pt}{0.964pt}}
\multiput(738.51,331.00)(2.000,2.000){2}{\rule{1.200pt}{0.482pt}}
\put(742.01,335){\rule{1.200pt}{0.964pt}}
\multiput(740.51,335.00)(3.000,2.000){2}{\rule{1.200pt}{0.482pt}}
\put(744.51,339){\rule{1.200pt}{0.964pt}}
\multiput(743.51,339.00)(2.000,2.000){2}{\rule{1.200pt}{0.482pt}}
\put(747.01,343){\rule{1.200pt}{0.964pt}}
\multiput(745.51,343.00)(3.000,2.000){2}{\rule{1.200pt}{0.482pt}}
\put(749.51,347){\rule{1.200pt}{0.964pt}}
\multiput(748.51,347.00)(2.000,2.000){2}{\rule{1.200pt}{0.482pt}}
\put(752.01,351){\rule{1.200pt}{0.964pt}}
\multiput(750.51,351.00)(3.000,2.000){2}{\rule{1.200pt}{0.482pt}}
\put(754.51,355){\rule{1.200pt}{1.204pt}}
\multiput(753.51,355.00)(2.000,2.500){2}{\rule{1.200pt}{0.602pt}}
\put(757.01,360){\rule{1.200pt}{0.964pt}}
\multiput(755.51,360.00)(3.000,2.000){2}{\rule{1.200pt}{0.482pt}}
\put(759.51,364){\rule{1.200pt}{0.964pt}}
\multiput(758.51,364.00)(2.000,2.000){2}{\rule{1.200pt}{0.482pt}}
\put(762.01,368){\rule{1.200pt}{0.964pt}}
\multiput(760.51,368.00)(3.000,2.000){2}{\rule{1.200pt}{0.482pt}}
\put(764.51,372){\rule{1.200pt}{1.204pt}}
\multiput(763.51,372.00)(2.000,2.500){2}{\rule{1.200pt}{0.602pt}}
\put(767.01,377){\rule{1.200pt}{0.964pt}}
\multiput(765.51,377.00)(3.000,2.000){2}{\rule{1.200pt}{0.482pt}}
\put(769.51,381){\rule{1.200pt}{0.964pt}}
\multiput(768.51,381.00)(2.000,2.000){2}{\rule{1.200pt}{0.482pt}}
\put(772.01,385){\rule{1.200pt}{1.204pt}}
\multiput(770.51,385.00)(3.000,2.500){2}{\rule{1.200pt}{0.602pt}}
\put(774.51,390){\rule{1.200pt}{0.964pt}}
\multiput(773.51,390.00)(2.000,2.000){2}{\rule{1.200pt}{0.482pt}}
\put(777.01,394){\rule{1.200pt}{0.964pt}}
\multiput(775.51,394.00)(3.000,2.000){2}{\rule{1.200pt}{0.482pt}}
\put(780.01,398){\rule{1.200pt}{1.204pt}}
\multiput(778.51,398.00)(3.000,2.500){2}{\rule{1.200pt}{0.602pt}}
\put(782.51,403){\rule{1.200pt}{0.964pt}}
\multiput(781.51,403.00)(2.000,2.000){2}{\rule{1.200pt}{0.482pt}}
\put(785.01,407){\rule{1.200pt}{1.204pt}}
\multiput(783.51,407.00)(3.000,2.500){2}{\rule{1.200pt}{0.602pt}}
\put(787.51,412){\rule{1.200pt}{0.964pt}}
\multiput(786.51,412.00)(2.000,2.000){2}{\rule{1.200pt}{0.482pt}}
\put(790.01,416){\rule{1.200pt}{1.204pt}}
\multiput(788.51,416.00)(3.000,2.500){2}{\rule{1.200pt}{0.602pt}}
\put(792.51,421){\rule{1.200pt}{0.964pt}}
\multiput(791.51,421.00)(2.000,2.000){2}{\rule{1.200pt}{0.482pt}}
\put(795.01,425){\rule{1.200pt}{1.204pt}}
\multiput(793.51,425.00)(3.000,2.500){2}{\rule{1.200pt}{0.602pt}}
\put(797.51,430){\rule{1.200pt}{0.964pt}}
\multiput(796.51,430.00)(2.000,2.000){2}{\rule{1.200pt}{0.482pt}}
\put(800.01,434){\rule{1.200pt}{1.204pt}}
\multiput(798.51,434.00)(3.000,2.500){2}{\rule{1.200pt}{0.602pt}}
\put(802.51,439){\rule{1.200pt}{0.964pt}}
\multiput(801.51,439.00)(2.000,2.000){2}{\rule{1.200pt}{0.482pt}}
\put(805.01,443){\rule{1.200pt}{1.204pt}}
\multiput(803.51,443.00)(3.000,2.500){2}{\rule{1.200pt}{0.602pt}}
\put(807.51,448){\rule{1.200pt}{0.964pt}}
\multiput(806.51,448.00)(2.000,2.000){2}{\rule{1.200pt}{0.482pt}}
\put(810.01,452){\rule{1.200pt}{1.204pt}}
\multiput(808.51,452.00)(3.000,2.500){2}{\rule{1.200pt}{0.602pt}}
\put(812.51,457){\rule{1.200pt}{0.964pt}}
\multiput(811.51,457.00)(2.000,2.000){2}{\rule{1.200pt}{0.482pt}}
\put(815.01,461){\rule{1.200pt}{1.204pt}}
\multiput(813.51,461.00)(3.000,2.500){2}{\rule{1.200pt}{0.602pt}}
\put(817.51,466){\rule{1.200pt}{0.964pt}}
\multiput(816.51,466.00)(2.000,2.000){2}{\rule{1.200pt}{0.482pt}}
\put(820.01,470){\rule{1.200pt}{1.204pt}}
\multiput(818.51,470.00)(3.000,2.500){2}{\rule{1.200pt}{0.602pt}}
\put(822.51,475){\rule{1.200pt}{0.964pt}}
\multiput(821.51,475.00)(2.000,2.000){2}{\rule{1.200pt}{0.482pt}}
\put(825.01,479){\rule{1.200pt}{1.204pt}}
\multiput(823.51,479.00)(3.000,2.500){2}{\rule{1.200pt}{0.602pt}}
\put(827.51,484){\rule{1.200pt}{0.964pt}}
\multiput(826.51,484.00)(2.000,2.000){2}{\rule{1.200pt}{0.482pt}}
\put(830.01,488){\rule{1.200pt}{1.204pt}}
\multiput(828.51,488.00)(3.000,2.500){2}{\rule{1.200pt}{0.602pt}}
\put(832.51,493){\rule{1.200pt}{0.964pt}}
\multiput(831.51,493.00)(2.000,2.000){2}{\rule{1.200pt}{0.482pt}}
\put(835.01,497){\rule{1.200pt}{1.204pt}}
\multiput(833.51,497.00)(3.000,2.500){2}{\rule{1.200pt}{0.602pt}}
\put(838.01,502){\rule{1.200pt}{0.964pt}}
\multiput(836.51,502.00)(3.000,2.000){2}{\rule{1.200pt}{0.482pt}}
\put(840.51,506){\rule{1.200pt}{0.964pt}}
\multiput(839.51,506.00)(2.000,2.000){2}{\rule{1.200pt}{0.482pt}}
\put(843.01,510){\rule{1.200pt}{1.204pt}}
\multiput(841.51,510.00)(3.000,2.500){2}{\rule{1.200pt}{0.602pt}}
\put(845.51,515){\rule{1.200pt}{0.964pt}}
\multiput(844.51,515.00)(2.000,2.000){2}{\rule{1.200pt}{0.482pt}}
\put(848.01,519){\rule{1.200pt}{0.964pt}}
\multiput(846.51,519.00)(3.000,2.000){2}{\rule{1.200pt}{0.482pt}}
\put(850.51,523){\rule{1.200pt}{1.204pt}}
\multiput(849.51,523.00)(2.000,2.500){2}{\rule{1.200pt}{0.602pt}}
\put(853.01,528){\rule{1.200pt}{0.964pt}}
\multiput(851.51,528.00)(3.000,2.000){2}{\rule{1.200pt}{0.482pt}}
\put(855.51,532){\rule{1.200pt}{0.964pt}}
\multiput(854.51,532.00)(2.000,2.000){2}{\rule{1.200pt}{0.482pt}}
\put(858.01,536){\rule{1.200pt}{0.964pt}}
\multiput(856.51,536.00)(3.000,2.000){2}{\rule{1.200pt}{0.482pt}}
\put(860.51,540){\rule{1.200pt}{1.204pt}}
\multiput(859.51,540.00)(2.000,2.500){2}{\rule{1.200pt}{0.602pt}}
\put(863.01,545){\rule{1.200pt}{0.964pt}}
\multiput(861.51,545.00)(3.000,2.000){2}{\rule{1.200pt}{0.482pt}}
\put(865.51,549){\rule{1.200pt}{0.964pt}}
\multiput(864.51,549.00)(2.000,2.000){2}{\rule{1.200pt}{0.482pt}}
\put(868.01,553){\rule{1.200pt}{0.964pt}}
\multiput(866.51,553.00)(3.000,2.000){2}{\rule{1.200pt}{0.482pt}}
\put(870.51,557){\rule{1.200pt}{0.964pt}}
\multiput(869.51,557.00)(2.000,2.000){2}{\rule{1.200pt}{0.482pt}}
\put(873.01,561){\rule{1.200pt}{0.964pt}}
\multiput(871.51,561.00)(3.000,2.000){2}{\rule{1.200pt}{0.482pt}}
\put(875.51,565){\rule{1.200pt}{0.964pt}}
\multiput(874.51,565.00)(2.000,2.000){2}{\rule{1.200pt}{0.482pt}}
\put(878.01,569){\rule{1.200pt}{0.964pt}}
\multiput(876.51,569.00)(3.000,2.000){2}{\rule{1.200pt}{0.482pt}}
\put(880.51,573){\rule{1.200pt}{0.964pt}}
\multiput(879.51,573.00)(2.000,2.000){2}{\rule{1.200pt}{0.482pt}}
\put(884,576.01){\rule{0.723pt}{1.200pt}}
\multiput(884.00,574.51)(1.500,3.000){2}{\rule{0.361pt}{1.200pt}}
\put(885.51,580){\rule{1.200pt}{0.964pt}}
\multiput(884.51,580.00)(2.000,2.000){2}{\rule{1.200pt}{0.482pt}}
\put(888.01,584){\rule{1.200pt}{0.964pt}}
\multiput(886.51,584.00)(3.000,2.000){2}{\rule{1.200pt}{0.482pt}}
\put(890.51,588){\rule{1.200pt}{0.964pt}}
\multiput(889.51,588.00)(2.000,2.000){2}{\rule{1.200pt}{0.482pt}}
\put(894,591.01){\rule{0.723pt}{1.200pt}}
\multiput(894.00,589.51)(1.500,3.000){2}{\rule{0.361pt}{1.200pt}}
\put(895.51,595){\rule{1.200pt}{0.964pt}}
\multiput(894.51,595.00)(2.000,2.000){2}{\rule{1.200pt}{0.482pt}}
\put(899,598.01){\rule{0.723pt}{1.200pt}}
\multiput(899.00,596.51)(1.500,3.000){2}{\rule{0.361pt}{1.200pt}}
\put(901.01,602){\rule{1.200pt}{0.964pt}}
\multiput(899.51,602.00)(3.000,2.000){2}{\rule{1.200pt}{0.482pt}}
\put(903.51,606){\rule{1.200pt}{0.723pt}}
\multiput(902.51,606.00)(2.000,1.500){2}{\rule{1.200pt}{0.361pt}}
\put(906.01,609){\rule{1.200pt}{0.964pt}}
\multiput(904.51,609.00)(3.000,2.000){2}{\rule{1.200pt}{0.482pt}}
\put(908.51,613){\rule{1.200pt}{0.723pt}}
\multiput(907.51,613.00)(2.000,1.500){2}{\rule{1.200pt}{0.361pt}}
\put(911.01,616){\rule{1.200pt}{0.964pt}}
\multiput(909.51,616.00)(3.000,2.000){2}{\rule{1.200pt}{0.482pt}}
\put(913.51,620){\rule{1.200pt}{0.723pt}}
\multiput(912.51,620.00)(2.000,1.500){2}{\rule{1.200pt}{0.361pt}}
\put(917,622.01){\rule{0.723pt}{1.200pt}}
\multiput(917.00,620.51)(1.500,3.000){2}{\rule{0.361pt}{1.200pt}}
\put(918.51,626){\rule{1.200pt}{0.723pt}}
\multiput(917.51,626.00)(2.000,1.500){2}{\rule{1.200pt}{0.361pt}}
\put(922,628.01){\rule{0.723pt}{1.200pt}}
\multiput(922.00,626.51)(1.500,3.000){2}{\rule{0.361pt}{1.200pt}}
\put(923.51,632){\rule{1.200pt}{0.964pt}}
\multiput(922.51,632.00)(2.000,2.000){2}{\rule{1.200pt}{0.482pt}}
\put(927,635.01){\rule{0.723pt}{1.200pt}}
\multiput(927.00,633.51)(1.500,3.000){2}{\rule{0.361pt}{1.200pt}}
\put(928.51,639){\rule{1.200pt}{0.723pt}}
\multiput(927.51,639.00)(2.000,1.500){2}{\rule{1.200pt}{0.361pt}}
\put(932,641.01){\rule{0.723pt}{1.200pt}}
\multiput(932.00,639.51)(1.500,3.000){2}{\rule{0.361pt}{1.200pt}}
\put(935,643.51){\rule{0.482pt}{1.200pt}}
\multiput(935.00,642.51)(1.000,2.000){2}{\rule{0.241pt}{1.200pt}}
\put(937,646.01){\rule{0.723pt}{1.200pt}}
\multiput(937.00,644.51)(1.500,3.000){2}{\rule{0.361pt}{1.200pt}}
\put(938.51,650){\rule{1.200pt}{0.723pt}}
\multiput(937.51,650.00)(2.000,1.500){2}{\rule{1.200pt}{0.361pt}}
\put(942,652.01){\rule{0.723pt}{1.200pt}}
\multiput(942.00,650.51)(1.500,3.000){2}{\rule{0.361pt}{1.200pt}}
\put(943.51,656){\rule{1.200pt}{0.723pt}}
\multiput(942.51,656.00)(2.000,1.500){2}{\rule{1.200pt}{0.361pt}}
\put(947,657.51){\rule{0.723pt}{1.200pt}}
\multiput(947.00,656.51)(1.500,2.000){2}{\rule{0.361pt}{1.200pt}}
\put(948.51,661){\rule{1.200pt}{0.723pt}}
\multiput(947.51,661.00)(2.000,1.500){2}{\rule{1.200pt}{0.361pt}}
\put(952,662.51){\rule{0.723pt}{1.200pt}}
\multiput(952.00,661.51)(1.500,2.000){2}{\rule{0.361pt}{1.200pt}}
\put(953.51,666){\rule{1.200pt}{0.723pt}}
\multiput(952.51,666.00)(2.000,1.500){2}{\rule{1.200pt}{0.361pt}}
\put(957,667.51){\rule{0.723pt}{1.200pt}}
\multiput(957.00,666.51)(1.500,2.000){2}{\rule{0.361pt}{1.200pt}}
\put(960,670.01){\rule{0.723pt}{1.200pt}}
\multiput(960.00,668.51)(1.500,3.000){2}{\rule{0.361pt}{1.200pt}}
\put(963,672.51){\rule{0.482pt}{1.200pt}}
\multiput(963.00,671.51)(1.000,2.000){2}{\rule{0.241pt}{1.200pt}}
\put(965,675.01){\rule{0.723pt}{1.200pt}}
\multiput(965.00,673.51)(1.500,3.000){2}{\rule{0.361pt}{1.200pt}}
\put(968,677.51){\rule{0.482pt}{1.200pt}}
\multiput(968.00,676.51)(1.000,2.000){2}{\rule{0.241pt}{1.200pt}}
\put(970,679.51){\rule{0.723pt}{1.200pt}}
\multiput(970.00,678.51)(1.500,2.000){2}{\rule{0.361pt}{1.200pt}}
\put(973,681.51){\rule{0.482pt}{1.200pt}}
\multiput(973.00,680.51)(1.000,2.000){2}{\rule{0.241pt}{1.200pt}}
\put(975,683.51){\rule{0.723pt}{1.200pt}}
\multiput(975.00,682.51)(1.500,2.000){2}{\rule{0.361pt}{1.200pt}}
\put(976.51,687){\rule{1.200pt}{0.723pt}}
\multiput(975.51,687.00)(2.000,1.500){2}{\rule{1.200pt}{0.361pt}}
\put(980,688.51){\rule{0.723pt}{1.200pt}}
\multiput(980.00,687.51)(1.500,2.000){2}{\rule{0.361pt}{1.200pt}}
\put(983,690.51){\rule{0.482pt}{1.200pt}}
\multiput(983.00,689.51)(1.000,2.000){2}{\rule{0.241pt}{1.200pt}}
\put(985,692.51){\rule{0.723pt}{1.200pt}}
\multiput(985.00,691.51)(1.500,2.000){2}{\rule{0.361pt}{1.200pt}}
\put(988,694.51){\rule{0.482pt}{1.200pt}}
\multiput(988.00,693.51)(1.000,2.000){2}{\rule{0.241pt}{1.200pt}}
\put(990,696.51){\rule{0.723pt}{1.200pt}}
\multiput(990.00,695.51)(1.500,2.000){2}{\rule{0.361pt}{1.200pt}}
\put(993,698.01){\rule{0.482pt}{1.200pt}}
\multiput(993.00,697.51)(1.000,1.000){2}{\rule{0.241pt}{1.200pt}}
\put(995,699.51){\rule{0.723pt}{1.200pt}}
\multiput(995.00,698.51)(1.500,2.000){2}{\rule{0.361pt}{1.200pt}}
\put(998,701.51){\rule{0.482pt}{1.200pt}}
\multiput(998.00,700.51)(1.000,2.000){2}{\rule{0.241pt}{1.200pt}}
\put(1000,703.51){\rule{0.723pt}{1.200pt}}
\multiput(1000.00,702.51)(1.500,2.000){2}{\rule{0.361pt}{1.200pt}}
\put(1003,705.51){\rule{0.482pt}{1.200pt}}
\multiput(1003.00,704.51)(1.000,2.000){2}{\rule{0.241pt}{1.200pt}}
\put(1005,707.01){\rule{0.723pt}{1.200pt}}
\multiput(1005.00,706.51)(1.500,1.000){2}{\rule{0.361pt}{1.200pt}}
\put(1008,708.51){\rule{0.482pt}{1.200pt}}
\multiput(1008.00,707.51)(1.000,2.000){2}{\rule{0.241pt}{1.200pt}}
\put(1010,710.01){\rule{0.723pt}{1.200pt}}
\multiput(1010.00,709.51)(1.500,1.000){2}{\rule{0.361pt}{1.200pt}}
\put(1013,711.51){\rule{0.482pt}{1.200pt}}
\multiput(1013.00,710.51)(1.000,2.000){2}{\rule{0.241pt}{1.200pt}}
\put(1015,713.51){\rule{0.723pt}{1.200pt}}
\multiput(1015.00,712.51)(1.500,2.000){2}{\rule{0.361pt}{1.200pt}}
\put(1018,715.01){\rule{0.723pt}{1.200pt}}
\multiput(1018.00,714.51)(1.500,1.000){2}{\rule{0.361pt}{1.200pt}}
\put(1021,716.51){\rule{0.482pt}{1.200pt}}
\multiput(1021.00,715.51)(1.000,2.000){2}{\rule{0.241pt}{1.200pt}}
\put(1023,718.01){\rule{0.723pt}{1.200pt}}
\multiput(1023.00,717.51)(1.500,1.000){2}{\rule{0.361pt}{1.200pt}}
\put(1026,719.01){\rule{0.482pt}{1.200pt}}
\multiput(1026.00,718.51)(1.000,1.000){2}{\rule{0.241pt}{1.200pt}}
\put(1028,720.51){\rule{0.723pt}{1.200pt}}
\multiput(1028.00,719.51)(1.500,2.000){2}{\rule{0.361pt}{1.200pt}}
\put(1031,722.01){\rule{0.482pt}{1.200pt}}
\multiput(1031.00,721.51)(1.000,1.000){2}{\rule{0.241pt}{1.200pt}}
\put(1033,723.01){\rule{0.723pt}{1.200pt}}
\multiput(1033.00,722.51)(1.500,1.000){2}{\rule{0.361pt}{1.200pt}}
\put(1036,724.51){\rule{0.482pt}{1.200pt}}
\multiput(1036.00,723.51)(1.000,2.000){2}{\rule{0.241pt}{1.200pt}}
\put(1038,726.01){\rule{0.723pt}{1.200pt}}
\multiput(1038.00,725.51)(1.500,1.000){2}{\rule{0.361pt}{1.200pt}}
\put(1041,727.01){\rule{0.482pt}{1.200pt}}
\multiput(1041.00,726.51)(1.000,1.000){2}{\rule{0.241pt}{1.200pt}}
\put(1043,728.01){\rule{0.723pt}{1.200pt}}
\multiput(1043.00,727.51)(1.500,1.000){2}{\rule{0.361pt}{1.200pt}}
\put(1046,729.01){\rule{0.482pt}{1.200pt}}
\multiput(1046.00,728.51)(1.000,1.000){2}{\rule{0.241pt}{1.200pt}}
\put(1048,730.51){\rule{0.723pt}{1.200pt}}
\multiput(1048.00,729.51)(1.500,2.000){2}{\rule{0.361pt}{1.200pt}}
\put(1051,732.01){\rule{0.482pt}{1.200pt}}
\multiput(1051.00,731.51)(1.000,1.000){2}{\rule{0.241pt}{1.200pt}}
\put(1053,733.01){\rule{0.723pt}{1.200pt}}
\multiput(1053.00,732.51)(1.500,1.000){2}{\rule{0.361pt}{1.200pt}}
\put(1056,734.01){\rule{0.482pt}{1.200pt}}
\multiput(1056.00,733.51)(1.000,1.000){2}{\rule{0.241pt}{1.200pt}}
\put(1058,735.01){\rule{0.723pt}{1.200pt}}
\multiput(1058.00,734.51)(1.500,1.000){2}{\rule{0.361pt}{1.200pt}}
\put(1061,736.01){\rule{0.482pt}{1.200pt}}
\multiput(1061.00,735.51)(1.000,1.000){2}{\rule{0.241pt}{1.200pt}}
\put(1063,737.01){\rule{0.723pt}{1.200pt}}
\multiput(1063.00,736.51)(1.500,1.000){2}{\rule{0.361pt}{1.200pt}}
\put(1066,738.01){\rule{0.482pt}{1.200pt}}
\multiput(1066.00,737.51)(1.000,1.000){2}{\rule{0.241pt}{1.200pt}}
\put(1068,739.01){\rule{0.723pt}{1.200pt}}
\multiput(1068.00,738.51)(1.500,1.000){2}{\rule{0.361pt}{1.200pt}}
\put(1071,740.01){\rule{0.482pt}{1.200pt}}
\multiput(1071.00,739.51)(1.000,1.000){2}{\rule{0.241pt}{1.200pt}}
\put(1073,741.01){\rule{0.723pt}{1.200pt}}
\multiput(1073.00,740.51)(1.500,1.000){2}{\rule{0.361pt}{1.200pt}}
\put(542.0,156.0){\usebox{\plotpoint}}
\put(1078,742.01){\rule{0.723pt}{1.200pt}}
\multiput(1078.00,741.51)(1.500,1.000){2}{\rule{0.361pt}{1.200pt}}
\put(1081,743.01){\rule{0.723pt}{1.200pt}}
\multiput(1081.00,742.51)(1.500,1.000){2}{\rule{0.361pt}{1.200pt}}
\put(1084,744.01){\rule{0.482pt}{1.200pt}}
\multiput(1084.00,743.51)(1.000,1.000){2}{\rule{0.241pt}{1.200pt}}
\put(1086,745.01){\rule{0.723pt}{1.200pt}}
\multiput(1086.00,744.51)(1.500,1.000){2}{\rule{0.361pt}{1.200pt}}
\put(1076.0,744.0){\usebox{\plotpoint}}
\put(1091,746.01){\rule{0.723pt}{1.200pt}}
\multiput(1091.00,745.51)(1.500,1.000){2}{\rule{0.361pt}{1.200pt}}
\put(1094,747.01){\rule{0.482pt}{1.200pt}}
\multiput(1094.00,746.51)(1.000,1.000){2}{\rule{0.241pt}{1.200pt}}
\put(1096,748.01){\rule{0.723pt}{1.200pt}}
\multiput(1096.00,747.51)(1.500,1.000){2}{\rule{0.361pt}{1.200pt}}
\put(1089.0,748.0){\usebox{\plotpoint}}
\put(1101,749.01){\rule{0.723pt}{1.200pt}}
\multiput(1101.00,748.51)(1.500,1.000){2}{\rule{0.361pt}{1.200pt}}
\put(1104,750.01){\rule{0.482pt}{1.200pt}}
\multiput(1104.00,749.51)(1.000,1.000){2}{\rule{0.241pt}{1.200pt}}
\put(1099.0,751.0){\usebox{\plotpoint}}
\put(1109,751.01){\rule{0.482pt}{1.200pt}}
\multiput(1109.00,750.51)(1.000,1.000){2}{\rule{0.241pt}{1.200pt}}
\put(1111,752.01){\rule{0.723pt}{1.200pt}}
\multiput(1111.00,751.51)(1.500,1.000){2}{\rule{0.361pt}{1.200pt}}
\put(1106.0,753.0){\usebox{\plotpoint}}
\put(1116,753.01){\rule{0.723pt}{1.200pt}}
\multiput(1116.00,752.51)(1.500,1.000){2}{\rule{0.361pt}{1.200pt}}
\put(1114.0,755.0){\usebox{\plotpoint}}
\put(1121,754.01){\rule{0.723pt}{1.200pt}}
\multiput(1121.00,753.51)(1.500,1.000){2}{\rule{0.361pt}{1.200pt}}
\put(1119.0,756.0){\usebox{\plotpoint}}
\put(1126,755.01){\rule{0.723pt}{1.200pt}}
\multiput(1126.00,754.51)(1.500,1.000){2}{\rule{0.361pt}{1.200pt}}
\put(1124.0,757.0){\usebox{\plotpoint}}
\put(1131,756.01){\rule{0.723pt}{1.200pt}}
\multiput(1131.00,755.51)(1.500,1.000){2}{\rule{0.361pt}{1.200pt}}
\put(1129.0,758.0){\usebox{\plotpoint}}
\put(1136,757.01){\rule{0.723pt}{1.200pt}}
\multiput(1136.00,756.51)(1.500,1.000){2}{\rule{0.361pt}{1.200pt}}
\put(1134.0,759.0){\usebox{\plotpoint}}
\put(1142,758.01){\rule{0.482pt}{1.200pt}}
\multiput(1142.00,757.51)(1.000,1.000){2}{\rule{0.241pt}{1.200pt}}
\put(1139.0,760.0){\usebox{\plotpoint}}
\put(1147,759.01){\rule{0.482pt}{1.200pt}}
\multiput(1147.00,758.51)(1.000,1.000){2}{\rule{0.241pt}{1.200pt}}
\put(1144.0,761.0){\usebox{\plotpoint}}
\put(1154,760.01){\rule{0.723pt}{1.200pt}}
\multiput(1154.00,759.51)(1.500,1.000){2}{\rule{0.361pt}{1.200pt}}
\put(1149.0,762.0){\rule[-0.600pt]{1.204pt}{1.200pt}}
\put(1159,761.01){\rule{0.723pt}{1.200pt}}
\multiput(1159.00,760.51)(1.500,1.000){2}{\rule{0.361pt}{1.200pt}}
\put(1157.0,763.0){\usebox{\plotpoint}}
\put(1167,762.01){\rule{0.482pt}{1.200pt}}
\multiput(1167.00,761.51)(1.000,1.000){2}{\rule{0.241pt}{1.200pt}}
\put(1162.0,764.0){\rule[-0.600pt]{1.204pt}{1.200pt}}
\put(1174,763.01){\rule{0.723pt}{1.200pt}}
\multiput(1174.00,762.51)(1.500,1.000){2}{\rule{0.361pt}{1.200pt}}
\put(1169.0,765.0){\rule[-0.600pt]{1.204pt}{1.200pt}}
\put(1184,764.01){\rule{0.723pt}{1.200pt}}
\multiput(1184.00,763.51)(1.500,1.000){2}{\rule{0.361pt}{1.200pt}}
\put(1177.0,766.0){\rule[-0.600pt]{1.686pt}{1.200pt}}
\put(1192,765.01){\rule{0.482pt}{1.200pt}}
\multiput(1192.00,764.51)(1.000,1.000){2}{\rule{0.241pt}{1.200pt}}
\put(1187.0,767.0){\rule[-0.600pt]{1.204pt}{1.200pt}}
\put(1202,766.01){\rule{0.723pt}{1.200pt}}
\multiput(1202.00,765.51)(1.500,1.000){2}{\rule{0.361pt}{1.200pt}}
\put(1194.0,768.0){\rule[-0.600pt]{1.927pt}{1.200pt}}
\put(1215,767.01){\rule{0.482pt}{1.200pt}}
\multiput(1215.00,766.51)(1.000,1.000){2}{\rule{0.241pt}{1.200pt}}
\put(1205.0,769.0){\rule[-0.600pt]{2.409pt}{1.200pt}}
\put(1227,768.01){\rule{0.723pt}{1.200pt}}
\multiput(1227.00,767.51)(1.500,1.000){2}{\rule{0.361pt}{1.200pt}}
\put(1217.0,770.0){\rule[-0.600pt]{2.409pt}{1.200pt}}
\put(1242,769.01){\rule{0.723pt}{1.200pt}}
\multiput(1242.00,768.51)(1.500,1.000){2}{\rule{0.361pt}{1.200pt}}
\put(1230.0,771.0){\rule[-0.600pt]{2.891pt}{1.200pt}}
\put(1263,770.01){\rule{0.482pt}{1.200pt}}
\multiput(1263.00,769.51)(1.000,1.000){2}{\rule{0.241pt}{1.200pt}}
\put(1245.0,772.0){\rule[-0.600pt]{4.336pt}{1.200pt}}
\put(1285,771.01){\rule{0.723pt}{1.200pt}}
\multiput(1285.00,770.51)(1.500,1.000){2}{\rule{0.361pt}{1.200pt}}
\put(1265.0,773.0){\rule[-0.600pt]{4.818pt}{1.200pt}}
\put(1315,772.01){\rule{0.723pt}{1.200pt}}
\multiput(1315.00,771.51)(1.500,1.000){2}{\rule{0.361pt}{1.200pt}}
\put(1288.0,774.0){\rule[-0.600pt]{6.504pt}{1.200pt}}
\put(1361,773.01){\rule{0.482pt}{1.200pt}}
\multiput(1361.00,772.51)(1.000,1.000){2}{\rule{0.241pt}{1.200pt}}
\put(1318.0,775.0){\rule[-0.600pt]{10.359pt}{1.200pt}}
\put(1363.0,776.0){\rule[-0.600pt]{18.308pt}{1.200pt}}
\end{picture}

%% file: FOM_Plot_Constant_N2.tex
% GNUPLOT: LaTeX picture
\setlength{\unitlength}{0.240900pt}
\ifx\plotpoint\undefined\newsavebox{\plotpoint}\fi
\sbox{\plotpoint}{\rule[-0.200pt]{0.400pt}{0.400pt}}%
\begin{picture}(1500,900)(0,0)
\font\gnuplot=cmr10 at 10pt
\gnuplot
\sbox{\plotpoint}{\rule[-0.200pt]{0.400pt}{0.400pt}}%
\put(181.0,123.0){\rule[-0.200pt]{4.818pt}{0.400pt}}
\put(161,123){\makebox(0,0)[r]{ 0}}
\put(1419.0,123.0){\rule[-0.200pt]{4.818pt}{0.400pt}}
\put(181.0,236.0){\rule[-0.200pt]{4.818pt}{0.400pt}}
\put(161,236){\makebox(0,0)[r]{ 0.2}}
\put(1419.0,236.0){\rule[-0.200pt]{4.818pt}{0.400pt}}
\put(181.0,350.0){\rule[-0.200pt]{4.818pt}{0.400pt}}
\put(161,350){\makebox(0,0)[r]{ 0.4}}
\put(1419.0,350.0){\rule[-0.200pt]{4.818pt}{0.400pt}}
\put(181.0,463.0){\rule[-0.200pt]{4.818pt}{0.400pt}}
\put(161,463){\makebox(0,0)[r]{ 0.6}}
\put(1419.0,463.0){\rule[-0.200pt]{4.818pt}{0.400pt}}
\put(181.0,577.0){\rule[-0.200pt]{4.818pt}{0.400pt}}
\put(161,577){\makebox(0,0)[r]{ 0.8}}
\put(1419.0,577.0){\rule[-0.200pt]{4.818pt}{0.400pt}}
\put(181.0,690.0){\rule[-0.200pt]{4.818pt}{0.400pt}}
\put(161,690){\makebox(0,0)[r]{ 1}}
\put(1419.0,690.0){\rule[-0.200pt]{4.818pt}{0.400pt}}
\put(181.0,803.0){\rule[-0.200pt]{4.818pt}{0.400pt}}
\put(161,803){\makebox(0,0)[r]{ 1.2}}
\put(1419.0,803.0){\rule[-0.200pt]{4.818pt}{0.400pt}}
\put(181.0,123.0){\rule[-0.200pt]{0.400pt}{4.818pt}}
\put(181,82){\makebox(0,0){ 0}}
\put(181.0,840.0){\rule[-0.200pt]{0.400pt}{4.818pt}}
\put(433.0,123.0){\rule[-0.200pt]{0.400pt}{4.818pt}}
\put(433,82){\makebox(0,0){ 5}}
\put(433.0,840.0){\rule[-0.200pt]{0.400pt}{4.818pt}}
\put(684.0,123.0){\rule[-0.200pt]{0.400pt}{4.818pt}}
\put(684,82){\makebox(0,0){ 10}}
\put(684.0,840.0){\rule[-0.200pt]{0.400pt}{4.818pt}}
\put(936.0,123.0){\rule[-0.200pt]{0.400pt}{4.818pt}}
\put(936,82){\makebox(0,0){ 15}}
\put(936.0,840.0){\rule[-0.200pt]{0.400pt}{4.818pt}}
\put(1187.0,123.0){\rule[-0.200pt]{0.400pt}{4.818pt}}
\put(1187,82){\makebox(0,0){ 20}}
\put(1187.0,840.0){\rule[-0.200pt]{0.400pt}{4.818pt}}
\put(1439.0,123.0){\rule[-0.200pt]{0.400pt}{4.818pt}}
\put(1439,82){\makebox(0,0){ 25}}
\put(1439.0,840.0){\rule[-0.200pt]{0.400pt}{4.818pt}}
\put(181.0,123.0){\rule[-0.200pt]{303.052pt}{0.400pt}}
\put(1439.0,123.0){\rule[-0.200pt]{0.400pt}{177.543pt}}
\put(181.0,860.0){\rule[-0.200pt]{303.052pt}{0.400pt}}
\put(40,491){\makebox(0,0){\rotatebox{90}{$N/N_0$}}}
\put(810,21){\makebox(0,0){$\tau\,/\,\tau_*$}}
\put(181.0,123.0){\rule[-0.200pt]{0.400pt}{177.543pt}}
\put(1279,820){\makebox(0,0)[r]{$N/N_0$}}
\put(1299.0,820.0){\rule[-0.200pt]{24.090pt}{0.400pt}}
\put(181,690){\usebox{\plotpoint}}
\put(181.0,690.0){\rule[-0.200pt]{303.052pt}{0.400pt}}
\sbox{\plotpoint}{\rule[-0.600pt]{1.200pt}{1.200pt}}%
\put(1279,779){\makebox(0,0)[r]{$\mathcal{P}$}}
\put(1299.0,779.0){\rule[-0.600pt]{24.090pt}{1.200pt}}
\put(181,123){\usebox{\plotpoint}}
\put(180.01,123){\rule{1.200pt}{0.964pt}}
\multiput(178.51,123.00)(3.000,2.000){2}{\rule{1.200pt}{0.482pt}}
\put(182.51,127){\rule{1.200pt}{0.723pt}}
\multiput(181.51,127.00)(2.000,1.500){2}{\rule{1.200pt}{0.361pt}}
\put(185.01,130){\rule{1.200pt}{0.964pt}}
\multiput(183.51,130.00)(3.000,2.000){2}{\rule{1.200pt}{0.482pt}}
\put(187.51,134){\rule{1.200pt}{0.723pt}}
\multiput(186.51,134.00)(2.000,1.500){2}{\rule{1.200pt}{0.361pt}}
\put(191,136.01){\rule{0.723pt}{1.200pt}}
\multiput(191.00,134.51)(1.500,3.000){2}{\rule{0.361pt}{1.200pt}}
\put(192.51,140){\rule{1.200pt}{0.964pt}}
\multiput(191.51,140.00)(2.000,2.000){2}{\rule{1.200pt}{0.482pt}}
\put(196,143.01){\rule{0.723pt}{1.200pt}}
\multiput(196.00,141.51)(1.500,3.000){2}{\rule{0.361pt}{1.200pt}}
\put(197.51,147){\rule{1.200pt}{0.964pt}}
\multiput(196.51,147.00)(2.000,2.000){2}{\rule{1.200pt}{0.482pt}}
\put(201,150.01){\rule{0.723pt}{1.200pt}}
\multiput(201.00,148.51)(1.500,3.000){2}{\rule{0.361pt}{1.200pt}}
\put(202.51,154){\rule{1.200pt}{0.723pt}}
\multiput(201.51,154.00)(2.000,1.500){2}{\rule{1.200pt}{0.361pt}}
\put(205.01,157){\rule{1.200pt}{0.964pt}}
\multiput(203.51,157.00)(3.000,2.000){2}{\rule{1.200pt}{0.482pt}}
\put(207.51,161){\rule{1.200pt}{0.723pt}}
\multiput(206.51,161.00)(2.000,1.500){2}{\rule{1.200pt}{0.361pt}}
\put(211,163.01){\rule{0.723pt}{1.200pt}}
\multiput(211.00,161.51)(1.500,3.000){2}{\rule{0.361pt}{1.200pt}}
\put(212.51,167){\rule{1.200pt}{0.964pt}}
\multiput(211.51,167.00)(2.000,2.000){2}{\rule{1.200pt}{0.482pt}}
\put(216,170.01){\rule{0.723pt}{1.200pt}}
\multiput(216.00,168.51)(1.500,3.000){2}{\rule{0.361pt}{1.200pt}}
\put(217.51,174){\rule{1.200pt}{0.723pt}}
\multiput(216.51,174.00)(2.000,1.500){2}{\rule{1.200pt}{0.361pt}}
\put(221,176.01){\rule{0.723pt}{1.200pt}}
\multiput(221.00,174.51)(1.500,3.000){2}{\rule{0.361pt}{1.200pt}}
\put(222.51,180){\rule{1.200pt}{0.723pt}}
\multiput(221.51,180.00)(2.000,1.500){2}{\rule{1.200pt}{0.361pt}}
\put(225.01,183){\rule{1.200pt}{0.964pt}}
\multiput(223.51,183.00)(3.000,2.000){2}{\rule{1.200pt}{0.482pt}}
\put(227.51,187){\rule{1.200pt}{0.723pt}}
\multiput(226.51,187.00)(2.000,1.500){2}{\rule{1.200pt}{0.361pt}}
\put(231,189.01){\rule{0.723pt}{1.200pt}}
\multiput(231.00,187.51)(1.500,3.000){2}{\rule{0.361pt}{1.200pt}}
\put(232.51,193){\rule{1.200pt}{0.723pt}}
\multiput(231.51,193.00)(2.000,1.500){2}{\rule{1.200pt}{0.361pt}}
\put(236,195.01){\rule{0.723pt}{1.200pt}}
\multiput(236.00,193.51)(1.500,3.000){2}{\rule{0.361pt}{1.200pt}}
\put(239,198.01){\rule{0.723pt}{1.200pt}}
\multiput(239.00,196.51)(1.500,3.000){2}{\rule{0.361pt}{1.200pt}}
\put(240.51,202){\rule{1.200pt}{0.723pt}}
\multiput(239.51,202.00)(2.000,1.500){2}{\rule{1.200pt}{0.361pt}}
\put(244,204.01){\rule{0.723pt}{1.200pt}}
\multiput(244.00,202.51)(1.500,3.000){2}{\rule{0.361pt}{1.200pt}}
\put(245.51,208){\rule{1.200pt}{0.723pt}}
\multiput(244.51,208.00)(2.000,1.500){2}{\rule{1.200pt}{0.361pt}}
\put(249,210.01){\rule{0.723pt}{1.200pt}}
\multiput(249.00,208.51)(1.500,3.000){2}{\rule{0.361pt}{1.200pt}}
\put(250.51,214){\rule{1.200pt}{0.723pt}}
\multiput(249.51,214.00)(2.000,1.500){2}{\rule{1.200pt}{0.361pt}}
\put(254,216.01){\rule{0.723pt}{1.200pt}}
\multiput(254.00,214.51)(1.500,3.000){2}{\rule{0.361pt}{1.200pt}}
\put(255.51,220){\rule{1.200pt}{0.723pt}}
\multiput(254.51,220.00)(2.000,1.500){2}{\rule{1.200pt}{0.361pt}}
\put(259,222.01){\rule{0.723pt}{1.200pt}}
\multiput(259.00,220.51)(1.500,3.000){2}{\rule{0.361pt}{1.200pt}}
\put(260.51,226){\rule{1.200pt}{0.723pt}}
\multiput(259.51,226.00)(2.000,1.500){2}{\rule{1.200pt}{0.361pt}}
\put(264,228.01){\rule{0.723pt}{1.200pt}}
\multiput(264.00,226.51)(1.500,3.000){2}{\rule{0.361pt}{1.200pt}}
\put(265.51,232){\rule{1.200pt}{0.723pt}}
\multiput(264.51,232.00)(2.000,1.500){2}{\rule{1.200pt}{0.361pt}}
\put(269,233.51){\rule{0.723pt}{1.200pt}}
\multiput(269.00,232.51)(1.500,2.000){2}{\rule{0.361pt}{1.200pt}}
\put(270.51,237){\rule{1.200pt}{0.723pt}}
\multiput(269.51,237.00)(2.000,1.500){2}{\rule{1.200pt}{0.361pt}}
\put(274,239.01){\rule{0.723pt}{1.200pt}}
\multiput(274.00,237.51)(1.500,3.000){2}{\rule{0.361pt}{1.200pt}}
\put(275.51,243){\rule{1.200pt}{0.723pt}}
\multiput(274.51,243.00)(2.000,1.500){2}{\rule{1.200pt}{0.361pt}}
\put(279,245.01){\rule{0.723pt}{1.200pt}}
\multiput(279.00,243.51)(1.500,3.000){2}{\rule{0.361pt}{1.200pt}}
\put(282,247.51){\rule{0.482pt}{1.200pt}}
\multiput(282.00,246.51)(1.000,2.000){2}{\rule{0.241pt}{1.200pt}}
\put(284,250.01){\rule{0.723pt}{1.200pt}}
\multiput(284.00,248.51)(1.500,3.000){2}{\rule{0.361pt}{1.200pt}}
\put(285.51,254){\rule{1.200pt}{0.723pt}}
\multiput(284.51,254.00)(2.000,1.500){2}{\rule{1.200pt}{0.361pt}}
\put(289,256.01){\rule{0.723pt}{1.200pt}}
\multiput(289.00,254.51)(1.500,3.000){2}{\rule{0.361pt}{1.200pt}}
\put(292,258.51){\rule{0.482pt}{1.200pt}}
\multiput(292.00,257.51)(1.000,2.000){2}{\rule{0.241pt}{1.200pt}}
\put(294,261.01){\rule{0.723pt}{1.200pt}}
\multiput(294.00,259.51)(1.500,3.000){2}{\rule{0.361pt}{1.200pt}}
\put(295.51,265){\rule{1.200pt}{0.723pt}}
\multiput(294.51,265.00)(2.000,1.500){2}{\rule{1.200pt}{0.361pt}}
\put(299,266.51){\rule{0.723pt}{1.200pt}}
\multiput(299.00,265.51)(1.500,2.000){2}{\rule{0.361pt}{1.200pt}}
\put(302,269.01){\rule{0.723pt}{1.200pt}}
\multiput(302.00,267.51)(1.500,3.000){2}{\rule{0.361pt}{1.200pt}}
\put(305,271.51){\rule{0.482pt}{1.200pt}}
\multiput(305.00,270.51)(1.000,2.000){2}{\rule{0.241pt}{1.200pt}}
\put(307,274.01){\rule{0.723pt}{1.200pt}}
\multiput(307.00,272.51)(1.500,3.000){2}{\rule{0.361pt}{1.200pt}}
\put(308.51,278){\rule{1.200pt}{0.723pt}}
\multiput(307.51,278.00)(2.000,1.500){2}{\rule{1.200pt}{0.361pt}}
\put(312,279.51){\rule{0.723pt}{1.200pt}}
\multiput(312.00,278.51)(1.500,2.000){2}{\rule{0.361pt}{1.200pt}}
\put(313.51,283){\rule{1.200pt}{0.723pt}}
\multiput(312.51,283.00)(2.000,1.500){2}{\rule{1.200pt}{0.361pt}}
\put(317,284.51){\rule{0.723pt}{1.200pt}}
\multiput(317.00,283.51)(1.500,2.000){2}{\rule{0.361pt}{1.200pt}}
\put(318.51,288){\rule{1.200pt}{0.723pt}}
\multiput(317.51,288.00)(2.000,1.500){2}{\rule{1.200pt}{0.361pt}}
\put(322,289.51){\rule{0.723pt}{1.200pt}}
\multiput(322.00,288.51)(1.500,2.000){2}{\rule{0.361pt}{1.200pt}}
\put(323.51,293){\rule{1.200pt}{0.723pt}}
\multiput(322.51,293.00)(2.000,1.500){2}{\rule{1.200pt}{0.361pt}}
\put(327,294.51){\rule{0.723pt}{1.200pt}}
\multiput(327.00,293.51)(1.500,2.000){2}{\rule{0.361pt}{1.200pt}}
\put(328.51,298){\rule{1.200pt}{0.723pt}}
\multiput(327.51,298.00)(2.000,1.500){2}{\rule{1.200pt}{0.361pt}}
\put(332,299.51){\rule{0.723pt}{1.200pt}}
\multiput(332.00,298.51)(1.500,2.000){2}{\rule{0.361pt}{1.200pt}}
\put(335,301.51){\rule{0.482pt}{1.200pt}}
\multiput(335.00,300.51)(1.000,2.000){2}{\rule{0.241pt}{1.200pt}}
\put(337,304.01){\rule{0.723pt}{1.200pt}}
\multiput(337.00,302.51)(1.500,3.000){2}{\rule{0.361pt}{1.200pt}}
\put(340,306.51){\rule{0.482pt}{1.200pt}}
\multiput(340.00,305.51)(1.000,2.000){2}{\rule{0.241pt}{1.200pt}}
\put(342,309.01){\rule{0.723pt}{1.200pt}}
\multiput(342.00,307.51)(1.500,3.000){2}{\rule{0.361pt}{1.200pt}}
\put(345,311.51){\rule{0.482pt}{1.200pt}}
\multiput(345.00,310.51)(1.000,2.000){2}{\rule{0.241pt}{1.200pt}}
\put(347,313.51){\rule{0.723pt}{1.200pt}}
\multiput(347.00,312.51)(1.500,2.000){2}{\rule{0.361pt}{1.200pt}}
\put(348.51,317){\rule{1.200pt}{0.723pt}}
\multiput(347.51,317.00)(2.000,1.500){2}{\rule{1.200pt}{0.361pt}}
\put(352,318.51){\rule{0.723pt}{1.200pt}}
\multiput(352.00,317.51)(1.500,2.000){2}{\rule{0.361pt}{1.200pt}}
\put(355,320.51){\rule{0.482pt}{1.200pt}}
\multiput(355.00,319.51)(1.000,2.000){2}{\rule{0.241pt}{1.200pt}}
\put(357,322.51){\rule{0.723pt}{1.200pt}}
\multiput(357.00,321.51)(1.500,2.000){2}{\rule{0.361pt}{1.200pt}}
\put(360,325.01){\rule{0.723pt}{1.200pt}}
\multiput(360.00,323.51)(1.500,3.000){2}{\rule{0.361pt}{1.200pt}}
\put(363,327.51){\rule{0.482pt}{1.200pt}}
\multiput(363.00,326.51)(1.000,2.000){2}{\rule{0.241pt}{1.200pt}}
\put(365,329.51){\rule{0.723pt}{1.200pt}}
\multiput(365.00,328.51)(1.500,2.000){2}{\rule{0.361pt}{1.200pt}}
\put(368,331.51){\rule{0.482pt}{1.200pt}}
\multiput(368.00,330.51)(1.000,2.000){2}{\rule{0.241pt}{1.200pt}}
\put(370,334.01){\rule{0.723pt}{1.200pt}}
\multiput(370.00,332.51)(1.500,3.000){2}{\rule{0.361pt}{1.200pt}}
\put(373,336.51){\rule{0.482pt}{1.200pt}}
\multiput(373.00,335.51)(1.000,2.000){2}{\rule{0.241pt}{1.200pt}}
\put(375,338.51){\rule{0.723pt}{1.200pt}}
\multiput(375.00,337.51)(1.500,2.000){2}{\rule{0.361pt}{1.200pt}}
\put(378,340.51){\rule{0.482pt}{1.200pt}}
\multiput(378.00,339.51)(1.000,2.000){2}{\rule{0.241pt}{1.200pt}}
\put(380,342.51){\rule{0.723pt}{1.200pt}}
\multiput(380.00,341.51)(1.500,2.000){2}{\rule{0.361pt}{1.200pt}}
\put(381.51,346){\rule{1.200pt}{0.723pt}}
\multiput(380.51,346.00)(2.000,1.500){2}{\rule{1.200pt}{0.361pt}}
\put(385,347.51){\rule{0.723pt}{1.200pt}}
\multiput(385.00,346.51)(1.500,2.000){2}{\rule{0.361pt}{1.200pt}}
\put(388,349.51){\rule{0.482pt}{1.200pt}}
\multiput(388.00,348.51)(1.000,2.000){2}{\rule{0.241pt}{1.200pt}}
\put(390,351.51){\rule{0.723pt}{1.200pt}}
\multiput(390.00,350.51)(1.500,2.000){2}{\rule{0.361pt}{1.200pt}}
\put(393,353.51){\rule{0.482pt}{1.200pt}}
\multiput(393.00,352.51)(1.000,2.000){2}{\rule{0.241pt}{1.200pt}}
\put(395,355.51){\rule{0.723pt}{1.200pt}}
\multiput(395.00,354.51)(1.500,2.000){2}{\rule{0.361pt}{1.200pt}}
\put(398,357.51){\rule{0.482pt}{1.200pt}}
\multiput(398.00,356.51)(1.000,2.000){2}{\rule{0.241pt}{1.200pt}}
\put(400,359.51){\rule{0.723pt}{1.200pt}}
\multiput(400.00,358.51)(1.500,2.000){2}{\rule{0.361pt}{1.200pt}}
\put(403,361.51){\rule{0.482pt}{1.200pt}}
\multiput(403.00,360.51)(1.000,2.000){2}{\rule{0.241pt}{1.200pt}}
\put(405,363.51){\rule{0.723pt}{1.200pt}}
\multiput(405.00,362.51)(1.500,2.000){2}{\rule{0.361pt}{1.200pt}}
\put(408,365.51){\rule{0.482pt}{1.200pt}}
\multiput(408.00,364.51)(1.000,2.000){2}{\rule{0.241pt}{1.200pt}}
\put(410,367.51){\rule{0.723pt}{1.200pt}}
\multiput(410.00,366.51)(1.500,2.000){2}{\rule{0.361pt}{1.200pt}}
\put(413,369.51){\rule{0.482pt}{1.200pt}}
\multiput(413.00,368.51)(1.000,2.000){2}{\rule{0.241pt}{1.200pt}}
\put(415,371.51){\rule{0.723pt}{1.200pt}}
\multiput(415.00,370.51)(1.500,2.000){2}{\rule{0.361pt}{1.200pt}}
\put(418,373.51){\rule{0.482pt}{1.200pt}}
\multiput(418.00,372.51)(1.000,2.000){2}{\rule{0.241pt}{1.200pt}}
\put(420,375.51){\rule{0.723pt}{1.200pt}}
\multiput(420.00,374.51)(1.500,2.000){2}{\rule{0.361pt}{1.200pt}}
\put(423,377.51){\rule{0.723pt}{1.200pt}}
\multiput(423.00,376.51)(1.500,2.000){2}{\rule{0.361pt}{1.200pt}}
\put(426,379.51){\rule{0.482pt}{1.200pt}}
\multiput(426.00,378.51)(1.000,2.000){2}{\rule{0.241pt}{1.200pt}}
\put(428,381.51){\rule{0.723pt}{1.200pt}}
\multiput(428.00,380.51)(1.500,2.000){2}{\rule{0.361pt}{1.200pt}}
\put(431,383.51){\rule{0.482pt}{1.200pt}}
\multiput(431.00,382.51)(1.000,2.000){2}{\rule{0.241pt}{1.200pt}}
\put(433,385.51){\rule{0.723pt}{1.200pt}}
\multiput(433.00,384.51)(1.500,2.000){2}{\rule{0.361pt}{1.200pt}}
\put(436,387.51){\rule{0.482pt}{1.200pt}}
\multiput(436.00,386.51)(1.000,2.000){2}{\rule{0.241pt}{1.200pt}}
\put(438,389.01){\rule{0.723pt}{1.200pt}}
\multiput(438.00,388.51)(1.500,1.000){2}{\rule{0.361pt}{1.200pt}}
\put(441,390.51){\rule{0.482pt}{1.200pt}}
\multiput(441.00,389.51)(1.000,2.000){2}{\rule{0.241pt}{1.200pt}}
\put(443,392.51){\rule{0.723pt}{1.200pt}}
\multiput(443.00,391.51)(1.500,2.000){2}{\rule{0.361pt}{1.200pt}}
\put(446,394.51){\rule{0.482pt}{1.200pt}}
\multiput(446.00,393.51)(1.000,2.000){2}{\rule{0.241pt}{1.200pt}}
\put(448,396.51){\rule{0.723pt}{1.200pt}}
\multiput(448.00,395.51)(1.500,2.000){2}{\rule{0.361pt}{1.200pt}}
\put(451,398.51){\rule{0.482pt}{1.200pt}}
\multiput(451.00,397.51)(1.000,2.000){2}{\rule{0.241pt}{1.200pt}}
\put(453,400.01){\rule{0.723pt}{1.200pt}}
\multiput(453.00,399.51)(1.500,1.000){2}{\rule{0.361pt}{1.200pt}}
\put(456,401.51){\rule{0.482pt}{1.200pt}}
\multiput(456.00,400.51)(1.000,2.000){2}{\rule{0.241pt}{1.200pt}}
\put(458,403.51){\rule{0.723pt}{1.200pt}}
\multiput(458.00,402.51)(1.500,2.000){2}{\rule{0.361pt}{1.200pt}}
\put(461,405.51){\rule{0.482pt}{1.200pt}}
\multiput(461.00,404.51)(1.000,2.000){2}{\rule{0.241pt}{1.200pt}}
\put(463,407.51){\rule{0.723pt}{1.200pt}}
\multiput(463.00,406.51)(1.500,2.000){2}{\rule{0.361pt}{1.200pt}}
\put(466,409.01){\rule{0.482pt}{1.200pt}}
\multiput(466.00,408.51)(1.000,1.000){2}{\rule{0.241pt}{1.200pt}}
\put(468,410.51){\rule{0.723pt}{1.200pt}}
\multiput(468.00,409.51)(1.500,2.000){2}{\rule{0.361pt}{1.200pt}}
\put(471,412.51){\rule{0.482pt}{1.200pt}}
\multiput(471.00,411.51)(1.000,2.000){2}{\rule{0.241pt}{1.200pt}}
\put(473,414.01){\rule{0.723pt}{1.200pt}}
\multiput(473.00,413.51)(1.500,1.000){2}{\rule{0.361pt}{1.200pt}}
\put(476,415.51){\rule{0.482pt}{1.200pt}}
\multiput(476.00,414.51)(1.000,2.000){2}{\rule{0.241pt}{1.200pt}}
\put(478,417.51){\rule{0.723pt}{1.200pt}}
\multiput(478.00,416.51)(1.500,2.000){2}{\rule{0.361pt}{1.200pt}}
\put(481,419.51){\rule{0.723pt}{1.200pt}}
\multiput(481.00,418.51)(1.500,2.000){2}{\rule{0.361pt}{1.200pt}}
\put(484,421.01){\rule{0.482pt}{1.200pt}}
\multiput(484.00,420.51)(1.000,1.000){2}{\rule{0.241pt}{1.200pt}}
\put(486,422.51){\rule{0.723pt}{1.200pt}}
\multiput(486.00,421.51)(1.500,2.000){2}{\rule{0.361pt}{1.200pt}}
\put(489,424.51){\rule{0.482pt}{1.200pt}}
\multiput(489.00,423.51)(1.000,2.000){2}{\rule{0.241pt}{1.200pt}}
\put(491,426.01){\rule{0.723pt}{1.200pt}}
\multiput(491.00,425.51)(1.500,1.000){2}{\rule{0.361pt}{1.200pt}}
\put(494,427.51){\rule{0.482pt}{1.200pt}}
\multiput(494.00,426.51)(1.000,2.000){2}{\rule{0.241pt}{1.200pt}}
\put(496,429.01){\rule{0.723pt}{1.200pt}}
\multiput(496.00,428.51)(1.500,1.000){2}{\rule{0.361pt}{1.200pt}}
\put(499,430.51){\rule{0.482pt}{1.200pt}}
\multiput(499.00,429.51)(1.000,2.000){2}{\rule{0.241pt}{1.200pt}}
\put(501,432.51){\rule{0.723pt}{1.200pt}}
\multiput(501.00,431.51)(1.500,2.000){2}{\rule{0.361pt}{1.200pt}}
\put(504,434.01){\rule{0.482pt}{1.200pt}}
\multiput(504.00,433.51)(1.000,1.000){2}{\rule{0.241pt}{1.200pt}}
\put(506,435.51){\rule{0.723pt}{1.200pt}}
\multiput(506.00,434.51)(1.500,2.000){2}{\rule{0.361pt}{1.200pt}}
\put(509,437.01){\rule{0.482pt}{1.200pt}}
\multiput(509.00,436.51)(1.000,1.000){2}{\rule{0.241pt}{1.200pt}}
\put(511,438.51){\rule{0.723pt}{1.200pt}}
\multiput(511.00,437.51)(1.500,2.000){2}{\rule{0.361pt}{1.200pt}}
\put(514,440.01){\rule{0.482pt}{1.200pt}}
\multiput(514.00,439.51)(1.000,1.000){2}{\rule{0.241pt}{1.200pt}}
\put(516,441.51){\rule{0.723pt}{1.200pt}}
\multiput(516.00,440.51)(1.500,2.000){2}{\rule{0.361pt}{1.200pt}}
\put(519,443.51){\rule{0.482pt}{1.200pt}}
\multiput(519.00,442.51)(1.000,2.000){2}{\rule{0.241pt}{1.200pt}}
\put(521,445.01){\rule{0.723pt}{1.200pt}}
\multiput(521.00,444.51)(1.500,1.000){2}{\rule{0.361pt}{1.200pt}}
\put(524,446.51){\rule{0.482pt}{1.200pt}}
\multiput(524.00,445.51)(1.000,2.000){2}{\rule{0.241pt}{1.200pt}}
\put(526,448.01){\rule{0.723pt}{1.200pt}}
\multiput(526.00,447.51)(1.500,1.000){2}{\rule{0.361pt}{1.200pt}}
\put(529,449.51){\rule{0.482pt}{1.200pt}}
\multiput(529.00,448.51)(1.000,2.000){2}{\rule{0.241pt}{1.200pt}}
\put(531,451.01){\rule{0.723pt}{1.200pt}}
\multiput(531.00,450.51)(1.500,1.000){2}{\rule{0.361pt}{1.200pt}}
\put(534,452.01){\rule{0.482pt}{1.200pt}}
\multiput(534.00,451.51)(1.000,1.000){2}{\rule{0.241pt}{1.200pt}}
\put(536,453.51){\rule{0.723pt}{1.200pt}}
\multiput(536.00,452.51)(1.500,2.000){2}{\rule{0.361pt}{1.200pt}}
\put(539,455.01){\rule{0.723pt}{1.200pt}}
\multiput(539.00,454.51)(1.500,1.000){2}{\rule{0.361pt}{1.200pt}}
\put(542,456.51){\rule{0.482pt}{1.200pt}}
\multiput(542.00,455.51)(1.000,2.000){2}{\rule{0.241pt}{1.200pt}}
\put(544,458.01){\rule{0.723pt}{1.200pt}}
\multiput(544.00,457.51)(1.500,1.000){2}{\rule{0.361pt}{1.200pt}}
\put(547,459.51){\rule{0.482pt}{1.200pt}}
\multiput(547.00,458.51)(1.000,2.000){2}{\rule{0.241pt}{1.200pt}}
\put(549,461.01){\rule{0.723pt}{1.200pt}}
\multiput(549.00,460.51)(1.500,1.000){2}{\rule{0.361pt}{1.200pt}}
\put(552,462.51){\rule{0.482pt}{1.200pt}}
\multiput(552.00,461.51)(1.000,2.000){2}{\rule{0.241pt}{1.200pt}}
\put(554,464.01){\rule{0.723pt}{1.200pt}}
\multiput(554.00,463.51)(1.500,1.000){2}{\rule{0.361pt}{1.200pt}}
\put(557,465.01){\rule{0.482pt}{1.200pt}}
\multiput(557.00,464.51)(1.000,1.000){2}{\rule{0.241pt}{1.200pt}}
\put(559,466.51){\rule{0.723pt}{1.200pt}}
\multiput(559.00,465.51)(1.500,2.000){2}{\rule{0.361pt}{1.200pt}}
\put(562,468.01){\rule{0.482pt}{1.200pt}}
\multiput(562.00,467.51)(1.000,1.000){2}{\rule{0.241pt}{1.200pt}}
\put(564,469.01){\rule{0.723pt}{1.200pt}}
\multiput(564.00,468.51)(1.500,1.000){2}{\rule{0.361pt}{1.200pt}}
\put(567,470.51){\rule{0.482pt}{1.200pt}}
\multiput(567.00,469.51)(1.000,2.000){2}{\rule{0.241pt}{1.200pt}}
\put(569,472.01){\rule{0.723pt}{1.200pt}}
\multiput(569.00,471.51)(1.500,1.000){2}{\rule{0.361pt}{1.200pt}}
\put(572,473.51){\rule{0.482pt}{1.200pt}}
\multiput(572.00,472.51)(1.000,2.000){2}{\rule{0.241pt}{1.200pt}}
\put(574,475.01){\rule{0.723pt}{1.200pt}}
\multiput(574.00,474.51)(1.500,1.000){2}{\rule{0.361pt}{1.200pt}}
\put(577,476.01){\rule{0.482pt}{1.200pt}}
\multiput(577.00,475.51)(1.000,1.000){2}{\rule{0.241pt}{1.200pt}}
\put(579,477.01){\rule{0.723pt}{1.200pt}}
\multiput(579.00,476.51)(1.500,1.000){2}{\rule{0.361pt}{1.200pt}}
\put(582,478.51){\rule{0.482pt}{1.200pt}}
\multiput(582.00,477.51)(1.000,2.000){2}{\rule{0.241pt}{1.200pt}}
\put(584,480.01){\rule{0.723pt}{1.200pt}}
\multiput(584.00,479.51)(1.500,1.000){2}{\rule{0.361pt}{1.200pt}}
\put(587,481.01){\rule{0.482pt}{1.200pt}}
\multiput(587.00,480.51)(1.000,1.000){2}{\rule{0.241pt}{1.200pt}}
\put(589,482.51){\rule{0.723pt}{1.200pt}}
\multiput(589.00,481.51)(1.500,2.000){2}{\rule{0.361pt}{1.200pt}}
\put(592,484.01){\rule{0.482pt}{1.200pt}}
\multiput(592.00,483.51)(1.000,1.000){2}{\rule{0.241pt}{1.200pt}}
\put(594,485.01){\rule{0.723pt}{1.200pt}}
\multiput(594.00,484.51)(1.500,1.000){2}{\rule{0.361pt}{1.200pt}}
\put(597,486.01){\rule{0.482pt}{1.200pt}}
\multiput(597.00,485.51)(1.000,1.000){2}{\rule{0.241pt}{1.200pt}}
\put(599,487.51){\rule{0.723pt}{1.200pt}}
\multiput(599.00,486.51)(1.500,2.000){2}{\rule{0.361pt}{1.200pt}}
\put(602,489.01){\rule{0.723pt}{1.200pt}}
\multiput(602.00,488.51)(1.500,1.000){2}{\rule{0.361pt}{1.200pt}}
\put(605,490.01){\rule{0.482pt}{1.200pt}}
\multiput(605.00,489.51)(1.000,1.000){2}{\rule{0.241pt}{1.200pt}}
\put(607,491.01){\rule{0.723pt}{1.200pt}}
\multiput(607.00,490.51)(1.500,1.000){2}{\rule{0.361pt}{1.200pt}}
\put(610,492.51){\rule{0.482pt}{1.200pt}}
\multiput(610.00,491.51)(1.000,2.000){2}{\rule{0.241pt}{1.200pt}}
\put(612,494.01){\rule{0.723pt}{1.200pt}}
\multiput(612.00,493.51)(1.500,1.000){2}{\rule{0.361pt}{1.200pt}}
\put(615,495.01){\rule{0.482pt}{1.200pt}}
\multiput(615.00,494.51)(1.000,1.000){2}{\rule{0.241pt}{1.200pt}}
\put(617,496.01){\rule{0.723pt}{1.200pt}}
\multiput(617.00,495.51)(1.500,1.000){2}{\rule{0.361pt}{1.200pt}}
\put(620,497.01){\rule{0.482pt}{1.200pt}}
\multiput(620.00,496.51)(1.000,1.000){2}{\rule{0.241pt}{1.200pt}}
\put(622,498.51){\rule{0.723pt}{1.200pt}}
\multiput(622.00,497.51)(1.500,2.000){2}{\rule{0.361pt}{1.200pt}}
\put(625,500.01){\rule{0.482pt}{1.200pt}}
\multiput(625.00,499.51)(1.000,1.000){2}{\rule{0.241pt}{1.200pt}}
\put(627,501.01){\rule{0.723pt}{1.200pt}}
\multiput(627.00,500.51)(1.500,1.000){2}{\rule{0.361pt}{1.200pt}}
\put(630,502.01){\rule{0.482pt}{1.200pt}}
\multiput(630.00,501.51)(1.000,1.000){2}{\rule{0.241pt}{1.200pt}}
\put(632,503.01){\rule{0.723pt}{1.200pt}}
\multiput(632.00,502.51)(1.500,1.000){2}{\rule{0.361pt}{1.200pt}}
\put(635,504.01){\rule{0.482pt}{1.200pt}}
\multiput(635.00,503.51)(1.000,1.000){2}{\rule{0.241pt}{1.200pt}}
\put(637,505.51){\rule{0.723pt}{1.200pt}}
\multiput(637.00,504.51)(1.500,2.000){2}{\rule{0.361pt}{1.200pt}}
\put(640,507.01){\rule{0.482pt}{1.200pt}}
\multiput(640.00,506.51)(1.000,1.000){2}{\rule{0.241pt}{1.200pt}}
\put(642,508.01){\rule{0.723pt}{1.200pt}}
\multiput(642.00,507.51)(1.500,1.000){2}{\rule{0.361pt}{1.200pt}}
\put(645,509.01){\rule{0.482pt}{1.200pt}}
\multiput(645.00,508.51)(1.000,1.000){2}{\rule{0.241pt}{1.200pt}}
\put(647,510.01){\rule{0.723pt}{1.200pt}}
\multiput(647.00,509.51)(1.500,1.000){2}{\rule{0.361pt}{1.200pt}}
\put(650,511.01){\rule{0.482pt}{1.200pt}}
\multiput(650.00,510.51)(1.000,1.000){2}{\rule{0.241pt}{1.200pt}}
\put(652,512.01){\rule{0.723pt}{1.200pt}}
\multiput(652.00,511.51)(1.500,1.000){2}{\rule{0.361pt}{1.200pt}}
\put(655,513.01){\rule{0.482pt}{1.200pt}}
\multiput(655.00,512.51)(1.000,1.000){2}{\rule{0.241pt}{1.200pt}}
\put(657,514.01){\rule{0.723pt}{1.200pt}}
\multiput(657.00,513.51)(1.500,1.000){2}{\rule{0.361pt}{1.200pt}}
\put(660,515.51){\rule{0.723pt}{1.200pt}}
\multiput(660.00,514.51)(1.500,2.000){2}{\rule{0.361pt}{1.200pt}}
\put(663,517.01){\rule{0.482pt}{1.200pt}}
\multiput(663.00,516.51)(1.000,1.000){2}{\rule{0.241pt}{1.200pt}}
\put(665,518.01){\rule{0.723pt}{1.200pt}}
\multiput(665.00,517.51)(1.500,1.000){2}{\rule{0.361pt}{1.200pt}}
\put(668,519.01){\rule{0.482pt}{1.200pt}}
\multiput(668.00,518.51)(1.000,1.000){2}{\rule{0.241pt}{1.200pt}}
\put(670,520.01){\rule{0.723pt}{1.200pt}}
\multiput(670.00,519.51)(1.500,1.000){2}{\rule{0.361pt}{1.200pt}}
\put(673,521.01){\rule{0.482pt}{1.200pt}}
\multiput(673.00,520.51)(1.000,1.000){2}{\rule{0.241pt}{1.200pt}}
\put(675,522.01){\rule{0.723pt}{1.200pt}}
\multiput(675.00,521.51)(1.500,1.000){2}{\rule{0.361pt}{1.200pt}}
\put(678,523.01){\rule{0.482pt}{1.200pt}}
\multiput(678.00,522.51)(1.000,1.000){2}{\rule{0.241pt}{1.200pt}}
\put(680,524.01){\rule{0.723pt}{1.200pt}}
\multiput(680.00,523.51)(1.500,1.000){2}{\rule{0.361pt}{1.200pt}}
\put(683,525.01){\rule{0.482pt}{1.200pt}}
\multiput(683.00,524.51)(1.000,1.000){2}{\rule{0.241pt}{1.200pt}}
\put(685,526.01){\rule{0.723pt}{1.200pt}}
\multiput(685.00,525.51)(1.500,1.000){2}{\rule{0.361pt}{1.200pt}}
\put(688,527.01){\rule{0.482pt}{1.200pt}}
\multiput(688.00,526.51)(1.000,1.000){2}{\rule{0.241pt}{1.200pt}}
\put(690,528.01){\rule{0.723pt}{1.200pt}}
\multiput(690.00,527.51)(1.500,1.000){2}{\rule{0.361pt}{1.200pt}}
\put(693,529.01){\rule{0.482pt}{1.200pt}}
\multiput(693.00,528.51)(1.000,1.000){2}{\rule{0.241pt}{1.200pt}}
\put(695,530.01){\rule{0.723pt}{1.200pt}}
\multiput(695.00,529.51)(1.500,1.000){2}{\rule{0.361pt}{1.200pt}}
\put(698,531.01){\rule{0.482pt}{1.200pt}}
\multiput(698.00,530.51)(1.000,1.000){2}{\rule{0.241pt}{1.200pt}}
\put(700,532.01){\rule{0.723pt}{1.200pt}}
\multiput(700.00,531.51)(1.500,1.000){2}{\rule{0.361pt}{1.200pt}}
\put(703,533.01){\rule{0.482pt}{1.200pt}}
\multiput(703.00,532.51)(1.000,1.000){2}{\rule{0.241pt}{1.200pt}}
\put(705,534.01){\rule{0.723pt}{1.200pt}}
\multiput(705.00,533.51)(1.500,1.000){2}{\rule{0.361pt}{1.200pt}}
\put(708,535.01){\rule{0.482pt}{1.200pt}}
\multiput(708.00,534.51)(1.000,1.000){2}{\rule{0.241pt}{1.200pt}}
\put(710,536.01){\rule{0.723pt}{1.200pt}}
\multiput(710.00,535.51)(1.500,1.000){2}{\rule{0.361pt}{1.200pt}}
\put(713,537.01){\rule{0.482pt}{1.200pt}}
\multiput(713.00,536.51)(1.000,1.000){2}{\rule{0.241pt}{1.200pt}}
\put(715,538.01){\rule{0.723pt}{1.200pt}}
\multiput(715.00,537.51)(1.500,1.000){2}{\rule{0.361pt}{1.200pt}}
\put(718,539.01){\rule{0.723pt}{1.200pt}}
\multiput(718.00,538.51)(1.500,1.000){2}{\rule{0.361pt}{1.200pt}}
\put(723,540.01){\rule{0.723pt}{1.200pt}}
\multiput(723.00,539.51)(1.500,1.000){2}{\rule{0.361pt}{1.200pt}}
\put(726,541.01){\rule{0.482pt}{1.200pt}}
\multiput(726.00,540.51)(1.000,1.000){2}{\rule{0.241pt}{1.200pt}}
\put(728,542.01){\rule{0.723pt}{1.200pt}}
\multiput(728.00,541.51)(1.500,1.000){2}{\rule{0.361pt}{1.200pt}}
\put(731,543.01){\rule{0.482pt}{1.200pt}}
\multiput(731.00,542.51)(1.000,1.000){2}{\rule{0.241pt}{1.200pt}}
\put(733,544.01){\rule{0.723pt}{1.200pt}}
\multiput(733.00,543.51)(1.500,1.000){2}{\rule{0.361pt}{1.200pt}}
\put(736,545.01){\rule{0.482pt}{1.200pt}}
\multiput(736.00,544.51)(1.000,1.000){2}{\rule{0.241pt}{1.200pt}}
\put(738,546.01){\rule{0.723pt}{1.200pt}}
\multiput(738.00,545.51)(1.500,1.000){2}{\rule{0.361pt}{1.200pt}}
\put(741,547.01){\rule{0.482pt}{1.200pt}}
\multiput(741.00,546.51)(1.000,1.000){2}{\rule{0.241pt}{1.200pt}}
\put(743,548.01){\rule{0.723pt}{1.200pt}}
\multiput(743.00,547.51)(1.500,1.000){2}{\rule{0.361pt}{1.200pt}}
\put(721.0,542.0){\usebox{\plotpoint}}
\put(748,549.01){\rule{0.723pt}{1.200pt}}
\multiput(748.00,548.51)(1.500,1.000){2}{\rule{0.361pt}{1.200pt}}
\put(751,550.01){\rule{0.482pt}{1.200pt}}
\multiput(751.00,549.51)(1.000,1.000){2}{\rule{0.241pt}{1.200pt}}
\put(753,551.01){\rule{0.723pt}{1.200pt}}
\multiput(753.00,550.51)(1.500,1.000){2}{\rule{0.361pt}{1.200pt}}
\put(756,552.01){\rule{0.482pt}{1.200pt}}
\multiput(756.00,551.51)(1.000,1.000){2}{\rule{0.241pt}{1.200pt}}
\put(758,553.01){\rule{0.723pt}{1.200pt}}
\multiput(758.00,552.51)(1.500,1.000){2}{\rule{0.361pt}{1.200pt}}
\put(746.0,551.0){\usebox{\plotpoint}}
\put(763,554.01){\rule{0.723pt}{1.200pt}}
\multiput(763.00,553.51)(1.500,1.000){2}{\rule{0.361pt}{1.200pt}}
\put(766,555.01){\rule{0.482pt}{1.200pt}}
\multiput(766.00,554.51)(1.000,1.000){2}{\rule{0.241pt}{1.200pt}}
\put(768,556.01){\rule{0.723pt}{1.200pt}}
\multiput(768.00,555.51)(1.500,1.000){2}{\rule{0.361pt}{1.200pt}}
\put(771,557.01){\rule{0.482pt}{1.200pt}}
\multiput(771.00,556.51)(1.000,1.000){2}{\rule{0.241pt}{1.200pt}}
\put(773,558.01){\rule{0.723pt}{1.200pt}}
\multiput(773.00,557.51)(1.500,1.000){2}{\rule{0.361pt}{1.200pt}}
\put(761.0,556.0){\usebox{\plotpoint}}
\put(778,559.01){\rule{0.723pt}{1.200pt}}
\multiput(778.00,558.51)(1.500,1.000){2}{\rule{0.361pt}{1.200pt}}
\put(781,560.01){\rule{0.723pt}{1.200pt}}
\multiput(781.00,559.51)(1.500,1.000){2}{\rule{0.361pt}{1.200pt}}
\put(784,561.01){\rule{0.482pt}{1.200pt}}
\multiput(784.00,560.51)(1.000,1.000){2}{\rule{0.241pt}{1.200pt}}
\put(786,562.01){\rule{0.723pt}{1.200pt}}
\multiput(786.00,561.51)(1.500,1.000){2}{\rule{0.361pt}{1.200pt}}
\put(776.0,561.0){\usebox{\plotpoint}}
\put(791,563.01){\rule{0.723pt}{1.200pt}}
\multiput(791.00,562.51)(1.500,1.000){2}{\rule{0.361pt}{1.200pt}}
\put(794,564.01){\rule{0.482pt}{1.200pt}}
\multiput(794.00,563.51)(1.000,1.000){2}{\rule{0.241pt}{1.200pt}}
\put(796,565.01){\rule{0.723pt}{1.200pt}}
\multiput(796.00,564.51)(1.500,1.000){2}{\rule{0.361pt}{1.200pt}}
\put(789.0,565.0){\usebox{\plotpoint}}
\put(801,566.01){\rule{0.723pt}{1.200pt}}
\multiput(801.00,565.51)(1.500,1.000){2}{\rule{0.361pt}{1.200pt}}
\put(804,567.01){\rule{0.482pt}{1.200pt}}
\multiput(804.00,566.51)(1.000,1.000){2}{\rule{0.241pt}{1.200pt}}
\put(806,568.01){\rule{0.723pt}{1.200pt}}
\multiput(806.00,567.51)(1.500,1.000){2}{\rule{0.361pt}{1.200pt}}
\put(799.0,568.0){\usebox{\plotpoint}}
\put(811,569.01){\rule{0.723pt}{1.200pt}}
\multiput(811.00,568.51)(1.500,1.000){2}{\rule{0.361pt}{1.200pt}}
\put(814,570.01){\rule{0.482pt}{1.200pt}}
\multiput(814.00,569.51)(1.000,1.000){2}{\rule{0.241pt}{1.200pt}}
\put(816,571.01){\rule{0.723pt}{1.200pt}}
\multiput(816.00,570.51)(1.500,1.000){2}{\rule{0.361pt}{1.200pt}}
\put(809.0,571.0){\usebox{\plotpoint}}
\put(821,572.01){\rule{0.723pt}{1.200pt}}
\multiput(821.00,571.51)(1.500,1.000){2}{\rule{0.361pt}{1.200pt}}
\put(824,573.01){\rule{0.482pt}{1.200pt}}
\multiput(824.00,572.51)(1.000,1.000){2}{\rule{0.241pt}{1.200pt}}
\put(826,574.01){\rule{0.723pt}{1.200pt}}
\multiput(826.00,573.51)(1.500,1.000){2}{\rule{0.361pt}{1.200pt}}
\put(819.0,574.0){\usebox{\plotpoint}}
\put(831,575.01){\rule{0.723pt}{1.200pt}}
\multiput(831.00,574.51)(1.500,1.000){2}{\rule{0.361pt}{1.200pt}}
\put(834,576.01){\rule{0.482pt}{1.200pt}}
\multiput(834.00,575.51)(1.000,1.000){2}{\rule{0.241pt}{1.200pt}}
\put(829.0,577.0){\usebox{\plotpoint}}
\put(839,577.01){\rule{0.723pt}{1.200pt}}
\multiput(839.00,576.51)(1.500,1.000){2}{\rule{0.361pt}{1.200pt}}
\put(842,578.01){\rule{0.482pt}{1.200pt}}
\multiput(842.00,577.51)(1.000,1.000){2}{\rule{0.241pt}{1.200pt}}
\put(836.0,579.0){\usebox{\plotpoint}}
\put(847,579.01){\rule{0.482pt}{1.200pt}}
\multiput(847.00,578.51)(1.000,1.000){2}{\rule{0.241pt}{1.200pt}}
\put(849,580.01){\rule{0.723pt}{1.200pt}}
\multiput(849.00,579.51)(1.500,1.000){2}{\rule{0.361pt}{1.200pt}}
\put(844.0,581.0){\usebox{\plotpoint}}
\put(854,581.01){\rule{0.723pt}{1.200pt}}
\multiput(854.00,580.51)(1.500,1.000){2}{\rule{0.361pt}{1.200pt}}
\put(857,582.01){\rule{0.482pt}{1.200pt}}
\multiput(857.00,581.51)(1.000,1.000){2}{\rule{0.241pt}{1.200pt}}
\put(852.0,583.0){\usebox{\plotpoint}}
\put(862,583.01){\rule{0.482pt}{1.200pt}}
\multiput(862.00,582.51)(1.000,1.000){2}{\rule{0.241pt}{1.200pt}}
\put(864,584.01){\rule{0.723pt}{1.200pt}}
\multiput(864.00,583.51)(1.500,1.000){2}{\rule{0.361pt}{1.200pt}}
\put(859.0,585.0){\usebox{\plotpoint}}
\put(869,585.01){\rule{0.723pt}{1.200pt}}
\multiput(869.00,584.51)(1.500,1.000){2}{\rule{0.361pt}{1.200pt}}
\put(872,586.01){\rule{0.482pt}{1.200pt}}
\multiput(872.00,585.51)(1.000,1.000){2}{\rule{0.241pt}{1.200pt}}
\put(867.0,587.0){\usebox{\plotpoint}}
\put(877,587.01){\rule{0.482pt}{1.200pt}}
\multiput(877.00,586.51)(1.000,1.000){2}{\rule{0.241pt}{1.200pt}}
\put(879,588.01){\rule{0.723pt}{1.200pt}}
\multiput(879.00,587.51)(1.500,1.000){2}{\rule{0.361pt}{1.200pt}}
\put(874.0,589.0){\usebox{\plotpoint}}
\put(884,589.01){\rule{0.723pt}{1.200pt}}
\multiput(884.00,588.51)(1.500,1.000){2}{\rule{0.361pt}{1.200pt}}
\put(882.0,591.0){\usebox{\plotpoint}}
\put(889,590.01){\rule{0.723pt}{1.200pt}}
\multiput(889.00,589.51)(1.500,1.000){2}{\rule{0.361pt}{1.200pt}}
\put(892,591.01){\rule{0.482pt}{1.200pt}}
\multiput(892.00,590.51)(1.000,1.000){2}{\rule{0.241pt}{1.200pt}}
\put(887.0,592.0){\usebox{\plotpoint}}
\put(897,592.01){\rule{0.482pt}{1.200pt}}
\multiput(897.00,591.51)(1.000,1.000){2}{\rule{0.241pt}{1.200pt}}
\put(894.0,594.0){\usebox{\plotpoint}}
\put(902,593.01){\rule{0.723pt}{1.200pt}}
\multiput(902.00,592.51)(1.500,1.000){2}{\rule{0.361pt}{1.200pt}}
\put(905,594.01){\rule{0.482pt}{1.200pt}}
\multiput(905.00,593.51)(1.000,1.000){2}{\rule{0.241pt}{1.200pt}}
\put(899.0,595.0){\usebox{\plotpoint}}
\put(910,595.01){\rule{0.482pt}{1.200pt}}
\multiput(910.00,594.51)(1.000,1.000){2}{\rule{0.241pt}{1.200pt}}
\put(907.0,597.0){\usebox{\plotpoint}}
\put(915,596.01){\rule{0.482pt}{1.200pt}}
\multiput(915.00,595.51)(1.000,1.000){2}{\rule{0.241pt}{1.200pt}}
\put(912.0,598.0){\usebox{\plotpoint}}
\put(920,597.01){\rule{0.482pt}{1.200pt}}
\multiput(920.00,596.51)(1.000,1.000){2}{\rule{0.241pt}{1.200pt}}
\put(922,598.01){\rule{0.723pt}{1.200pt}}
\multiput(922.00,597.51)(1.500,1.000){2}{\rule{0.361pt}{1.200pt}}
\put(917.0,599.0){\usebox{\plotpoint}}
\put(927,599.01){\rule{0.723pt}{1.200pt}}
\multiput(927.00,598.51)(1.500,1.000){2}{\rule{0.361pt}{1.200pt}}
\put(925.0,601.0){\usebox{\plotpoint}}
\put(932,600.01){\rule{0.723pt}{1.200pt}}
\multiput(932.00,599.51)(1.500,1.000){2}{\rule{0.361pt}{1.200pt}}
\put(930.0,602.0){\usebox{\plotpoint}}
\put(937,601.01){\rule{0.723pt}{1.200pt}}
\multiput(937.00,600.51)(1.500,1.000){2}{\rule{0.361pt}{1.200pt}}
\put(935.0,603.0){\usebox{\plotpoint}}
\put(942,602.01){\rule{0.723pt}{1.200pt}}
\multiput(942.00,601.51)(1.500,1.000){2}{\rule{0.361pt}{1.200pt}}
\put(940.0,604.0){\usebox{\plotpoint}}
\put(947,603.01){\rule{0.723pt}{1.200pt}}
\multiput(947.00,602.51)(1.500,1.000){2}{\rule{0.361pt}{1.200pt}}
\put(950,604.01){\rule{0.482pt}{1.200pt}}
\multiput(950.00,603.51)(1.000,1.000){2}{\rule{0.241pt}{1.200pt}}
\put(945.0,605.0){\usebox{\plotpoint}}
\put(955,605.01){\rule{0.482pt}{1.200pt}}
\multiput(955.00,604.51)(1.000,1.000){2}{\rule{0.241pt}{1.200pt}}
\put(952.0,607.0){\usebox{\plotpoint}}
\put(960,606.01){\rule{0.723pt}{1.200pt}}
\multiput(960.00,605.51)(1.500,1.000){2}{\rule{0.361pt}{1.200pt}}
\put(957.0,608.0){\usebox{\plotpoint}}
\put(965,607.01){\rule{0.723pt}{1.200pt}}
\multiput(965.00,606.51)(1.500,1.000){2}{\rule{0.361pt}{1.200pt}}
\put(963.0,609.0){\usebox{\plotpoint}}
\put(970,608.01){\rule{0.723pt}{1.200pt}}
\multiput(970.00,607.51)(1.500,1.000){2}{\rule{0.361pt}{1.200pt}}
\put(968.0,610.0){\usebox{\plotpoint}}
\put(975,609.01){\rule{0.723pt}{1.200pt}}
\multiput(975.00,608.51)(1.500,1.000){2}{\rule{0.361pt}{1.200pt}}
\put(973.0,611.0){\usebox{\plotpoint}}
\put(980,610.01){\rule{0.723pt}{1.200pt}}
\multiput(980.00,609.51)(1.500,1.000){2}{\rule{0.361pt}{1.200pt}}
\put(978.0,612.0){\usebox{\plotpoint}}
\put(985,611.01){\rule{0.723pt}{1.200pt}}
\multiput(985.00,610.51)(1.500,1.000){2}{\rule{0.361pt}{1.200pt}}
\put(983.0,613.0){\usebox{\plotpoint}}
\put(993,612.01){\rule{0.482pt}{1.200pt}}
\multiput(993.00,611.51)(1.000,1.000){2}{\rule{0.241pt}{1.200pt}}
\put(988.0,614.0){\rule[-0.600pt]{1.204pt}{1.200pt}}
\put(998,613.01){\rule{0.482pt}{1.200pt}}
\multiput(998.00,612.51)(1.000,1.000){2}{\rule{0.241pt}{1.200pt}}
\put(995.0,615.0){\usebox{\plotpoint}}
\put(1003,614.01){\rule{0.482pt}{1.200pt}}
\multiput(1003.00,613.51)(1.000,1.000){2}{\rule{0.241pt}{1.200pt}}
\put(1000.0,616.0){\usebox{\plotpoint}}
\put(1008,615.01){\rule{0.482pt}{1.200pt}}
\multiput(1008.00,614.51)(1.000,1.000){2}{\rule{0.241pt}{1.200pt}}
\put(1005.0,617.0){\usebox{\plotpoint}}
\put(1013,616.01){\rule{0.482pt}{1.200pt}}
\multiput(1013.00,615.51)(1.000,1.000){2}{\rule{0.241pt}{1.200pt}}
\put(1010.0,618.0){\usebox{\plotpoint}}
\put(1021,617.01){\rule{0.482pt}{1.200pt}}
\multiput(1021.00,616.51)(1.000,1.000){2}{\rule{0.241pt}{1.200pt}}
\put(1015.0,619.0){\rule[-0.600pt]{1.445pt}{1.200pt}}
\put(1026,618.01){\rule{0.482pt}{1.200pt}}
\multiput(1026.00,617.51)(1.000,1.000){2}{\rule{0.241pt}{1.200pt}}
\put(1023.0,620.0){\usebox{\plotpoint}}
\put(1031,619.01){\rule{0.482pt}{1.200pt}}
\multiput(1031.00,618.51)(1.000,1.000){2}{\rule{0.241pt}{1.200pt}}
\put(1028.0,621.0){\usebox{\plotpoint}}
\put(1036,620.01){\rule{0.482pt}{1.200pt}}
\multiput(1036.00,619.51)(1.000,1.000){2}{\rule{0.241pt}{1.200pt}}
\put(1033.0,622.0){\usebox{\plotpoint}}
\put(1043,621.01){\rule{0.723pt}{1.200pt}}
\multiput(1043.00,620.51)(1.500,1.000){2}{\rule{0.361pt}{1.200pt}}
\put(1038.0,623.0){\rule[-0.600pt]{1.204pt}{1.200pt}}
\put(1048,622.01){\rule{0.723pt}{1.200pt}}
\multiput(1048.00,621.51)(1.500,1.000){2}{\rule{0.361pt}{1.200pt}}
\put(1046.0,624.0){\usebox{\plotpoint}}
\put(1056,623.01){\rule{0.482pt}{1.200pt}}
\multiput(1056.00,622.51)(1.000,1.000){2}{\rule{0.241pt}{1.200pt}}
\put(1051.0,625.0){\rule[-0.600pt]{1.204pt}{1.200pt}}
\put(1061,624.01){\rule{0.482pt}{1.200pt}}
\multiput(1061.00,623.51)(1.000,1.000){2}{\rule{0.241pt}{1.200pt}}
\put(1058.0,626.0){\usebox{\plotpoint}}
\put(1068,625.01){\rule{0.723pt}{1.200pt}}
\multiput(1068.00,624.51)(1.500,1.000){2}{\rule{0.361pt}{1.200pt}}
\put(1063.0,627.0){\rule[-0.600pt]{1.204pt}{1.200pt}}
\put(1073,626.01){\rule{0.723pt}{1.200pt}}
\multiput(1073.00,625.51)(1.500,1.000){2}{\rule{0.361pt}{1.200pt}}
\put(1071.0,628.0){\usebox{\plotpoint}}
\put(1081,627.01){\rule{0.723pt}{1.200pt}}
\multiput(1081.00,626.51)(1.500,1.000){2}{\rule{0.361pt}{1.200pt}}
\put(1076.0,629.0){\rule[-0.600pt]{1.204pt}{1.200pt}}
\put(1089,628.01){\rule{0.482pt}{1.200pt}}
\multiput(1089.00,627.51)(1.000,1.000){2}{\rule{0.241pt}{1.200pt}}
\put(1084.0,630.0){\rule[-0.600pt]{1.204pt}{1.200pt}}
\put(1094,629.01){\rule{0.482pt}{1.200pt}}
\multiput(1094.00,628.51)(1.000,1.000){2}{\rule{0.241pt}{1.200pt}}
\put(1091.0,631.0){\usebox{\plotpoint}}
\put(1101,630.01){\rule{0.723pt}{1.200pt}}
\multiput(1101.00,629.51)(1.500,1.000){2}{\rule{0.361pt}{1.200pt}}
\put(1096.0,632.0){\rule[-0.600pt]{1.204pt}{1.200pt}}
\put(1109,631.01){\rule{0.482pt}{1.200pt}}
\multiput(1109.00,630.51)(1.000,1.000){2}{\rule{0.241pt}{1.200pt}}
\put(1104.0,633.0){\rule[-0.600pt]{1.204pt}{1.200pt}}
\put(1116,632.01){\rule{0.723pt}{1.200pt}}
\multiput(1116.00,631.51)(1.500,1.000){2}{\rule{0.361pt}{1.200pt}}
\put(1111.0,634.0){\rule[-0.600pt]{1.204pt}{1.200pt}}
\put(1124,633.01){\rule{0.482pt}{1.200pt}}
\multiput(1124.00,632.51)(1.000,1.000){2}{\rule{0.241pt}{1.200pt}}
\put(1119.0,635.0){\rule[-0.600pt]{1.204pt}{1.200pt}}
\put(1131,634.01){\rule{0.723pt}{1.200pt}}
\multiput(1131.00,633.51)(1.500,1.000){2}{\rule{0.361pt}{1.200pt}}
\put(1126.0,636.0){\rule[-0.600pt]{1.204pt}{1.200pt}}
\put(1139,635.01){\rule{0.723pt}{1.200pt}}
\multiput(1139.00,634.51)(1.500,1.000){2}{\rule{0.361pt}{1.200pt}}
\put(1134.0,637.0){\rule[-0.600pt]{1.204pt}{1.200pt}}
\put(1147,636.01){\rule{0.482pt}{1.200pt}}
\multiput(1147.00,635.51)(1.000,1.000){2}{\rule{0.241pt}{1.200pt}}
\put(1142.0,638.0){\rule[-0.600pt]{1.204pt}{1.200pt}}
\put(1154,637.01){\rule{0.723pt}{1.200pt}}
\multiput(1154.00,636.51)(1.500,1.000){2}{\rule{0.361pt}{1.200pt}}
\put(1149.0,639.0){\rule[-0.600pt]{1.204pt}{1.200pt}}
\put(1162,638.01){\rule{0.482pt}{1.200pt}}
\multiput(1162.00,637.51)(1.000,1.000){2}{\rule{0.241pt}{1.200pt}}
\put(1157.0,640.0){\rule[-0.600pt]{1.204pt}{1.200pt}}
\put(1169,639.01){\rule{0.723pt}{1.200pt}}
\multiput(1169.00,638.51)(1.500,1.000){2}{\rule{0.361pt}{1.200pt}}
\put(1164.0,641.0){\rule[-0.600pt]{1.204pt}{1.200pt}}
\put(1179,640.01){\rule{0.723pt}{1.200pt}}
\multiput(1179.00,639.51)(1.500,1.000){2}{\rule{0.361pt}{1.200pt}}
\put(1172.0,642.0){\rule[-0.600pt]{1.686pt}{1.200pt}}
\put(1187,641.01){\rule{0.482pt}{1.200pt}}
\multiput(1187.00,640.51)(1.000,1.000){2}{\rule{0.241pt}{1.200pt}}
\put(1182.0,643.0){\rule[-0.600pt]{1.204pt}{1.200pt}}
\put(1197,642.01){\rule{0.723pt}{1.200pt}}
\multiput(1197.00,641.51)(1.500,1.000){2}{\rule{0.361pt}{1.200pt}}
\put(1189.0,644.0){\rule[-0.600pt]{1.927pt}{1.200pt}}
\put(1205,643.01){\rule{0.482pt}{1.200pt}}
\multiput(1205.00,642.51)(1.000,1.000){2}{\rule{0.241pt}{1.200pt}}
\put(1200.0,645.0){\rule[-0.600pt]{1.204pt}{1.200pt}}
\put(1215,644.01){\rule{0.482pt}{1.200pt}}
\multiput(1215.00,643.51)(1.000,1.000){2}{\rule{0.241pt}{1.200pt}}
\put(1207.0,646.0){\rule[-0.600pt]{1.927pt}{1.200pt}}
\put(1222,645.01){\rule{0.723pt}{1.200pt}}
\multiput(1222.00,644.51)(1.500,1.000){2}{\rule{0.361pt}{1.200pt}}
\put(1217.0,647.0){\rule[-0.600pt]{1.204pt}{1.200pt}}
\put(1232,646.01){\rule{0.723pt}{1.200pt}}
\multiput(1232.00,645.51)(1.500,1.000){2}{\rule{0.361pt}{1.200pt}}
\put(1225.0,648.0){\rule[-0.600pt]{1.686pt}{1.200pt}}
\put(1242,647.01){\rule{0.723pt}{1.200pt}}
\multiput(1242.00,646.51)(1.500,1.000){2}{\rule{0.361pt}{1.200pt}}
\put(1235.0,649.0){\rule[-0.600pt]{1.686pt}{1.200pt}}
\put(1252,648.01){\rule{0.723pt}{1.200pt}}
\multiput(1252.00,647.51)(1.500,1.000){2}{\rule{0.361pt}{1.200pt}}
\put(1245.0,650.0){\rule[-0.600pt]{1.686pt}{1.200pt}}
\put(1263,649.01){\rule{0.482pt}{1.200pt}}
\multiput(1263.00,648.51)(1.000,1.000){2}{\rule{0.241pt}{1.200pt}}
\put(1255.0,651.0){\rule[-0.600pt]{1.927pt}{1.200pt}}
\put(1275,650.01){\rule{0.723pt}{1.200pt}}
\multiput(1275.00,649.51)(1.500,1.000){2}{\rule{0.361pt}{1.200pt}}
\put(1265.0,652.0){\rule[-0.600pt]{2.409pt}{1.200pt}}
\put(1285,651.01){\rule{0.723pt}{1.200pt}}
\multiput(1285.00,650.51)(1.500,1.000){2}{\rule{0.361pt}{1.200pt}}
\put(1278.0,653.0){\rule[-0.600pt]{1.686pt}{1.200pt}}
\put(1295,652.01){\rule{0.723pt}{1.200pt}}
\multiput(1295.00,651.51)(1.500,1.000){2}{\rule{0.361pt}{1.200pt}}
\put(1288.0,654.0){\rule[-0.600pt]{1.686pt}{1.200pt}}
\put(1308,653.01){\rule{0.482pt}{1.200pt}}
\multiput(1308.00,652.51)(1.000,1.000){2}{\rule{0.241pt}{1.200pt}}
\put(1298.0,655.0){\rule[-0.600pt]{2.409pt}{1.200pt}}
\put(1321,654.01){\rule{0.482pt}{1.200pt}}
\multiput(1321.00,653.51)(1.000,1.000){2}{\rule{0.241pt}{1.200pt}}
\put(1310.0,656.0){\rule[-0.600pt]{2.650pt}{1.200pt}}
\put(1331,655.01){\rule{0.482pt}{1.200pt}}
\multiput(1331.00,654.51)(1.000,1.000){2}{\rule{0.241pt}{1.200pt}}
\put(1323.0,657.0){\rule[-0.600pt]{1.927pt}{1.200pt}}
\put(1343,656.01){\rule{0.723pt}{1.200pt}}
\multiput(1343.00,655.51)(1.500,1.000){2}{\rule{0.361pt}{1.200pt}}
\put(1333.0,658.0){\rule[-0.600pt]{2.409pt}{1.200pt}}
\put(1358,657.01){\rule{0.723pt}{1.200pt}}
\multiput(1358.00,656.51)(1.500,1.000){2}{\rule{0.361pt}{1.200pt}}
\put(1346.0,659.0){\rule[-0.600pt]{2.891pt}{1.200pt}}
\put(1371,658.01){\rule{0.482pt}{1.200pt}}
\multiput(1371.00,657.51)(1.000,1.000){2}{\rule{0.241pt}{1.200pt}}
\put(1361.0,660.0){\rule[-0.600pt]{2.409pt}{1.200pt}}
\put(1384,659.01){\rule{0.482pt}{1.200pt}}
\multiput(1384.00,658.51)(1.000,1.000){2}{\rule{0.241pt}{1.200pt}}
\put(1373.0,661.0){\rule[-0.600pt]{2.650pt}{1.200pt}}
\put(1399,660.01){\rule{0.482pt}{1.200pt}}
\multiput(1399.00,659.51)(1.000,1.000){2}{\rule{0.241pt}{1.200pt}}
\put(1386.0,662.0){\rule[-0.600pt]{3.132pt}{1.200pt}}
\put(1414,661.01){\rule{0.482pt}{1.200pt}}
\multiput(1414.00,660.51)(1.000,1.000){2}{\rule{0.241pt}{1.200pt}}
\put(1401.0,663.0){\rule[-0.600pt]{3.132pt}{1.200pt}}
\put(1429,662.01){\rule{0.482pt}{1.200pt}}
\multiput(1429.00,661.51)(1.000,1.000){2}{\rule{0.241pt}{1.200pt}}
\put(1416.0,664.0){\rule[-0.600pt]{3.132pt}{1.200pt}}
\put(1431.0,665.0){\rule[-0.600pt]{1.927pt}{1.200pt}}
\sbox{\plotpoint}{\rule[-0.500pt]{1.000pt}{1.000pt}}%
\put(1279,738){\makebox(0,0)[r]{FOM}}
\multiput(1299,738)(20.756,0.000){5}{\usebox{\plotpoint}}
\put(1399,738){\usebox{\plotpoint}}
\put(181,123){\usebox{\plotpoint}}
\put(181.00,123.00){\usebox{\plotpoint}}
\put(201.56,124.19){\usebox{\plotpoint}}
\put(221.59,128.20){\usebox{\plotpoint}}
\put(241.37,133.79){\usebox{\plotpoint}}
\put(260.57,141.52){\usebox{\plotpoint}}
\put(279.22,150.07){\usebox{\plotpoint}}
\put(297.53,159.26){\usebox{\plotpoint}}
\put(315.67,168.67){\usebox{\plotpoint}}
\put(332.86,179.29){\usebox{\plotpoint}}
\put(350.08,190.04){\usebox{\plotpoint}}
\put(367.76,200.84){\usebox{\plotpoint}}
\put(384.56,212.56){\usebox{\plotpoint}}
\put(401.42,223.95){\usebox{\plotpoint}}
\put(418.42,235.42){\usebox{\plotpoint}}
\put(435.20,247.47){\usebox{\plotpoint}}
\put(452.18,259.18){\usebox{\plotpoint}}
\put(468.70,271.23){\usebox{\plotpoint}}
\put(485.63,282.63){\usebox{\plotpoint}}
\put(502.30,294.43){\usebox{\plotpoint}}
\put(519.08,306.08){\usebox{\plotpoint}}
\put(535.73,317.73){\usebox{\plotpoint}}
\put(552.84,328.84){\usebox{\plotpoint}}
\put(569.70,340.47){\usebox{\plotpoint}}
\put(586.77,351.85){\usebox{\plotpoint}}
\put(604.21,362.48){\usebox{\plotpoint}}
\put(621.38,373.69){\usebox{\plotpoint}}
\put(638.70,384.57){\usebox{\plotpoint}}
\put(656.00,395.00){\usebox{\plotpoint}}
\put(673.80,405.40){\usebox{\plotpoint}}
\put(691.57,416.05){\usebox{\plotpoint}}
\put(709.48,425.74){\usebox{\plotpoint}}
\put(727.54,435.77){\usebox{\plotpoint}}
\put(745.86,444.95){\usebox{\plotpoint}}
\put(763.95,454.63){\usebox{\plotpoint}}
\put(782.25,463.83){\usebox{\plotpoint}}
\put(800.67,472.84){\usebox{\plotpoint}}
\put(819.52,481.26){\usebox{\plotpoint}}
\put(838.34,489.78){\usebox{\plotpoint}}
\put(857.10,498.05){\usebox{\plotpoint}}
\put(876.31,505.77){\usebox{\plotpoint}}
\put(894.95,514.32){\usebox{\plotpoint}}
\put(914.50,520.83){\usebox{\plotpoint}}
\put(933.71,528.57){\usebox{\plotpoint}}
\put(953.07,535.36){\usebox{\plotpoint}}
\put(972.62,541.87){\usebox{\plotpoint}}
\put(991.98,548.66){\usebox{\plotpoint}}
\put(1011.48,555.00){\usebox{\plotpoint}}
\put(1031.04,561.02){\usebox{\plotpoint}}
\put(1050.55,566.85){\usebox{\plotpoint}}
\put(1070.33,572.00){\usebox{\plotpoint}}
\put(1090.00,577.50){\usebox{\plotpoint}}
\put(1109.82,582.41){\usebox{\plotpoint}}
\put(1129.57,587.29){\usebox{\plotpoint}}
\put(1149.42,592.14){\usebox{\plotpoint}}
\put(1169.24,597.00){\usebox{\plotpoint}}
\put(1189.11,601.04){\usebox{\plotpoint}}
\put(1209.08,605.00){\usebox{\plotpoint}}
\put(1228.89,609.00){\usebox{\plotpoint}}
\put(1248.70,613.00){\usebox{\plotpoint}}
\put(1268.74,616.37){\usebox{\plotpoint}}
\put(1288.78,620.00){\usebox{\plotpoint}}
\put(1308.90,623.00){\usebox{\plotpoint}}
\put(1329.11,626.37){\usebox{\plotpoint}}
\put(1349.30,629.43){\usebox{\plotpoint}}
\put(1369.56,632.00){\usebox{\plotpoint}}
\put(1389.61,635.00){\usebox{\plotpoint}}
\put(1409.86,637.43){\usebox{\plotpoint}}
\put(1430.09,640.00){\usebox{\plotpoint}}
\put(1439,641){\usebox{\plotpoint}}
\end{picture}